\documentclass[aps,prd,reprint,preprintnumbers,superscriptaddress,amsmath,amssymb,floatfix,longbibliography]{revtex4-2}
\pdfoutput=1
\usepackage[colorlinks,citecolor=blue,urlcolor=blue,linkcolor=red]{hyperref}
\usepackage{subfigure}
\usepackage{graphicx}
\usepackage{dcolumn}
\usepackage{amsmath}
\usepackage{bm}

\begin{document}
\title{$\eta_c\eta_c$ and $J/\psi J/\psi$ scattering from lattice QCD}

%%%%%%%%%%%%%%%%%%%%%%%%%%%
\author{Geng Li}
\email{ligeng@ihep.ac.cn}
\affiliation{Institute of High Energy Physics, Chinese Academy of Sciences, Beijing 100049, People's Republic of China}
\affiliation{Center for High Energy Physics, Henan Academy of Sciences, Zhengzhou 450046, People's Republic of China}

\author{Chunjiang Shi}
%\email{shichunjiang@ihep.ac.cn}
\affiliation{Institute of High Energy Physics, Chinese Academy of Sciences, Beijing 100049, People's Republic of China}
\affiliation{School of Physical Sciences, University of Chinese Academy of Sciences, Beijing 100049, People's Republic of China}

\author{Ying Chen}
\email{cheny@ihep.ac.cn}
\affiliation{Institute of High Energy Physics, Chinese Academy of Sciences, Beijing 100049, People's Republic of China}
\affiliation{Center for High Energy Physics, Henan Academy of Sciences, Zhengzhou 450046, People's Republic of China}
\affiliation{School of Physical Sciences, University of Chinese Academy of Sciences, Beijing 100049, People's Republic of China}

\author{Wei Sun}
%\email{sunwei@ihep.ac.cn}
\affiliation{Institute of High Energy Physics, Chinese Academy of Sciences, Beijing 100049, People's Republic of China}
%%%%%%%%%%%%%%%%%%%%%%%%%%%

\begin{abstract}
	We investigate the $S$-wave $\eta_c\eta_c$ and $J/\psi J/\psi$ scattering in the $J^{PC}=(0,2)^{++}$ channels up to a center-of-mass energy of 6.6~GeV. The calculations are carried out at two unphysical pion masses, $m_\pi\approx 420$~MeV and 250~MeV, in $N_f=2$ lattice QCD. For each $m_\pi$, we extract the finite-volume energy levels on two lattices with an identical lattice spacing ($a\simeq 0.136~\mathrm{fm}$) but different spatial volumes. Since the coupled-channel effects between the $\eta_c\eta_c$ and $J/\psi J/\psi$ channels are found to be negligible, we analyze the corresponding scattering properties using the single-channel L"{u}scher method. We find that the interactions in these dicharmonium systems are dominated by the quark rearrangement effect. In the $0^{++}$ channel, the near-threshold attraction in $J/\psi J/\psi$ and repulsion in $\eta_c\eta_c$ can be explained through the Fierz rearrangement. The attractive interaction in the ${}^1S_0$ $J/\psi J/\psi$ channel allows for the existence of a near-threshold scalar structure, which may correspond to the $X(6200)$. However, possible left-hand-cut effects due to light-hadron exchange prevent us from precisely determining the pole positions. In the $2^{++}$ channel, while the ${}^5S_2$ $J/\psi J/\psi$ system exhibits a repulsive interaction near threshold, the scattering amplitude has a Castillejo-Dalitz-Dyson zero at $\sqrt{s}=6.45~\mathrm{GeV}$ and a resonance pole at $\sqrt{s}=\big[6.543(10)-i,0.548(34)/2\big]~\mathrm{GeV}$ for $m_\pi\approx 420~\mathrm{MeV}$ and $\big[6.538(13)-i,0.537(56)/2\big]~\mathrm{GeV}$ for $m_\pi\approx 250~\mathrm{MeV}$, where the uncertainties are statistical. This resonance may correspond to the $X(6600)$ (or $X(6400)$) reported by the ATLAS and CMS Collaborations. Our result supports its $2^{++}$ assignment, in agreement with the latest spin-parity determination by CMS. Despite the observed dominance of the quark rearrangement effect, the light-hadron dynamics underlying dicharmonium scattering need to be explored at lighter pion masses. Furthermore, since our lattices are coarse and the lattice volumes are small, the associated systematic uncertainties should be controlled in future studies using more sophisticated lattice setups.
\end{abstract}

\maketitle

%%%%%%%%%%%%%%%%%%%%%%%%%%%
\section{Introduction}\label{sec:intro}
%%%%%%%%%%%%%%%%%%%%%%%%%%%
In 2020, LHCb analyzed the $J/\psi J/\psi$ invariant mass spectrum and observed a narrow structure around 6.9~GeV, named $X(6900)$, as well as a broad structure near the $J/\psi J/\psi$ threshold~\cite{LHCb:2020bwg}. 
ATLAS also performed an analysis of the $J/\psi J/\psi$ mass spectra in the $4\mu$ system~\cite{ATLAS:2023bft}.
Assuming three interfering resonances, ATLAS observed two additional states with masses around 6.4~GeV and 6.6~GeV, labeled as $X(6400)$ and $X(6600)$, in addition to $X(6900)$.
Shortly after, CMS reported the observation of $X(6600)$, $X(6900)$, and a new structure, $X(7100)$~\cite{CMS:2023owd, Wang:2024koq, CMS:2025xwt}. 
These states are now named as $T_{cc\bar c\bar c}$ states.   
Recently, CMS refined the analysis of $J/\psi J/\psi$ spectrum in proton-proton collisions with 3.6 times more statistics~\cite{CMS:2025fpt}, which confirms these three structures with the statistical significance well above $5\sigma$ based on quantum interference among structures, and indicates that they may have the same $J^{PC}$ quantum numbers. CMS also performed an analysis on the angular distribution of the decay products of these states~\cite{CMS:2025fpt}, which prefers the quantum numbers $2^{++}$ while $0^{++}$ is disfavored at 95\% confidence level. These states may be candidates for $cc\bar{c}\bar{c}$ tetraquark states, $T_{cc\bar{c}\bar{c}}$, and have stimulated extensive theoretical studies~\cite{Guo:2020pvt,liu:2020eha,Chen:2020xwe,Maiani:2020pur,Dong:2020nwy,Chao:2020dml,Zhu:2020xni,Gong:2020bmg,Liang:2021fzr,Liu:2021rtn,Dong:2021lkh,Niu:2022jqp,Dong:2022sef,Kuang:2023vac,Huang:2024jin,Wu:2024euj,Zhang:2024qkg} (see Ref.~\cite{Chen:2022asf} for a review). 
The fully-heavy tetraquark states were first investigated from the 1980s~\cite{Iwasaki:1975pv,Chao:1980dv,Ader:1981db,Li:1983ru,Heller:1985cb,Badalian:1985es,Brambilla:2015rqa}. 
Prior to the experimental observations of these structures, there were theoretical studies on fully-charmed tetraquark states using numerous diquark-diquark interpolation fields within the framework of QCD sum rules, which predicted $T_{cc\bar{c}\bar{c}}$ spectra for various $J^{PC}$ states~\cite{Chen:2016jxd,Wang:2017jtz,Wang:2018poa,Wang:2022xja}.
Recently, a formalism is developped for the production of $T_{cc\bar c \bar c}$ states~\cite{Belov:2024qyi}, which shows that the production cross section of $X(6900)$ as a fully-charmed tetraquark is compatible with a $J^{PC}=2^{++}$ state. 
Since these $T_{cc\bar{c}\bar{c}}$ structures are observed in the $J/\psi J/\psi$ system, it is intriguing to understand the dynamics between the two $J/\psi$ particles (or, more generally, between two charmonium states) that contribute to the formation of these structures.
Some phenomenological studies attribute the dynamics to pomeron exchanges (gluonic interactions) or light meson exchanges between the two charmonia~\cite{Gong:2020bmg}. 

In this exploratory study, we investigate the dicharmonium interaction and its scattering properties from lattice QCD.
As the first step, we examine the coupled-channel scattering between $\eta_c \eta_c$ and $J/\psi J/\psi$ in the $J^{PC} = 0^{++}$ and $2^{++}$ channels with center-of-mass energies up to 6.6~GeV, which is below the $J/\psi\psi(2S)$ threshold. 
We restrict our analysis of the $0^{++}$ channel to the ${}^1S_0$ $S$-wave scattering. 
For the $2^{++}$ channel, we focus on the ${}^5S_2$ $J/\psi J/\psi$ scattering, including ${}^1D_2$ $\eta_c \eta_c$ operators to complete the lattice operator basis and ensure a reliable extraction of finite-volume energies (FVEs).
We perform lattice calculations on $N_f=2$ anisotropic gauge ensembles at two pion masses, $m_\pi\approx 420$ and $250$~MeV, to study the possible $m_\pi$-dependence of dicharmonium scattering properties, which are extracted from the finite-volume energy levels via Lüscher's formalism~\cite{Luscher:1986pf,Luscher:1990ux,Luscher:1991cf}.  
For each pion mass, we employ both $12^3 \times 96$ and $16^3 \times 128$ lattice volumes. 
In practical calculation, we adopt the distillation method~\cite{Peardon:2009gh}, which provides a systematic framework to handle simultaneously the operator construction, the quark annihilation effect, and the calculation of huge correlation matrices.
Many efforts have been devoted to the reliable extraction of the energy levels of the $\eta_c \eta_c-J/\psi J/\psi$ systems.
%We further investigate the dependence of the lattice energy levels on the dimension of the Laplacian-Heaviside subspace.

The paper is organized as follows.
In Sec.~\ref{sectionII}, we provide the numerical details of the lattice calculation, including lattice configuration, operator setup, correlation computation, and energy-level extraction. 
In Sec.~\ref{sectionIII}, we analyze the extracted energy levels within the framework of scattering theory and identify the possible existence of new states. 
In Sec.~\ref{sectionIV}, we make discussion of the results. 
In Sec.~\ref{sectionV}, a summary is presented.

%%%%%%%%%%%%%%%%%%%%%%%%%%%
\section{Numerical details}\label{sectionII}
%%%%%%%%%%%%%%%%%%%%%%%%%%%

%%%%%%%%%%%%%%%%%%%%%%%%%%%
\begin{table*}[t]
	\renewcommand\arraystretch{1.5}
	\caption{The detailed parameters of our improved ensembles (IE), labeled as `Lxx' for the lattice size and `Mxxx' for the pion mass, are summarized below.}
	\label{tab:config}
	\begin{ruledtabular}
		\begin{tabular}{ccccccccccc}
			IE      & $N_s^3\times N_t$ & $m_\pi$(MeV) & $m_\pi L_s$ & $M_{\eta_c}$(MeV) & $M_{J/\psi}$(MeV) & $a_t^{-1}$(GeV) & $\xi_{\eta_c}$ & $\xi_{J/\psi}$ & $N_V$ & $N_\mathrm{cfg}$ \\
			\hline
			L12M420 & $12^3\times$96    & 426(4)       & $\sim 3.4$  & 2995.97(66)       & 3088.50(86)       & 7.219           & 5.100(6)       & 5.087(8)       & 170   & 400              \\
			L16M420 & $16^3\times$128   & 417(6)       & $\sim 4.6$  & 2995.25(34)       & 3087.14(48)       & 7.219           & 5.081(8)       & 5.063(10)      & 120   & 400             \\
			L12M250 & $12^3\times$96    & $\sim 250$& $\sim 2.1$  & 2995.67(69)       & 3081.66(91)       & 7.276           & 5.148(7)       & 5.118(7)       & 170   & 400              \\
			L16M250 & $16^3\times$128   & 250(3)       & $\sim 2.8$  & 2998.32(34)       & 3085.29(45)       & 7.276           & 5.120(9)       & 5.103(11)      & 120   & 400              \\
		\end{tabular}
	\end{ruledtabular}
\end{table*}
%%%%%%%%%%%%%%%%%%%%%%%%%%%

%%%%%%%%%%%%%%%%%%%%%%%%%%%
\subsection{Lattice configuration}\label{secII:configuration}
%%%%%%%%%%%%%%%%%%%%%%%%%%%
We use $N_f = 2$ dynamical gauge configurations with degenerate $u$ and $d$ quarks at pion masses of $420~(250)$~MeV (labeled as `M420' and `M250'). 
These ensembles are generated on anisotropic lattices with aspect ratio $\xi \equiv a_s / a_t \approx 5$, where $a_s$ and $a_t$ denote the lattice spacings in the spatial and temporal directions, respectively.
The tadpole-improved Symanzik gauge action~\cite{Morningstar:1997ff,Chen:2005mg} and the tadpole-improved anisotropic clover fermion action~\cite{Zhang:2001in,Su:2004sc,CLQCD:2009nvn} are utilized for both the sea light quarks and the valence charm quark. 
For each pion mass, we have two distinct lattice volumes that share the same spatial lattice spacing~\cite{Li:2024pfg}, namely, $N_s^3 \times N_t = 12^3 \times 96$ and $16^3 \times 128$ (labeled as `L12' and `L16'). 
The spatial lattice spacing $a_s$ is determined to be $a_s = 0.136(2)$~fm through the Wilson flow method~\cite{Luscher:2010iy,BMW:2012hcm}. 
The spatial lattice extent is given by $L_s = N_s \times a_s$. 
Consequently, we calculate the pion dispersion relation to determine the physical anisotropy parameter $\xi$, obtaining temporal lattice spacings of $a_t^{-1} = 7.219~(7.276)$~GeV for the M420~(M250) ensembles, respectively. 
The charm quark mass parameter is tuned to reproduce the spin-averaged mass of the $1S$ charmonium state, $\left(M_{\eta_c} + 3M_{J/\psi} \right) / 4 = 3069$~MeV. 
The parameters of the gauge ensembles are collected in Table~\ref{tab:config}.
The hyperfine splitting $M_{J/\psi}-M_{\eta_c} \sim 90$~MeV obtained from our lattice setup deviates noticeably from the experimental value (114~MeV), indicating that this quantity is subject to discretization uncertainties, as also observed in previous lattice studies.

The distillation method~\cite{Peardon:2009gh} is adopted to calculate the correlation functions. 
The eigenspace of the gauge-invariant Laplacian operator is truncated by retaining only the eigenmodes corresponding to the $N_V$ smallest eigenvalues.
The corresponding $N_V$ eigenvectors span a Laplacian Heaviside subspace (LHS) and provide a smearing scheme for quark fields named by LHS smearing. 
The single particle interpolation operators of charmonia and dicharmonium operators are built in term of the LHS smeared charm quark fields throughout this work. 
The encapsulated all-to-all propagators (perambulators) of charm quark are calculated in this subspace and are used to compute the correlation functions of charmonium and dicharmonium operators. 
One can refer to the original Ref.~\cite{Peardon:2009gh} for details. 

In this work, we aim to study the $\eta_c\eta_c$ and $J/\psi J/\psi$ scattering up to the center-of-mass energy around 6.6~GeV. 
Since the $\eta_c$ and $J/\psi$ mesons are heavy, the maximum relative momentum of the two particles is set to $k^2 \sim 1.9~\mathrm{GeV}^2$.
However, similar to Gaussian smearing, the LHS-smeared operators also suffer from momentum suppression~\cite{Bali:2016lva}, such that their couplings to higher-momentum states (after momentum projection) are suppressed by $e^{-k^2/(4\sigma)}$, where $k$ and $\sigma$ denote the momentum and the smearing size, respectively.
Therefore, the smearing sizes need to be sufficiently small to allow proper access to higher-momentum states.
For LHS smearing, a smaller smearing size requires a larger $N_V$ on the same lattice.
On the other hand, to achieve the same smearing size, $N_V$ must increase with the spatial volume of the lattice.
In practice, $N_V$ is set to be 170 (120) for $N_s=12$ ($N_s=16$) lattices, and the perambulators are calculated with the source time slice running over all the time range of each lattice. 
These choices are based on a balance between our physical goals and the available computing resources, as larger lattice sizes and higher values of $N_V$ require significantly more computational power and storage space.
The detailed discussion can be found in Sec.~\ref{secII:NV-dep}.  
It should be noted that our lattice sizes yield $m_\pi L_s \sim 2.1$ or 2.8 on the `M250' ensembles, which may lead to appreciable finite-volume effects for light hadrons.
However, for fully-charmed systems, light hadrons contribute only through vacuum polarization diagrams of light quarks, which are expected to be suppressed according to the Okubo-Zweig-Iizuka (OZI) rule.
We provide further details on this issue in Sec.~\ref{sectionIV}.      

%%%%%%%%%%%%%%%%%%%%%%%%%%%
\begin{figure*}[t]
	\centering
	\includegraphics[width=0.45\linewidth]{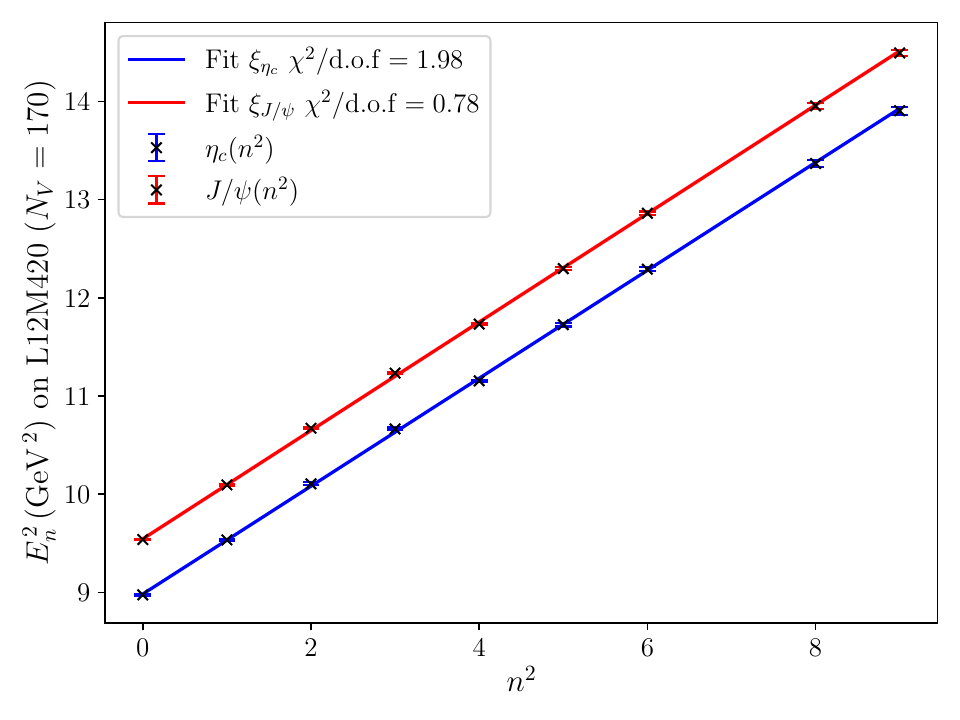}
	\includegraphics[width=0.45\linewidth]{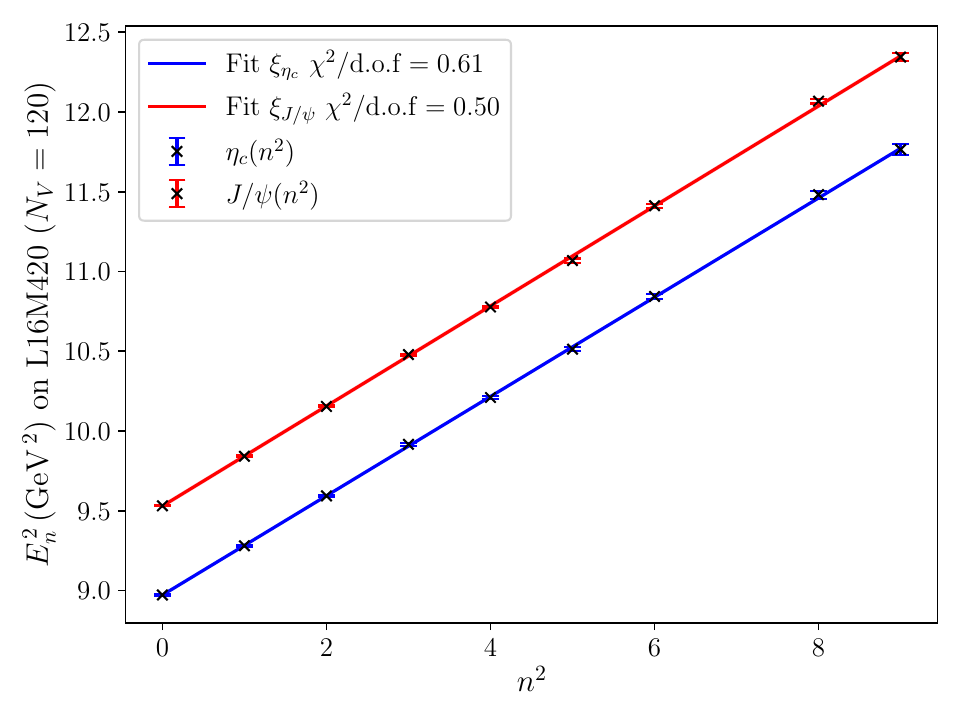}\\
	\includegraphics[width=0.45\linewidth]{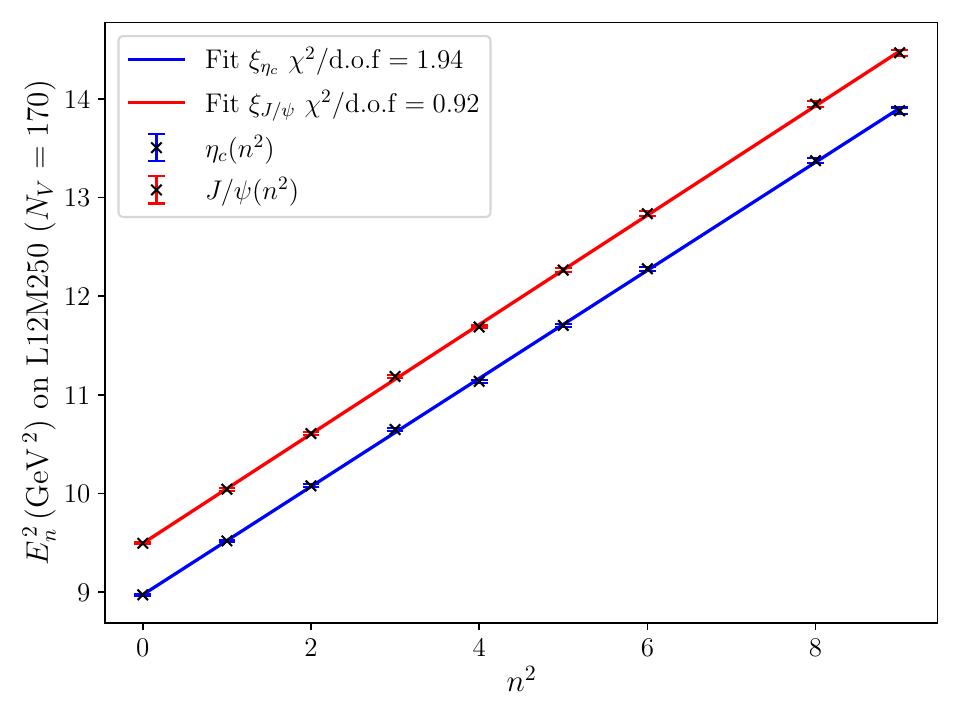}
	\includegraphics[width=0.45\linewidth]{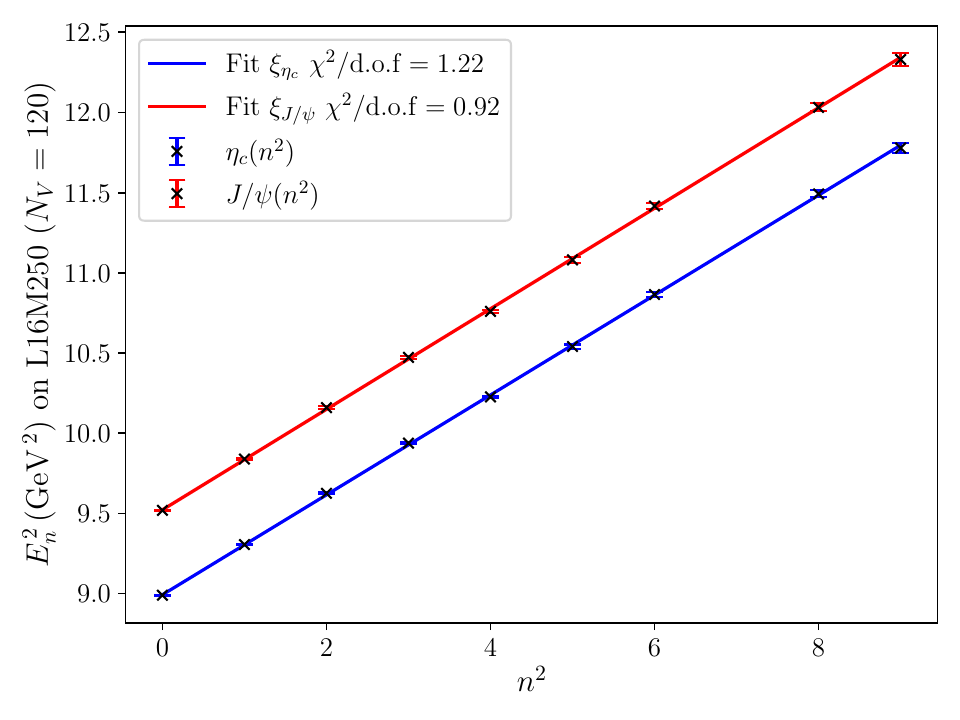}
	\caption{The dispersion relations of the charmonium states $J/\psi$ and $\eta_c$ with momentum $n^2$ are shown for L12 and L16 ensembles at two pion masses, where the statistical error is estimated using the jackknife method.
		The aspect ratios $\xi_{\eta_c}$ and $\xi_{J/\psi}$ are extracted by fitting Eq.~\eqref{eq:dispersion}, with all fits yielding $\chi^2/\mathrm{d.o.f.} < 2$.}
	\label{fig:ksi}
\end{figure*}
%%%%%%%%%%%%%%%%%%%%%%%%%%%

\begin{figure}[t]
	\centering
	\includegraphics[width=0.99\linewidth]{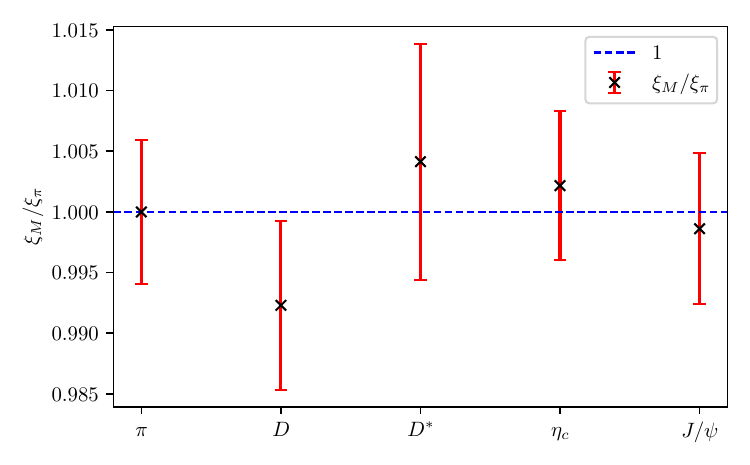}
	\caption{The relative aspect ratio $\xi_M/\xi_\pi$ for various mesons ($\pi, D,D^*,\eta_c, J/\psi$) on the L16M420 ensemble.}
	\label{fig:ksi-ratio}
\end{figure}

As a first step, we examine the dispersion relations of the $J/\psi$ and $\eta_c$ mesons on all four lattices used, which is crucial for the subsequent scattering analysis. 
The results are shown in Fig.~\ref{fig:ksi}.
The data points correspond to the squared energies $E_n^2$ of the $\eta_c$ and $J/\psi$ mesons at momenta $k^2 = (2\pi/L_s)^2 n^2$.
The straight lines represent the continuum dispersion relation,
\begin{equation}\label{eq:dispersion}
	E_n^2 = m_M^2 + \frac{1}{\xi_M^2} \left(\frac{2\pi}{a_t N_s}\right)^2 n^2,
\end{equation}
where $M$ denotes $\eta_c$ or $J/\psi$ meson, $m_M$ is the mass of the particle $M$, and $\xi_M$ is the aspect ratio for each particle, obtained by fitting the data.
The values of $\xi_{\eta_c,J/\psi}$ are also collected in Table~\ref{tab:config} and are applied to the analysis of the related scattering. 
Obviously, Eq.~\eqref{eq:dispersion} describes the data fairly well. 
We also check the dispersion relation of other mesons ($M$), such as $\pi, D, D^*$, on the L16M420 ensemble, and observe the same phenomenon. 
Figure~\ref{fig:ksi-ratio} shows the ratios of the obtained aspect ratios $\xi_M$ of indiviual particles to $\xi_\pi=5.07(3)$, where one can see that they are consistent with each other within 1\%. 
This continuum dispersion relation has also been observed in other lattice QCD studies using anisotropic lattices; see, for example, Ref.~\cite{Wilson:2023anv}.  

%%%%%%%%%%%%%%%%%%%%%%%%%%%
%%%%%%%%%%%%%%%%%%%%%%%%%%%

%%%%%%%%%%%%%%%%%%%%%%%%%%%
\subsection{Two-particle operators}\label{secII:oper}
%%%%%%%%%%%%%%%%%%%%%%%%%%%

%%%%%%%%%%%%%%%%%%%%%%%%%%%%%%%%%%%%%%%%%%%%%%%%%%%%%%
\begin{table}[t]
	\caption{Decomposition of subduced representations of SU(2) with respect to the octahedral group O, for angular momenta up to $J=5$.}
	\label{table:representations}
	\begin{ruledtabular}
		\begin{tabular}{ccccccc}
			O$\backslash J$  &   $0$     & $1$ &   $2$ & $3$   &  $4$    &  $5$ \\
			\hline
			$A_1$           &   1    &0     &0    &0       &1    & 0 \\
			$A_2$           &   0    &0     &0    &1       &0    & 0 \\
			$E$             &   0    &0     &1    &0       &1    & 1 \\
			$T_1$           &   0    &1     &0    &1       &1    & 2 \\
			$T_2$           &   0    &0     &1    &1       &1    & 1 \\
		\end{tabular}
	\end{ruledtabular}
\end{table}
%%%%%%%%%%%%%%%%%%%%%%%%%%%%%%%%%%%%%%%%%%%%%%%%%%%%%%%

%%%%%%%%%%%%%%%%%%%%%%%%%%%
\begin{figure}[t]
	\centering
	\subfigure[Direct]{
		\includegraphics[width=0.45\linewidth]{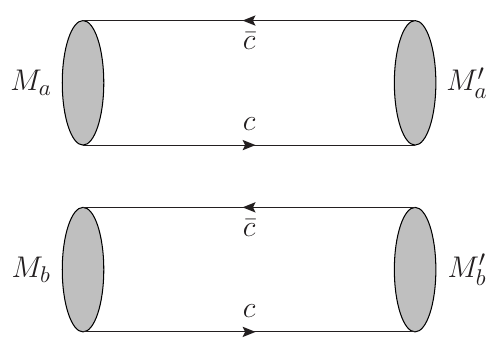}}
	\subfigure[Quark rearrangement]{
		\includegraphics[width=0.45\linewidth]{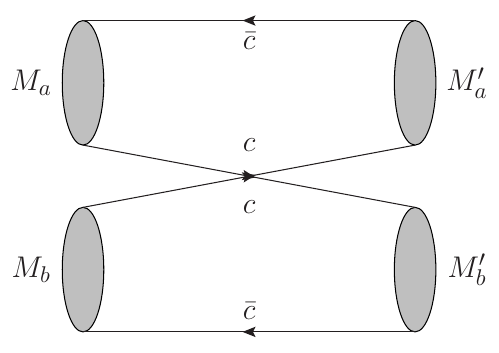}}\\
	\subfigure[Single annihilation]{
		\includegraphics[width=0.45\linewidth]{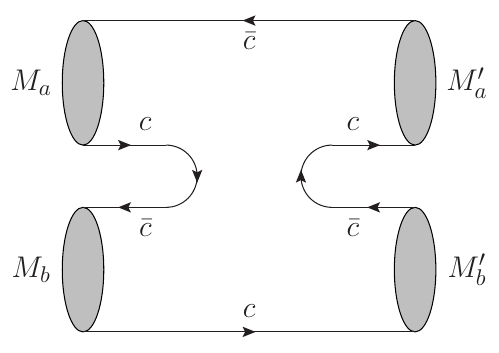}}
	\subfigure[Double annihilation]{
		\includegraphics[width=0.45\linewidth]{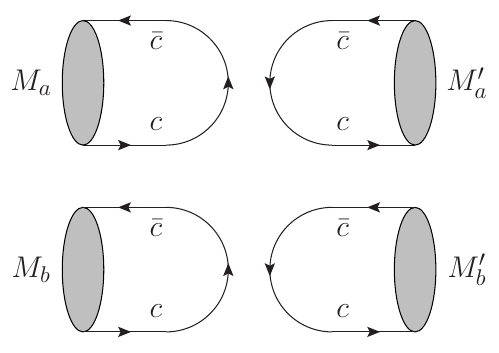}}\\
	\subfigure[$c\bar{c}$ and two-particle operators]{
		\includegraphics[width=0.45\linewidth]{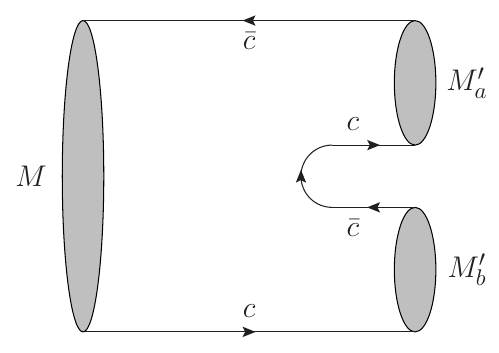}}
	\subfigure[$c\bar{c}$ operator]{
		\includegraphics[width=0.45\linewidth]{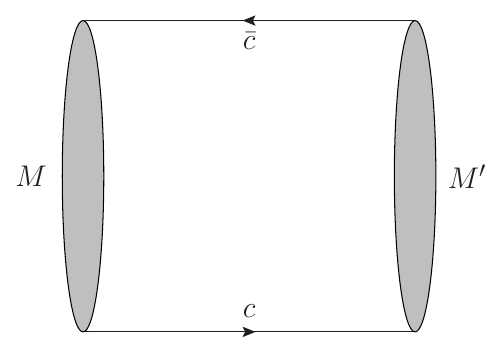}}
	\caption{Schematic illustrations of quark Wick contractions. }
	\label{fig:diagrams}
\end{figure}
%%%%%%%%%%%%%%%%%%%%%%%%%%%

We investigate the $0^{++}$ and $2^{++}$ channels of the $c\bar{c}c\bar{c}$ system with meson-meson type operator construction, with center-of-mass energies up to 6.6~GeV. 
Within this energy range, only the coupled-channel scattering involving $\eta_c \eta_c$ and $J/\psi J/\psi$ is considered, excluding contributions from excited charmonium states such as $\psi(3770)$ and $\eta_c(2S)$. 

The operator sets are constructed in the rest frame of the system, where each operator $\mathcal{O}_\alpha$ is a meson-meson operator defined as $\mathcal{O}_{MM}(n^2(k)) = \mathcal{O}_M(\vec{k}) \circ \mathcal{O}_M(-\vec{k})$, with $M$ denoting either $\eta_c$ or $J/\psi$ meson, and the symbol `$\circ$' stands for the specific combinations of the different orientations of $\vec{k}$ and the indices of $\mathcal{O}_\psi^i$~\cite{Cheung:2017tnt}.
Here, $\mathcal{O}_M^{(i)}(\vec{k})=[\bar{c}\Gamma_M^{(i)} c](\vec{k})$ represents a single-meson operator with $M$ referring to either $\eta_c$ ($\Gamma_M=\gamma_5$) or $J/\psi$ ($\Gamma_M^i=\gamma_i$), and $n^2(k)=n_1^2+n_2^2+n_3^2$ labels the magnitude of the relative momentum mode $\vec{k}=\frac{2\pi}{L_s}(n_1,n_2,n_3)$. 

The lattice version of the ${}^1S_0$ operators belongs to the $A_1$ irreducible representation (irrep) of the octahedral group O and is built as
\begin{equation} 
	\begin{aligned}
		\label{eq:operator-A1}
		\mathcal{O}^{A_1}_{\eta_c\eta_c}(n^2(k))&=\sum\limits_{R\in \mathrm{O}} \mathcal{O}_{\eta_c}(R\vec{k})\mathcal{O}_{\eta_c}(-R \vec{k}),\\
		\mathcal{O}^{A_1}_{\psi\psi}(n^2(k))&=\sum\limits_{i,R\in \mathrm{O}} \mathcal{O}_{\psi}^i(R\vec{k})\mathcal{O}_{\psi}^i(-R \vec{k}),
	\end{aligned}
\end{equation} 
where $R\in \mathrm{O}$ denotes an operation of O acting on the momentum $\vec{k}$.
The explicit expressions of all the $\eta_c \eta_c$ operators are presented in Appendix~\ref{appendix:A}. 
The $J/\psi J/\psi$ operators are summed over momentum space in the same way as the $\eta_c \eta_c$ operators, with the additional sum over the index $i$.
Note that the $0^{++}$ system includes ${}^1S_0$ $\eta_c \eta_c$ scattering states and (${}^1S_0$, ${}^5D_0$) $J/\psi J/\psi$ scattering states. 
While the momentum configurations of the operators belong to the $A_1$ irrep, which correspond to $S$-wave ($l=0$) and $G$-wave ($l=4$) in the continuum limit, as summarized in Table~\ref{table:representations}, where the correspondence between the continuum angular momentum ($J$) and the irreps of O is given.
Therefore, our $S$($D$)-wave operators possibly do not couple to $D$($S$)-wave states, such that the coupled-channel effects from the $D$-wave can be neglected.
In principle, the correlation functions of these dicharmonium operators receive contributions from charm quark-antiquark annihilation diagrams, which are illustrated in Fig.~\ref{fig:diagrams}.
The double annihilation diagram~(d), where both $c\bar{c}$ pairs of the operator annihilate at the sink and source time slices, results in the propagation of glueballs and (multiple) light hadron states in the presence of light sea quarks. 
These contributions are expected to be doubly suppressed by the OZI rule and can thus be neglected. 
The single annihilation diagram~(c), on the other hand, leads to the propagation of hadron systems containing a single charm quark-antiquark pair. 
This contribution is not OZI suppressed and should be taken into account. 
For diagrams~(e,f), we tentatively include the single-particle scalar $\bar{c}c$ operator $\mathcal{O}_{\chi_{c0}}$ in the $0^{++}$ channel, and illustrate the additional difficulties it causes in extracting the $\eta_c\eta_c$ and $J/\psi J/\psi$ energies.

In the $2^{++}$ channel, the finite-volume energies of the ${}^1D_2$ $\eta_c\eta_c$ scattering lie below those of the corresponding ${}^5S_2$ $J/\psi J/\psi$ states. 
To reliably extract the energy levels of the $J/\psi J/\psi$ system, we include $D$-wave $\eta_c\eta_c$ operators to complete the set of interpolating fields.
The construction of the $2^{++}$ operators involves additional complexity. 
The continuum $J=2$ is subduced into the irreps $E$ and $T_2$ of O on the lattice, namely, $J=2\to E\oplus T_2$, as shown in Table~\ref{table:representations}. 
In practice, the $2^{++}$ $\mathcal{O}_{MM}$ operators are constructed to be the first component $E(1)$ of the irrep $E$.
As $\eta_c$ has spin zero, the total angular momentum of the $\eta_c\eta_c$ pair is determined solely by their relative orbital angular momentum. Thus the ${}^1D_2$ $\eta_c\eta_c$ operators $\mathcal{O}_{\eta_c\eta_c}^{E(1)}$ belong to the representation $E^{++} \leftarrow A_1^{(s)} \otimes E^{(l)}$, where the superscripts $s$ and $l$ denote spin and orbital configurations, respectively, as given by
\begin{equation}
	\label{eq:operator-E1}
	\mathcal{O}_{\eta_c\eta_c}^{E(1)}=\sum\limits_{R\in O} c_R^{E(1)} \mathcal{O}_{\eta_c}(R\vec{k})\mathcal{O}_{\eta_c}(-R \vec{k}),
\end{equation}
where $\{c_R^{E(1)},R\in O\}$ is the character of $E(1)$. 
The explicit expressions can be found in Appendix~\ref{appendix:A}.
For the ${}^5S_2$ $J/\psi J/\psi$ channel, the operators corresponding to $E^{++} \leftarrow E^{(s)} \otimes A_1^{(l)}$ representation has
\begin{equation} 
	\begin{aligned}
		\mathcal{O}&_{\psi\psi}^{E(1)}({}^5S_2)=\\
		&\sum\limits_{R\in O}\left[ \mathcal{O}_{\psi}^1 (R\vec{k})\mathcal{O}_{\psi}^1(-R \vec{k})-\mathcal{O}_{\psi}^2 (R\vec{k})\mathcal{O}_{\psi}^2(-R \vec{k})\right].
	\end{aligned}
\end{equation}
We also construct the ${}^1D_2$ $J/\psi J/\psi$ operators corresponding to the $E^{++} \leftarrow A_1^{(s)} \otimes E^{(l)}$ representation, 
\begin{equation} 
	\begin{aligned}
		\mathcal{O}_{\psi\psi}^{E(1)} ({}^1D_2)=
		\sum\limits_{i, R\in O}\bigg[ c_R^{E(1)}\mathcal{O}_{\psi}^i (R\vec{k})\mathcal{O}_{\psi}^i (-R \vec{k})\bigg],
	\end{aligned}
\end{equation} 
to explore the possible the coupled-channel effects between the $S$- and $D$-waves.

Eventually, the $0^{++}$,~$2^{++}$ operator sets $\mathcal{S}_{0,2}$ for each ensemble are chosen to be 
\begin{equation} 
	\begin{aligned}
		\label{eq:op-set}
		\mathcal{S}_0&=\{\mathcal{O}^{A_1}_{\eta_c\eta_c}(n^2),\mathcal{O}^{A_1}_{\psi\psi}(n^2), \mathcal{O}_{\chi_{c0}}; n^2=0,1,2,3,4\},  \\
		\mathcal{S}_2&=\{\mathcal{O}^{E(1)}_{\eta_c\eta_c}(n^2),\mathcal{O}^{E(1)}_{\psi\psi}(n^2),\mathcal{O}_{\chi_{c2}}; n^2=0,1,2,3,4\}.
	\end{aligned}
\end{equation}
For each operator set we compute the correlation matrix,
\begin{equation}
	C_{\alpha\beta}(t)=\langle \mathcal{O}_\alpha (t) \mathcal{O}_\beta^\dagger(0)\rangle,
\end{equation}
with $\mathcal{O}_\alpha$ referring to the operators in the operator set. 

%%%%%%%%%%%%%%%%%%%%%%%%%%%
\subsection{$0^{++}$ Correlation functions}\label{secII:corr0pp}
%%%%%%%%%%%%%%%%%%%%%%%%%%%

%%%%%%%%%%%%%%%%%%%%%%%%%%%
\begin{figure}[t]
	\centering
	\includegraphics[width=0.99\linewidth]{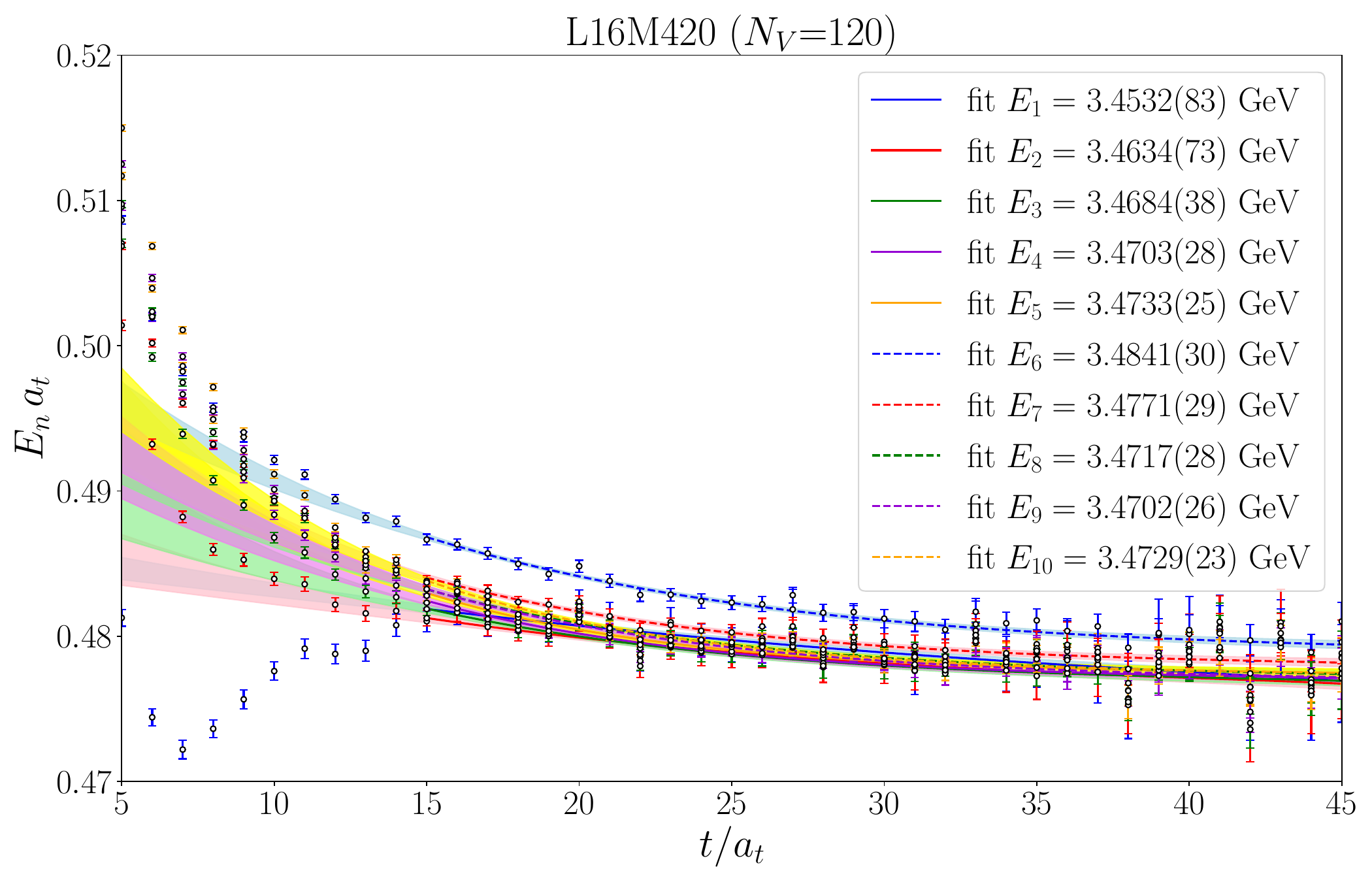}
	\caption{The effective mass extracted from diagram~(c) is approximately 3.45~GeV, which lies well below 6~GeV.}
	\label{fig:diagram-c-effmass}
\end{figure}
%%%%%%%%%%%%%%%%%%%%%%%%%%%

The correlation matrix $C_{\alpha\beta}(t)$ contains quite a number of components like $\langle \mathcal{O}_{M_1M_1}(\vec{k},t) ~ \mathcal{O}_{M_2M_2}(\vec{k}',0)\rangle$ and $\langle \mathcal{O}_{M_1M_1}(\vec{k},t)\mathcal{O}_{\chi_{c0}}(0)\rangle$ in the $0^{++}$ channel. Meanwhile, each element of $C_{\alpha\beta}(t)$ has contributions from various quark diagrams after Wick's contraction, which are illustrated in Fig.~\ref{fig:diagrams}.
The $\langle \mathcal{O}_{M_1M_1}(\vec{k},t) ~ \mathcal{O}_{M_2M_2}(\vec{k}',0)\rangle$-type component receives contributions from diagrams~(a--d), the $\langle \mathcal{O}_{M_1M_1}(\vec{k},t) ~ \mathcal{O}_{\chi_{c0}}(0)\rangle$-type component from diagram~(e), and the $\langle \mathcal{O}_{\chi_{c0}}(t)\mathcal{O}_{\chi_{c0}}(0)\rangle$ component from diagram~(f). 
Diagram~(d) is doubly suppressed by the OZI rule and is neglected in the practical calculation. 
Actually, annihilation diagram is usually ignored even in the calculation of charmonium correlation function. 
Figure~\ref{fig:diagram-c-effmass} shows the effective mass functions obtained from diagram~(c),
\begin{equation}
	E^{(\mathrm{eff})}(t) ~ a_t =\cosh^{-1}\left[\frac{C(t+a_t)+C(t-a_t)}{2C(t)}\right].
\end{equation}
The effective energies are obtained by two-mass-term fits. 
It is seen that the ground state is very close to the $\chi_{c0}$ mass $m_{\chi_{c0}}= 3.453(2)$~GeV determined on the same ensemble. 
Since the charm quark is heavy, the propagation of the $c\bar{c}c\bar{c}$ system is exponentially suppressed relative to that of a single $c\bar{c}$ pair, roughly by a factor of $e^{-m_{\eta_c} t}$.
Consequently, in the large-$t$ region, this diagram is dominated by the propagator of the $\chi_{c0}$ state.
Similarly, diagram~(e) also reflects the $\chi_{c0}$ propagator at large $t$.
When the $\mathcal{O}_{\chi_{c0}}$ operator is included, contributions from diagrams~(c--e) in Fig.~\ref{fig:diagrams} should be considered. 
As a result, the correlation matrix $C_{\alpha\beta}(t)$ becomes dominated by the $\chi_{c0}$ contribution, making it difficult to reliably extract the energies of dicharmonium states lying far above $m_{\chi_{c0}}$. 
Since the mass of $\chi_{c0}$ is well-below the energy region of interest, and under the assumption that it is largely disentangled from the dicharmonium systems, we omit $\mathcal{O}_{\chi_{c0}}$ from the operator set $\mathcal{S}_0$ and neglect the contribution from diagram~(c) in practical calculations. 
From now on, $\mathcal{S}_0$ refers to the operator set containing only two-particle operators, with the corresponding correlation matrix still denoted as $C_{\alpha\beta}(t)$. 
For the same reason, the $\chi_{c2}$ operator is also omitted in the $2^{++}$ channel.

%%%%%%%%%%%%%%%%%%%%%%%%%%%
\begin{figure*}[t]
	\centering
	\includegraphics[width=0.45\linewidth]{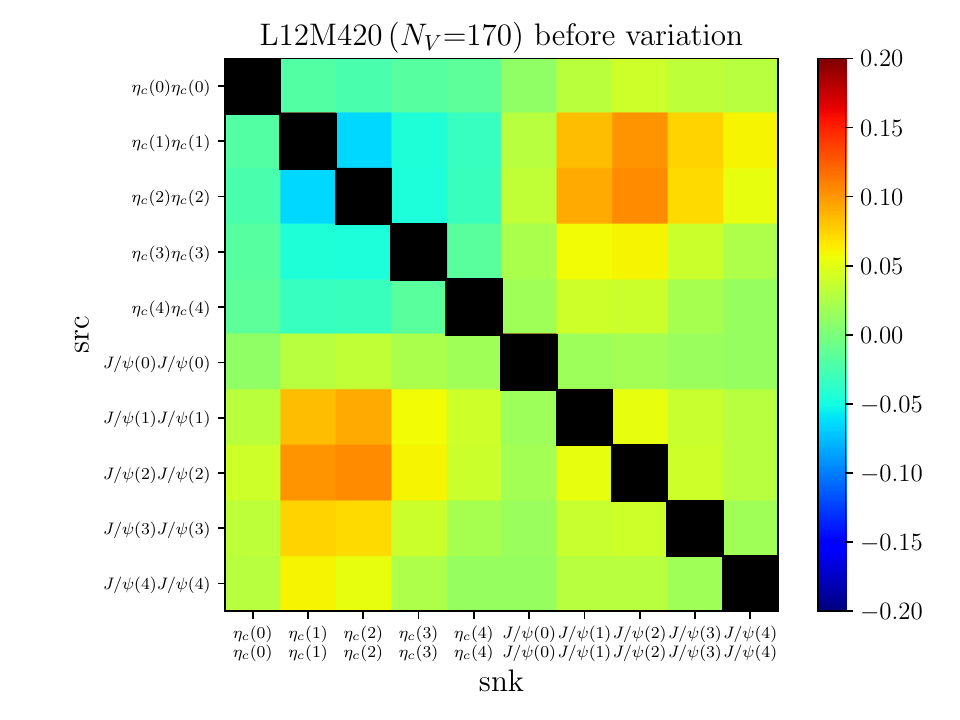}
	\includegraphics[width=0.45\linewidth]{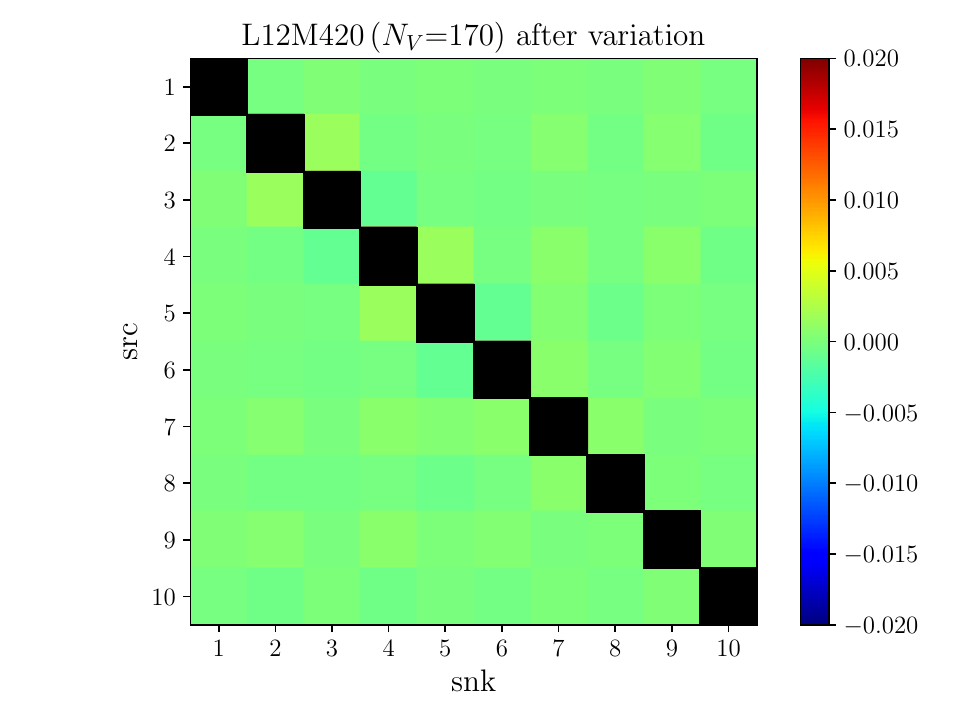}\\
	\includegraphics[width=0.45\linewidth]{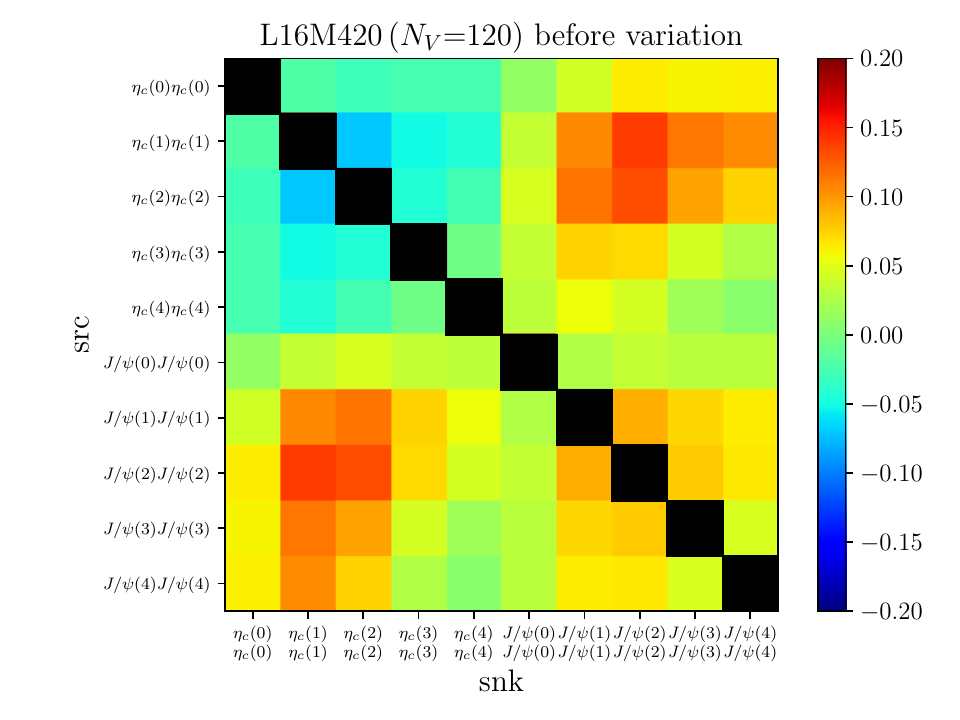}
	\includegraphics[width=0.45\linewidth]{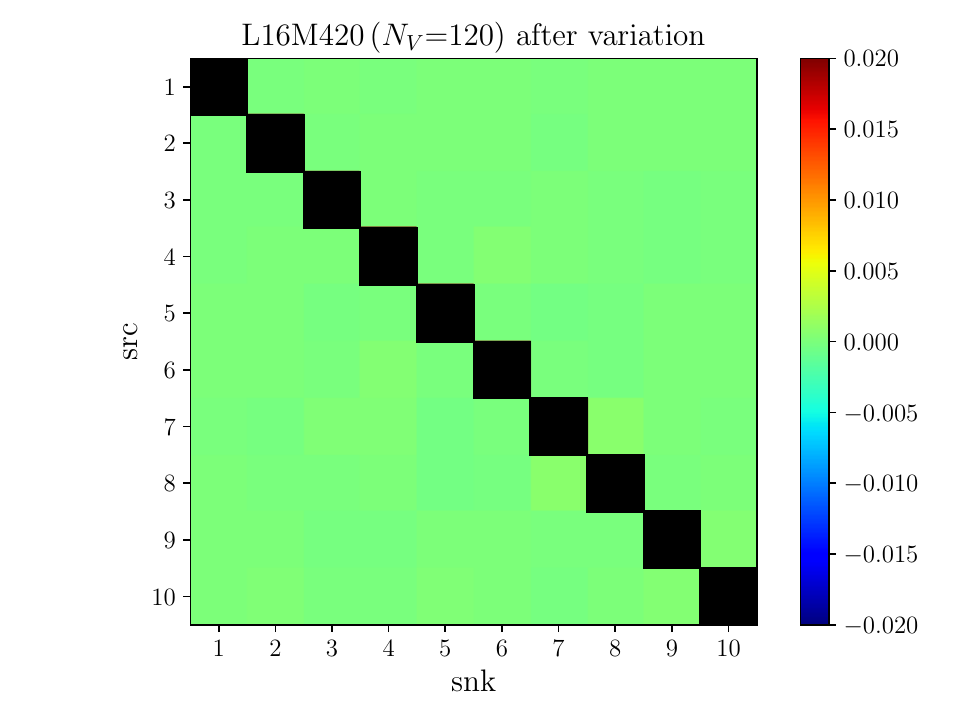}
	\caption{Weight functions $w_{\alpha\beta}(t)$ and $w_{mn}(t)$ in the $0^{++}$ channel. 
		The upper two panels are $w_{\alpha\beta}(t=20~a_t)$ (left) and $w_{mn}(t=20~a_t)$ (right) on the L12M420 ensemble. The lower panels are $w_{\alpha\beta}(t=35~a_t)$ (left) and $w_{mn}(t=35~a_t)$ (right) on the L16420 ensemble. 
		The axis labels are the contents of operators $\mathcal{O}_\alpha$ for $w_{\alpha\beta}(t)$, and refer to the $n$th optimized operator $\mathcal{O}^{(n)}$ for $w_{mn}(t)$. 
		In each panel, the diagonal elements are equal to one and are shown in black. 
		The sizeable values of $w_{\alpha\beta}$ for $\alpha\ne\beta$ indicate the substantial correlation between different operators $\mathcal{O}_\alpha$, while the tiny values (uniformly $(1\sim 2)\times 10^{-3}$) of $w_{mn}$ for $m\ne n$ manifest that the optimized operators $\mathcal{O}^{(n)}$ are almost orthogonal. 
		Similar patterns are observed for the M250 ensembles.}  
	\label{fig:wt-0pp}
\end{figure*}
%%%%%%%%%%%%%%%%%%%%%%%%%%%

We introduce the following weight function to visualize the correlation strength between different two-particle operators $\mathcal{O}_{MM}(k)$ ($\mathcal{O}_\alpha$), 
\begin{equation}\label{eq:wta}
	w_{\alpha\beta}(t)=\frac{C_{\alpha\beta}(t)}{\sqrt{C_{\alpha\alpha}(t)C_{\beta\beta}(t)}}.
\end{equation}
The left two panels in Fig.~\ref{fig:wt-0pp} illustrate the 2d histograms of $w_{\alpha\beta}(t)$ at $t=20~a_t$ for L12M420 ensemble and $t=35~a_t$ for L16M420 ensemble. 
When $\alpha = \beta$, the sink and source operators have the same particle content and the same relative momentum configuration $k$. 
In this case, $C_{\alpha\alpha}(t)$ is dominated by the contribution from diagram (a) and reflects the propagation of the two-particle system corresponding to the operator $\mathcal{O}_\alpha(k)$.
The diagonal elements of $w_{\alpha\beta}(t)$ equal to one exactly by definition, while the off-diagonal elements are nonzero and have magnitudes in the range from $\mathcal{O}(10^{-2})$ to $\mathcal{O}(10^{-1})$. 
The different amplitudes of $w_{\alpha\beta}(t)$ at $t=20~a_t$ ($t=35~a_t$) indicates the different composition of all the relevant energy eigenstates at this time slice. Let $|n\rangle$ be the $n$th eigenstate of the lattice Hamiltonian with the eigenenergy $E_n$, then $C_{\alpha\beta}(t)$ can be decomposed as 
\begin{equation}
	C_{\alpha\beta}(t)=\sum\limits_{n} Z_\alpha^{(n)} Z_\beta^{(n)*} e^{-E_n t} +(t\to T-t),
\end{equation}
where $Z_n=\langle 0|\mathcal{O}_\alpha|n\rangle$ is the coupling of $\mathcal{O}_\alpha$ to the $n$th energy eigenstate (here we use the normalization $\langle n|n\rangle =1$). So sizeable values of $w_{\alpha\beta}$ for $\alpha\ne \beta$ indicate the substantial correlations between operators of different particle constituents and different relative momentum modes.

%%%%%%%%%%%%%%%%%%%%%%%%%%%
\begin{figure}[t]
	\centering
	\includegraphics[width=0.9\linewidth]{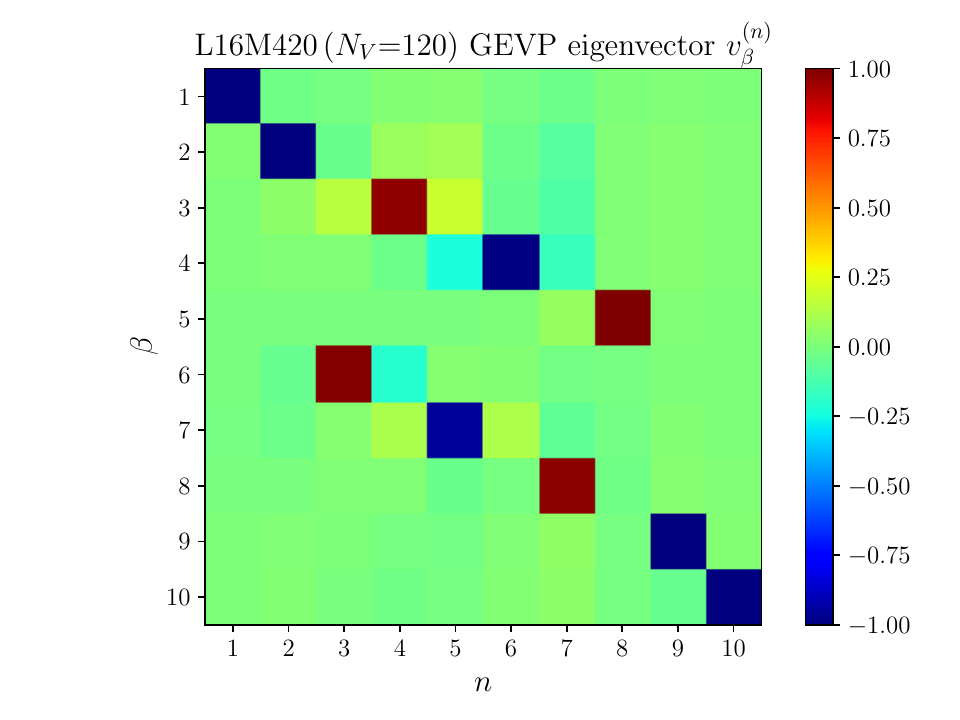}
	\caption{The eigenvector $v_\beta^{(n)}$ by solving the GEVP, where the main components determine the correspondence between energy levels before and after the variation.}
	\label{fig:GEVP_vector}
\end{figure}
%%%%%%%%%%%%%%%%%%%%%%%%%%%

In order to extract reliably the finite-volume energy levels relevant to $\eta_c\eta_c-J/\psi J/\psi$ scattering, we need the optimized operator $\mathcal{O}^{(n)}$ that couples predominantly to the $n$th energy eigenstate. 
In doing so, we solve the generalized eigenvalue problem (GEVP), also known as the variation method, for the correlation matrix $C_{\alpha\beta}(t)$ with given $t_0$ and $t_1$, namely,
\begin{equation}
	\label{eq:gevp}
	C_{\alpha\beta}(t_1)v_\beta^{(n)}=\lambda^{(n)}(t_1,t_0) C_{\alpha\beta}(t_0)v_\beta^{(n)},
\end{equation}
where $v_\alpha^{(n)}$ is eigenvector of the $n$th largest eigenvalue $\lambda^{(n)}(t_1,t_0)$. 
Subsequently, we obtain the optimized operator $\mathcal{O}^{(n)}=v_\alpha^{(n)}\mathcal{O}_\alpha$.

Using the operator set $\mathcal{S}_0$ (excluding $\mathcal{O}_{\chi_{c0}}$) and the choices of $(t_0,t_1)/a_t=(15,25)$ on L12 and $(t_0,t_1)/a_t=(30,40)$ on L16, we obtain a set of optimized operators $\{\mathcal{O}^{(n)}, n=1,2,\ldots,10\}$, whose correlation matrix is calculated as 
\begin{equation}
	C_{mn}(t)=\langle \mathcal{O}^{(m)}(t)\mathcal{O}^{(n)}(0)\rangle\equiv v_\alpha^{(m)}  C_{\alpha\beta}(t)v_\beta^{(n)}.
\end{equation}
It is seen that $C_{mn}(t)$ can be calculated directly from $C_{\alpha\beta}(t)$ and   
the magnitudes of $v_{\alpha}^{(m)}$ and $v_{\beta}^{(n)}$ determine the contribution of $C_{\alpha\beta}(t)$ to $C_{mn}(t)$.
As an example, the eigenvector $v_\beta^{(n)}$ is shown in Fig.~\ref{fig:GEVP_vector}, while $v_{\alpha}^{(m)}$ corresponds to its transpose.
It can be seen that each column in Fig.~\ref{fig:GEVP_vector} (with a fixed $n$-index, i.e., each optimized correlation) is predominantly contributed by a single original correlation (with the $\beta$-index).
It establishes the correspondence between the energy levels before and after solving GEVP.

In order to check the orthogonality of the optimized operators $\mathcal{O}^{(n)}$, we also introduce a weight function $w_{mn}(t)$ similar to Eq.~\eqref{eq:wta},
\begin{equation}
	\label{eq:wtn}
	w_{mn}(t)=\frac{C_{mn}(t)}{\sqrt{C_{mm}(t)C_{nn}(t)}}.
\end{equation}
As shown in the right panels in Fig.~\ref{fig:wt-0pp}, the off-diagonal elements of $w_{mn}(t)$ ($m\ne n$) have very tiny values ( $(1\sim 2) \times 10^{-3}$) uniformly in comparison with the diagonal elements $w_{nn}=1$. 
This indicates the fairly good orthogonality of the optimized operators $\mathcal{O}^{(n)}$. 
This is encouraging since the $n$th eigenenergy $E^{(n)}$ can be reliably extracted from $C_{nn}(t)$. 
Actually, Fig.~\ref{fig:wt-0pp} implies
\begin{equation}
	\begin{aligned}
		\mathcal{O}^{(n)}|0\rangle &=Z_n\left(|n\rangle+\sum\limits_{m\ne n} \epsilon_{m}|m\rangle\right),\\
		C_{nn}(t)&= Z_n^2\left(e^{E_n t}+\sum\limits_{m\ne n}\epsilon_m^2 e^{-E_m t}\right),
	\end{aligned}
\end{equation}
with $\epsilon_m \lesssim \mathcal{O}(10^{-3})$. For simplicity, we assume there is only one state $|m\rangle$ in the summation of the above equations and let $\Delta E=E_m-E_n$. 
If $C_{nn}(t)$ is fitted by a single exponential function (the $t\to T-t$ term owing to the periodic boundary condition is omitted temporarily), then one has 
\begin{equation}
	\begin{aligned}
		E_n^{(\mathrm{eff})}(t)&=\frac{1}{a_t}\ln \frac{C_{nn}(t)}{C_{nn}(t+a_t)}\\
		&\approx E_n\left(1+\epsilon_m^2 \frac{\Delta E}{E_n} e^{-\Delta E t}\right),
	\end{aligned}
\end{equation}
where the second term in the brackets is due to the contamination from the state $|m\rangle$. For the typical values $|\Delta E|/E_n \in (0.01,0.1)$ in this study, the relative deviation of $E_n^{(\mathrm{eff})}(t)$ from $E_n$ is of order $\mathcal{O}(10^{-8}-10^{-6})$ in the time range $t\sim 40 a_t$ and is much smaller than the statistical error.

%%%%%%%%%%%%%%%%%%%%%%%%%%%
\begin{figure*}[t]
	\centering
	\includegraphics[width=0.45\linewidth]{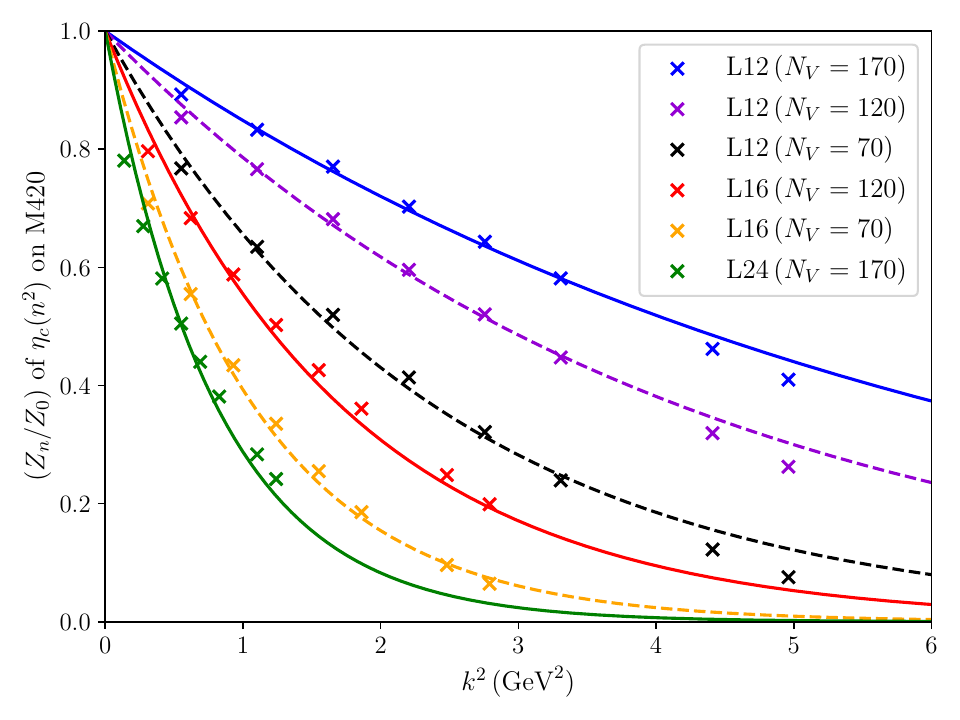}
	\includegraphics[width=0.45\linewidth]{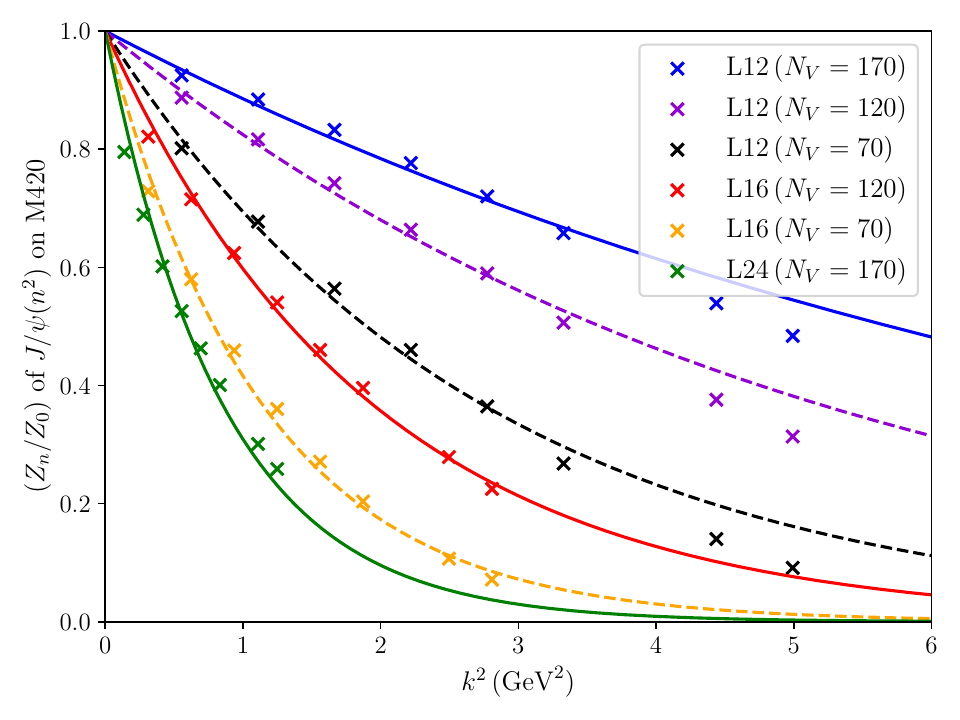}
	\caption{The spectral weight $Z_n/Z_0$ for $C^{(n)}(t) = |Z_n|^2 \frac{{N_s}^3}{2E_n^M} e^{-E^M_n t}$ of $\eta_c(n^2)$ (left) and $J/\psi(n^2)$ (right) varying $N_V$ values, on three-size M420 ensembles.
		As the momentum $k^2$ increases, the spectral weight exhibits exponential suppression, given by $Z_n \sim Z_0 ~ e^{-\frac{k^2}{4\sigma}}$.}
	\label{fig:suppression-NV}
\end{figure*}
%%%%%%%%%%%%%%%%%%%%%%%%%%%

\subsection{$N_V$ dependence of LHS smearing}\label{secII:NV}
The distillation method provides a quark field smearing scheme, namely, LHS smearing
\begin{equation}
	c(\vec{x})=\sum\limits_{\vec{y}}\sum\limits_{i=1}^{N_V} V_i^\dagger(\vec{x})V_i(\vec{y})\tilde{c}(\vec{y})\equiv \int d^3\vec{y} \Phi(\vec{x},\vec{y})\tilde{c}(\vec{y}),
\end{equation}
where $V_i(\vec{x})$ is the $i$th eigenvector of the lattice gauge invariant Laplacian operator and $\tilde{c}$ stands for the unsmeared charm quark field. 
Previous lattice studies~\cite{HadronSpectrum:2009krc,ZhangRenQiang:2021gnn} have observed that the smearing function behaves similarly to the Gaussian smearing $\Phi(\vec{x},\vec{y})\sim e^{-\sigma r^2/2}$ where $r=|\vec{x}-\vec{y}|$ and $\sigma$ varying with $N_V$.
The LHS smeared charm quark field $c(\vec{x})$ has a Gaussian spatial distribution around $\vec{x}$. 
Consequently, the operator $\mathcal{O}_M(\vec{x})=[\bar{c}\Gamma_M c]$ has the following property,
\begin{equation}
	\mathcal{O}_M(\vec{x})\sim \int d^3 \vec{y} e^{-\sigma|\vec{y}-\vec{x}|^2}\mathcal{O}^0_M(\vec{y}),
\end{equation}
where $\mathcal{O}_M^0(\vec{y})=[\bar{\tilde{c}}\Gamma\tilde{c}](\vec{y})$ is the local operator at $\vec{y}$. 
The operator $\mathcal{O}_M(\vec{k})$ with a definite momentum $\vec{k}$ is obtained from the Fourier transformation,
\begin{equation}\label{eq:suppression}
	\mathcal{O}_M(\vec{k})=\frac{1}{V_3}\int d^3\vec{x} e^{-i\vec{k}\cdot\vec{x}} \mathcal{O}_M(\vec{x})\propto e^{-\frac{k^2}{4\sigma}}\mathcal{O}_M^0(\vec{k}). 
\end{equation}
where $V_3$ is the spatial volume. 
For single-particle states of $\eta_c$ and $J/\psi$, the operator couplings are written as 
\begin{eqnarray}
	\langle 0|\mathcal{O}_{\eta_c} (\vec{k})|\eta_c,\vec{k}\rangle&\equiv& Z_P\nonumber\\
	\langle 0|\mathcal{O}_{\psi}^i(\vec{k})|J/\psi,\vec{k},\lambda\rangle &\equiv & Z_V \epsilon_i(\vec{k},\lambda), 
\end{eqnarray}
where $\vec{\epsilon}(\vec{k},\lambda)$ is the polarization vector of $J/\psi$, and $Z_P$ and $Z_V$ are constant independent of $\vec{k}$. 
Thus, Eq.~\eqref{eq:suppression} indicates that the coupling of the LHS smeared operator $\mathcal{O}_M(\vec{k})$ to a state of momentum $\vec{k}$ is suppressed by a Gaussian factor $e^{-k^2/(4\sigma)}$. 
This has been verified to be correct in the actual calculations.
As usual, the correlation function of $\mathcal{O}_M(\vec{k})$ is parametrized as 
\begin{equation}
	C_M(\vec{k},t) = |Z_k|^2 \frac{{N_s}^3}{2E_k} e^{-E_k t}~~~(t/a_t \gg 1),
\end{equation}
where the factor $N_s^3$ is due to fact that both the sink and source operators are smeared.
By fitting the single-particle correlations in the region $t/a_t = 20{-}108$, we obtained the single-particle mass spectrum $E_k$, as shown in Fig.~\ref{fig:ksi}, along with the corresponding values of $|Z_k|^2$.
The ratios $Z_k/Z_0$ of $\eta_c$ and $J/\psi$ on ensembles L12M420 and L16M420 (and an ensemble L24M420 that has the similar parameters to those of L12M420 and L16M420 except for $L_s=24 a_s$) are plotted as points with respect to $k^2$ (in physical units) in the left ($\eta_c$) and right ($J/\psi$) panels of Fig.~\ref{fig:suppression-NV}, where the ratios are derived at different $N_V=(70,120,170)$ and the curves show the Gaussian function $e^{-k^2/(4\sigma)}$ with estimated values of $\sigma$ (not serious fits) at different $N_V$. 

From the plots one can get the following information:
\begin{itemize}
	\item For the same number of $N_V$ on the same lattice, the LHS smearing effect is the same for the operators of $\eta_c$ and $J/\psi$. The momentum suppression can be well approximated by a Gaussion profile $e^{-k^2/(4\sigma)}$.
	\item On the same lattice, the momentum suppression becomes stronger when $N_V$ decreases. 
	\item With the same $N_V$, the larger the lattice size, the stronger the momentum suppression. 
	For the L24M420 ensemble, even with $N_V=170$ the coupling to the state of $k^2\sim 2~\mathrm{GeV}^2$ is suppressed to be $Z_k/Z_0\sim 0.1$; 
	\item For the correlation functions $C_{\alpha\alpha}(t)$ (the diagonal elements of the correlation matrix), their magnitudes scale approximately as $|Z_{k_\alpha}/Z_0|^4$ with respect to $k^2$. 
	Since the upper bound of the energy range we are considering corresponds to the momentum $k_\alpha^2\sim 2~\mathrm{GeV}^2$, this scale factor takes the approximate values $\sim 0.7^4, 0.3^4, 0.1^4$ for $N_s(N_V)=12(170),16(120),24(170)$, respectively.   
\end{itemize}
Therefore, a large enough $N_V$ is required for a given lattice size in order for the coupling of operators to higher momentum states to be enhanced. 
Unfortunately, for a give smear size ($1/\sqrt{\sigma}$), the required $N_V$ increases with the lattice volume~\cite{HadronSpectrum:2009krc}. 
Due to the limitation of our computing resources and the storage space, we can only afford the calculation with $(N_s,N_V)=(12,170)$ or $(16,120)$, which are the practical parameters in this study. 
Actually, $N_V=120$ on our L16 lattices is not large enough for us to reliably extract the energy levels around 6.5~GeV, see Sec.~\ref{secII:NV-dep}. 

%%%%%%%%%%%%%%%%%%%%%%%%%%%%%%%%%%%%
\begin{figure*}[t]
	\centering
	\includegraphics[width=0.45\linewidth]{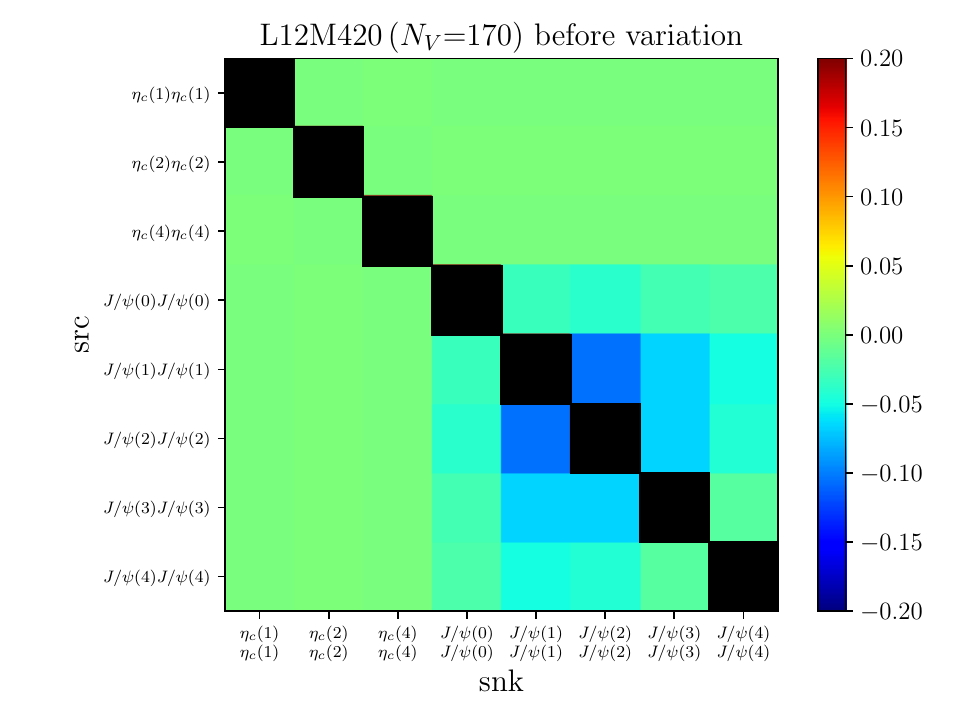}
	\includegraphics[width=0.45\linewidth]{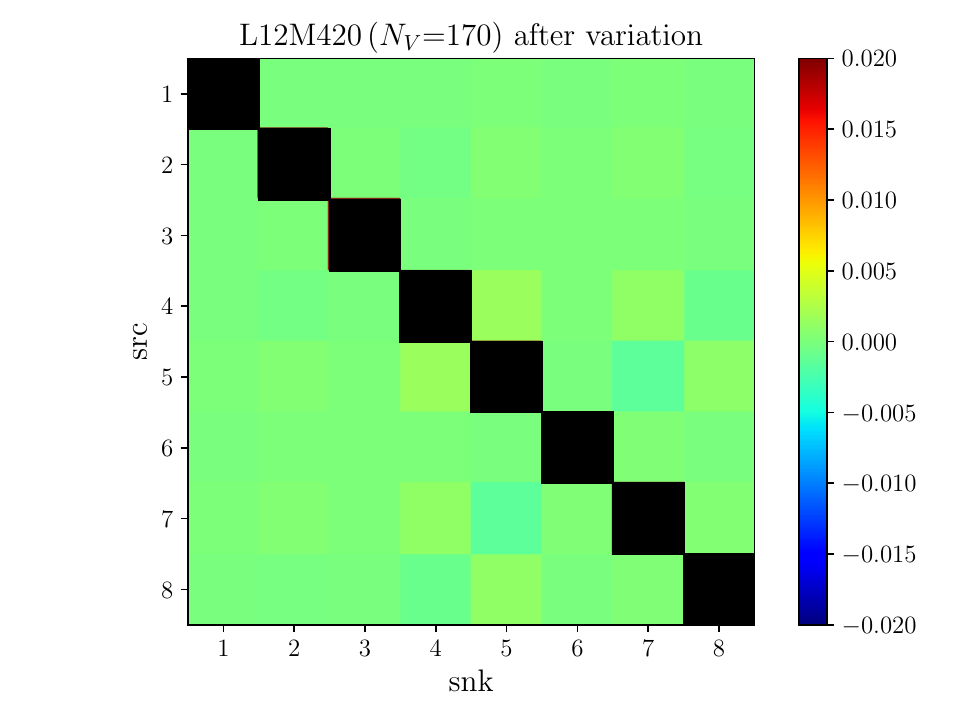}\\
	\includegraphics[width=0.45\linewidth]{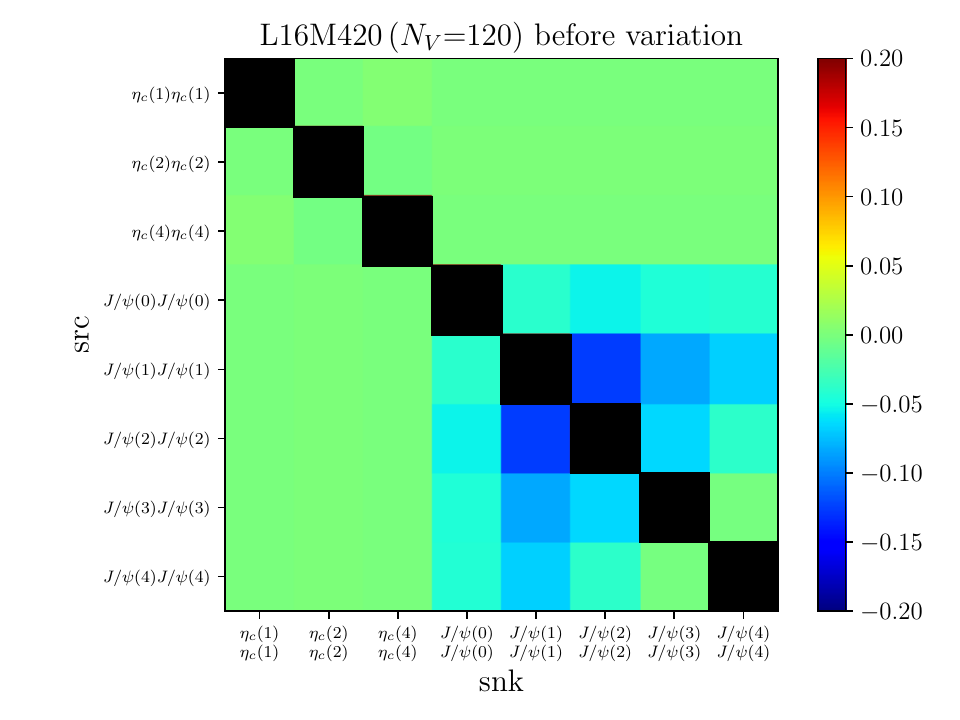}
	\includegraphics[width=0.45\linewidth]{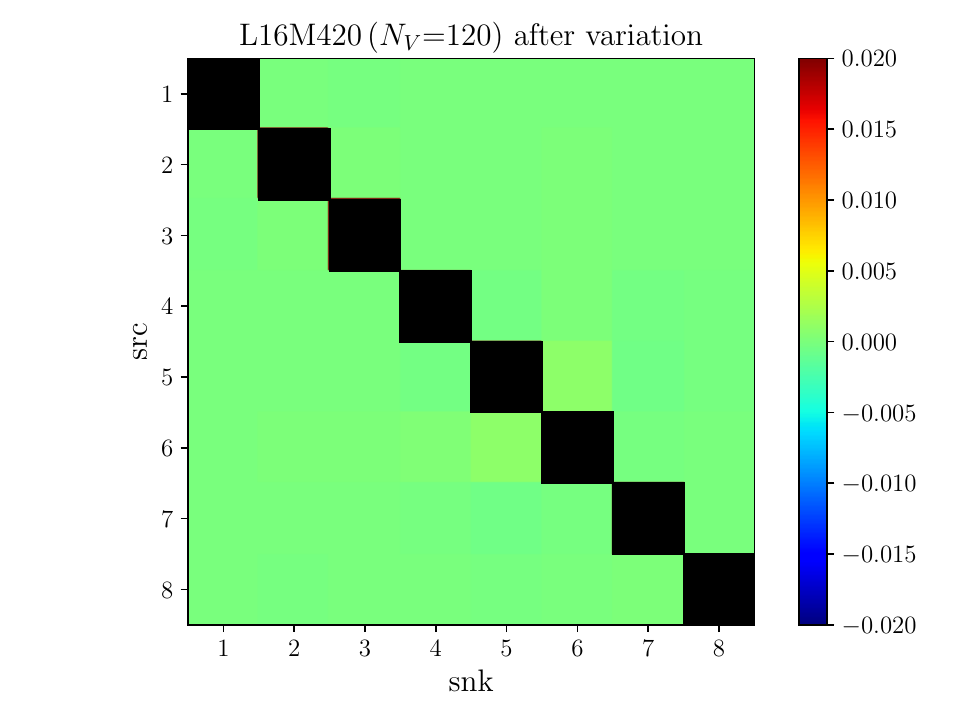}
	\caption{Weight functions $w_{\alpha\beta}(t)$ and $w_{mn}(t)$ in the $2^{++}$ channel. 
		The upper two panels are $w_{\alpha\beta}(t=20~a_t)$ (left) and $w_{mn}(t=20~a_t)$ (right) on the L12M420 ensemble. 
		The lower panels are $w_{\alpha\beta}(t=35~a_t)$ (left) and $w_{mn}(t=35~a_t)$ (right) on the L16420 ensemble. 
		The axis labels are the contents of operators $\mathcal{O}_\alpha$ for $w_{\alpha\beta}(t)$, and refer to the $n$th optimized operator $\mathcal{O}^{(n)}$ for $w_{mn}(t)$. 
		In each panel, the diagonal elements are equal to one and are shown in black. A striking observation is that the $\eta_c\eta_c$ operators almost do not couple with $J/\psi J/\psi$ operators, while $J/\psi J/\psi$ operators with different relative momenta have sizeable correlations. 
		The $w_{mn}$ is almost a unit matrix indicating the optimized operators $\mathcal{O}^{(n)}$ are nearly orthogonal. 
		Similar patterns are observed for the M250 ensembles.   }
	\label{fig:wt-2pp}
\end{figure*}
%%%%%%%%%%%%%%%%%%%%%%%%%%%

Based on the discussions above, we now explain why the weight matrix $w_{mn}$ for the L16 lattices is more diagonal than that for the L12 lattices (see Fig.~\ref{fig:wt-0pp}).
In general, if a correlation matrix $C_{\alpha\beta}(t)$ contains contributions from $N$ energy eigenstates $\{|n\rangle, n=1,2,\ldots,N\}$ of the lattice Hamiltonian of the system in a specific time range, then one needs only $N$ linearly independent operators that satisfy the relation,
\begin{equation}
	\label{eq:states}
	\mathcal{O}_\alpha|0\rangle=\sum\limits_n |n\rangle\langle n|\mathcal{O}_\alpha|0\rangle\equiv \sum\limits_n Z_\alpha^{(n)}|n\rangle,
\end{equation}
with $\det \mathbf{Z}\equiv\det \{Z^{(n)}_\alpha\}\ne 0$. These operator establish a complete operator set $\mathcal{S}=\{\mathcal{O}_\alpha, \alpha=1,2,\ldots, N\}$, such that one can obtain $N$ optimized operators $\mathcal{O}^{(n)}=(\mathbf{C}^{-1})_\alpha^n\mathcal{O}_\alpha$ with each operator $\mathcal{O}^{(n)}$ coupling exclusively to $|n\rangle$, namely,
\begin{equation}
	|n\rangle=(\mathbf{C}^{-1})_\alpha^n\mathcal{O}_\alpha|0\rangle=\mathcal{O}^{(n)}|0\rangle.
\end{equation}
One can see that the eigenvectors of GEVP are actually $v_\alpha^{(n)}=(\mathbf{C}^{-1})_\alpha^n$ and the optimized operators $\mathcal{O}^{(n)}$ are exactly orthogonal. 
If operators are less than the states, the operators $\mathcal{O}^{(n)}=v_\alpha^{(n)}\mathcal{O}_\alpha$ can not be exactly orthogonal because the matrix $\mathbf{Z}$ is not invertible. 
On the L16 lattices, with a smaller $N_V = 120$ compared to $N_V = 170$ on the smaller L12 lattices, higher-momentum states are more strongly suppressed. 
As a result, the number of operators in $\mathcal{S}_0$ is closer to the number of states, making the optimized operators $\mathcal{O}^{(n)}$ more orthogonal than those on the L12 lattices.

%%%%%%%%%%%%%%%%%%%%%%%%%%%%%%%%%%%%%%%%%%%%%%%%%%%%%%%%%%%%%%%%%%%%%%%
\subsection{$2^{++}$ Correlation functions}\label{secII:corr2pp}
%%%%%%%%%%%%%%%%%%%%%%%%%%%%%%%%%%%%%%%%%%%%%%%%%%%%%%%%%%%%%%%%%%%%%%%

With the prescriptions discussed in the previous sections, the operator sets used to determine the finite-volume energy levels in the $0^{++}$ and $2^{++}$ channels are reduced to the following:
\begin{equation}
	\begin{aligned}
		\label{eq:op-set-new}
		\mathcal{S}_0&=\{\mathcal{O}_{\eta_c\eta_c}(n^2),\mathcal{O}_{\psi\psi}(n^2); n^2=0,1,2,3,4\},\\
		\mathcal{S}_2&=\{\mathcal{O}_{\eta_c\eta_c}(n^2\ne 0,3),\mathcal{O}_{\psi\psi}(n^2); n^2=0,1,2,3,4\}.
	\end{aligned}
\end{equation} 
Based on the $^1D_2$ $\eta_c \eta_c$ and $^5S_2$ $J/\psi J/\psi$ operator set $\mathcal{S}_2$ in Eq.~\eqref{eq:op-set-new}, we calculate the correlation matrix $C_{\alpha\beta}(t)$, whose correlation weight matrix $w_{\alpha\beta}(t)$ is shown in the right two panels (upper left panel for L12M420 ($N_V=170$) and lower left panel for L16M420 ($N_V=120$)) of Fig.~\ref{fig:wt-2pp}. 

This striking observation is that $\mathcal{O}_{\eta_c\eta_c}$ operators almost do not couple with $\mathcal{O}_{\psi\psi}$ operators since $w_{\alpha\beta}\approx 0$ for $(\alpha,\beta)\to (\eta_c\eta_c,\psi\psi)$ (The correlations between $\eta_c\eta_c$ operators with different relative momentum is also very weak). 
This provides a direct and clear indication that the $\eta_c\eta_c-J/\psi J/\psi$ coupled channel effects is negligible. After solving the GEVP to $C_{\alpha\beta}(t)$, we obtain the correlation
matrix $C_{mn}(t)$ of optimized operators $\mathcal{O}^{(n)}$, whose correlation strengths $w_{mn}$ are also shown in Fig.~\ref{fig:wt-2pp}, where the upper~(lower) right panel is for the L12M420 (L16M420) ensemble. 
It is seen that $w_{mn}(t)$ of the L16M420 ensemble is more diagonal than that of L12M420 for the same reason as the case of the $0^{++}$ channel.
For the center-of-mass energy range below 6.6~GeV, we consider the $J/\psi J/\psi$ operators up to $n^2=4$. 
On the L12 ensembles, the $J/\psi J/\psi$ energy at $n^2=5$ is about 6.85~GeV, which lies well above the upper limit of our energy window, where scattering channels such as $J/\psi \psi'$ and $J/\psi \psi(3770)$ also become relevant.

%%%%%%%%%%%%%%%%%%%%%%%%%%%%%%%%%%%%%%%%%%%%%%%%%%%%%%%%%%%%%%%%%%%%%%%%%%%%%%%%%%%%%%%%
\subsection{Energy-level determination}\label{secII:fit}
%%%%%%%%%%%%%%%%%%%%%%%%%%%%%%%%%%%%%%%%%%%%%%%%%%%%%%%%%%%%%%%%%%%%%%%%%%%%%%%%%%%%%%%%
After the GEVP analysis to the correlation matrix $C_{\alpha\beta}(t)$ of these operator set on each gauge ensemble with specific values of $N_V$, we obtain the correlation functions $C_{nn}(t)$ of the optimized operators $\mathcal{O}^{(n)}$, each of which is expected have predominantly the contribution from the $n$th energy eigenstate. 

The correlation functions $C_{nn}(t)$ maintain a good signal-to-noise ratio throughout the entire time range, and facilitate the data analysis for extracting the energy eigenvalues 
$E_n$ over wide time windows.
In two-particle systems, the periodic boundary condition permits the two particles either to propagate alongside from the source time slice ($t=0$ for example) to a sink time slice $t<T/2$, or to propagate independently with one particle running from $0$ to $t$ and the other running in the opposite time direction from $0$ to $t$ around the time circle. 
The later propagation is commonly referred to as the thermal state.
Given that the two particles have energies $E$ and $E'$, the time evolution of the thermal state can be expressed as
\begin{equation}
	\begin{aligned}
		C_\mathrm{thm}(t)&=A\left[e^{-Et}e^{-E'(T-t)}+e^{-E(T-t)}e^{-E't}\right]  \\
		&= 2A e^{-(E+E')\frac{T}{2}} \cosh\left[(E-E')(t-\frac{T}{2})\right],
	\end{aligned}
\end{equation}
where $A$ is a overall factor from the operator coupling. For a system of two identical particles in its rest frame, the condition $E=E'$ holds, which leads to a constant
$C_\mathrm{thm}(t)$. 
This is the scenario under consideration.

Both the two ways of propagation contribute to $C_{nn}(t)$ and the contribution from the thermal state becomes more pronounced when $t\to T/2$. 
Since we need to extract the energy of the lowest state contributing to $C_{nn}(t)$ in the large $t$ range where the contamination of higer states can be negligible, the contribution from the thermal state should be considered. 
We adopt the following function form to parametrize $C_{nn}(t)$,
\begin{equation}
	\begin{aligned}
		\label{eq:3-state-fit}
		C^{(n)}(t) & =  W\cosh\left[E_n(t-\frac{T}{2})\right] + W_\mathrm{thm} \\
		&~ + W'\cosh\left[E_n'(t-\frac{T}{2})\right],
	\end{aligned}
\end{equation}
where the first term accounts for the contribution of the $n$th eigen state of the lattice Hamiltonian, the second term (a constant) comes from the thermal state, and the third term is added to absorb the residual contamination from higher states. 
This method is referred to as `3-state' fitting in the following.

\begin{figure*}[htbp]
	\centering
	\includegraphics[width=0.49\linewidth]{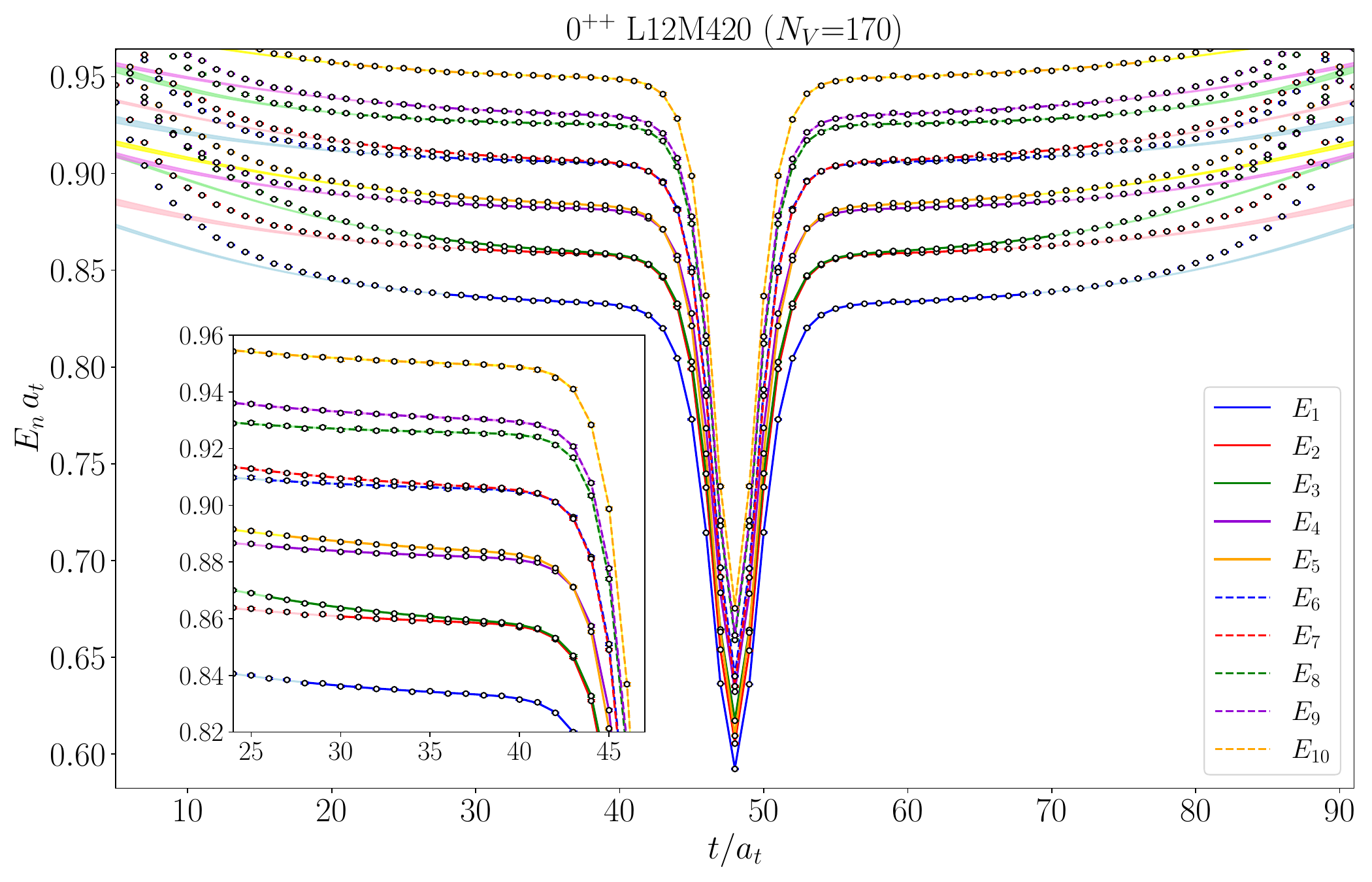}
	\includegraphics[width=0.49\linewidth]{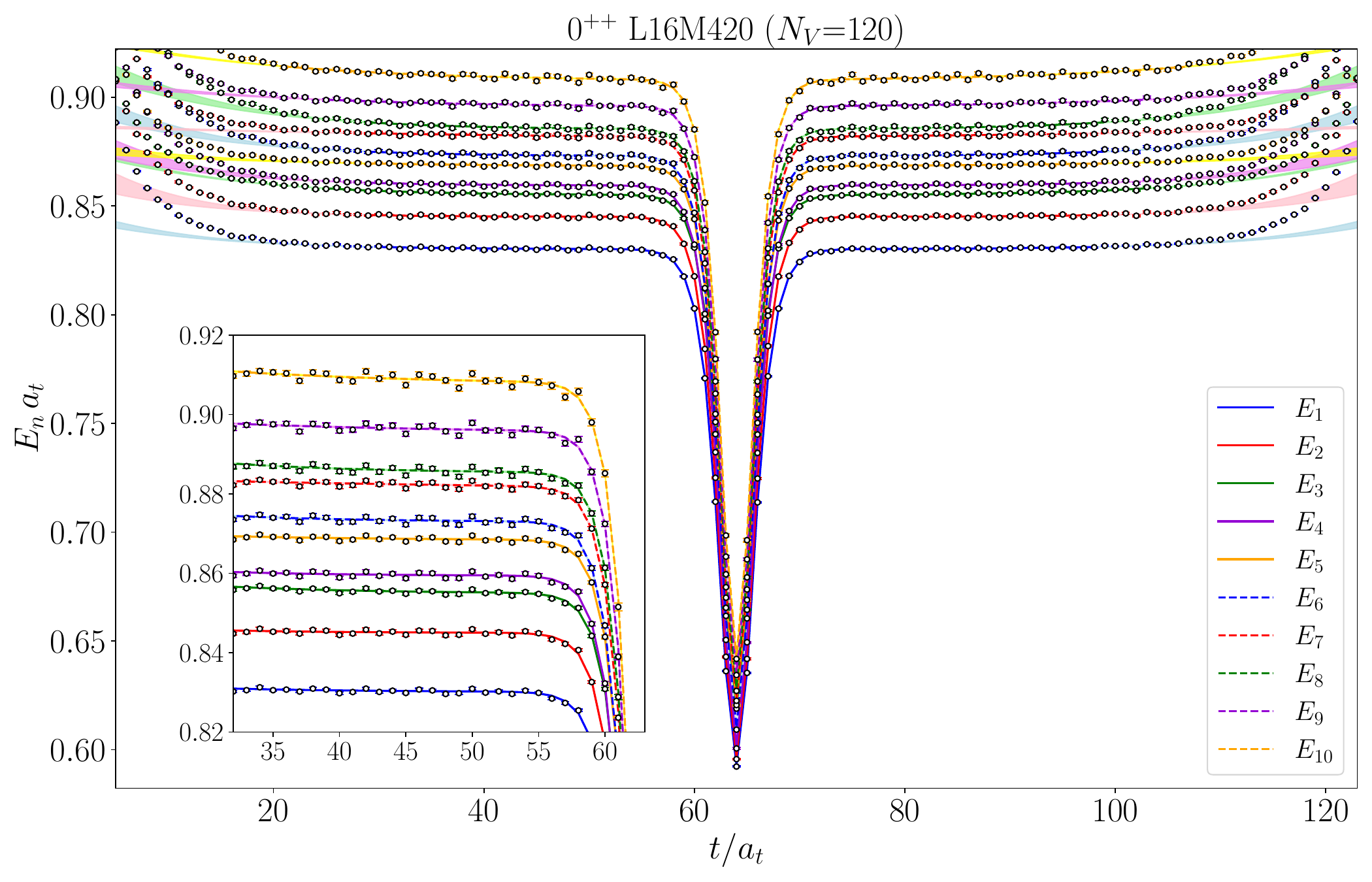}\\
	\includegraphics[width=0.49\linewidth]{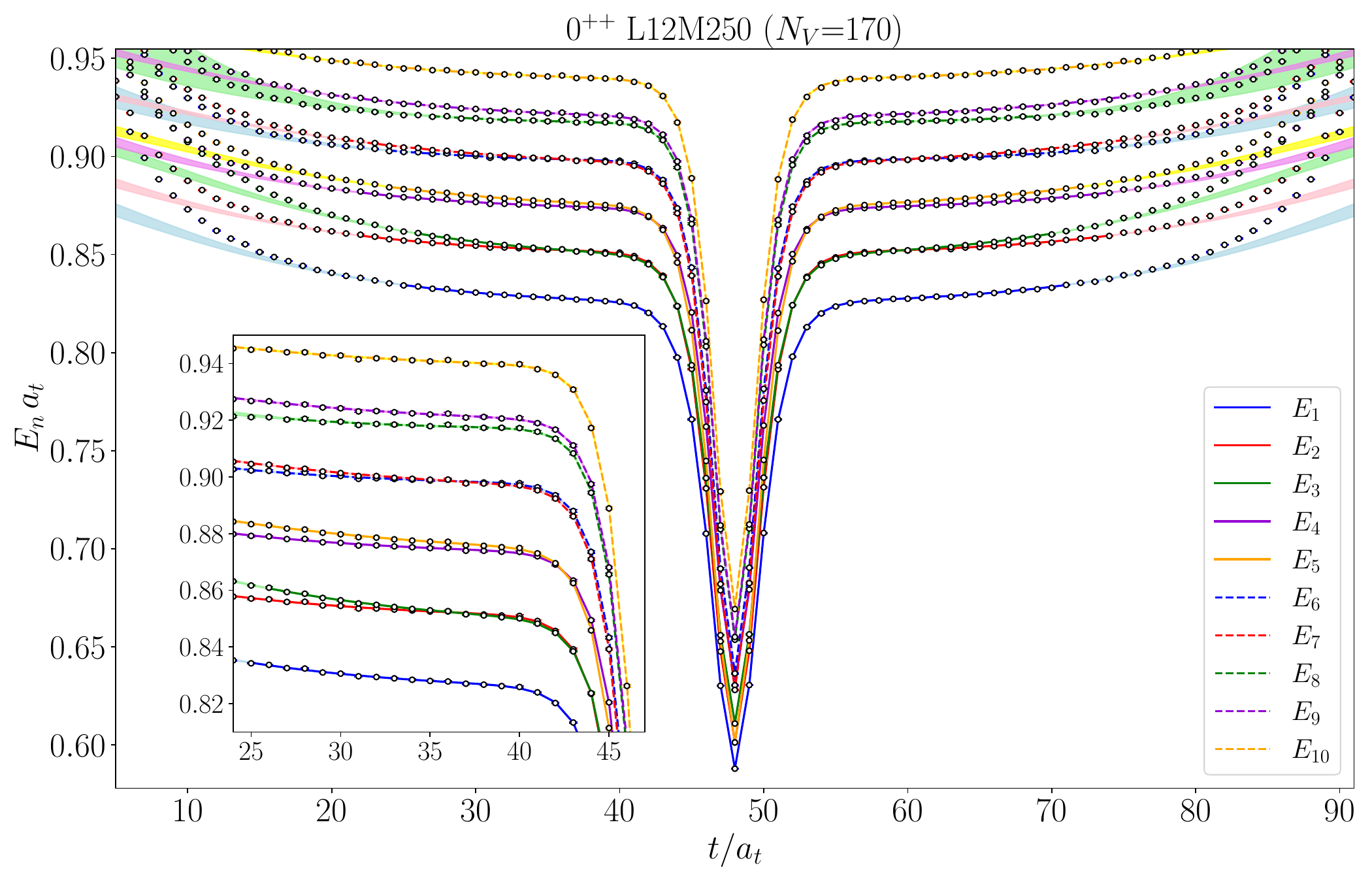}
	\includegraphics[width=0.49\linewidth]{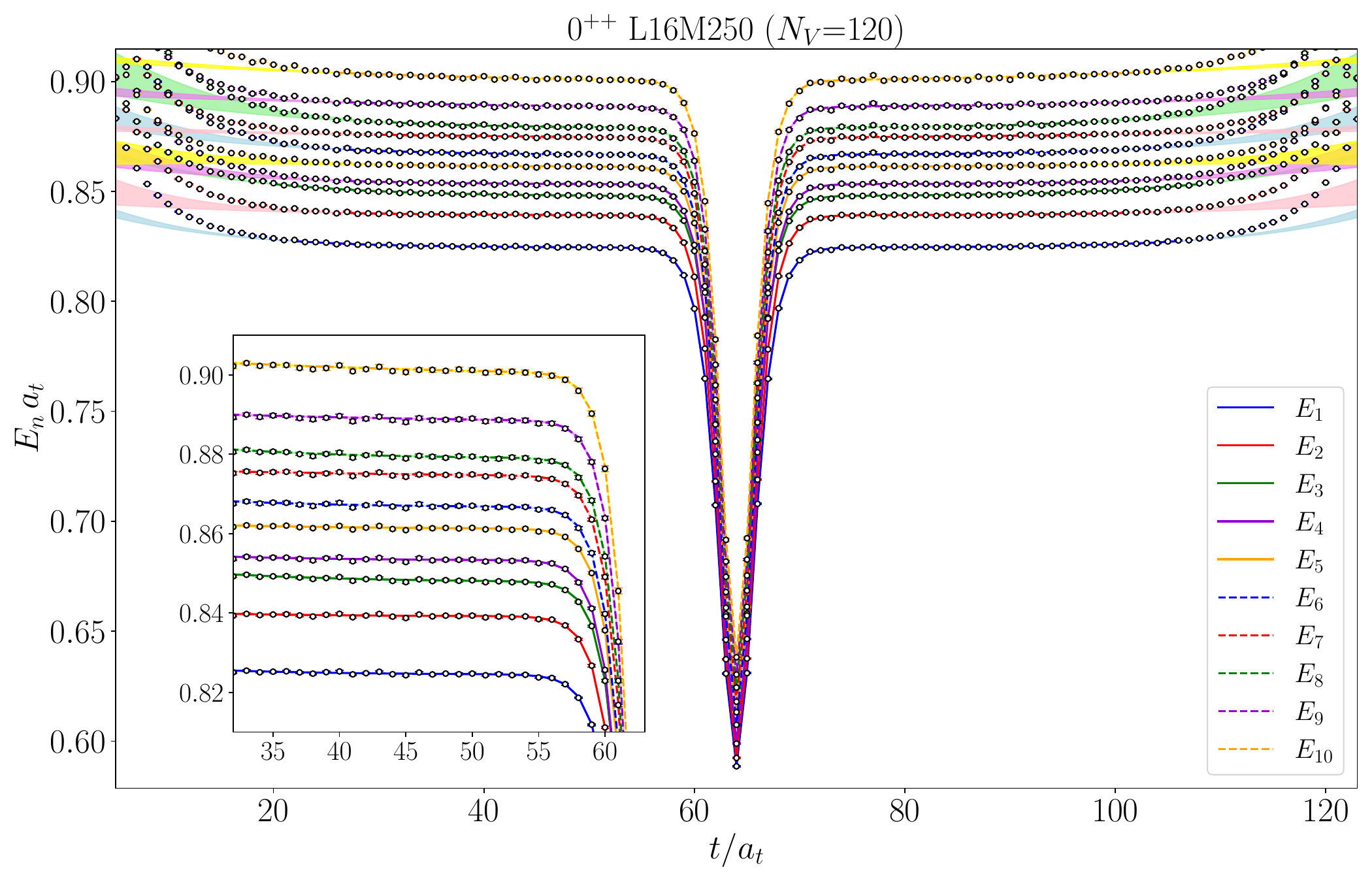}\\
	\includegraphics[width=0.49\linewidth]{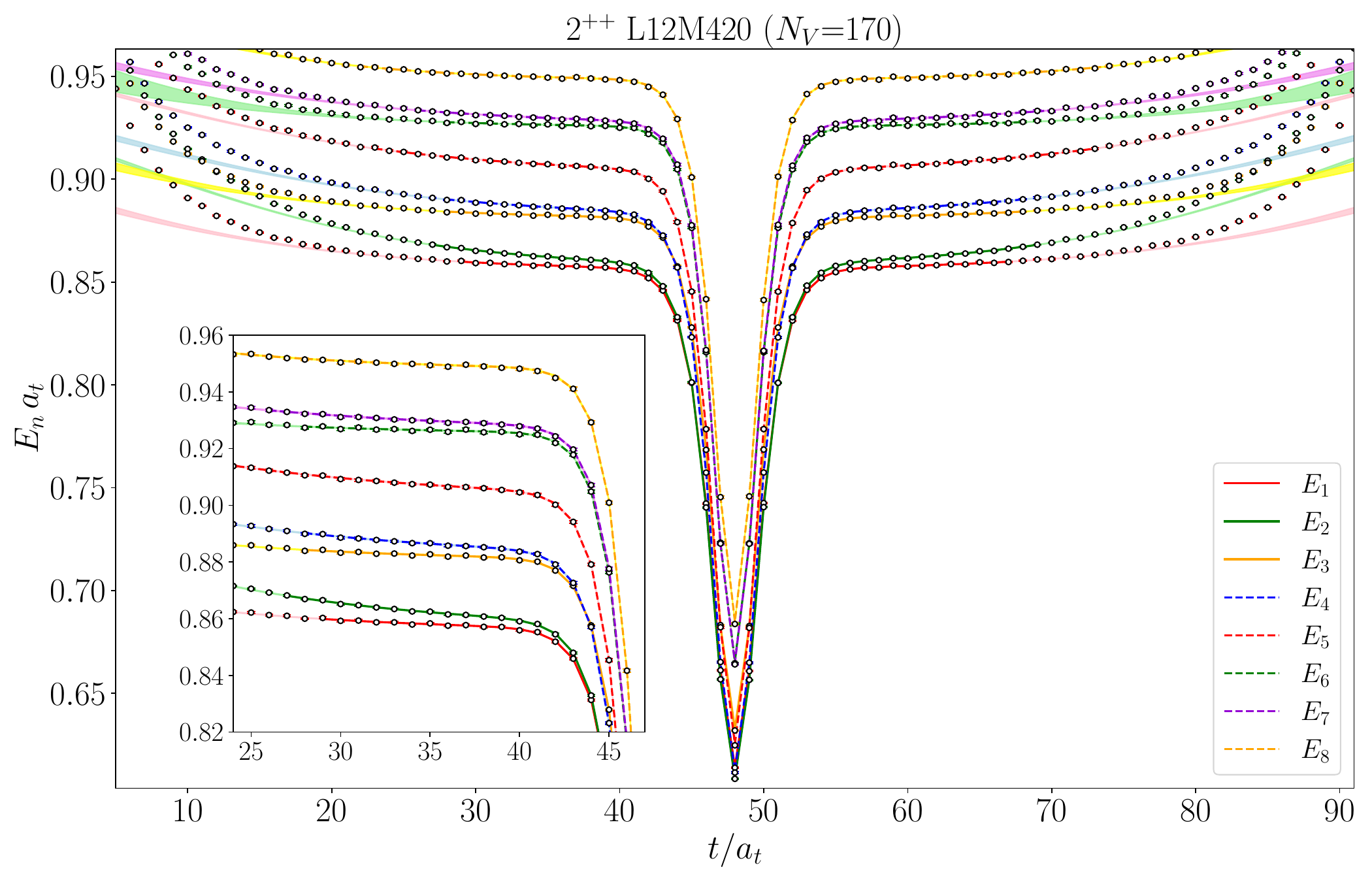}
	\includegraphics[width=0.49\linewidth]{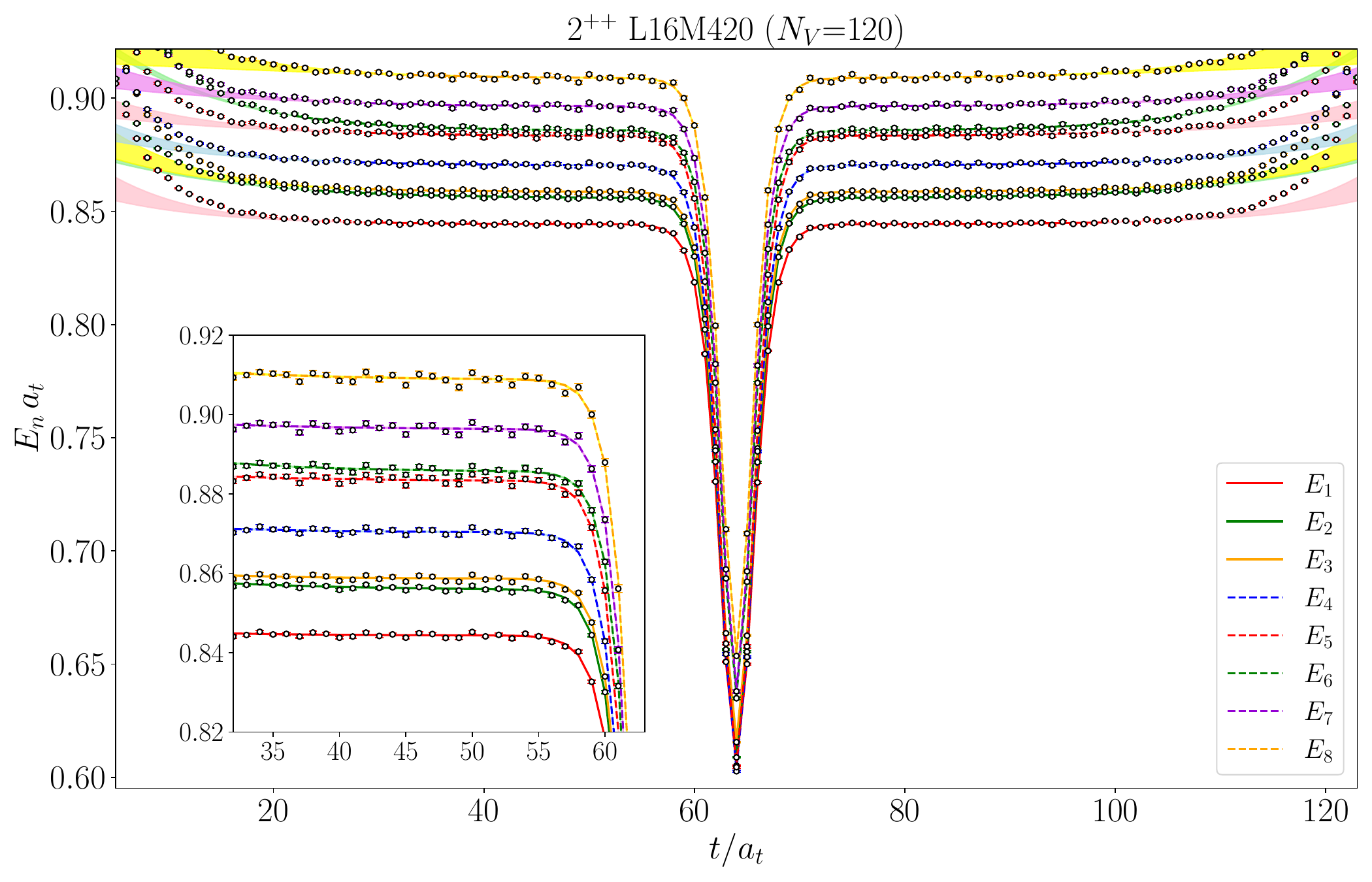}\\
	\includegraphics[width=0.49\linewidth]{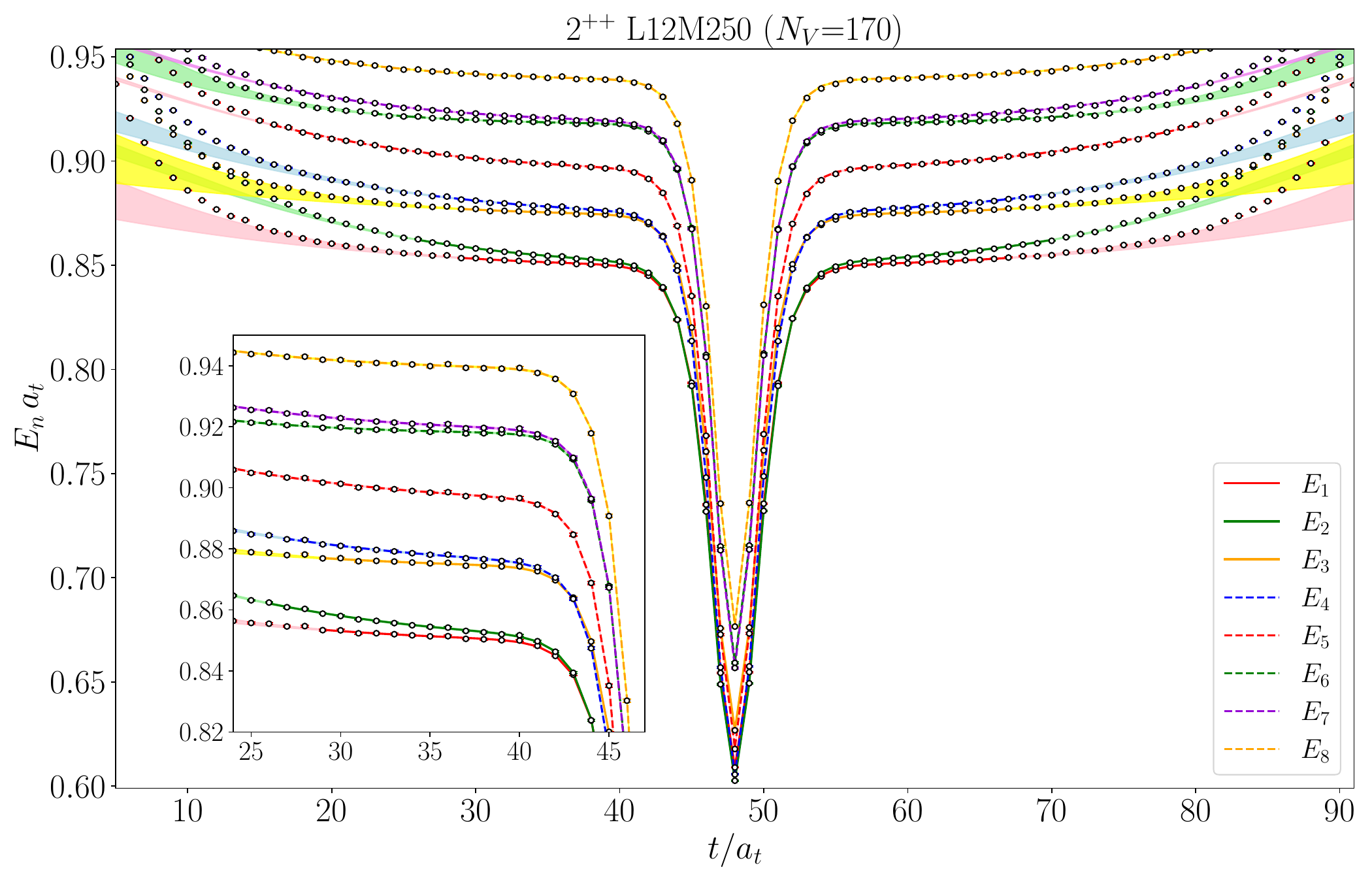}
	\includegraphics[width=0.49\linewidth]{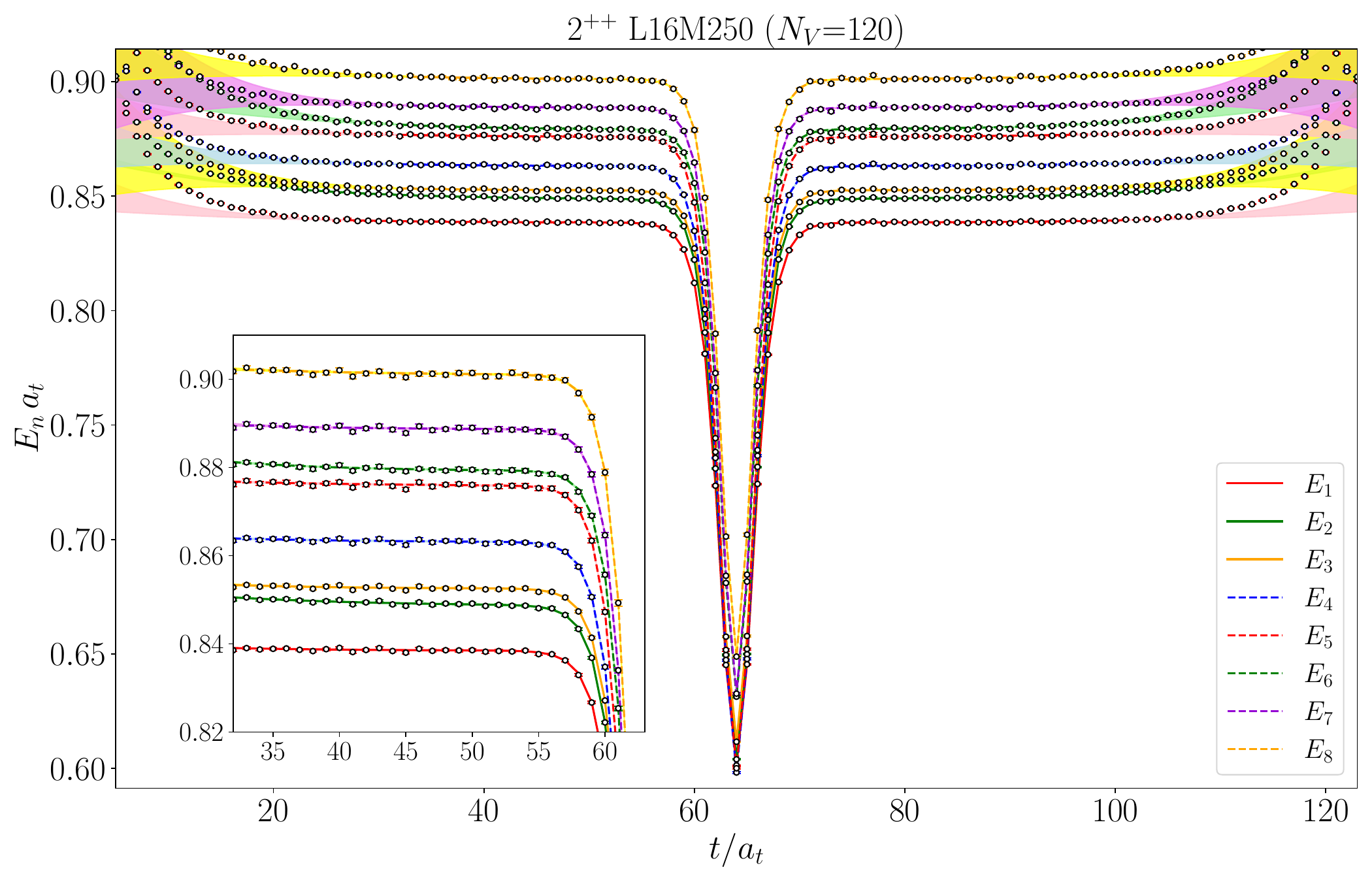}\\
	\caption{The effective energy levels $E_n$ of the $0^{++}$ and $2^{++}$ channels obtained from the `3-state' fitting after variation with the complete operator sets of $\mathcal S_0$ and $\mathcal S_2$ in Eq.~\eqref{eq:op-set-new}, respectively.}
	\label{fig:fit-energy-levels-3}
\end{figure*}

Figure~\ref{fig:fit-energy-levels-3} shows the plots of the effective energy functions of $\{C_{nn}(t),n=1,2,\ldots, 10\}$ in the ${}^1S_0$ channel and $\{C_{nn}(t),n=1,2,\ldots, 8\}$ in the ${}^5S_2$ channel, respectively, which is defined by 
\begin{equation}
	E_n^{(\mathrm{eff})}(t) ~ a_t=\cosh^{-1}\left[\frac{C_{nn}(t-a_t)+C_{nn}(t+a_t)}{2C_{nn}(t)}\right].
\end{equation}
The upper-left, upper-right, lower-left, lower-right panels of the $0^{++}$ or $2^{++}$ system are for the ensembles L12M420 ($N_V=170)$), L16M420 ($N_V=120$), L12M250 ($N_V=170$) and L16M250 ($N_V=120$), respectively. 
The data points correspond to the lattice simulation results, and the colored bands illustrate the fits based on Eq.~\eqref{eq:3-state-fit}. 
The dips around $t=T/2$ due to the thermal states manifest the importance of their contribution.
It is seen that Eq.~\eqref{eq:3-state-fit} describes all the correlation functions $C_{nn}(t)$ very well. 
The inset in each panel shows a magnified view of the $t$-dependence of $E_n^{(\mathrm{eff})}$, highlighting the robustness of the fits.
%%%%%%%%%%%%%%%%%%%%%%%%%%%%%%%%%%%%%%%%
%%%%%%%%%%%%%%%%%%%%%%%%%%%

%%%%%%%%%%%%%%%%%%%%%%%%%%%
\begin{figure*}[htbp]
	\centering
	\includegraphics[width=0.45\linewidth]{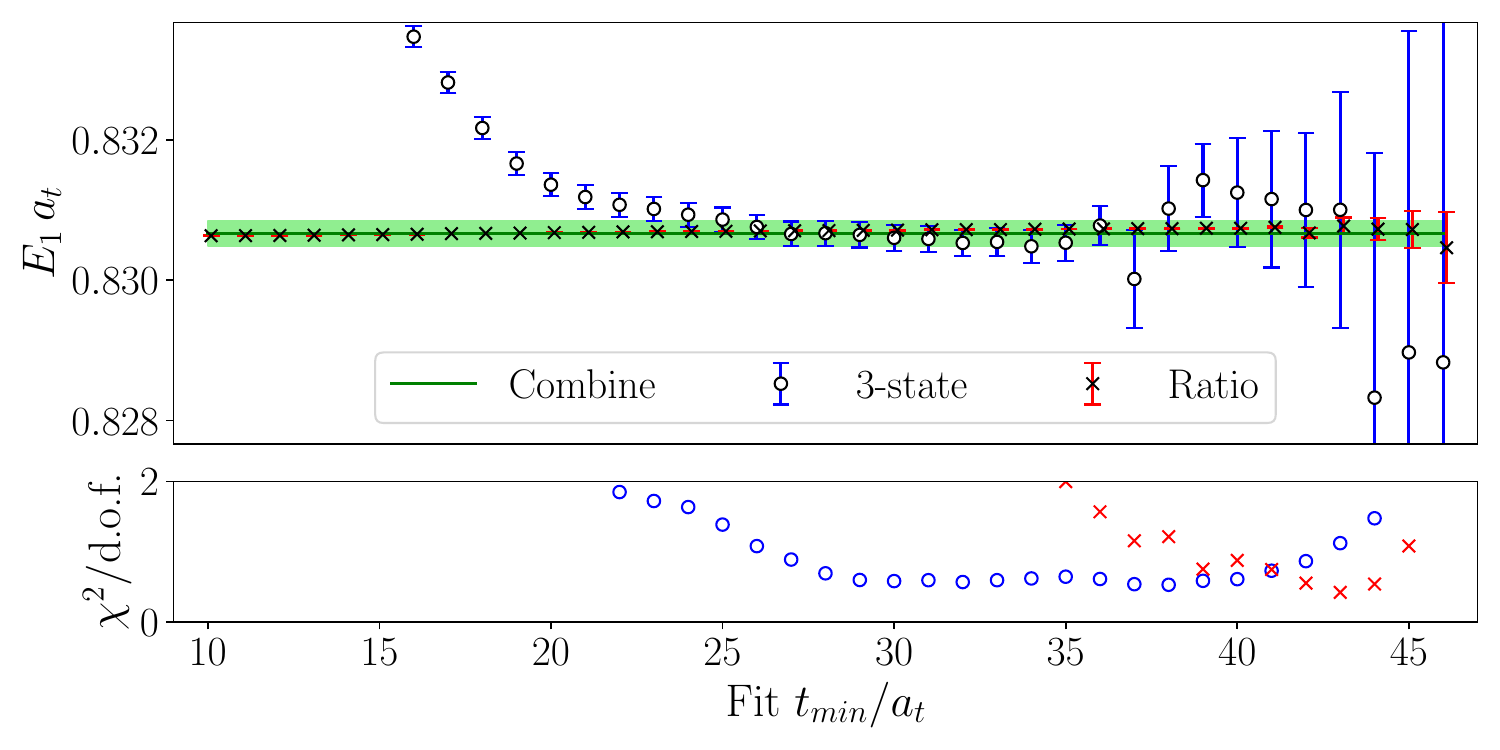}
	\includegraphics[width=0.45\linewidth]{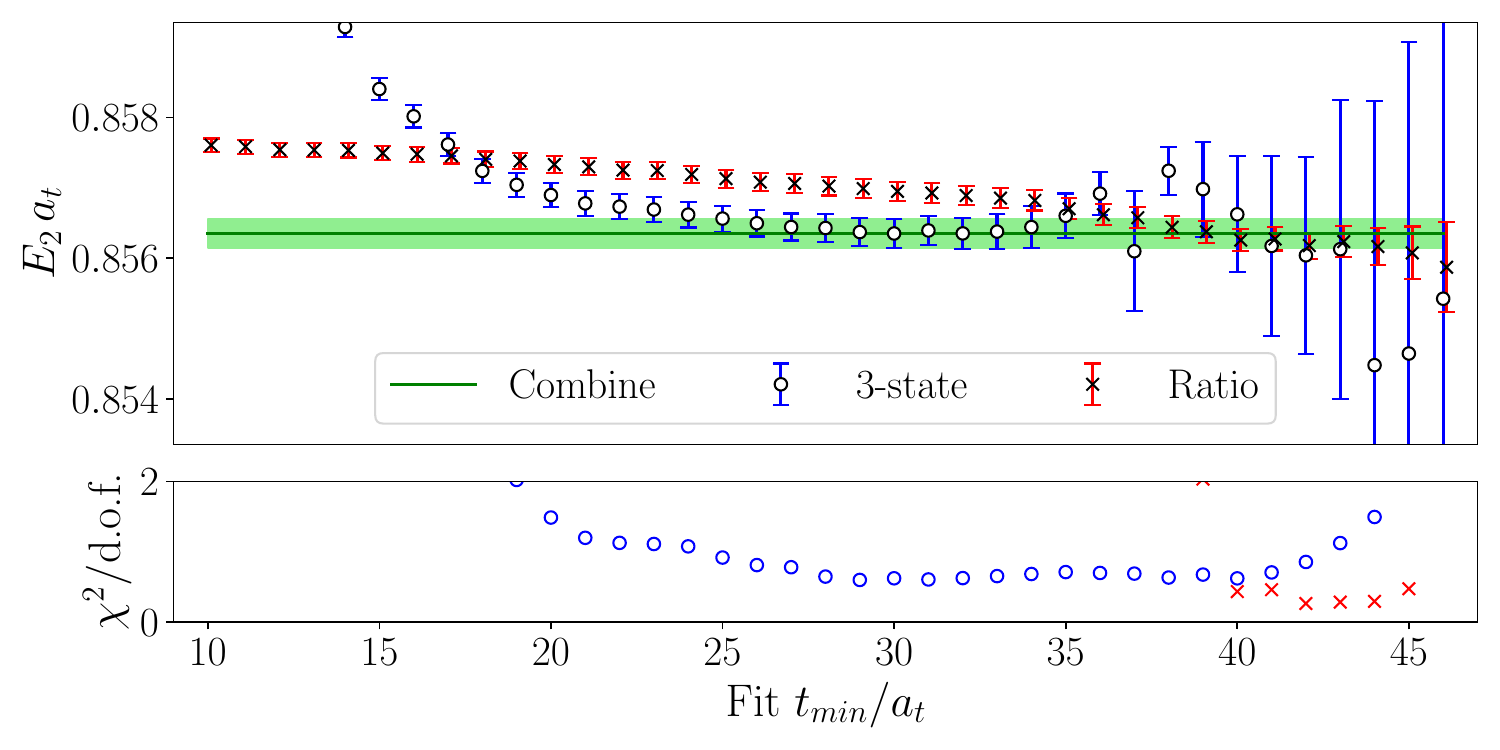}\\
	\includegraphics[width=0.45\linewidth]{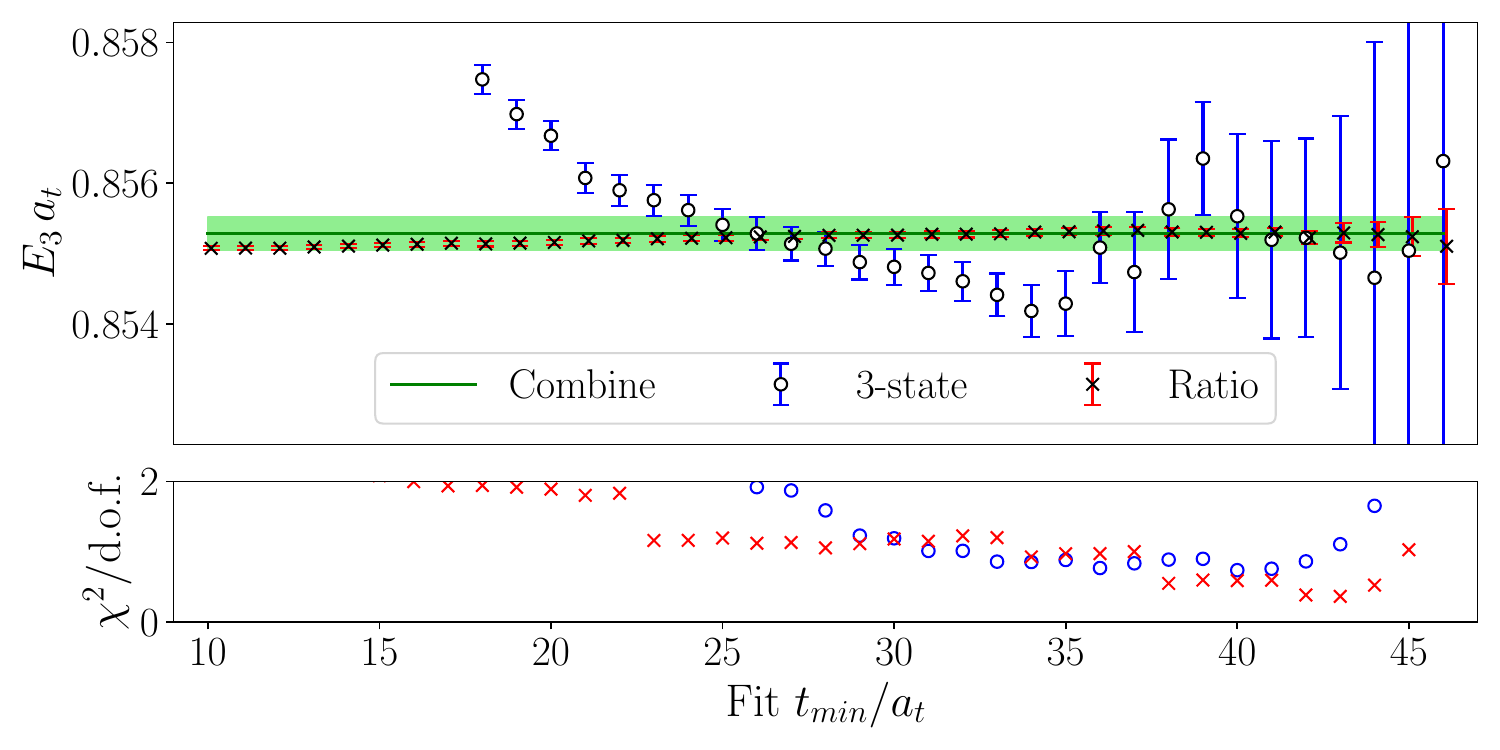}
	\includegraphics[width=0.45\linewidth]{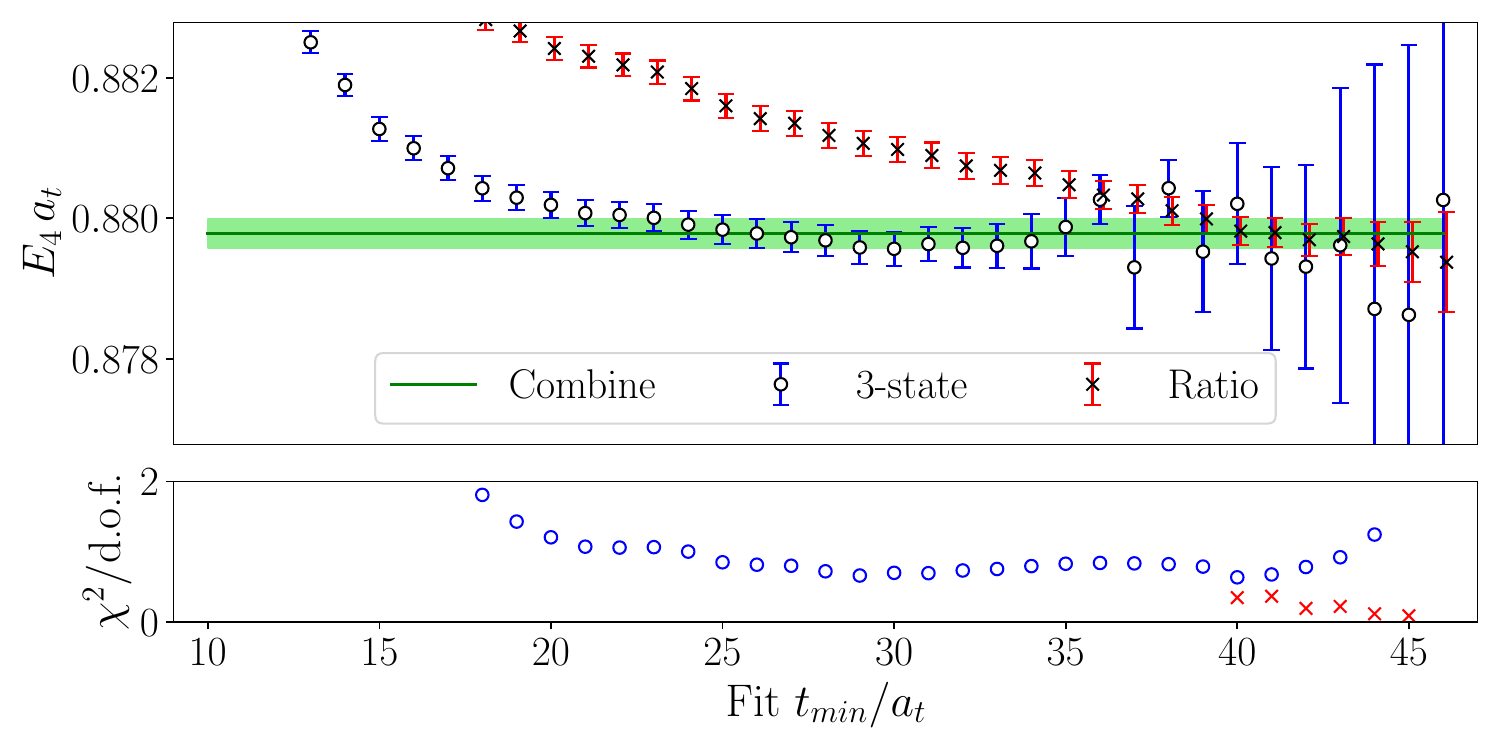}\\
	\includegraphics[width=0.45\linewidth]{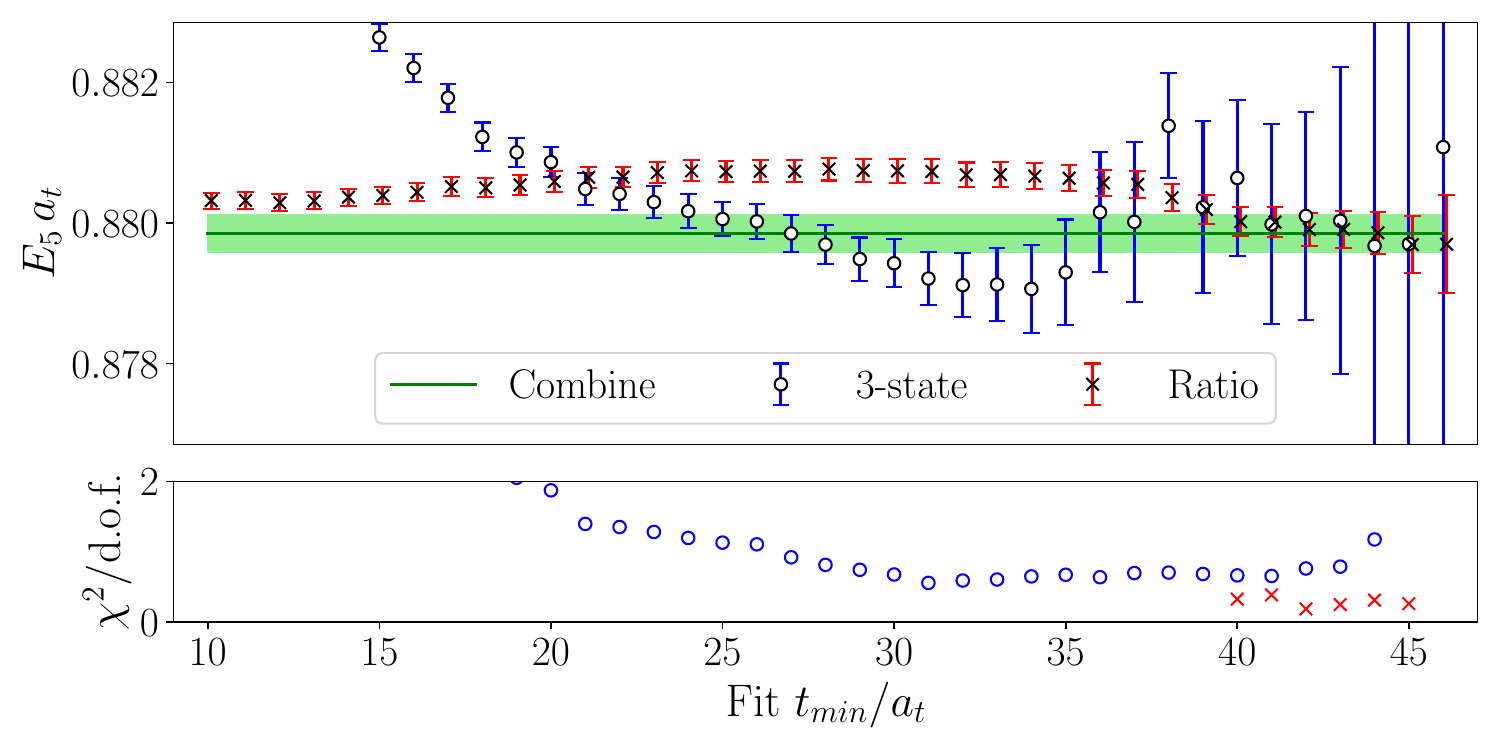}
	\includegraphics[width=0.45\linewidth]{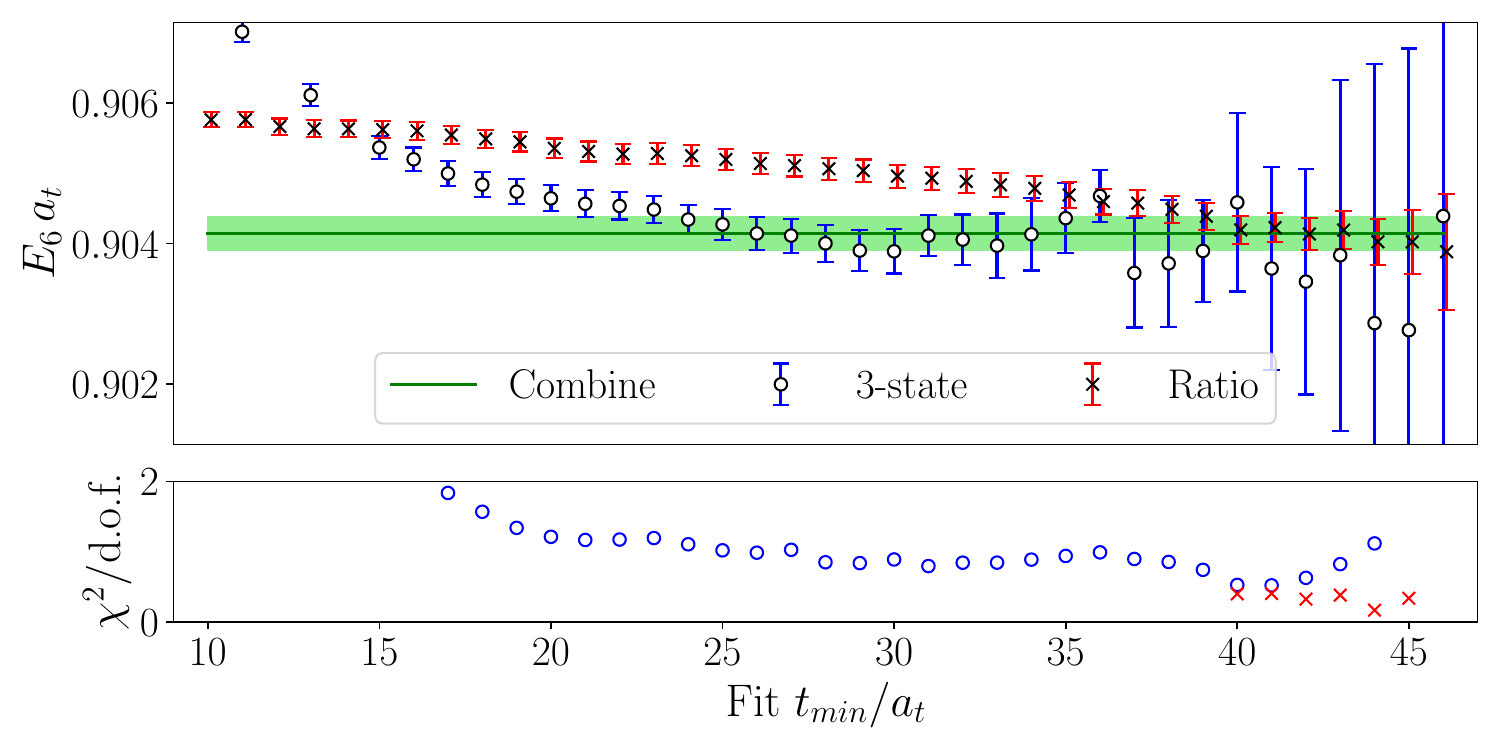}\\
	\includegraphics[width=0.45\linewidth]{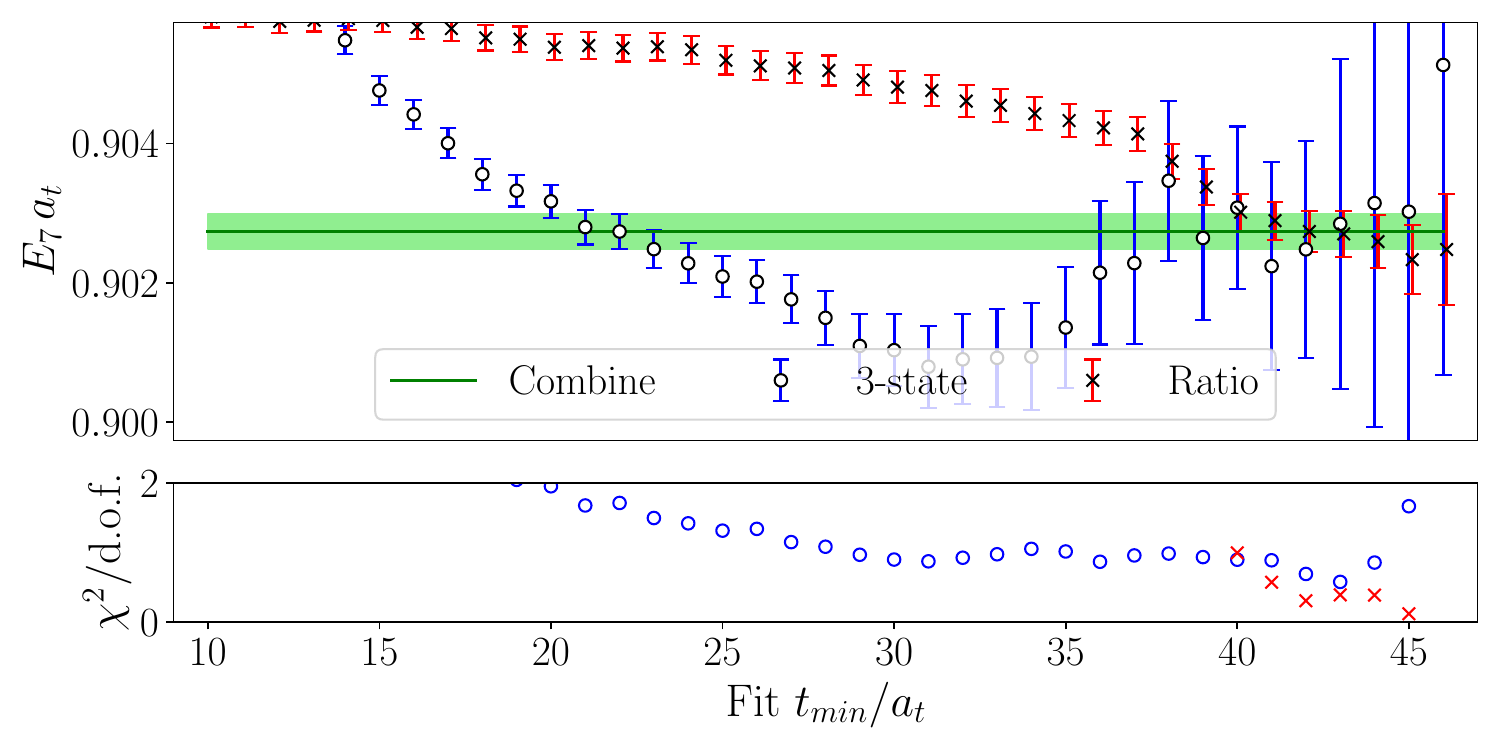}
	\includegraphics[width=0.45\linewidth]{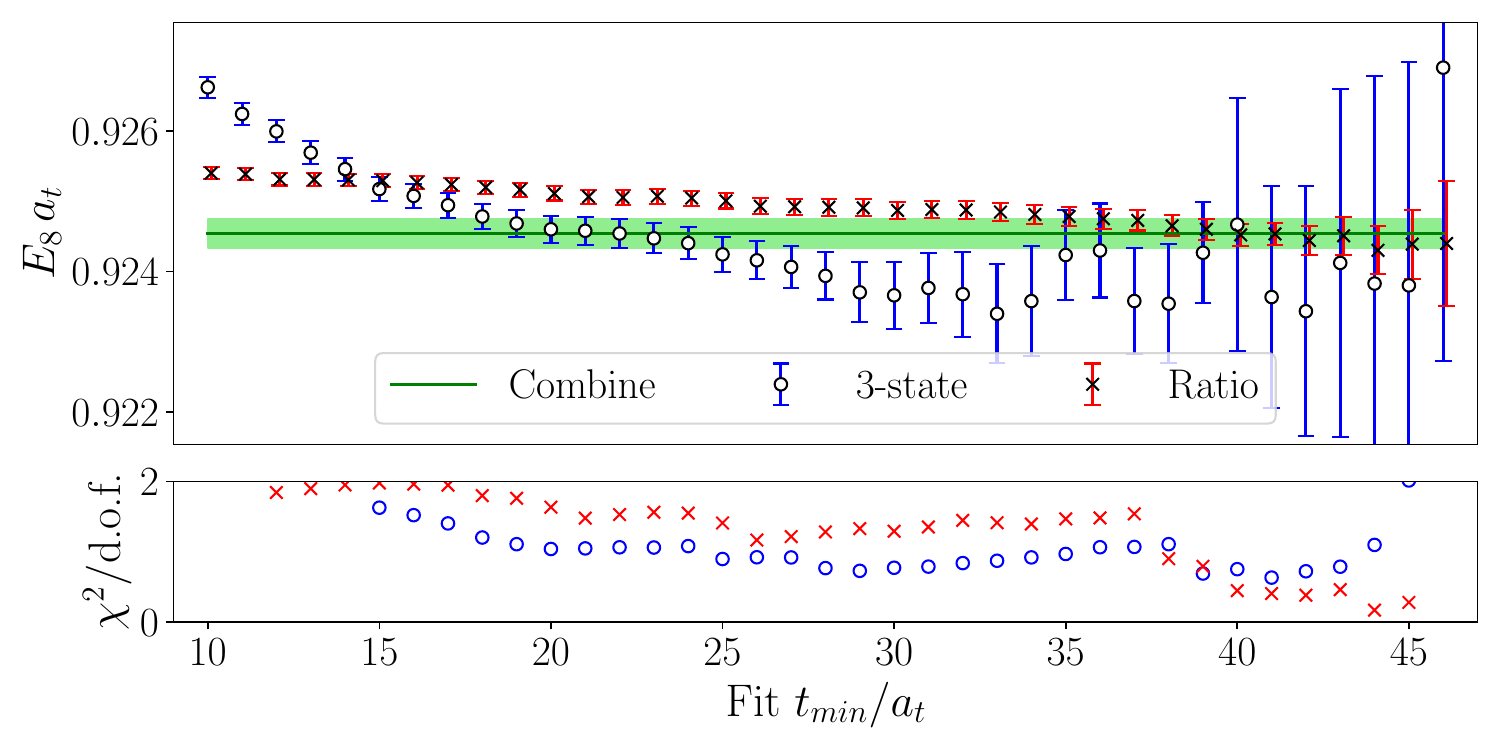}\\
	\includegraphics[width=0.45\linewidth]{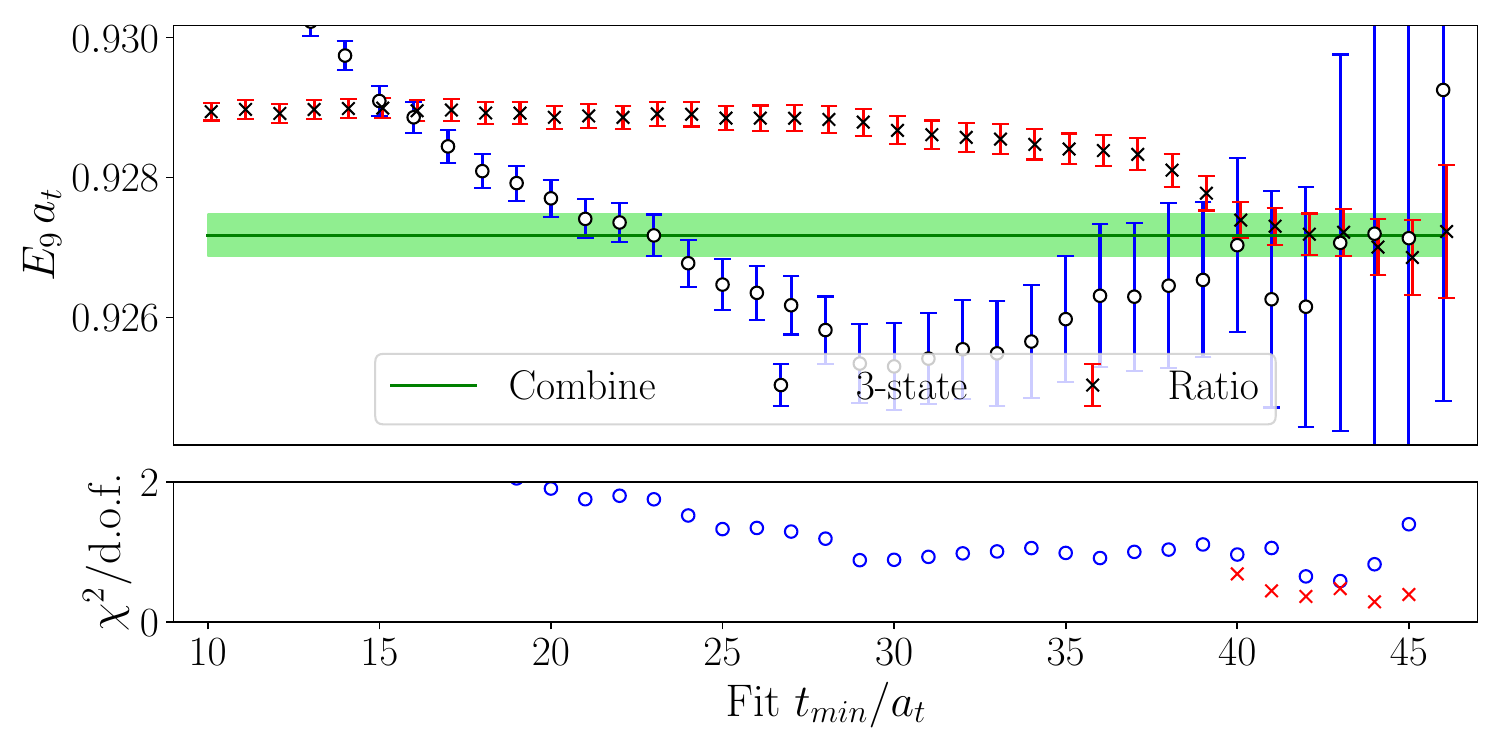}
	\includegraphics[width=0.45\linewidth]{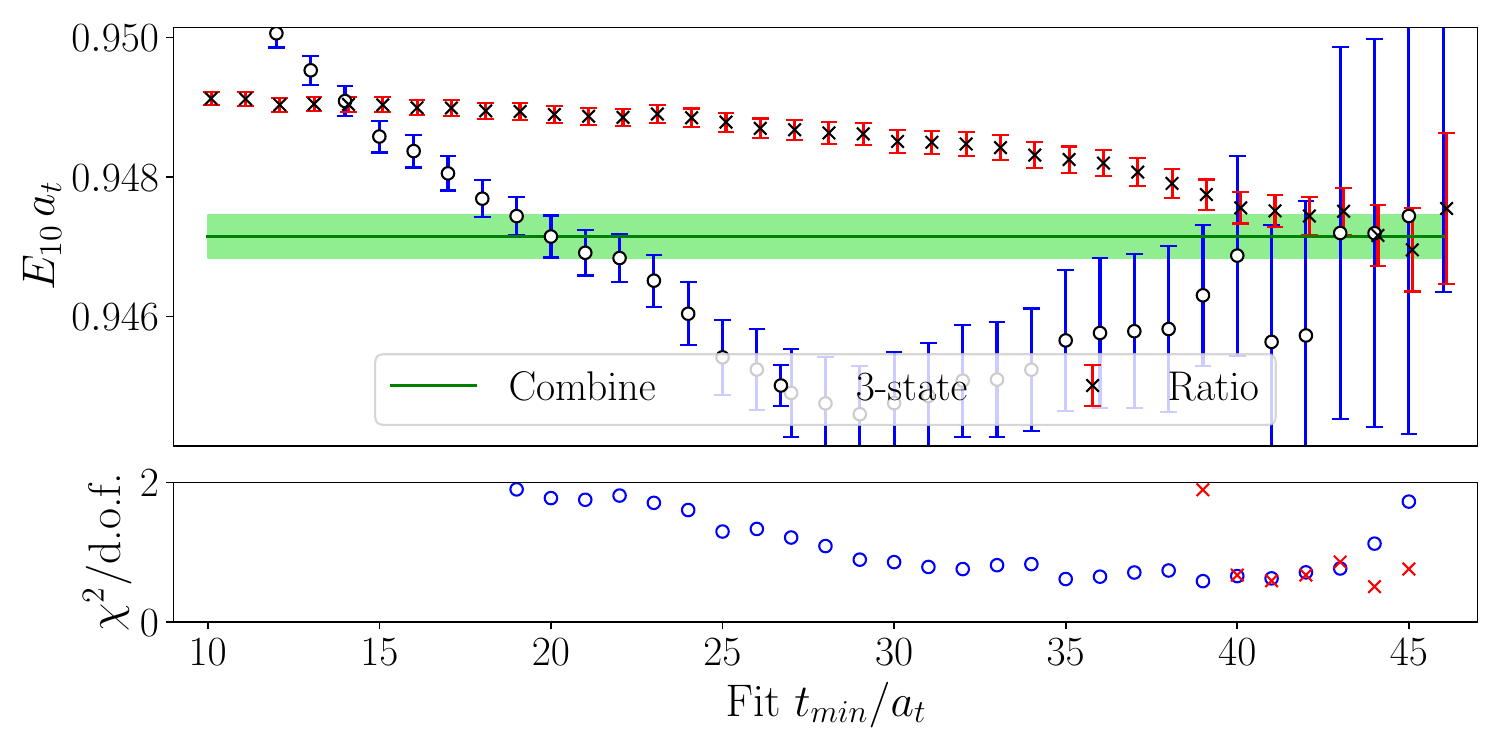}
	\caption{Stability of the two fitting methods on the L12M420 ensemble with $N_V=170$ is demonstrated by varying the minimum fitting time $t_{\rm min}$ in the $0^{++}$ system. 
		The blue dotted and red crossed points represent the results obtained using the `3-state' and `Ratio' fitting method, respectively. 
		The green bands represent our final determination of the energy levels, where both methods yield consistent results within a one-sigma confidence interval and are used in all subsequent analyses.
		The two fits for the higher-energy levels show discrepancies, and these levels are consequently excluded from the subsequent analysis.
		The lower block of each panel displays the corresponding $\chi^2/\mathrm{d.o.f.}$ values associated with each fit. The corresponding results for the other ensembles are shown in Figs.~\ref{fig:L16-fit-tmin}--\ref{fig:L16-2pp-pi-fit-tmin} of Appendix~\ref{appendix:B}.}
	\label{fig:L12-fit-tmin}
\end{figure*}
%%%%%%%%%%%%%%%%%%%%%%%%%%%

%%%%%%%%%%%%%%%%%%%%%%%%%%%
\begin{figure*}[htbp]
	\centering
	\includegraphics[width=0.45\linewidth]{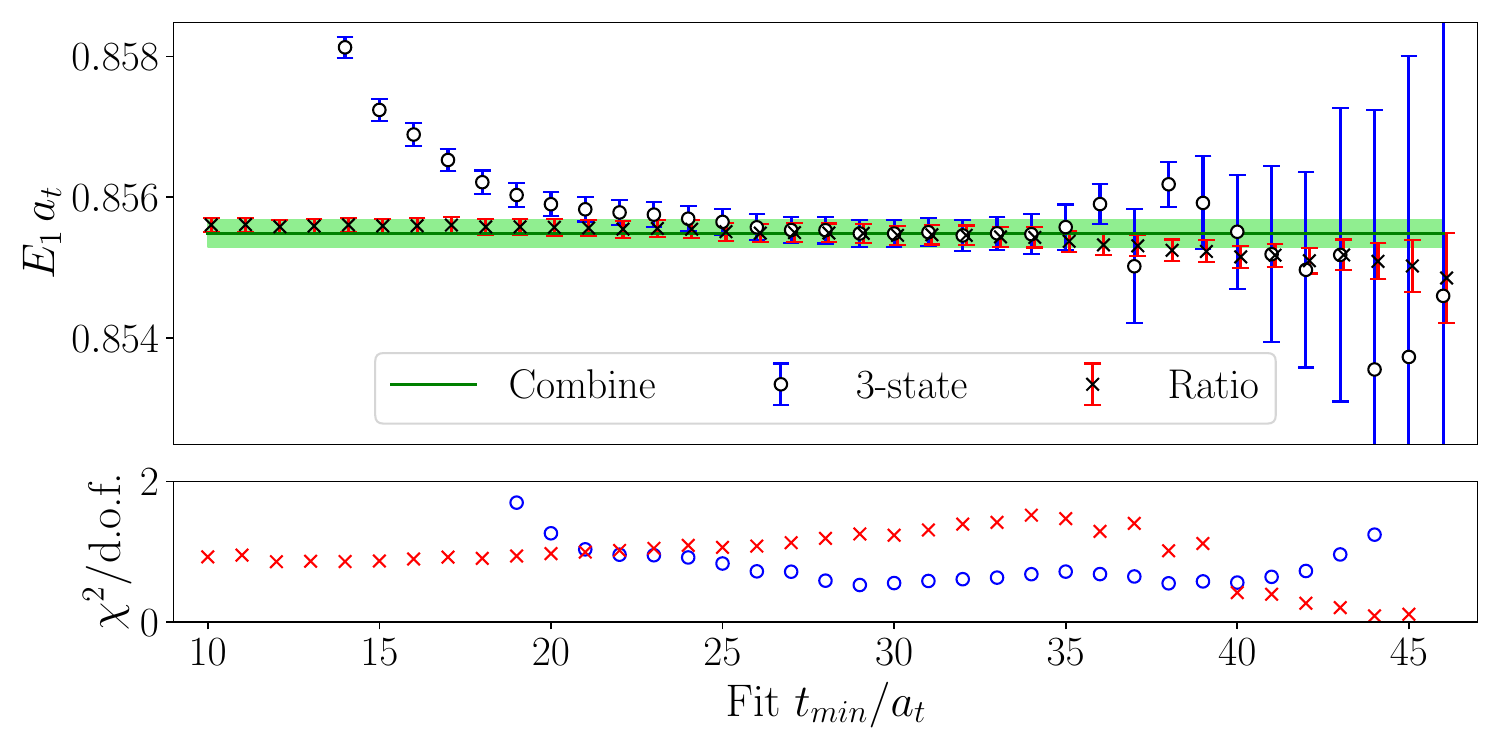}
	\includegraphics[width=0.45\linewidth]{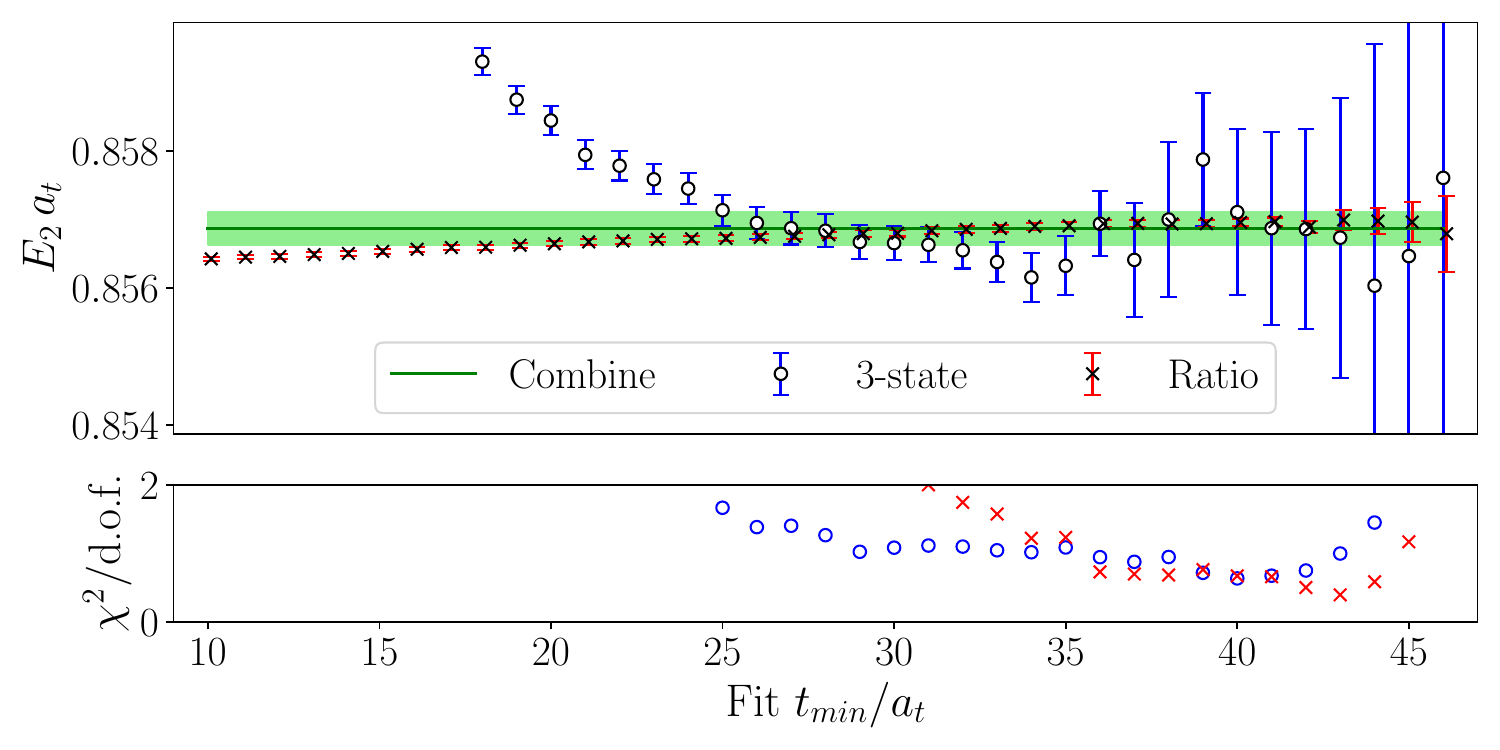}\\
	\includegraphics[width=0.45\linewidth]{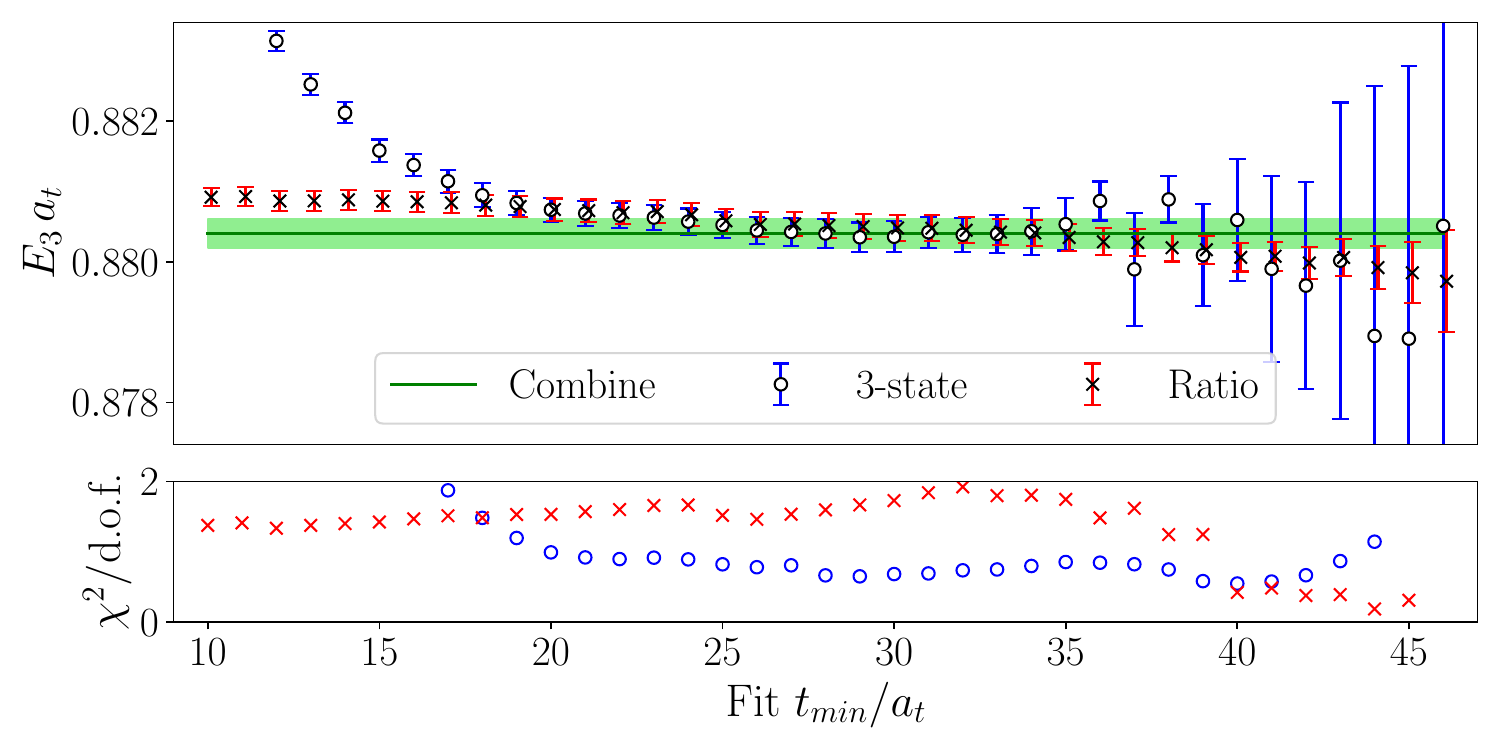}
	\includegraphics[width=0.45\linewidth]{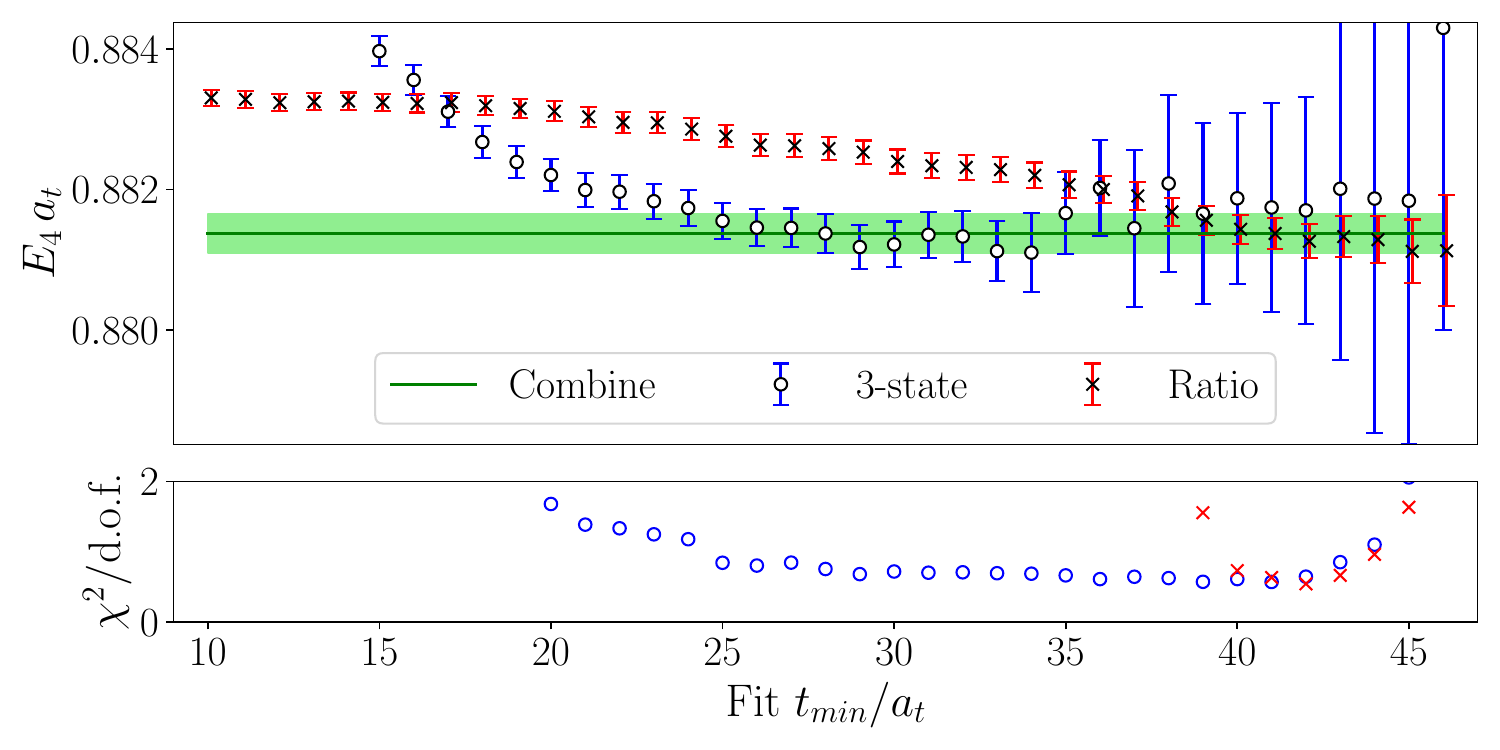}\\
	\includegraphics[width=0.45\linewidth]{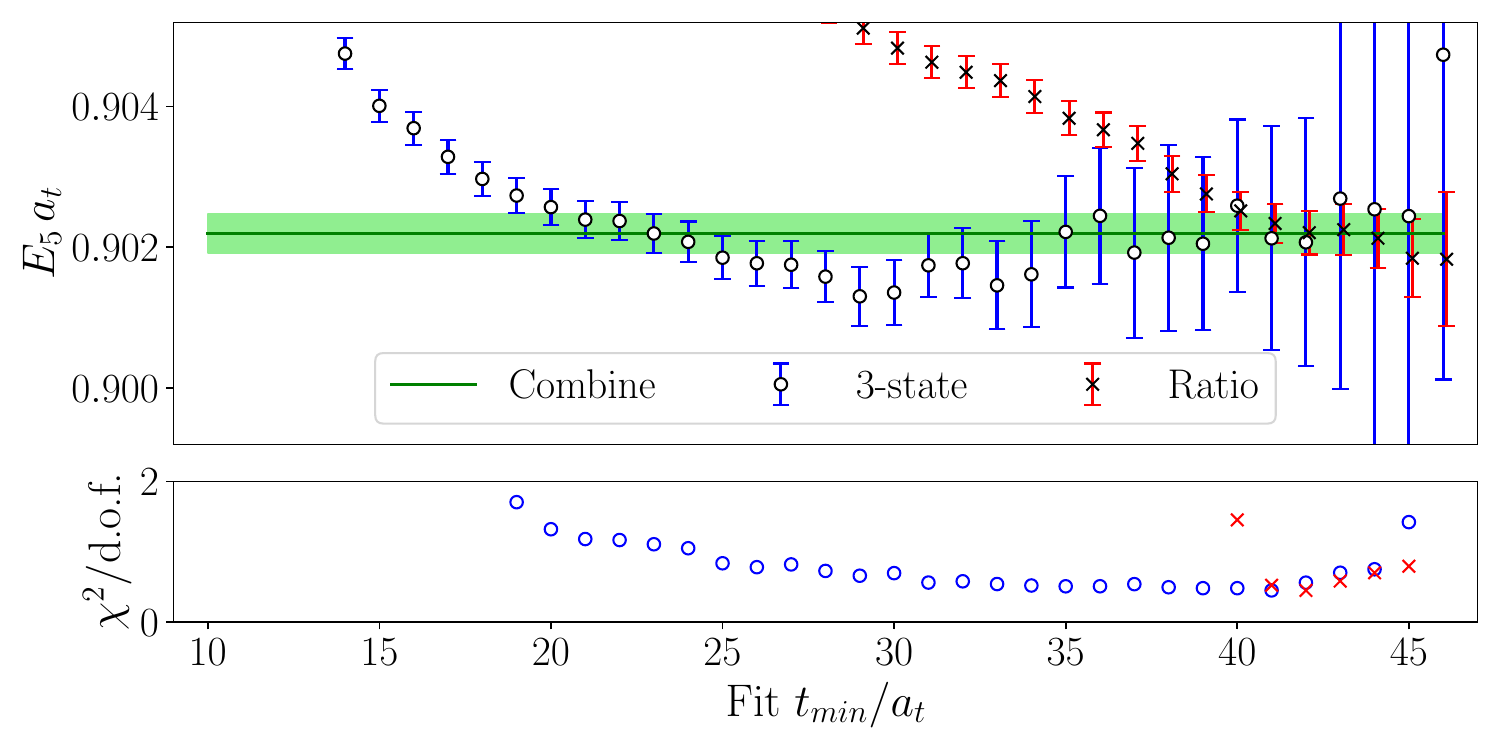}
	\includegraphics[width=0.45\linewidth]{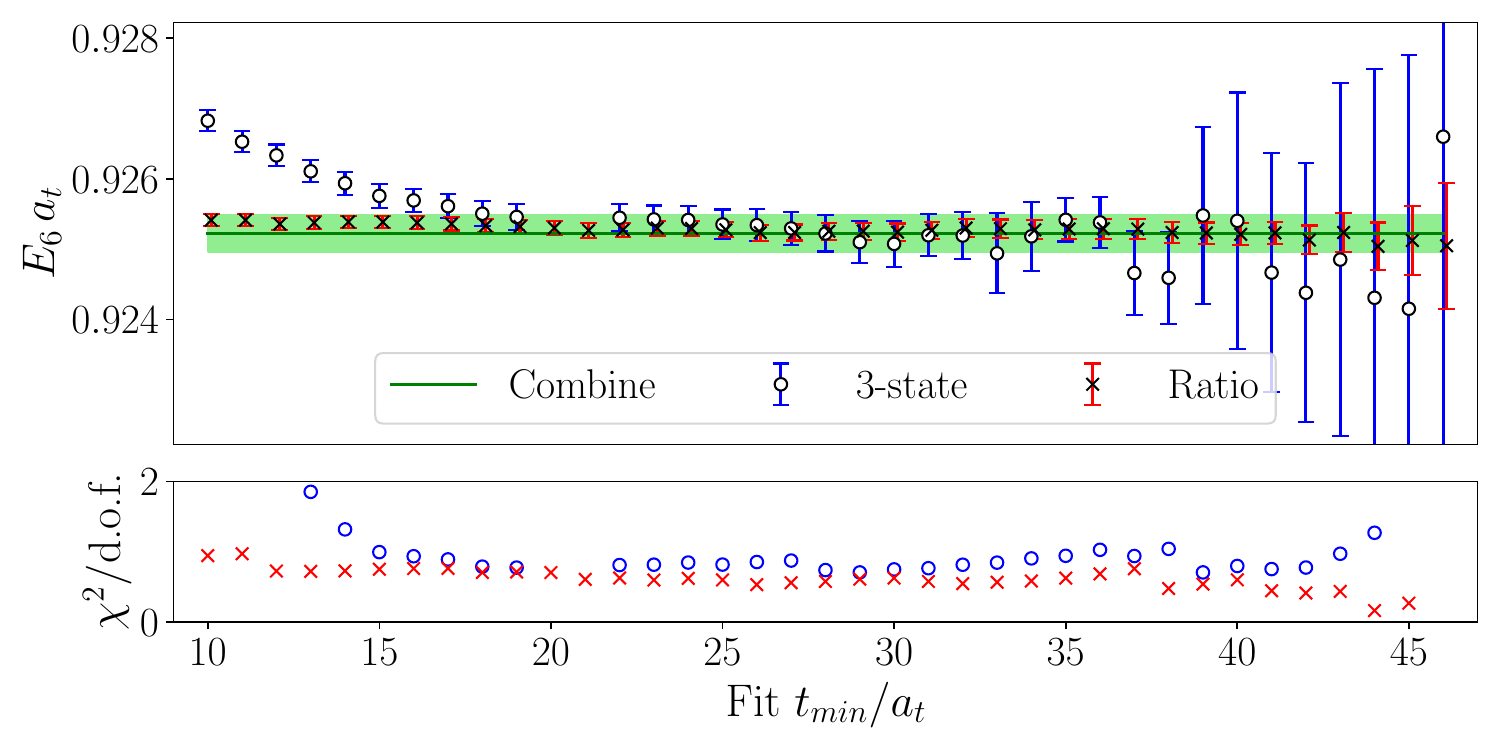}\\
	\includegraphics[width=0.45\linewidth]{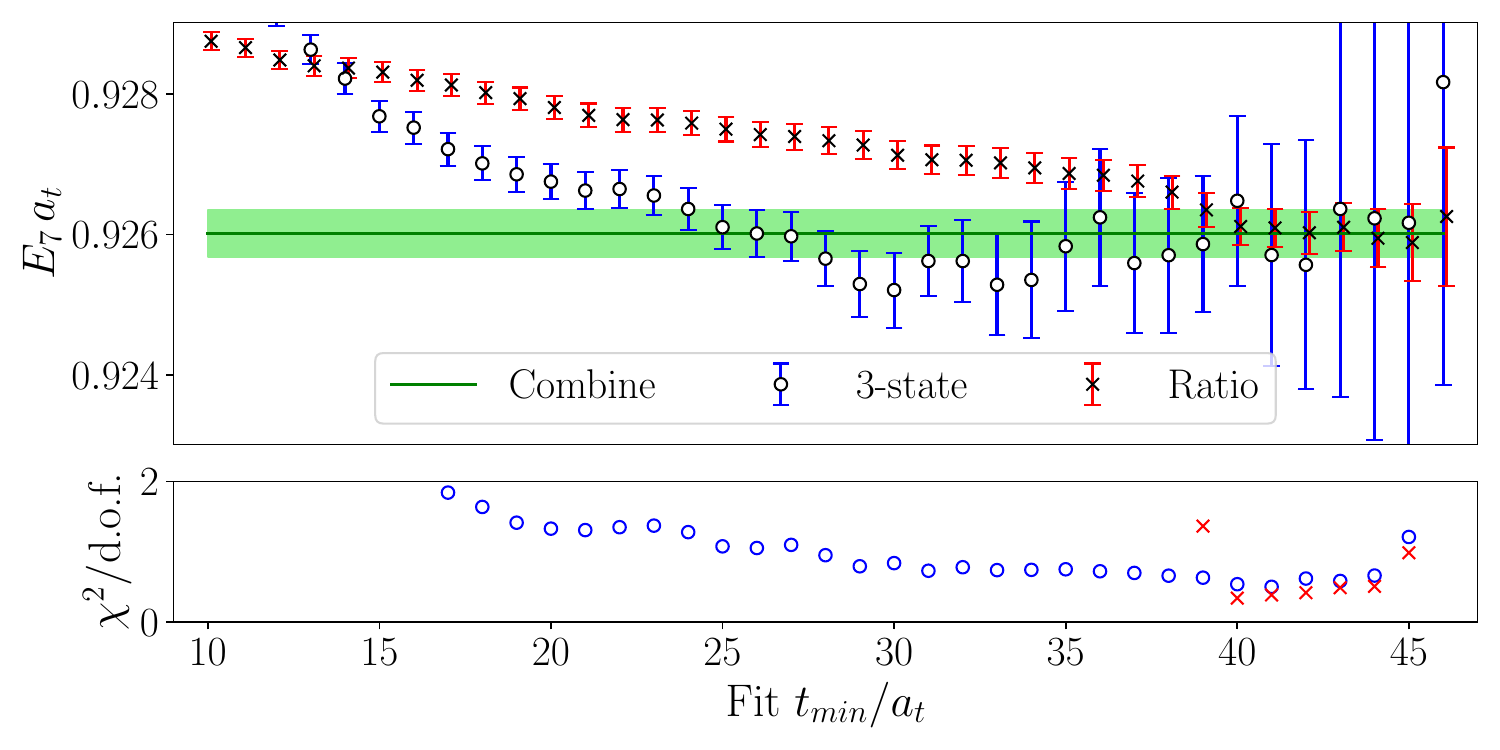}
	\includegraphics[width=0.45\linewidth]{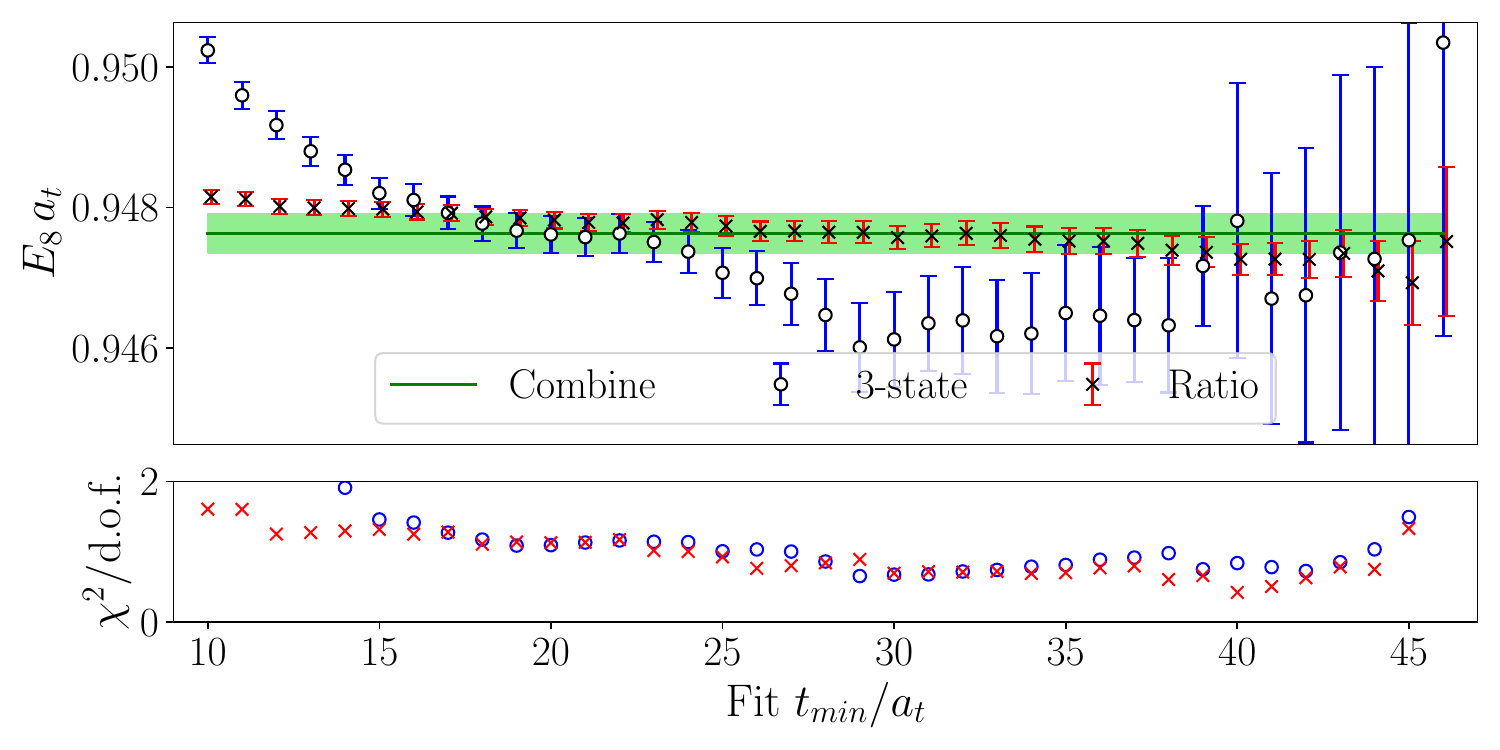}
	\caption{The stability of the two fitting methods on the L12M420 ensemble with $N_V=170$ is demonstrated by varying the minimum fitting time $t_{\rm min}$ in the $2^{++}$ system. 
	}
	\label{fig:L12-2pp-fit-tmin}
\end{figure*}
%%%%%%%%%%%%%%%%%%%%%%%%%%%

In the practical procedure of the data analysis, we set the fit time window to be $t\in [t_\mathrm{min},T/2]$ for each $C_{nn}(t)$ and vary the lower bound $t_\mathrm{min}$ to check the stability of the fit. 
The ten (eight) panels in Fig.~\ref{fig:L12-fit-tmin}~(\ref{fig:L12-2pp-fit-tmin}) present the fitting details of the $0^{++}$~($2^{++}$) energy levels $\{E_n, n=1,2,\ldots, 10~(8)\}$ on L12M420, respectively. 
The blue point in the upper block of each panel is the fitted $E_n$ at a specific $t_\mathrm{min}$ and the blue dot in the lower panel gives the $\chi^2/\mathrm{d.o.f}$ of the fit. 
It is seen that, when $t_\mathrm{min}/a_t>20$, although the fitted values of $E_n$ are consistent with each other within errors for different $t_\mathrm{min}$, the central values vary by roughly 0.1--0.2\%. 
The comparable $\chi^2/\mathrm{d.o.f}$ at these $t_\mathrm{min}$ cannot give us a criterion to choose the definite final value of $E_n$. 
Note that this 0.1--0.2\% variation of the central value will affect the derivation of the scattering properties. 

%%%%%%%%%%%%%%%%%%%%%%%%%%%
\begin{figure*}[t]
	\centering
	\includegraphics[width=0.49\linewidth]{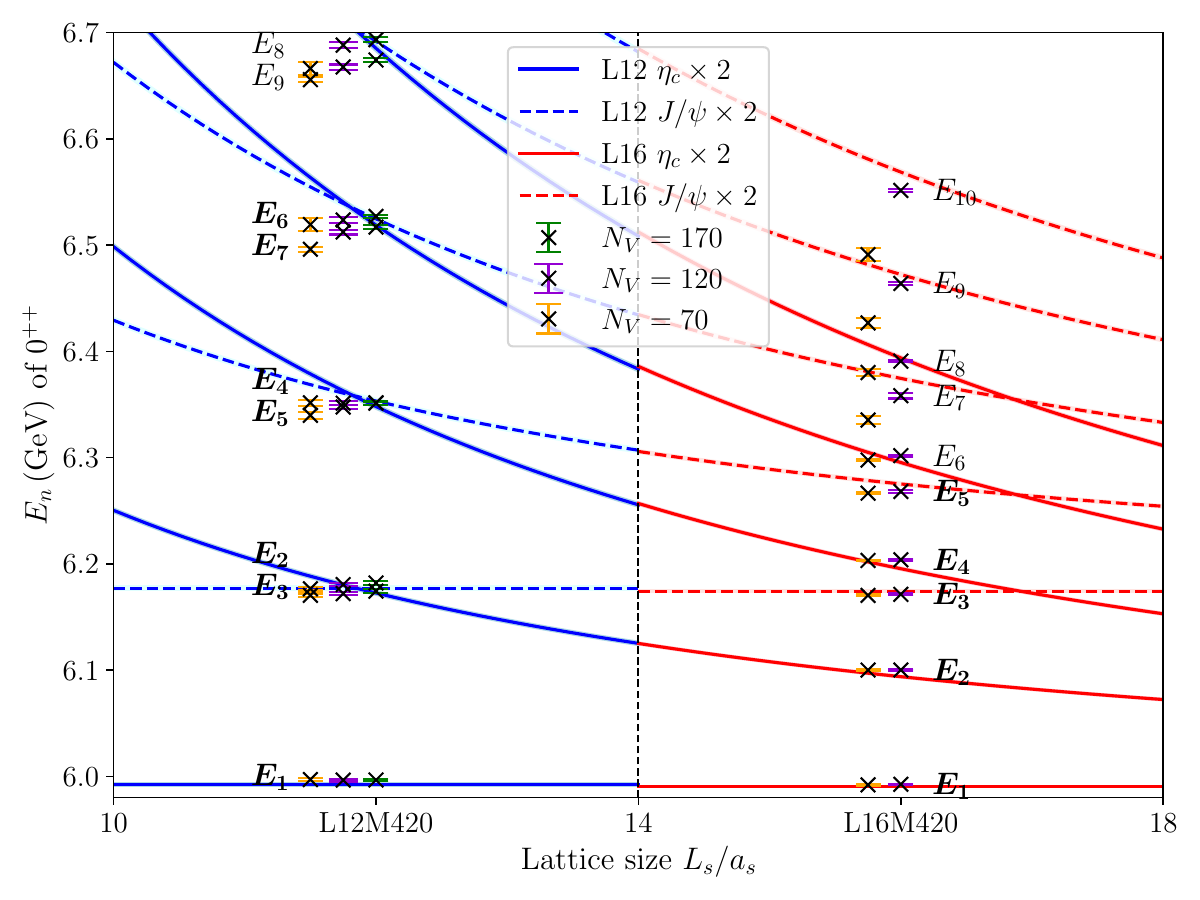}
	\includegraphics[width=0.49\linewidth]{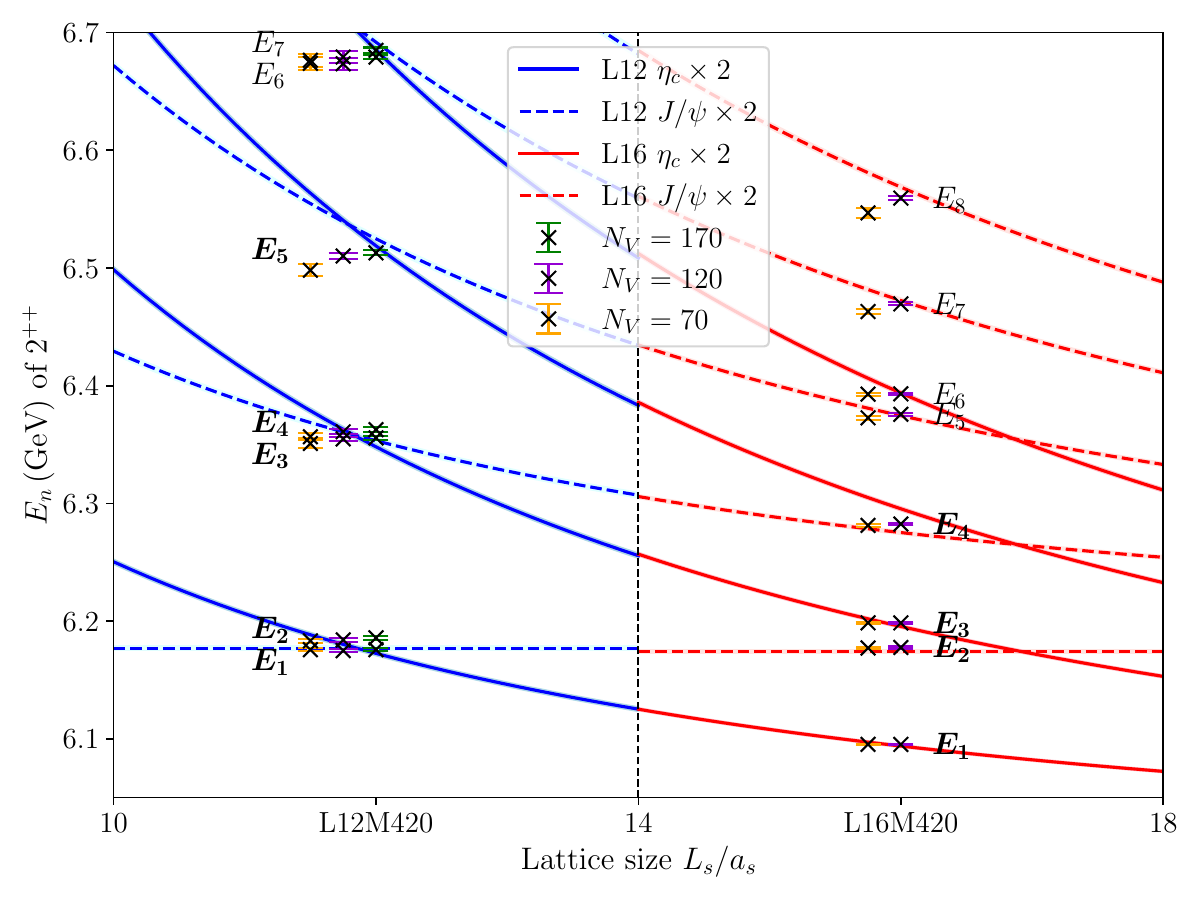}
	\caption{Energy-level determination of the $0^{++}$ (left) and $2^{++}$ (right) system with varying $N_V$ values on the L12M420 (blue) and L16M420 (red) ensembles. 
		Colored data points represent the extracted lattice energy levels, with results from different $N_V$ values slightly offset horizontally for visual clarity.
		Solid and dashed lines represent the noninteracting energy levels of di-$ \eta_c$ and di-$J/\psi$ channels, respectively. 
		Energy levels $E_n$ used for the scattering analysis are highlighted in bold.}
	\label{fig:spectrum}
\end{figure*}
%%%%%%%%%%%%%%%%%%%%%%%%%%%

An alternative way to extract the energy levels of a two-particle system is the so-called `ratio' method proposed by Refs.~\cite{Umeda:2007hy, Feng:2009ij}. 
In order to eliminate the constant term from the thermal states, one can introduce the ratio function,
\begin{equation}
	R_n(t+a_t)=\frac{C_{nn}(t)-C_{nn}(t+2a_t)}{C_M^2(k(n),t)-C_{M}^2(k(n),t+2a_t)},
\end{equation}
where $C_M(k(n),t)$ is the correlation function of the constituent particle $M$ with a momentum $k(n)=|\vec{k}(n)|={2\pi n}/{L_s}$. 
In the large-$t$ region where the higher-state contamination exists [the third term in Eq.~\eqref{eq:3-state-fit} can be neglected], the ratio function $R(t)$ can be approximated by 
\begin{equation}\label{eq:ratio}
	R_n(t+a_t)\approx A_R\frac{\sinh{(E_n t')}}{\sinh{(2E_M(n) t')}},
\end{equation}
where and $t'\equiv t+a_t-T/2$ has been defined and $E_M(n)=\sqrt{m_M^2+k^2(n)}$ is the energy of the $M$ particle with the momentum $k(n)$.
Then a joint fit to $R(t+a_t)$ and $C_M(k(n),t)$ in the time window $t\in[t_{\rm min},~T/2]$ will give an estimate of $E_n$. 
The fit details are also shown in Figs.~\ref{fig:L12-fit-tmin} and \ref{fig:L12-2pp-fit-tmin}; in each panel of the figure, the red point in the upper block shows the fitted value of $E_n$ 
using Eq.~\eqref{eq:ratio} at different $t_\mathrm{min}$ and the red cross in the lower block presents the $\chi^2/\mathrm{d.o.f}$ of the fit. 
Obviously, although smaller errors (owing to the ratios) of $E_n$, the $t_\mathrm{min}$-dependence of $E_n$ is more pronounced for higher $E_n$. 
Furthermore, the values of $\chi^2/\mathrm{d.o.f}$ becomes smaller than one only when $t_\mathrm{min}/a_t>40$ where the values of $E_n$
show more or less a plateau that merges with the values of $E_n$ obtained using Eq.~\eqref{eq:3-state-fit}. 
We attribute the reason for this phenomenon to the contamination from higher states [especially for $C_M(k(n),t)$], which renders Eq.~\eqref{eq:ratio} not so precisely as expected. 
Anyway, the plateau region of $E_n$ from Eq.~\eqref{eq:ratio} provide us a guide and a self-consistent check for the final determination of the values of $E_n$. 
Using the `ratio' method as a benchmark, we identify the optimal fitting range for the `3-state' method, whose final results—shown as green bands—capture the combined uncertainties of $E_n$ through their widths. 
Note that the `ratio' method works quite well for the studies on $\pi\pi$ scattering because the first excited state of $\pi$ is much higher in mass than $\pi$ (in experiments, the $\pi(1300)$ is taken as the first excited state of $\pi$). 
The same fitting procedure is performed for other ensembles in Appendix~\ref{appendix:B}, which show a similar pattern.

%%%%%%%%%%%%%%%%%%%%%%%%%%%%%%%%%%%%%%%%%%%%%%%%%%%%%%%%%%%%%%%%%%%%%%%%%%%%%%%%%%%%%%
\subsection{Check on the $N_V$ dependence of energy levels}\label{secII:NV-dep}
%%%%%%%%%%%%%%%%%%%%%%%%%%%%%%%%%%%%%%%%%%%%%%%%%%%%%%%%%%%%%%%%%%%%%%%%%%%%%%%%%%%%%%

In Sec.~\ref{secII:NV}, we have discussed that the LHS smearing results in the suppression of the couplings of operators to higher momentum states. 
Regarding this fact we consider the two-particle operators $\mathcal{O}_\alpha$ of relative momenta up to $n^2=4$. 
However, it is still intriguing if the obtained energy levels suffer from the truncated Laplacian Heaviside subspace defined by $N_V$. 
In doing so, we repeat the procedure of the energy determination discussed above on L12M420 ensemble with $N_V=70,120,N170$ and on L16M420 with $N_V=70,120$ (larger values of $N_V$ on this ensemble are nonaffordable for us with our available computing resources). 

We conduct this validation in the $0^{++}$ channel and anticipate analogous behavior in the $2^{++}$ channel.
The derived energy levels on the two ensembles at different values of $N_V$ are shown as data points in Fig.~\ref{fig:spectrum}, where curves indicate the noninteracting energies $E^{(0)}_n$ of $\eta_c\eta_c$ (solid lines) and $J/\psi J/\psi$ (dashed lines) and are plotted with respect to the lattice size $L_s$ through the relation,
\begin{equation}
	E_n^{(0)}=2\sqrt{m_M^2+k^2(n)}\equiv 2\sqrt{m_M^2+\left(\frac{2\pi}{L_s}\right)^2 n^2}.
\end{equation}
The tiny differences of the values of $m_M$ on the two ensemble (see in Table~\ref{tab:config}) makes the curves on L12M420 (blue) and those on L16M420 (red) not exactly connected. 
A clear $N_V$-dependence is observed for high-energy levels:
\begin{itemize}
	\item For $0^{++}$ system on the L12M420 ensemble, the energy levels with $N_V=70$ are obviously lower than those with $N_V=120$ and $170$ when $E_n>6.3$~GeV. The values of $E_n$ obtained with $N_V=120$ are consistent with those obtained with $N_V=170$ under 6.6~GeV. This indicate the values of $\{E_n, n=1,2,\ldots,7\}$ derived with $N_V=170$ are close to saturate the exact value at $N_V=3 N_s^3$ that is full dimension of the Laplacian space. While on L16M420, the $E_n$ values obtained with $N_V=70$ and $N_V=120$ are compatible with each other only when $E_n$ is lower than 6.3~GeV. So we will use the energy levels $\{E_n, n=1,2,\ldots,5\}$ in the analysis of the $\eta_c\eta_c$ and $J/\psi J/\psi$ scattering properties.;
	
	\item For the $2^{++}$ system, a similar analysis is performed. 
	As a result, we adopt the energy levels $\{E_n, n=1,2,\ldots,5~(4)\}$ on L12~(L16) ensembles, respectively, which are highlighted in bold in the right panel of Fig.~\ref{fig:spectrum}.
\end{itemize}

In summary, for ensembles L12M420 and L12M250, energy levels derived using $N_V = 170$ are reliable up to $E_n < 6.6$~GeV. 
In contrast, for the L16M420 and L16M250 ensembles (with $N_V = 120$), only the energy levels below 6.3~GeV are considered reliable.
The study of the scattering properties in the remainder of this article will be based on these energy levels.   

%%%%%%%%%%%%%%%%%%%%
\subsection{Check on the $S$- and $D$-wave coupling in $J/\psi J/\psi$ scattering}
\label{Sec:Dwave}
We focus on the $S$-wave $J/\psi J/\psi$ scattering in the $2^{++}$ channel, and therefore construct the $J/\psi J/\psi$ operators to transform according to the irrep 
$E^{++} \;\leftarrow\; E^{(s)} \otimes A_1^{(l)}$, which corresponds to ${}^5 S_2$ (and ${}^5 G_2$) quantum numbers in the continuum limit, as explained in Sec.~\ref{secII:corr2pp}. 
However, the ${}^1 D_2$ and ${}^5 D_2$ states can also give $2^{++}$ quantum number, so we check the possible coupling between $S$- and $D$-wave $J/\psi J/\psi$ scattering states. 
For simplicity, we only consider the ${}^1 D_2$ states whose lattice operators are in the irrep $E^{++}\leftarrow A_1^{(s)}\otimes E^{(l)}$ (see Eq.~\eqref{eq:op-set}). 
Since the energy levels employed in our scattering analysis reach at most $n^2=2$, we restrict the relative momentum mode to be $n^2\leq 3$. 

\begin{figure}[t]
	\centering
	\includegraphics[width=0.9\linewidth]{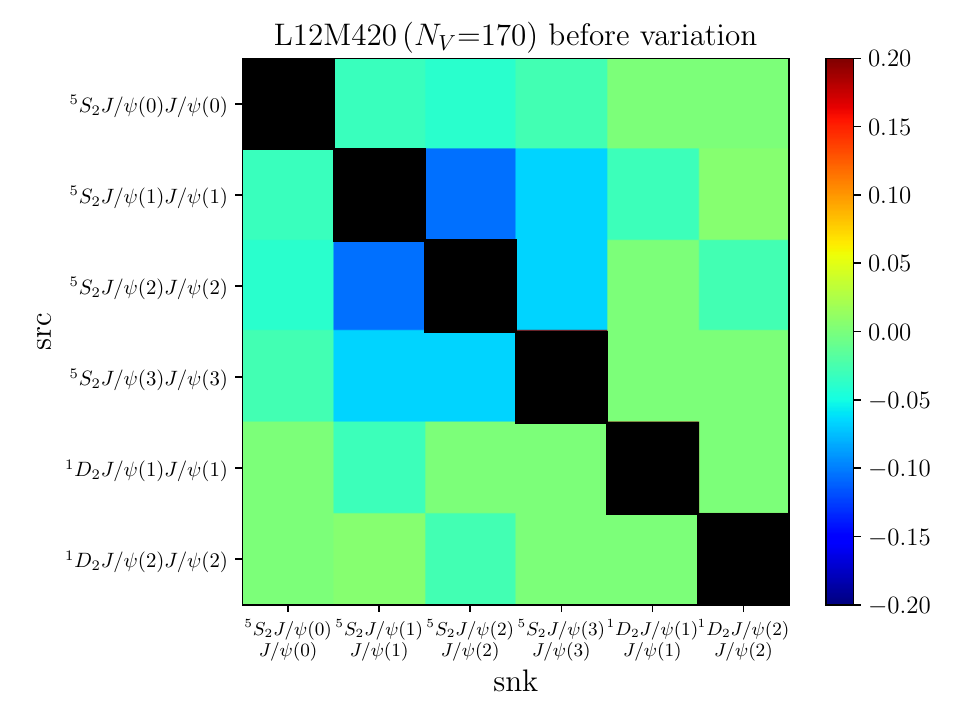}
	\caption{Correlation matrix $w_{\alpha\beta}(t)$ at $t = 20~a_t$ of the ${}^5S_2$ and ${}^1D_2$ $J/\psi J/\psi$ system on the L12M420 ensemble, where the ${}^1D_2$ $\eta_c \eta_c$ channel has been omitted, as supported by previous studies indicating that the coupling is rather weak.
		Diagonal elements are shown in black and normalized to $1$. }
	\label{fig:Dwave}
\end{figure}

We calculate the correlation matrix of these two types of operators on the L12M420 ensemble and extract the energy levels after solving the related GEVP. Figure~\ref{fig:Dwave} shows the weight functions $w_{\alpha\beta}(t)$ of the correlation matrix, where one can see little correlations between the two types ($S$- and $D$-wave) of operators.   

We compare the energy levels of the $S$- and $D$-wave states in Table~\ref{table:Dwave}. 
It is observed that the $S$-wave energies remain almost unchanged with or without the inclusion of $D$-wave operators. 
This indicates that the $S$–$D$ coupling in the $2^{++}$ $J/\psi J/\psi$ scattering channel is negligible. 
In addition, the lowest $D$-wave energy is consistent with the corresponding noninteracting level within uncertainties, implying that the $D$-wave $J/\psi J/\psi$ interaction is very weak.

%%%%%%%%%%%%%%%%%%%%%%%%%%%
%%%%%%%%%%%%%%%%%%%%%%%%%%%

%%%%%%%%%%%%%%%%%%%%%%%%%%%
\begin{table}[t]
	\caption{Comparison of energy levels (in GeV) from the noninteracting di-$J/\psi$ system (`$E_n^{(0)}$'), the $S$-wave energy levels used in the scattering analysis below (`$E_n$' highlighted in bold), and levels obtained via joint GEVP analyses of the ${}^5S_2$ and ${}^1D_2$ $J/\psi J/\psi$ system (`$E_n'''$').}
	\label{table:Dwave}
	\begin{ruledtabular}
		\begin{tabular}{llll}
			$J/\psi J/\psi$ channels& L12 $E_n^{(0)}$ & L12 $E_n$ & L12 $E_n'''$ \\
			\hline  
			${}^5S_2$ $(n^2=0)$ & 6.1770(17)&\bf{6.1858(17)}& 6.1855(17)            \\
			${}^5S_2$ $(n^2=1)$ & 6.3532(22)&\bf{6.3626(20)}& 6.3634(20)         \\
			${}^1D_2$ $(n^2=1)$ & 6.3532(22)&-						& 6.3528(20)          \\
			${}^5S_2$ $(n^2=2)$ & 6.5247(27)&\bf{6.5129(20)}& 6.5147(19)          \\
			${}^1D_2$ $(n^2=2)$ & 6.5247(27)&-						&  6.5304(19)        \\
			${}^5S_2$ $(n^2=3)$ & 6.6918(31)&    6.6849(24) & 6.6902(19)          \\
		\end{tabular}
	\end{ruledtabular}
\end{table}
%%%%%%%%%%%%%%%%%%%%%%%%%%%

%%%%%%%%%%%%%%%%%%%%%%%%%%%%%%%%%%%%%%%%%%%%%%%%%%%%%%%%%%%%%%%%%%%%%%%%%%%%%%%%%%
\section{Scattering Analysis}\label{sectionIII}
%%%%%%%%%%%%%%%%%%%%%%%%%%%%%%%%%%%%%%%%%%%%%%%%%%%%%%%%%%%%%%%%%%%%%%%%%%%%%%%%%%
A state-of-the-art technique for lattice QCD studies of hadron-hadron scattering is L\"uscher’s formalism, which establishes a connection between the finite-volume energy spectrum computed on a Euclidean spacetime lattice and the corresponding infinite-volume scattering matrix through quantization conditions,
\begin{equation}\label{eq:Luscher}
	\det \bigg[\mathbf{I}+i\boldsymbol{\rho}(E_\mathrm{cm})\mathbf{t}(E_\mathrm{cm})\big(\mathbf{I}-i\boldsymbol{\mathcal{M}}(E_\mathrm{cm},L_s)\big)\bigg]=0,
\end{equation}
where $\mathbf I$ denotes the identity matrix, $E_\mathrm{cm}$ is the center-of-mass energy of the system, 
the matrix $\boldsymbol{\rho}(E_{\mathrm{cm}})$ is diagonal in channel space, with each diagonal element given by the corresponding phase-space factor, $\mathbf{t}$ is the scattering matrix at $E_\mathrm{cm}$, and $\boldsymbol{\mathcal{M}}(E_{\mathrm{cm}},L_s)$ is a known matrix-valued function, which depends on the center-of-mass energy $E_{\mathrm{cm}}$ and the spatial lattice extent $L_s$.
Therefore, once an appropriate parametrization of $\mathbf{t}$-matrix is chosen, its parameters can be determined by fitting Eq.~\eqref{eq:Luscher} to the set of center-of-mass energies $E_{\mathrm{cm}}$ extracted from lattice QCD calculations.
The unitarity condition of $\mathbf{S}$-matrix requires $\operatorname{Im}~(\mathbf{t}^{-1})=-\boldsymbol{\rho}$ when $E_\mathrm{cm}$ is larger than the related threshold energy. 
After analytically continuing the invariant mass $\sqrt s$ into the complex plane, the $\mathbf{K}$-matrix provides a convenient and physically motivated parametrization,
\begin{equation}
	\mathbf{t}^{-1}(s)=\mathbf{K}^{-1}(s)-i\boldsymbol{\rho}(s),
\end{equation}
where the elements of the $\mathbf{K}$-matrix are real-valued functions of $s$, and their functional forms are not unique.

%%%%%%%%%%%%%%%%%%%%%%%%%%%
\begin{table*}[t]
	\caption{Energy levels (in GeV) extracted from noninteracting dicharmonium $E_n^{(0)}$, and from the ${}^1S_0$ $\eta_c \eta_c-J/\psi J/\psi$ system at two pion masses using different GEVP analyses;
		$E_n$ (using the complete set of operators from both channels),
		$E_n'$ (using only $\eta_c \eta_c$ operators), and
		$E_n''$ (using only $J/\psi J/\psi$ operators).
		Energy levels $E_n$ used for the scattering analysis are highlighted in bold.} 
	\label{tab:energys}
	\begin{ruledtabular}
		\begin{tabular}{lllllllll}
			$E_n$ & L12 $E_n^{(0)}$ & L12 $E_n$ & L12 $E_n'$ & L12 $E_n''$ & L16 $E_n^{(0)}$ & L16 $E_n$ & L16 $E_n'$ & L16 $E_n''$ \\ 
			\hline 
			&          &               &            &$m_\pi=420$~MeV&        &               &           &             \\
			\hline  
			$E_1$   & 5.9919(13)&\bf{5.9966(13)}& 5.9965(13) & -          & 5.9905(7)  &\bf{5.9925(7)} & 5.9926(7) & -           \\
			$E_2$   & 6.1726(17)&\bf{6.1820(15)}& 6.1840(16) & -          & 6.0938(9)  &\bf{6.1001(7)} & 6.1003(7) & -           \\
			$E_3$   & 6.1770(17)&\bf{6.1743(17)}& -          & 6.1721(16) & 6.1743(10) &\bf{6.1713(9)} & -         & 6.1715(9)   \\
			$E_4$   & 6.3481(21)&\bf{6.3512(15)}& 6.3566(16) & -          & 6.1954(11) &\bf{6.2039(7)} & 6.2033(7) & -           \\
			$E_5$   & 6.3532(22)&\bf{6.3516(19)}& -          & 6.3343(18) & 6.2752(13) &\bf{6.2681(11)}& -         & 6.2544(14)  \\
			$E_6$   & 6.5188(24)&\bf{6.5270(17)}& 6.5262(17) & -          & 6.2954(14) &    6.3018(8)  & 6.3000(8) & -           \\
			$E_7$   & 6.5247(27)&\bf{6.5169(18)}& -          & 6.5012(19) & 6.3745(16) &    6.3583(27) & -         & 6.3382(16)  \\
			$E_8$   & 6.6852(27)&    6.6742(15) & 6.6743(35) & -          & 6.3906(12) &    6.3908(10) & 6.3901(10)& -           \\
			$E_9$   & 6.6918(31)&    6.6933(22) & -          & 6.6743(22) & 6.4724(18) &    6.4539(15) & -         & 6.4399(20)  \\
			$E_{10}$& 6.8549(36)&    6.8374(22) & -          & 6.8161(24) & 6.5687(21) &    6.5515(15) & -         & 6.5225(24)  \\
			\hline
			&           &               &            &$m_\pi=250$~MeV&        &               &           &             \\
			\hline
			$E_1$   & 5.9913(14)&\bf{5.9968(20)}& 5.9958(26) & -          & 5.9966(7)  &\bf{5.9987(7)} & 5.9987(7) & -           \\
			$E_2$   & 6.1714(18)&\bf{6.1811(20)}& 6.1831(33) & -          & 6.0996(10) &\bf{6.1057(8)} & 6.1058(8) & -        \\
			$E_3$   & 6.1633(20)&\bf{6.1608(25)}& -          & 6.1575(25) & 6.1706(10) &\bf{6.1680(9)} & -         & 6.1674(10) \\
			$E_4$   & 6.3464(22)&\bf{6.3448(21)}& 6.3543(40) & -          & 6.2009(14) &\bf{6.2080(9)} & 6.2077(10)& -           \\
			$E_5$   & 6.3406(24)&\bf{6.3445(27)}& -          & 6.3228(31) & 6.2714(14) &\bf{6.2652(16)}& -         & 6.2539(15)  \\
			$E_6$   & 6.5167(26)&\bf{6.5246(24)}& 6.5235(27) & -          & 6.3005(17) &    6.3053(11) & 6.3036(11)& -           \\
			$E_7$   & 6.5130(28)&\bf{6.5025(26)}& -          & 6.4931(24) & 6.3706(18) &    6.3578(36) & -         & 6.3422(20)  \\
			$E_8$   & 6.6827(30)&    6.6692(24) & 6.6706(22) & -          & 6.3986(20) &    6.3945(14) & 6.3934(12)& -           \\
			$E_9$   & 6.6810(32)&    6.6805(25) & -          & 6.6644(28) & 6.4682(22) &    6.4600(31) & -         & 6.4445(24)  \\
			$E_{10}$& 6.8449(36)&    6.8206(30) & -          & 6.7996(38) & 6.5645(25) &    6.5429(39) & -         & 6.5287(26)   
		\end{tabular}
	\end{ruledtabular}
\end{table*}
%%%%%%%%%%%%%%%%%%%%%%%%%%%

In the following, we will analyze the properties of the ${}^1S_0$ $\eta_c\eta_c-J/\psi J/\psi$ scattering in the $0^{++}$ channel, as well as the ${}^1D_2$ $\eta_c\eta_c$ and ${}^5S_2$ $J/\psi J/\psi$ scattering in the $2^{++}$ channel. 
Since the charm quark-antiquark annihilation effect is not considered in our calculation, our scattering analysis
effectively ignores any coupling to light-hadron states with the same quantum
numbers.
As discussed in Secs.~\ref{secIII:0pp} and \ref{secIII:2pp}, our $S$-wave two-hadron operators are constructed using momentum configurations that belong to the $A_1$ irrep of the lattice symmetry group O in the two-hadron rest frame, which contains contributions from orbital angular momenta $l = 0, 4, \ldots$.
Therefore, the corresponding finite-volume energy levels encode information not only about the $S$-wave ($l=0$) scattering but also about higher partial waves ($l \ge 4$).
Fortunately, for sufficiently low-energy scattering, the phase shifts are expected to obey the hierarchy $\delta_0 \gg \delta_4$, such that contributions from partial waves with $l \ge 4$ are negligible~\cite{Luscher:1990ux}.
This assumption is usually adopted in lattice QCD studies of $S$-wave hadron--hadron scattering. 
On the other hand, for nonrelativistic two-body scattering, the largest orbital angular momentum of relevance at a given scattering momentum $k$ can be estimated as $l_{\mathrm{max}} \lesssim kR$, where $R$ denotes the interaction range.
For the $\eta_c\eta_c$ and $J/\psi J/\psi$ scattering processes considered in this work, the lightest meson that can be exchanged is the $\sigma$ meson, whose mass is approximately $m_\sigma \sim 2m_\pi$. 
The interaction range can therefore be estimated as $R \gtrsim 1/(2m_\pi)$.
At our highest center-of-mass energy, $E \simeq 6.6$~GeV, one finds $l_{\mathrm{max}} \lesssim 3$ for $m_\pi = 250$~MeV and $l_{\mathrm{max}} \lesssim 2$ for $m_\pi = 420$~MeV.
The contributions from partial waves with $l \ge 4$ can therefore be temporarily neglected.
As discussed in Sec.~\ref{Sec:Dwave}, the $D$-wave $J/\psi J/\psi$ interaction is found to be very weak, which provides further justification for neglecting the $l=4$ partial wave.
In this respect, we extract the $S$-wave scattering phase shift $\delta_0$ directly from the finite-volume energy levels using L\"uscher’s formalism.  

One can also compute finite-volume energy levels in moving frames to obtain more energy levels, so as to better constrain the scattering amplitudes. 
However, in moving frames, the finite-volume spectrum that contains the information of the $S$-wave scattering also receives contribution from high partial waves starting from $l=2$. 
As we have addressed above, for $\eta_c\eta_c$ and $J/\psi J/\psi$ scatterings at the center-of-mass energy near 6.6 GeV, the $l=2$ partial wave may not be neglected.  
The partial-wave mixing makes a clean determination of the $S$-wave amplitude highly nontrivial~\cite{Dudek:2012gj}.
To avoid uncontrolled systematic effects associated with partial-wave mixing, we restrict the present exploratory study to the  calculation in the rest frame.

%%%%%%%%%%%%%%%%%%%%%%%%%%%%%%%%%%%%%%%%%%%%%%%%%%%%%%%%%%%%%%%%%%%%%%%%%%
\subsection{$0^{++}$ $\eta_c\eta_c-J/\psi J/\psi$ scattering}\label{secIII:0pp}
%%%%%%%%%%%%%%%%%%%%%%%%%%%%%%%%%%%%%%%%%%%%%%%%%%%%%%%%%%%%%%%%%%%%%%%%%%
In the ${}^1S_0$ channel of the $\eta_c\eta_c-J/\psi J/\psi$ system, we obtain ten energy levels $\{E_n, n=1,2,\ldots,10\}$ on the four ensembles L12M420 ($N_V=170$), L16M420 ($N_V=120$), L12M250 ($N_V=170$) and L16M250 ($N_V=120$) employing the ten operators in $S_0$, which are collected in Table~\ref{tab:energys} under the label `$E_n$'. 
The association with a particular $E_n$ is made based on the observation that the corresponding state couples most strongly to the $\eta_c\eta_c$ operator and that $E_n$ is closest to the corresponding noninteracting energy under the label `$E_n^{(0)}$'.
The correspondence between the energy levels before and after the variation is illustrated in Fig.~\ref{fig:GEVP_vector}.
First, we examine the possible coupled-channel effects between $\eta_c\eta_c$ and $J/\psi J/\psi$ by performing the following steps:
\begin{itemize}
	\item 
	We first switch off the coupling to the $J/\psi J/\psi$ operators and perform a GEVP analysis on the correlation matrix of the five $\eta_c\eta_c$ operators with $n^2 \le 4$. 
	This yields five new energy levels, which are listed in Table~\ref{tab:energys} and are labeled by `$E_n'$'.
	Clearly, the $E_n'$ levels are compatible with the $E_n$ levels on a one-to-one basis within uncertainties, except for $E_4$ and $E_4'$ on L12M420, which differ slightly for unknown reasons.
	This indicates that the inclusion of $J/\psi J/\psi$ operators does not affect the energy levels of the $\eta_c\eta_c$ system.
	Since the higher excited states are well absorbed in the three-state fit, the extracted $\eta_c \eta_c$ energy levels are reliable, whether the $J/\psi J/\psi$ operators are included or not.
	Therefore, since the $\eta_c \eta_c$ energy levels remain unchanged when comparing single- and coupled-channel analyses, we conclude that the coupled-channel effects between the $S$-wave $\eta_c \eta_c$ and $J/\psi J/\psi$ channels are very weak.
	When comparing the $E_n$ (and $E_n'$) identified as $\eta_c\eta_c$ energies with the corresponding noninteracting energies $E_n^{(0)}$, one can see that $E_n(E_n')\gtrsim E_n^{(0)}$ in the energy region below 6.6~GeV, which hints at a repulsive interaction in the low-energy range near the $\eta_c\eta_c$ threshold.  
	\item Then we perform the GEVP analysis to the correlation matrix of the five $J/\psi J/\psi$ operators and obtain five new energy levels $E_n''$. 
	Their correspondence to specific $E_n$ values is based on the similar considerations discussed above. 
	There are small but non-negligible differences between $E_n$ and $E_n''$, which indicate the necessity of the inclusion of $\eta_c\eta_c$ operators to derive the energies of the $J/\psi J/\psi$ system. 
	In our calculation, both the direct diagram (labeled D)
	and the quark rearrangement diagram (labeled Q) in Fig.~\ref{fig:diagrams} are considered. 
	The Q diagram reflects, to some extent, the effect of Fierz rearrangement among quarks during the scattering process (see the related discussions in Sec.~\ref{sectionIV}). 
	Thus, both $\eta_c\eta_c$ and $J/\psi J/\psi$ operators couple to the same set of energy levels.
	For a given relative momentum mode, the energy of the $\eta_c\eta_c$ system is lower than that of the $J/\psi J/\psi$ system, and its energy can be reliably determined regardless of whether the corresponding $J/\psi J/\psi$ operator is included.
	In contrast, for the same momentum mode, the $J/\psi J/\psi$ energy lies above that of $\eta_c\eta_c$, while the $J/\psi J/\psi$ operator also couples to the $\eta_c\eta_c$ state via the Q diagram.
	Therefore, omitting the corresponding $\eta_c\eta_c$ operator may bias the determination of the $J/\psi J/\psi$ energy levels and result in a downward shift of the extracted energies, even though there is no direct coupled-channel effect between the $\eta_c\eta_c$ and $J/\psi J/\psi$ states.
	In this respect, the $J/\psi J/\psi$ spectrum obtained with the $\eta_c\eta_c$ operators included is more reliable and is used for the $J/\psi J/\psi$ scattering analysis. 
	In contrast with the $\eta_c\eta_c$ case, the $J/\psi J/\psi$ energies $E_n$ are uniformly lower than the corresponding noninteracting energies $E_n^{(0)}$ and hint at an attractive interaction.
\end{itemize}

Observing that the $\eta_c\eta_c$--$J/\psi J/\psi$ coupled-channel effect is weak, we perform a single-channel analysis of the $\eta_c\eta_c$ and $J/\psi J/\psi$ scattering. 
In the single-channel case, the scattering matrices reduce to one-dimensional quantities.
Throughout this work, we adopt the convention $\rho(s) = {k}/{\sqrt{s}}$, taking into account the factor of $1/2$ due to the presence of two identical particles in each channel.

For the $S$-wave single-channel two-body scattering, the scattering amplitude $t$ can be expressed in terms of the scattering phase $\delta_0$ as 
\begin{equation}
	t(s)=\frac{\sqrt{s}}{k\cot \delta_0(k)-ik},
\end{equation}
where $k$ is the magnitude of the scattering momentum defined through the center-of-mass $\sqrt{s}=E_\mathrm{cm}$, namely, 
\begin{equation}\label{eq:k-s}
	E_\mathrm{cm}=2\sqrt{m_M^2+k^2}
\end{equation}
with $m_M$ being the mass of $\eta_c$ or $J/\psi$. 
On the finite volume lattice with size $L_s$, one can introduce a dimensionless quantity $q^2=(L_s/2\pi)^2 k^2$ for convenience. 
The L\"{u}scher's quantization condition in Eq.~\eqref{eq:Luscher} is simplified as 
\begin{equation}\label{eq:phase}
	k\cot\delta_0(k) =k \mathcal{M}_{0000}(q^2,L_s)=\frac{2}{\sqrt{\pi}L_s}\mathcal{Z}_{00}(1,q^2),
\end{equation}
where the $l=0$ contribution of the L\"uscher zeta function,
\begin{equation}
	\mathcal{Z}_{00}=\frac{1}{\sqrt{4\pi}}\sum\limits_{\vec{n}\in Z^3} \frac{1}{\vec{n}^2-q^2}.
\end{equation}
Using Eq.~\eqref{eq:phase}, one can determine the scattering phase $\delta_0$ at each finite-volume energy derived from lattice QCD. 

%%%%%%%%%%%%%%%%%%%%%%%%%%%
\begin{figure*}[t]
	\centering
	\includegraphics[width=0.45\linewidth]{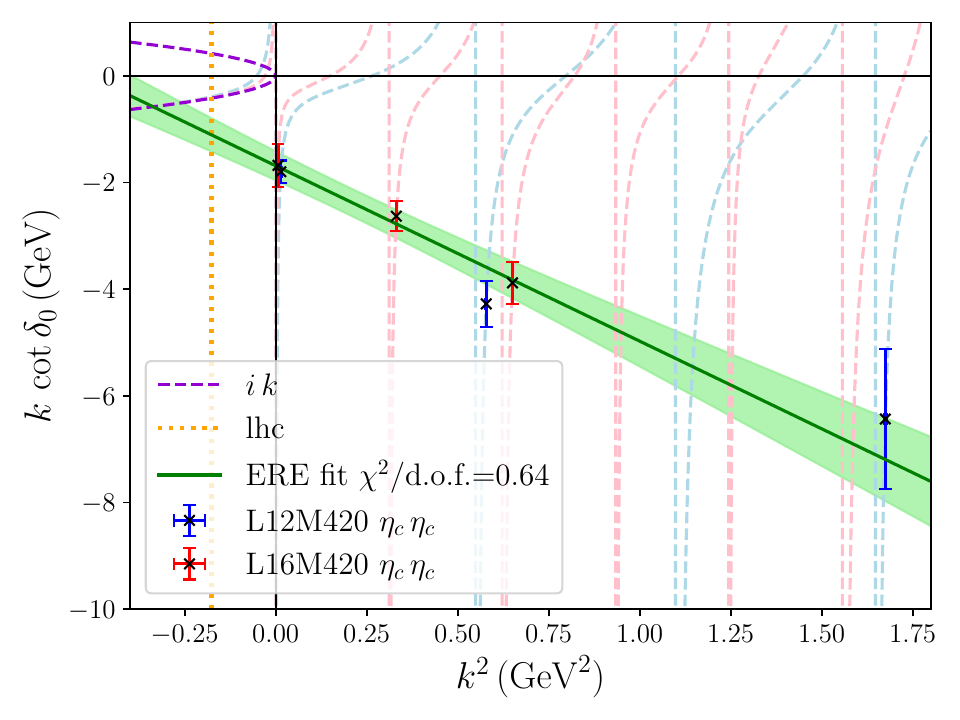}
	\includegraphics[width=0.45\linewidth]{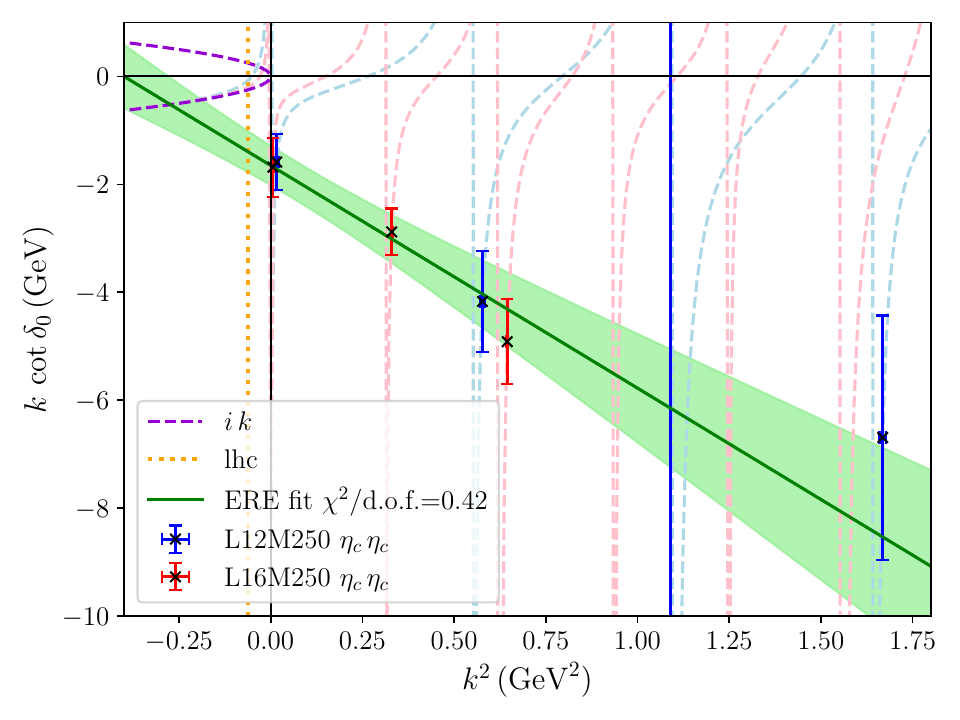}\\
	\includegraphics[width=0.45\linewidth]{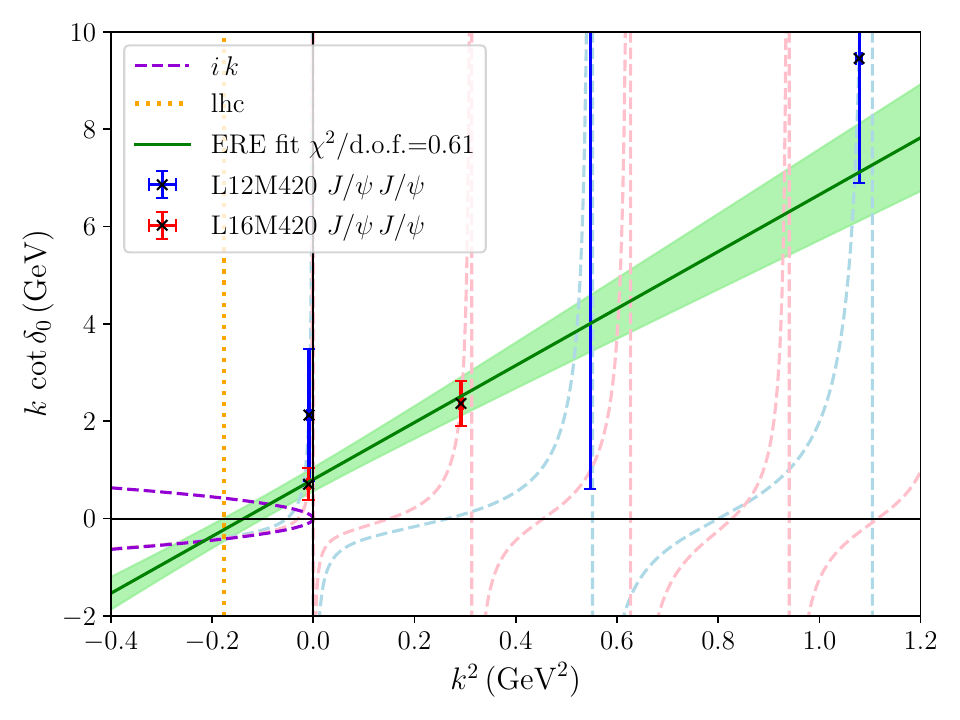}
	\includegraphics[width=0.45\linewidth]{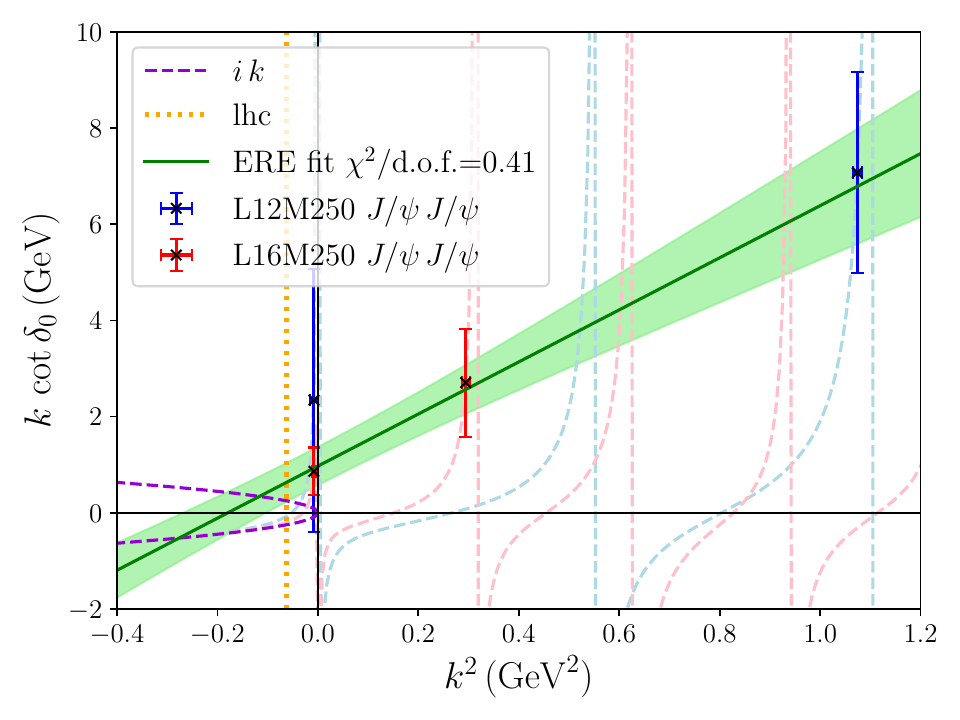}\\
	\includegraphics[width=0.45\linewidth]{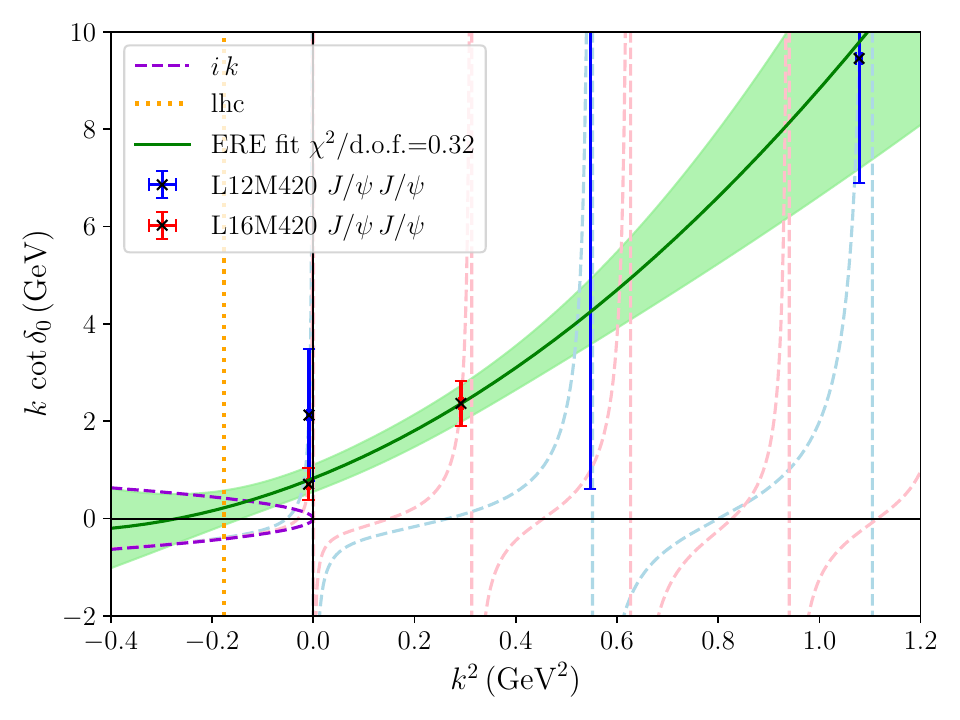}
	\includegraphics[width=0.45\linewidth]{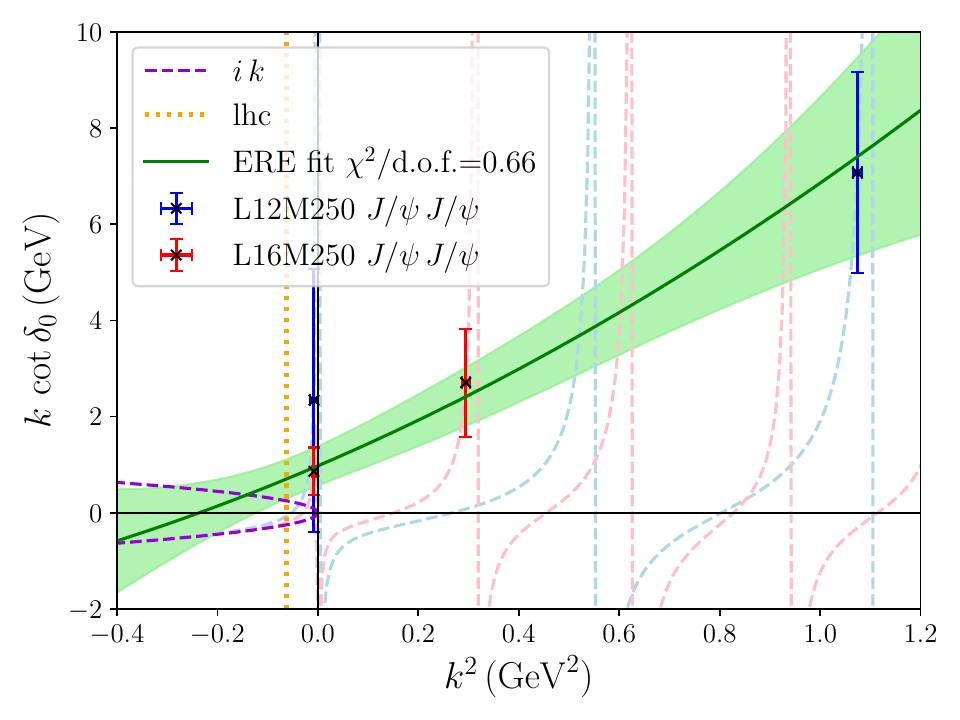}
	\caption{The scattering phases $k \cot \delta_0(k)$ for the $0^{++}$ system in the ${}^1S_0$ $\eta_c \eta_c$ (upper panels) and ${}^1S_0$ $J/\psi J/\psi$ (middle and lower panels) channels, for M420 (left) and M250 (right) ensembles.
		The middle panels show fit results using the lowest-order ERE in Eq.~\eqref{eq:ERE}, while the lower panels use results up to $k^4$ term in Eq.~\eqref{eq:erek4}.
		The blue and red points represent lattice energy levels accompanied by statistical errors from L12 and L16 lattices, respectively. 
		Light blue and red dashed curves indicate the Riemann zeta function.
		The green band represents the ERE fit applied on both lattice volumes.
		The purple dashed line represents the pole equation $k\cot \delta_0 (k)=i ~ k$ in Eq.~\eqref{eq:pole}.
		The orange dotted line denotes the left-hand cut (lhc) arising from $\sigma$-meson (two pion) exchange.}   
	\label{fig:phase-0pp}
\end{figure*}
%%%%%%%%%%%%%%%%%%%%%%%%%%%

The $0^{++}$ $\eta_c\eta_c$ scattering phase shifts are shown in the upper two panels of Fig.~\ref{fig:phase-0pp} (left for M420 and right for M250). 
The data points of $k\cot\delta_0(k)$ (shown as blue points for the L12 lattice and red points for the L16 lattice) exhibit an approximately linear dependence on $k^2$ up to $k^2 \sim 1.5~\mathrm{GeV}^2$, and happen to be described by the leading-order effective range expansion (ERE),
\begin{equation}\label{eq:ERE}
	k \cot\delta_0(k)=\frac{1}{a_0}+\frac{1}{2} r_0 k^2+\mathcal O(k^4),
\end{equation}
where $a_0$ is the $S$-wave scattering length and $r_0$ is the effective range.  
The green bands in the upper panels of Fig.~\ref{fig:phase-0pp} illustrate Eq.~(\ref{eq:ERE}) with the fitted parameters,
\begin{equation}\label{eq:di-etac}
	\begin{aligned}
		&{\rm M420:~}a_0= -0.117(19) ~{\rm fm}, ~ r_0=-1.30(20) ~{\rm fm},\\
		&{\rm M250:~}a_0= -0.120(26) ~{\rm fm}, ~ r_0=-1.63(40) ~{\rm fm}, 
	\end{aligned}
\end{equation}
which are determined mainly by the data points with $k^2<0.6~\rm{GeV}^2$. We notice that a lattice QCD study calculates the low-energy $S$-wave $\eta_c\eta_c$ scattering at zero relative momentum and obtains a similar scattering length $a_0^{0^{++}} = -0.104(9)$ in the continuum limit~\cite{Meng:2024czd}. 
As listed in Table~\ref{tab:energys}, the lowest $\eta_c\eta_c$ scattering energy on each ensemble lies above the $\eta_c\eta_c$ threshold.
Also the negative value of $a_0$ indicates a repulsive interaction in ${}^1S_0$ $\eta_c\eta_c$ system.

However, a puzzling observation is that, upon continuation into the $k^2<0$ region, 
the fitted function from Eq.~\eqref{eq:ERE} intersects the $ik$ curve on the lower branch (as shown in the upper panels of Fig.~\ref{fig:phase-0pp}), thereby implying a bound-state pole of the scattering amplitude $t(s)$. 
Such a bound state is incompatible with a repulsive interaction and is therefore clearly unphysical.
A possible explanation for this unphysical bound-state pole is that the functional form assumed in Eq.~\eqref{eq:ERE} is not valid over the entire $k^2<0$ region.
For instance, the exchange of light hadrons between the two scattering charmonia
can generate a left-hand cut (lhc) in the scattering amplitude~\cite{Du:2023hlu,Meng:2023bmz,Raposo:2023oru}.
For $\eta_c\eta_c$ and $J/\psi J/\psi$ scattering, the lightest hadron(s) that can be exchanged is (are) the $\sigma$ meson (two pions), and the corresponding lhc point is
\begin{equation}
	(k^{\sigma}_{\rm lhc})^2 \sim -\frac{m_\sigma^2}{4}\sim - m_\pi^2,
\end{equation}
where the $\sigma$ mass is estimated to be $m_\sigma\sim 2m_\pi$, indicated by the orange dotted line in each panel of Fig.~\ref{fig:phase-0pp}. 
Although the lattice data suggest that, within the energy window accessible to the simulations, the contribution from light-hadron exchange is weaker than that from charm-quark interchange (see Sec.~\ref{sectionIV}), light-hadron exchange nevertheless leads to nonanalytic structures associated with the left-hand cut, which can lie close to the physical region and significantly affect the amplitude.
Since Eq.~\eqref{eq:ERE} is valid only to the right of the lhc point, any intersection occurring on the left side is physically meaningless.
Therefore, our analysis does not support the presence of additional structures in the ${}^1S_0$ $\eta_c \eta_c$ channel.

\begin{figure}[t]
	\centering
	\includegraphics[width=0.99\linewidth]{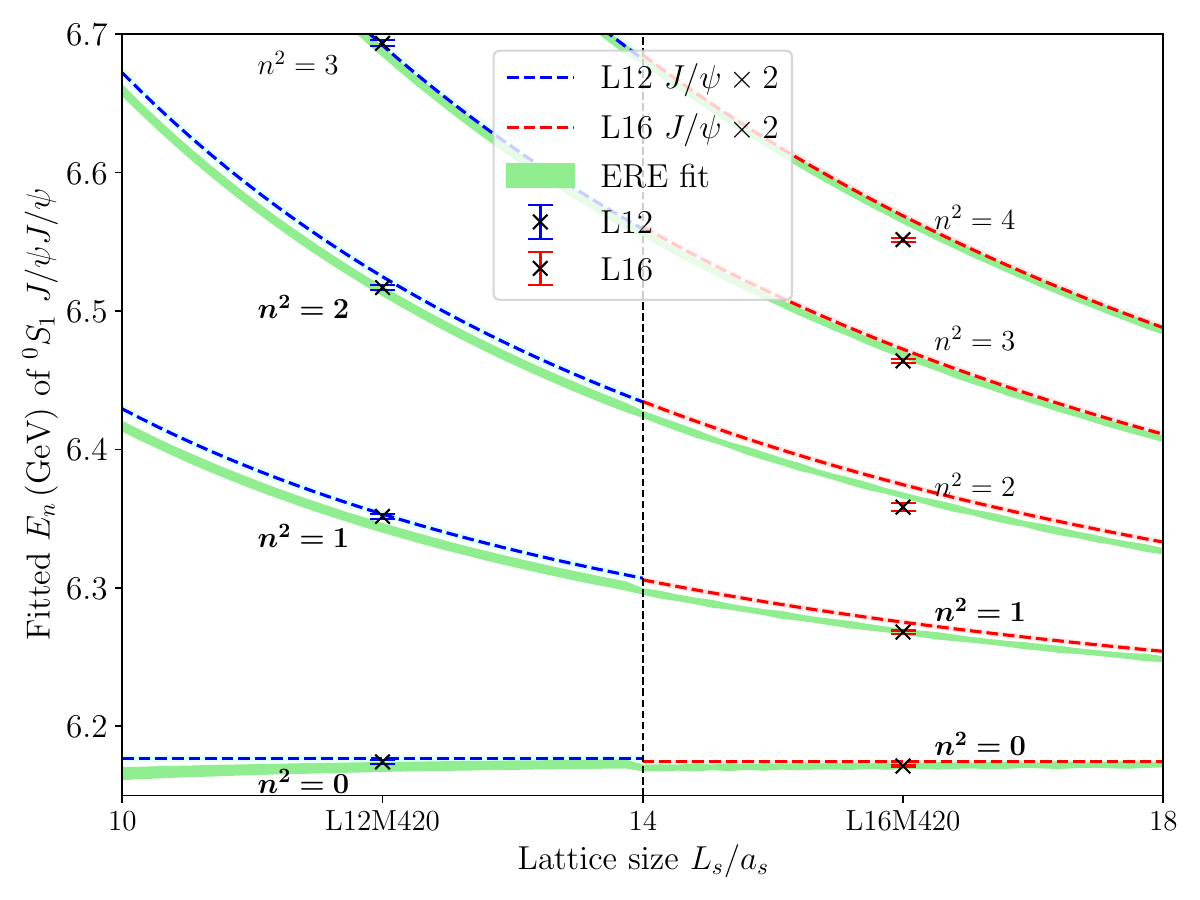}
	\caption{Energy levels with linear fits for the ${^1}S_0$ $J/\psi J/\psi$ system on M420. Energy levels $E_n(n^2)$ used for fitting are highlighted in bold.
		The green bands shows the theoretical predictions of FVEs based on the linear ERE fit.}
	\label{fig:fit-spectrum-0pp}
\end{figure}

The scattering phase of the ${}^1S_0$ $J/\psi J/\psi$ scattering is extracted in the same way. 
As shown in the middle two panels of Fig.~\ref{fig:phase-0pp} (left for M420 and right for M250), the data points of $k\cot\delta_0(k)$ approximately lie on a straight line over a wide momentum range, up to $k^2 \simeq 1.2~\mathrm{GeV}^2$ ($E_\mathrm{cm} \sim 6.5~\mathrm{GeV}$).
The linear fit using Eq.~\eqref{eq:ERE} yields the following parameter values:
\begin{equation}\label{eq:erek2}
	\begin{aligned}
		&{\rm M420:~}a_0= 0.25(7) ~{\rm fm}, ~~ r_0=2.31(33) ~{\rm fm},\\
		&{\rm M250:~}a_0= 0.20(8) ~{\rm fm}, ~~ r_0=2.14(41) ~{\rm fm},
	\end{aligned}
\end{equation} 
which is indicated by the green band.
It can be seen that the values of $a_0$ and $r_0$ at the two pion masses are consistent with each other.
The validity of the fit can be further assessed by comparing the energy levels obtained from lattice calculations with the theoretical predictions based on the parametrization of Eq.~\eqref{eq:ERE} using the fitted parameters $a_0$ and $r_0$, as shown in Fig.~\ref{fig:fit-spectrum-0pp}.
It is seen that the theoretical predictions based on the linear ERE fit (green bands) agree well with most of the lattice energy levels under $6.6~(6.3)$~GeV (except for the point with $n^2=1$ on the L12 lattice, which shows a slight deviation).

%%%%%%%%%%%%%%%%%%%%%%%%%%%
%%%%%%%%%%%%%%%%%%%%%%%%%%%

%%%%%%%%%%%%%%%%%%%%%%%%%%%
\begin{figure*}[t]
	\centering
	\includegraphics[width=0.45\linewidth]{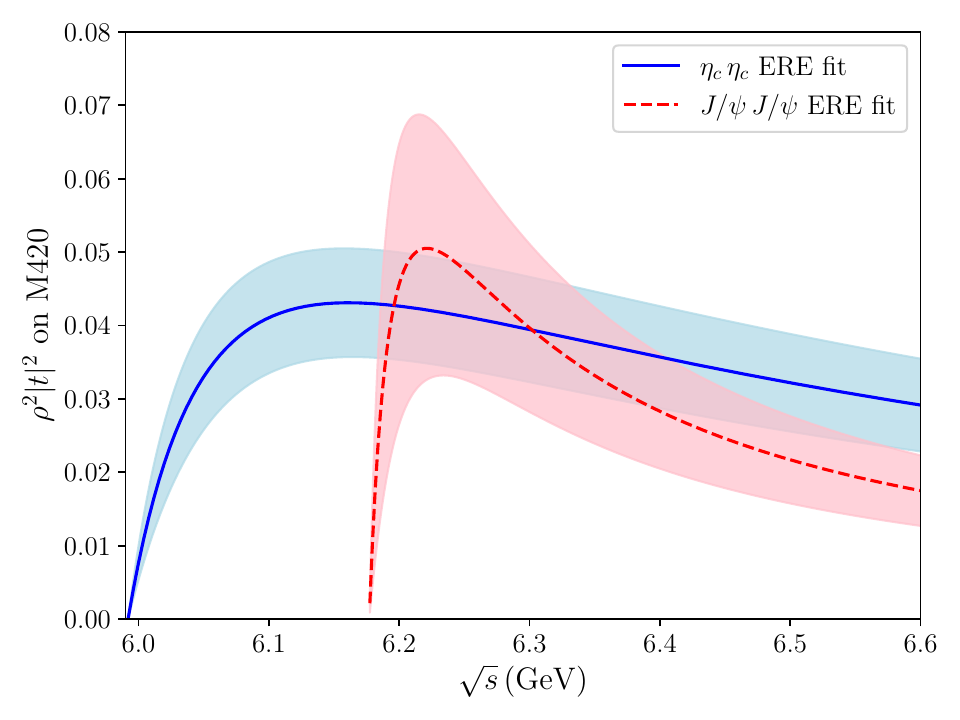}
	\includegraphics[width=0.45\linewidth]{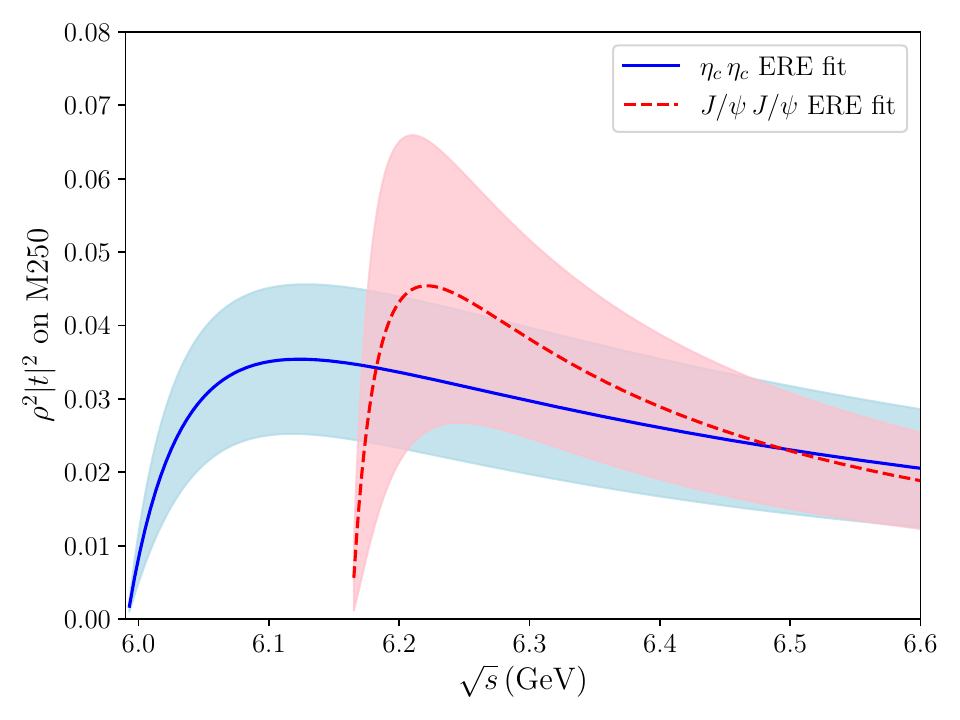}
	\caption{The cross section $\rho^2 |t|^2$ versus $\sqrt{s}$ on the M420~(left panels) and M250~(right panels) ensembles.
		The blue and red curves represent the results via ERE fitting for the ${}^1S_0$ $\eta_c \eta_c$ and ${}^1S_0$ $J/\psi J/\psi$ channels, respectively. }
	\label{fig:cross-section}
\end{figure*}
%%%%%%%%%%%%%%%%%%%%%%%%%%%

The positive values of the scattering length $a_0$ are consistent with the attractive interaction between $J/\psi J/\psi$ in this channel, as suggested by the observation that the finite-volume energies lie below the corresponding noninteracting energies.
With the obtained $a_0$ and $r_0$, the pole singularities of the scattering amplitude $t(s)$ can be investigated by solving the pole equation,
\begin{equation}\label{eq:pole}
	k\cot \delta_0 (k)=i ~ k,
\end{equation}
with the parametrization in Eq.~\eqref{eq:ERE}. 
The solutions of the pole equation are indicated in Fig.~\ref{fig:phase-0pp} by the intersection points between the $k\cot\delta_0(k)$ curve and the $ik$ curve (dashed parabola).
In each of the middle two panels, the intersection point closest to the threshold lies on the upper branch of the $ik$ curve and corresponds to a virtual-state pole at $k_V = -i|k_V|$.
The corresponding binding energy is $28(10)$~MeV on M420 and $38(20)$~MeV on M250, respectively. 
The other intersection, located on the lower branch of the $ik$ curve, appears to be a bound-state pole at $k_B = i|k_B|$.
However, this pole fails the consistency check for a bound state, which requires that the slope of $k\cot\delta_0(k)$ at $k^2 = k_B^2$ be smaller than that of $ik$, and is therefore essentially unphysical~\cite{Iritani:2017rlk}.
The appearance of this unphysical pole further indicates that Eq.~\eqref{eq:ERE} becomes invalid for large negative $k^2$.

Since our data points cover a wide range of $k^2$, we also attempt to fit $k\cot\delta_0(k)$ by including a $k^4$ term in Eq.~\eqref{eq:ERE}, namely,
\begin{equation}\label{eq:erek4}
	k \cot \delta_0(k) = \frac{1}{a_0} + \frac{1}{2} r_0 k^2 + c_0 k^4 + \mathcal{O}(k^6),
\end{equation}
and obtain the parameters
\begin{equation}
	\begin{aligned}
		&{\rm M420:~}a_0 = 0.24(8)~{\rm fm},~r_0= 1.6(4) ~{\rm fm},~c_0=0.03(2)~{\rm fm}^{-3},\\
		&{\rm M250:~}a_0 = 0.20(9)~{\rm fm},~r_0= 1.8(7) ~{\rm fm},~c_0=0.01(2)~{\rm fm}^{-3}.
	\end{aligned}
\end{equation}
Compared with the values in Eq.~\eqref{eq:erek2}, the scattering lengths $a_0$ remain nearly unchanged, the effective ranges $r_0$ have slightly smaller central values but are consistent within uncertainties, and the coefficient $c_0$ is small and compatible with zero.
As shown in the bottom two panels of Fig.~\ref{fig:phase-0pp}, the virtual-state pole persists with little change in its position, while the inclusion of the $k^4$ term either removes the unphysical bound-state pole (for M420) or shifts it further away from the threshold (for M250).

However, similar to the $\eta_c\eta_c$ scattering, the continuation of Eqs.~\eqref{eq:erek2} and \eqref{eq:erek4} into the $k^2<0$ region may also be affected by the left-hand cut, which is indicated by the orange vertical line in Fig.~\ref{fig:phase-0pp}.
In principle, the presence of the lhc will significantly alter the behavior of $k\cot\delta_0(k)$ as $k^2$ approaches the lhc point and can affect the pole positions~\cite{Du:2023hlu,Meng:2023bmz,Raposo:2023oru}.
Therefore, at the present stage, we cannot draw a definitive conclusion regarding the existence of a sub-threshold virtual state.

In any case, the finite-volume energies of the $\eta_c\eta_c$ and $J/\psi J/\psi$ systems all lie to the right of the lhc point, where L\"{u}scher's formalism is valid, and the same applies to the scattering phase $k\cot\delta_0(k)$.
Accordingly, we can use the parametrization of Eq.~\eqref{eq:erek2} to examine the energy dependence of the scattering cross section, defined as $\rho^2 |t|^2$.
Figure~\ref{fig:cross-section} shows $\rho^2 |t|^2$ for the $0^{++}$ $\eta_c\eta_c$ (blue band) and $J/\psi J/\psi$ (red) channels, based on the linear ERE fit, as a function of the center-of-mass energy $\sqrt{s} \equiv E_\mathrm{cm}$, where a rapid rise near the threshold is observed for the $J/\psi J/\psi$ system.
It is worth noting that the $J/\psi J/\psi$ spectra reported by LHCb~\cite{LHCb:2020bwg}, ATLAS~\cite{ATLAS:2023bft}, and CMS~\cite{CMS:2023owd,Wang:2024koq,CMS:2025fpt} also show a similar rapid rise near the $J/\psi J/\psi$ threshold.

%%%%%%%%%%%%%%%%%%%%%%%%%%%
\subsection{$2^{++}$ $\eta_c\eta_c-J/\psi J/\psi$ scattering}\label{secIII:2pp}
%%%%%%%%%%%%%%%%%%%%%%%%%%%
In the $2^{++}$ channel of the $\eta_c\eta_c-J/\psi J/\psi$ system, we obtain eight energy levels $\{E_n, n=1,2,\ldots,8\}$ on the four ensembles of L12M420 ($N_V=170$), L16M420 ($N_V=120$), L12M250 ($N_V=170$) and L16M250 ($N_V=120$) employing the eight operators in $S_2$, which are collected in Table~\ref{tab:energys_2pp} with the label `$E_n$'. 

%%%%%%%%%%%%%%%%%%%%%%%%%%%
\begin{table*}[t]
	\caption{Energy levels (in GeV) extracted from noninteracting dicharmonium $E_n^{(0)}$, and from the $2^{++}$ $\eta_c \eta_c-J/\psi J/\psi$ system at two pion masses using different GEVP analyses;
		$E_n$ (using the complete set of operators from both channels),
		$E_n'$ (using only $\eta_c \eta_c$ operators), and
		$E_n''$ (using only $J/\psi J/\psi$ operators).
		Energy levels $E_n$ used for the scattering analysis are highlighted in bold.} 
	\label{tab:energys_2pp}
	\begin{ruledtabular}
		\begin{tabular}{lllllllll}
			$E_n$ & L12 $E_n^{(0)}$ & L12 $E_n$ & L12 $E_n'$ & L12 $E_n''$ & L16 $E_n^{(0)}$ & L16 $E_n$ & L16 $E_n'$ & L16 $E_n''$ \\
			\hline 
			&          &               &            &$m_\pi=420$~MeV&       &               &           &             \\
			\hline  
			$E_1$ & 6.1726(17)&\bf{6.1757(14)}& 6.1761(14) & -          & 6.0938(9)  &\bf{6.0954(7)} & 6.0952(7) & -           \\
			$E_2$ & 6.1770(17)&\bf{6.1858(17)}& -          & 6.1863(17) & 6.1743(10) &\bf{6.1766(9)} & -         & 6.1774(9)   \\
			$E_3$ & 6.3481(21)&\bf{6.3556(15)}& 6.3553(16) & -          & 6.1954(11) &\bf{6.1985(7)} & 6.1984(7) & -           \\
			$E_4$ & 6.3532(22)&\bf{6.3626(20)}& -          & 6.3626(20) & 6.2752(13) &\bf{6.2824(10)}& -         & 6.2822(10)  \\
			$E_5$ & 6.5247(27)&\bf{6.5129(20)}& -          & 6.5121(21) & 6.3745(16) &    6.3757(12) & -         & 6.3757(12)  \\
			$E_6$ & 6.6852(27)&    6.6792(19) & 6.6792(19) & -          & 6.3906(12) &    6.3934(8)  & 6.3930(8) & -           \\
			$E_7$ & 6.6918(31)&    6.6849(24) & -          & 6.6849(24) & 6.4724(18) &    6.4701(11) & -         & 6.4695(13)  \\
			$E_8$ & 6.8549(36)&    6.8410(20) & -          & 6.8391(22) & 6.5687(21) &    6.5602(13) & -         & 6.5592(15)  \\
			\hline
			&           &               &            &$m_\pi=250$~MeV&      &               &           &             \\
			\hline
			$E_1$ & 6.1714(18)&\bf{6.1737(37)}& 6.1737(37) & -          & 6.0996(10) &\bf{6.1004(8)} & 6.1004(8) & -           \\
			$E_2$ & 6.1633(20)&\bf{6.1730(26)}& -          & 6.1730(26) & 6.1706(10) &\bf{6.1736(10)}& -         & 6.1736(10)  \\
			$E_3$ & 6.3464(22)&\bf{6.3523(40)}& 6.3525(39) & -          & 6.2009(14) &\bf{6.2030(9)} & 6.2029(9) & -           \\
			$E_4$ & 6.3406(24)&\bf{6.3509(37)}& -          & 6.3508(37) & 6.2714(14) &\bf{6.2791(15)}& -         & 6.2790(14)  \\
			$E_5$ & 6.5130(28)&\bf{6.5026(21)}& -          & 6.5026(21) & 6.3706(18) &    6.3718(20) & -         & 6.3719(17)  \\
			$E_6$ & 6.6827(30)&    6.6745(20) & 6.6745(20) & -          & 6.3986(20) &    6.3967(13) & 6.3966(11)& -           \\
			$E_7$ & 6.6810(32)&    6.6743(20) & -          & 6.6743(20) & 6.4682(22) &    6.4657(14) & -         & 6.4656(14)  \\
			$E_8$ & 6.8449(36)&    6.8246(20) & -          & 6.8223(25) & 6.5645(25) &    6.5543(27) & -         & 6.5551(19)   
		\end{tabular}
	\end{ruledtabular}
\end{table*}
%%%%%%%%%%%%%%%%%%%%%%%%%%%

Similar to the $0^{++}$ case, we follow the same steps to check for possible coupled-channel effects between the ${}^1D_2$ $\eta_c\eta_c$ and ${}^5S_2$ $J/\psi J/\psi$ scattering in the $2^{++}$ channel: 
\begin{itemize}
	\item We perform the GEVP analysis to the correlation matrix of the three $\eta_c\eta_c$ operators with $n^2=1,2,4$ ($n^2=3$ mode has no $E(1)$ representations of the octahedral group O), and obtain three new energy levels $E_n'$, which are almost the same as the corresponding $E_n$ within errors (see Table~\ref{tab:energys_2pp}). 
	This indicates that the inclusion of $J/\psi J/\psi$ operators does not affect the values of the energy levels of the $\eta_c\eta_c$ system. 
	The lowest two $\eta_c\eta_c$ energies, $E_n$ ($E_n'$), lie above the corresponding noninteracting energies $E_n^{(0)}$. 
	Note that, since only three $\eta_c\eta_c$ operators are used, the third energy level may be contaminated by higher states.
	This indicates a repulsive interaction between the ${}^1D_2$ $\eta_c\eta_c$ system near the threshold.
	
	\item Then we perform the GEVP analysis to the correlation matrix of the five $J/\psi J/\psi$ operators and obtain five new energy levels $E_n''$, which are in good agreement with the corresponding $E_n$ values and exhibit the independence of the $\eta_c\eta_c$ system.
\end{itemize}
These observations indicate that in the $2^{++}$ channel, the coupled-channel effects between the ${}^1D_2$ $\eta_c\eta_c$ and ${}^5S_2$ $J/\psi J/\psi$ systems are negligible.

\begin{figure*}[t]
	\centering
	\includegraphics[width=0.45\linewidth]{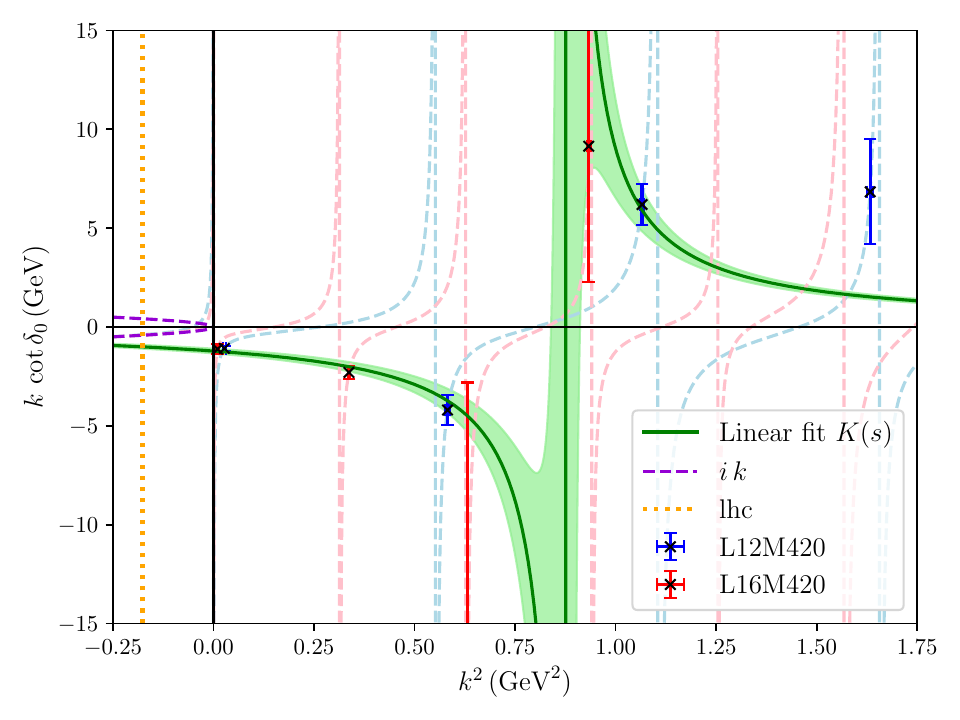}
	\includegraphics[width=0.45\linewidth]{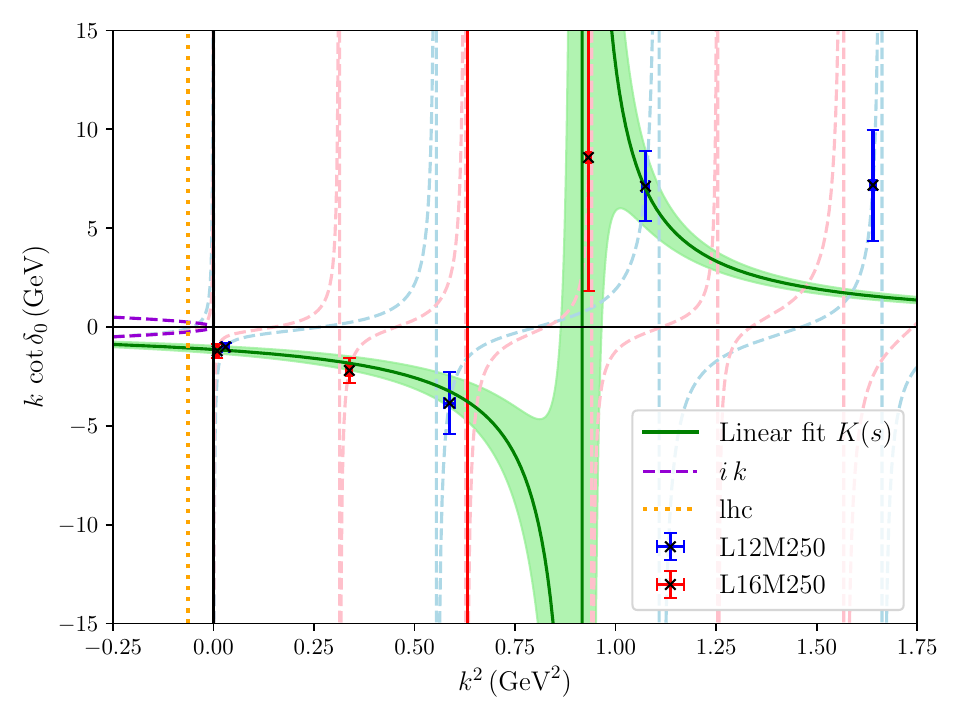}\\
	\includegraphics[width=0.45\linewidth]{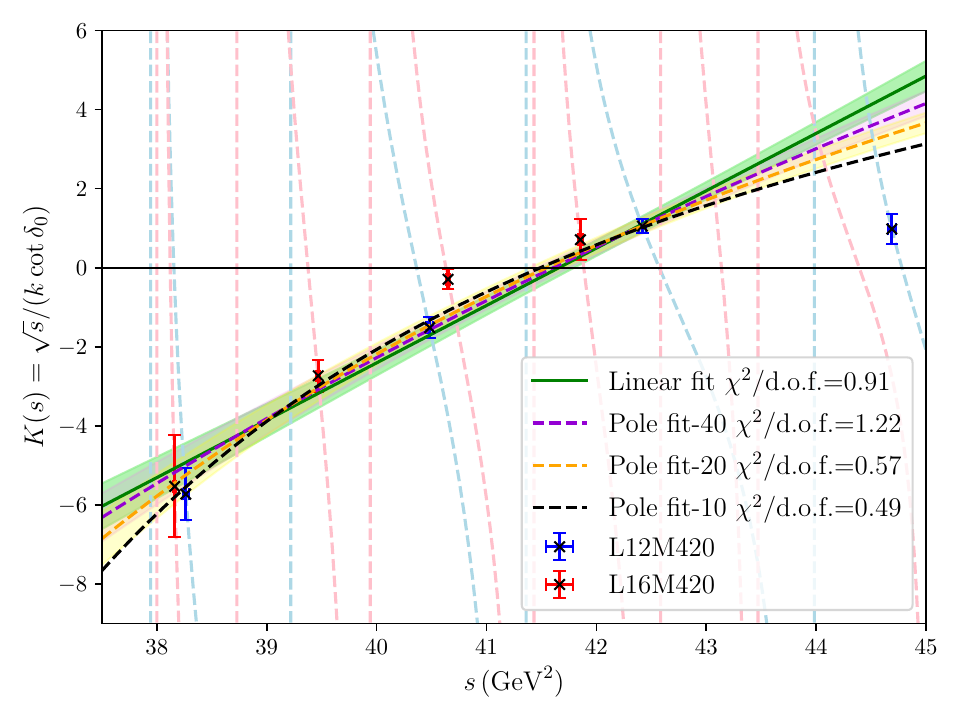}
	\includegraphics[width=0.45\linewidth]{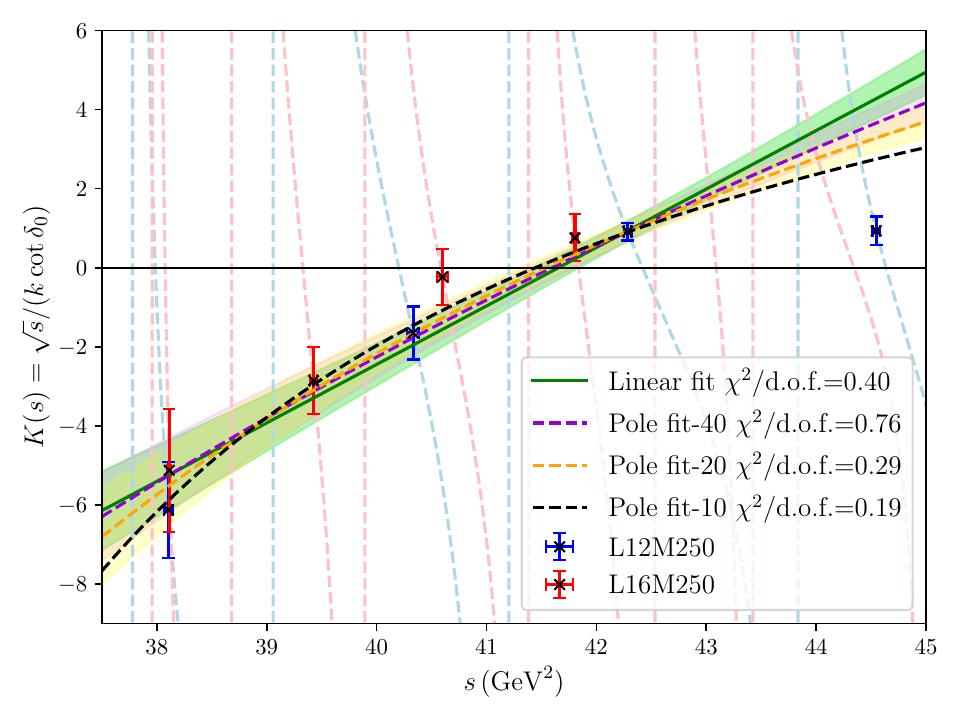}
	\caption{The scattering phases $k \cot \delta_0(k)$ (upper panels) and the $K(s)$ functions (lower panels) in the ${}^5S_2$ $J/\psi J/\psi$ channel on M420 (left) and M250 (right) ensembles. 
		The blue and red points represent the lattice measurements from the L12 and L16 ensembles, respectively. 
		The light dashed lines represent the Riemann zeta function. 
		The green band shows the linear fit in Eq.~\eqref{eq:linear}, and the purple, orange, and black curves correspond to the fits using Eq.~\eqref{eq:pole-term} with various initial $g_0$ values. The $\chi^2/{\rm d.o.f.}$ values for different fits are also given. 
		Note that the fits use only the leftmost three blue points and the leftmost two red points, which are considered to be reliable (see Sec.~\ref{secII:NV-dep}).}       
	\label{fig:kcot_jpsi}
\end{figure*}

Consequently, a single-channel scattering analysis is carried out for the ${}^5S_2$ $J/\psi J/\psi$ system. 
The values of the scattering phase $k\cot\delta_0(k)$ are extracted from the $J/\psi J/\psi$ energy levels in Table~\ref{tab:energys_2pp} through Eq.~\eqref{eq:phase}, which are plotted in the two upper two panels of Fig.~\ref{fig:kcot_jpsi} (the left panel for M420 and the right one for M250). 
In contrast to the previous cases, the values of $k\cot\delta_0(k)$ in this channel exhibits a very interesting behavior that $k\cot\delta_0(k)$ seems having two branches: $k\cot\delta_0(k)$ is negative and decreases monotonically with $k^2$ when $k^2$ is lower than $\sim 0.9~\mathrm{GeV}^2$, while it is positive and also decreases monotonically with $k^2$ when $k^2$ is beyond that value. 
Clearly, this behavior cannot be well-described by a polynomial function like Eq.~\eqref{eq:ERE} or Eq.~\eqref{eq:erek4}.
Thus, we employ the $K$-matrix formalism to parametrize the scattering amplitude $t(s)$, namely,
\begin{equation}
	t^{-1}(s)=K^{-1}(s)-i\rho(s)
\end{equation}
with $K(s)$ being a real function of $s$. 
Accordingly, the scattering phase shift can be related to the $K$-matrix as
\begin{equation}\label{eq:phase-th}
	k\cot\delta_0(k)=\sqrt{s}K^{-1}(s).
\end{equation}
An initial attempt is to adopt the single-pole functional form for $K(s)$, 
\begin{equation}\label{eq:pole-term}
	K(s)=\frac{g_0^2}{m_0^2-s}+\gamma,
\end{equation}
where the bare coupling $g_0^2>0$, the bare pole $m_0^2$ and $\gamma$ are parameters to be determined. 
It should be noted that the fit is performed using the energy levels under $6.6~(6.3)$~GeV on the L12~(L16) volume, as discussed in Sec.~\ref{secII:NV-dep}.
In the practical data analysis, it is found that the fitted parameters $g_0^2,m_0^2, \gamma$ are unstable and depend on their initial values. 
The fitting results with various initial values $g_0=(10, 20, 40)$~GeV (labeled as `Pole fit-(10,20,40)') are listed in the upper block of Table~\ref{tab:pole}. 
Although this functional form describes the data well, the parameter values differ significantly and exhibit strong correlations among $g_0$, $m_0^2$, and $\gamma$.
It indicates that the functional form in Eq.~\eqref{eq:pole-term} may be unsuitable and could contain a redundant parameter.

%%%%%%%%%%%%%%%%%%%%%%%%%%%
%%%%%%%%%%%%%%%%%%%%%%%%%%%

%%%%%%%%%%%%%%%%%%%%%%%%%%%
\begin{table*}[t]
	\caption{The $K$-matrix fitting parameters and the properties of the resonance~($R$) in the ${}^5S_2$ $J/\psi J/\psi$ scattering. 
		The linear fit results highlighted in bold are considered the most appropriate.}
	\label{tab:pole}
	\begin{ruledtabular}
		\begin{tabular}{lrrrrr}
			Parameters  & Pole fit-10 M420 & Pole fit-20 M420 & Pole fit-40 M420 & \textbf{Linear fit M420} & \textbf{Linear fit M250} \\
			\hline
			$g_0$ (GeV)                          & 11.5(1.5)     & 18.8(1.2)    & 38.6(0.7)       & -               & -               \\
			$m_0^2$ (GeV$^{2}$)                & 31.0(1.3)     & 24.9(1.4)    & 8.3(1.8)        & -               & -               \\
			$\gamma$                           & 12.5(1.8)     & 21.3(1.2)    & 44.9(1.0)       & -               & -               \\
			$a$ (GeV$^{-2}$)                & -            & -            & -                & 1.45(12)        & 1.48(20)        \\
			$b$                             & -            & -            & -                & -60.4(4.9)      &-61.6(8.4)       \\
			\hline
			$\operatorname{Re}~(k_R)$ (GeV)           & 1.073(25)    & 1.108(20)    & 1.130(18)        & 1.140(17)       & 1.148(24)       \\
			$\operatorname{Im}~(k_R)$ (GeV)           & -0.572(31)   & -0.499(18)   & -0.446(21)       & -0.393(20)      & -0.383(34)      \\
			$m_R$ (GeV)                        & 6.449(21)    & 6.495(12)    & 6.524(11)        & 6.543(10)       & 6.538(13)       \\
			$\Gamma_R$ (GeV)                   & 0.761(42)    & 0.681(32)    & 0.617(35)        & 0.548(34)       & 0.537(56)       \\
			$|c_R|^2$ (GeV$^2$)                & 22.2(1.0)    & 19.7(0.5)    & 18.0(0.6)        & 16.2(0.7)       & 15.9(1.1)       \\
			$|c_R|$ (GeV)                      & 4.71(10)     & 4.39(6)      & 4.44(6)          & 4.02(8)         & 3.99(14)        \\
			$\Gamma_{J/\psi J/\psi}$ (GeV)     & 0.494(19)    & 0.469(17)    & 0.443(19)        & 0.408(19)       & 0.406(31)       \\
			$\mathrm{Br}_{J/\psi J/\psi}$      & 65(2)\%      & 69(2)\%      & 72(2)\%          & 74(2)\%         & 75(3)\%         \\
		\end{tabular}
	\end{ruledtabular}
\end{table*}
%%%%%%%%%%%%%%%%%%%%%%%%%%%

%%%%%%%%%%%%%%%%%%%%%%%%%%%
\begin{figure}[t]
	\centering
	\includegraphics[width=0.99\linewidth]{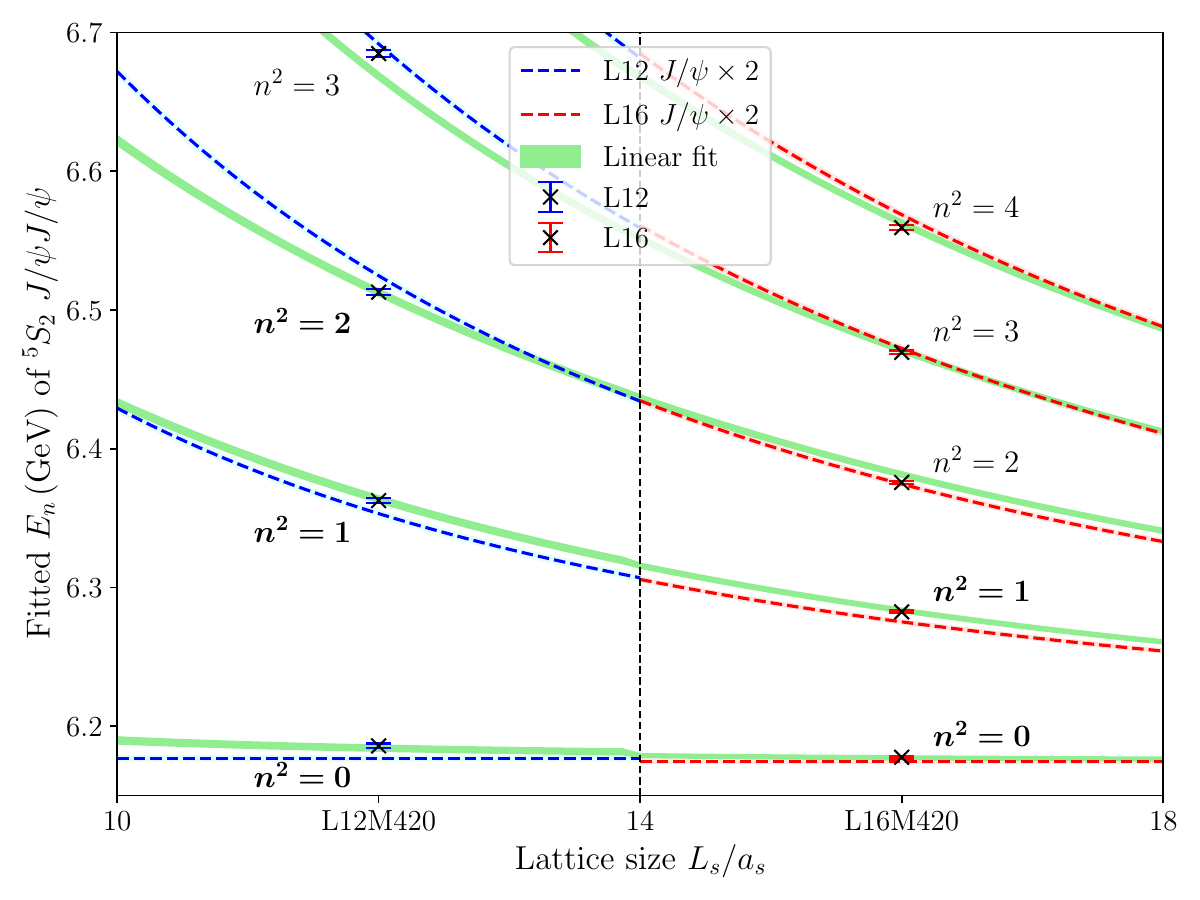}
	\caption{Energy levels with linear fits for the ${^5}S_2$ $J/\psi J/\psi$ system on M420. Energy levels $E_n(n^2)$ used for fitting are highlighted in bold.
		The green bands shows the theoretical predictions of FVEs based on the linear fit of $K$-matrix.}
	\label{fig:fit-spectrum}
\end{figure}
%%%%%%%%%%%%%%%%%%%%%%%%%%%

The pole behavior of $k\cot\delta_0(k)$ motivates us to find a proper function form for $K(s)$ by examining the $k^2$-dependence of $K(s)=\sqrt{s}(k\cot\delta_0(k))^{-1}$, which are plotted in the lower two panels of Fig.~\ref{fig:kcot_jpsi}.
The dashed lines with error bands illustrate the three fits using Eq.~\eqref{eq:pole-term} with different initial $g_0$ values. 
It is observed that the $(k \cot \delta_0(k))^{-1}$ data points from the lattice exhibit an almost linear dependence on $k^2$. 
This behavior also explains the strong correlations observed among the three parameters in Eq.~\eqref{eq:pole-term}.
The approximate linear behavior motivates us to paramterize $K(s)$ as
\begin{equation}\label{eq:linear}
	K(s)=a ~ s + b.
\end{equation}
The fitted values of the parameters $a$ and $b$ are listed in Table~\ref{tab:pole}.
The fit quality of the linear $K(s)$ function can be evaluated by comparing the lattice energy levels $E_n$ with the theoretical predictions obtained using L\"{u}scher's quantization condition of Eq.~\eqref{eq:Luscher}, or more specifically Eq.~\eqref{eq:phase} for single-channel $S$-wave scattering, based on the fitted parameters $a$ and $b$.
This comparison is illustrated in Fig.~\ref{fig:fit-spectrum}, where the dashed lines are the noninteracting energies of the di-$J/\psi$ system with respective to the lattice size $L_s$, the data points are the lattice energy levels, and the colored bands are the theoretical predictions through the L\"{u}scher's quantization condition and agree with the lattice data up to $6.6~(6.3)$~GeV very well. 

%%%%%%%%%%%%%%%%%%%%%%%%%%%
\begin{figure*}[t]
	\centering
	\includegraphics[width=0.45\linewidth]{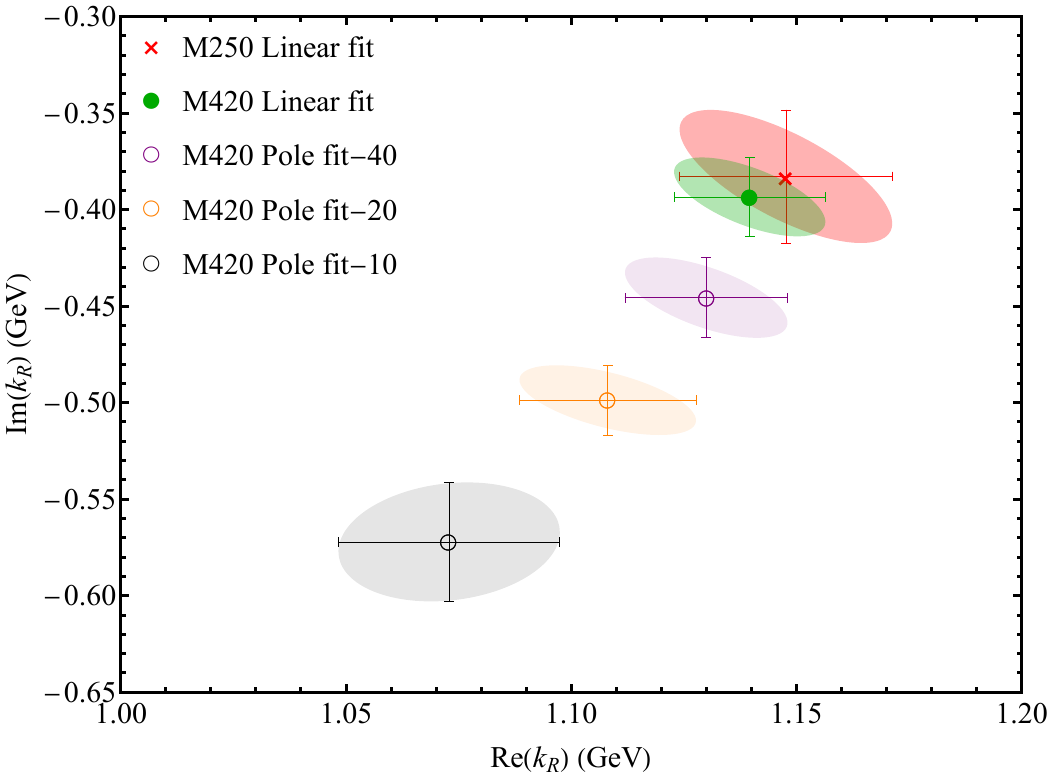}
	\includegraphics[width=0.45\linewidth]{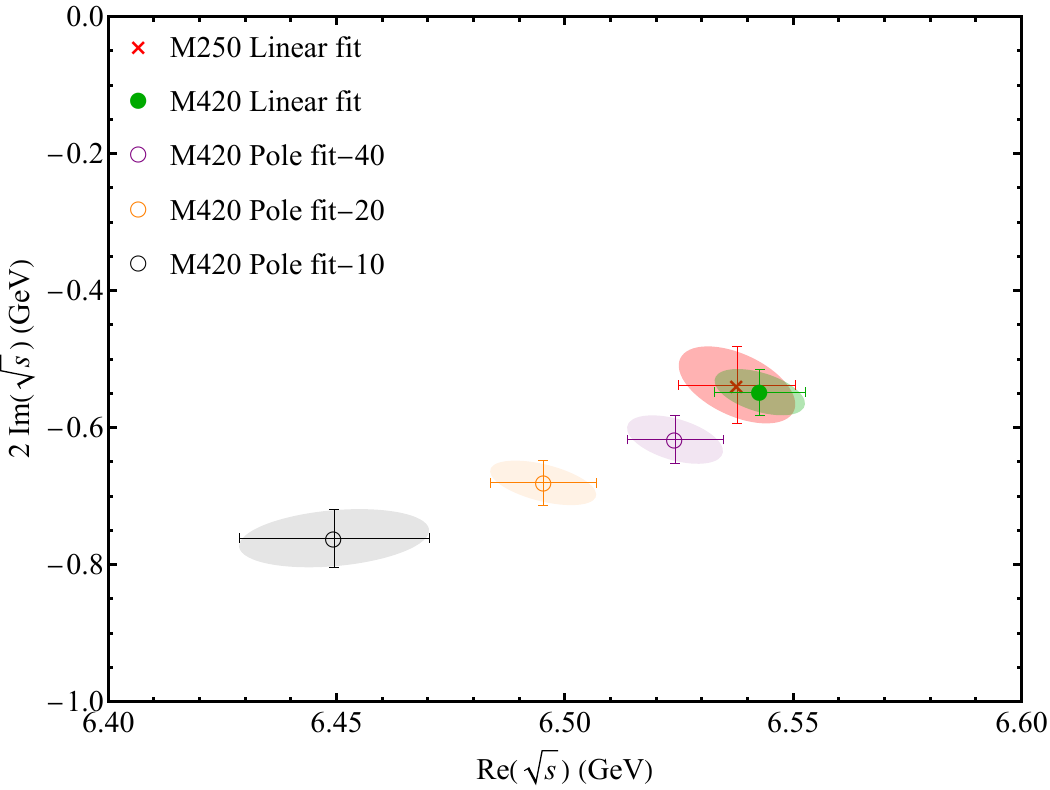}
	\caption{The resonance pole positions in the complex $k$-plane (left panel) and $\sqrt{s}$-plane (right panel).
		With increasing initial values of the pole fitting parameter $g_0$, the pole position converges toward the linear fit result (the green point for M420 and red one for M250).}
	\label{fig:pole-position}
\end{figure*}
%%%%%%%%%%%%%%%%%%%%%%%%%%%

%%%%%%%%%%%%%%%%%%%%%%%%%%%
\begin{figure*}[t]
	\centering
	\includegraphics[width=0.45\linewidth]{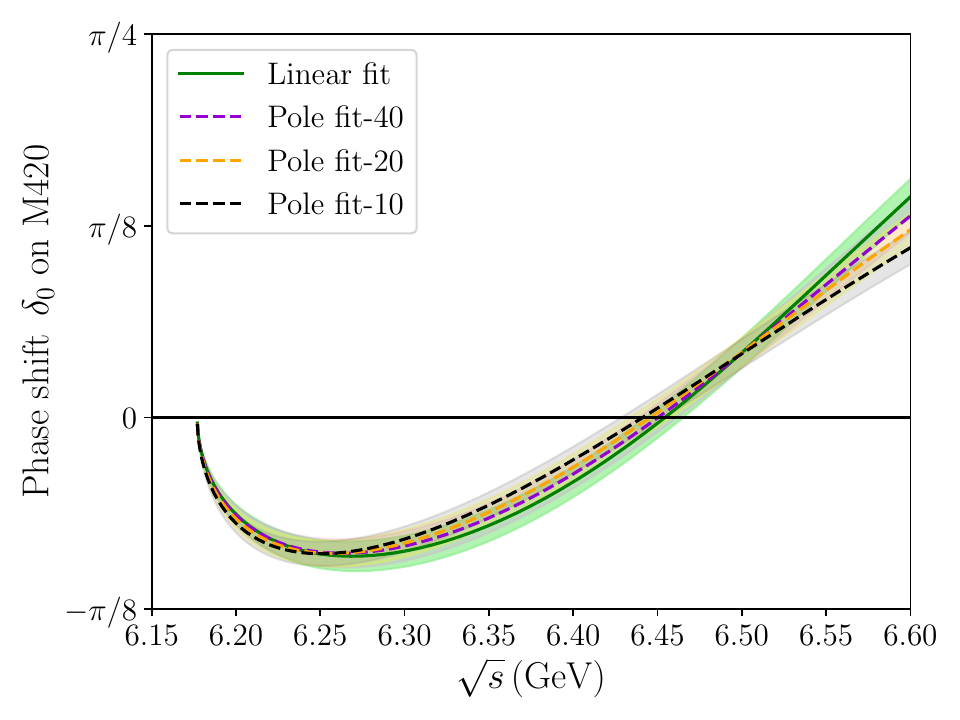}
	\includegraphics[width=0.45\linewidth]{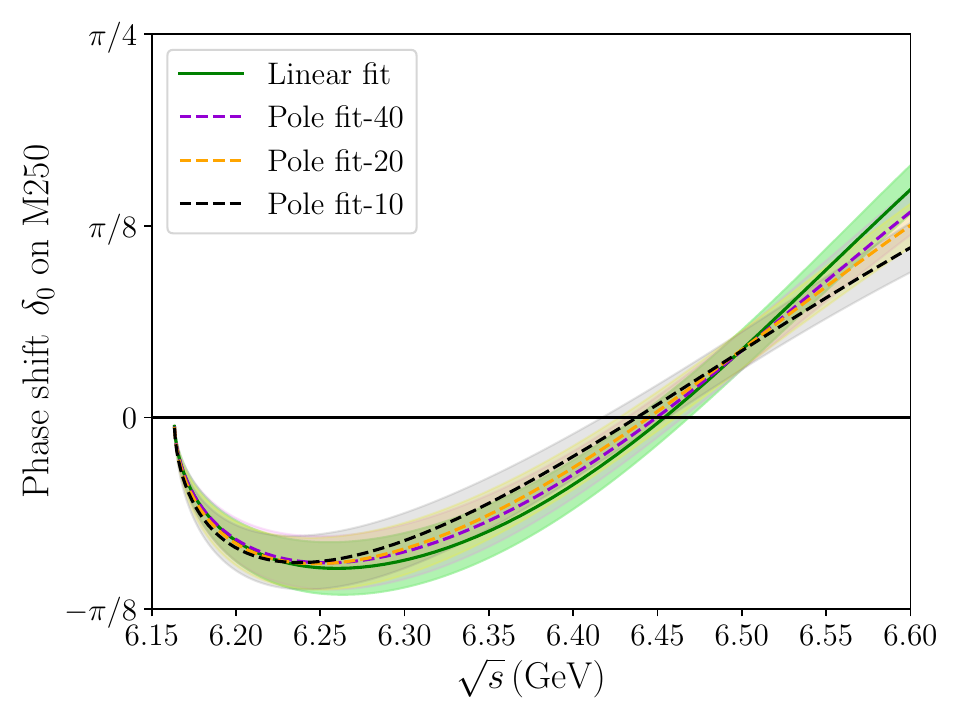}\\
	\includegraphics[width=0.45\linewidth]{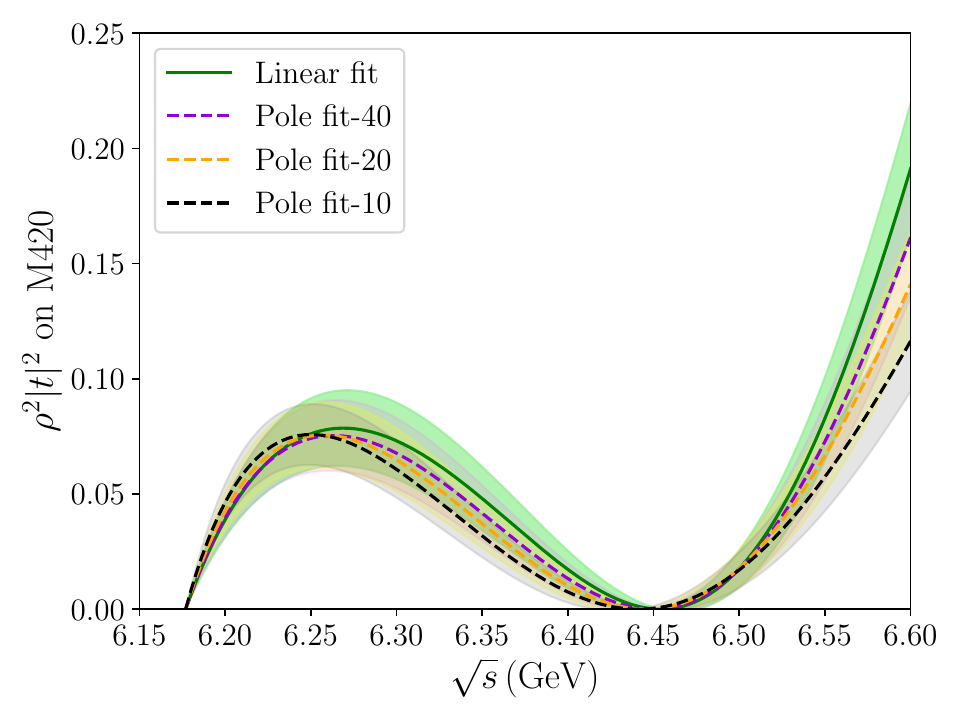}
	\includegraphics[width=0.45\linewidth]{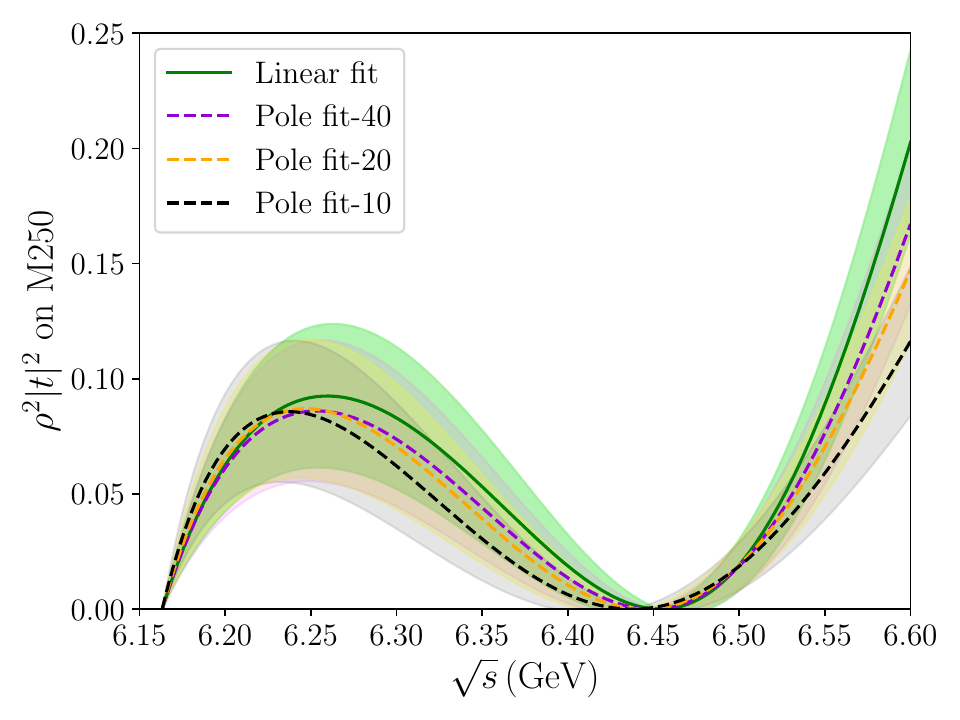}
	\caption{The scattering phase shift $\delta_0$ (upper panels) and the squared modulus of the transition matrix (lower panels) are shown as functions of $\sqrt{s}$ for the ${^5}S_2$ $J/\psi J/\psi$ channel on the M420 (left panels) and M250 (right panels) ensembles.
		The curves shown in different colors correspond to the results with various fits for $K(s)$. }
	\label{fig:phase-2pp}
\end{figure*}
%%%%%%%%%%%%%%%%%%%%%%%%%%%

The parametrization in Eq.~\eqref{eq:linear} carries significant physical implications.
Firstly, the scattering length $a_0$ in this channel can be obtained by 
\begin{equation}
	\frac{1}{a_0}=\lim\limits_{k\to 0} k\cot\delta_0(k)=\frac{ E_\mathrm{th}}{a E_\mathrm{th}^2+b},
\end{equation}
where $E_\mathrm{th}$ is di-$J/\psi$ threshold. With the obtained values of $a$ and $b$, one has 
\begin{equation} 
	\begin{aligned}
		&{\rm M420:~}a_0= -0.163(16) ~{\rm fm},\\ 
		&{\rm M250:~}a_0= -0.171(29) ~{\rm fm}, 
	\end{aligned} 
\end{equation}
which are consistent with that reported in Ref.~\cite{Meng:2024czd}, where the scattering length in the continuum limit is given as $a_0^{2^{++}} = -0.165(16)$~fm. 
Secondly, $K(s)$ has a zero point given by $s_0=-{b}/{a}$. 
At the two pion masses, we have the explicit values,
\begin{equation} 
	\begin{aligned}
		&{\rm M420:~}\sqrt{s_0}= 6.454(11) ~{\rm GeV}, ~(k_0^2 = 0.876(36)~{\rm GeV}^2),\\ 
		&{\rm M250:~}\sqrt{s_0}= 6.454(15) ~{\rm GeV}, ~(k_0^2 = 0.917(51)~{\rm GeV}^2). 
	\end{aligned} 
\end{equation}
This zero is responsible for the divergence of $k \cot \delta_0(k)$.
Note that $K(s)$ is related to the scattering amplitude $t(s)$ through
\begin{equation}\label{eq:ampK}
	t(s)=\frac{K(s)}{1-i\rho(s)K(s)}.
\end{equation}
Thus, the zero of $K(s)$ corresponds to the zero of $t(s)$, which is commonly referred to as a Castillejo-Dalitz-Dyson (CDD) zero~\cite{Castillejo:1955ed,Oller:2024lrk} and leads to a zero in the scattering cross section.
Though the origin of CDD zeros remains unclear, there are theoretical arguments suggesting that the CDD zeros are closely related to discrete bare states in a free theory~\cite{Dyson:1957rgq,Krivoruchenko:2010ft,Li:2021cue}. 
It is interesting that the value $\sqrt{s_0}\approx 6.45$~GeV is compatible with the phenomenological prediction of the lowest $2^{++}$ $cc\bar{c}\bar{c}$ tetraquark state~\cite{Chen:2016jxd,Lu:2020cns,Wang:2021kfv}.

With the parametrizations of $K(s)$ in Eqs.~\eqref{eq:pole-term} and \eqref{eq:linear} and the determined parameters, we obtain the scattering amplitude $t(s)$ through Eq.~\eqref{eq:ampK}. 
Then, the pole singularities of $t(s)$ can be investigated by solving the pole equation on the analytically continued complex $s$-plane (or equivalently, the complex $k$-plane)
\begin{equation}
	1 - i \rho(s) K(s) = 0.
\end{equation}
For each parametrization, we obtain the three poles in the complex $k$-plane, namely, a pure imaginary pole $k_B=i|k_B|$ and a pair of complex poles $k_R=\pm|\mathrm{Re}(k_R)|-i|\mathrm{Im}(k_R)|$. 
The imaginary pole, located in the region $k_B^2 < -0.34~\rm{GeV}^2$, lies far to the left of the left-hand cut at $k^2 \sim -m_\pi^2$ and is therefore considered unphysical.
The complex-conjugate pole pair lies in the lower half of the complex $k$-plane, corresponding to poles on the second Riemann sheet.
A physical resonance pole with $\mathrm{Re}(k_R)>0$ is located at $\sqrt{s_R} = m_R - i {\Gamma_R}/{2}$.
The parameters $\mathrm{Re}(k_R)>0, \mathrm{Im}(k_R), m_R$ and $\Gamma_R$ of each parametrization are listed in the lower block of Table~\ref{tab:pole}.
Figure~\ref{fig:pole-position} shows the pole position in the complex $k$-plane with $\operatorname{Re}~(k_R)>0$ (left panel) and the complex $\sqrt{s}$-plane (right panel). 
For the M420 ensemble, the pole positions obtained from the three fits (the grey, orange, and purple points) using the $K(s)$ function in Eq.~\eqref{eq:pole-term} vary within a certain region.
As shown in the lower two panels of Fig.~\ref{fig:kcot_jpsi}, the three fits exhibit slight differences in the energy region above $\sqrt{s} > 6.52~\mathrm{GeV}$, where no reliable lattice data are available.
The green (red) points in Fig.~\ref{fig:pole-position} are obtained via the linear $K(s)$ function Eq.~\eqref{eq:linear} on the M420 (M250) ensemble. 
One can see that the results at the two different pion masses are consistent with each other and exhibit little dependence on $m_\pi$.

When $s$ is close to the pole $s_R$, the scattering amplitude $t(s)$ has the asymptotic form given by the residue theorem
\begin{equation}\label{eq:near-pole}
	t(s)\approx \frac{c_R^2}{s_R-s},
\end{equation}
where the residue of the pole gives the effective coupling $c_R$. 
For each parametrization of $K(s)$ we obtain a complex $c_R^2$ whose modulus $|c_R|^2$ (and also $|c_R|$) are shown in Table~\ref{tab:pole}.
With the effective coupling $|c_R|^2$, we perform a self-consistent check by calculate the partial decay width $\Gamma(R\to J/\psi J/\psi)$ tentatively in the narrow resonance approximation.
Assuming the coupling $|c_R|^2$ at the resonance pole is insensitive to energy,  the partial decay width of the $S$-wave $J/\psi J/\psi$ decay mode is
\begin{equation}\label{eq:partial}
	\Gamma(J/\psi J/\psi) = |c_R|^2 \frac{k}{m_R^2} 
\end{equation}
where $k=\sqrt{m_R^2/4-m_{J/\psi}^2}$ is the decay momentum, and the symmetry factor $1/2$ for the di-$J/\psi$ system has been considered implicitly. 
The obtained partial width is about $65\%\sim 75\%$ of the total width $\Gamma$ determined by the pole position (see Table~\ref{tab:pole}). 
Although $|c_R|^2$ and $(m_R, \Gamma_R)$ are model-independent quantities associated with the resonance pole, the ratio $\mathrm{Br}_{J/\psi J/\psi} = \Gamma(J/\psi J/\psi)/\Gamma_R$ should not be directly identified with the physical branching fraction of the $J/\psi J/\psi$ decay mode.
Firstly, there do not exist other decay modes that can have totally a branching fraction of 20\%-30\%. 
Secondly, the narrow-width approximation in Eq.~\eqref{eq:partial} is not expected to be valid, since the total width is as large as $\sim 500~\mathrm{MeV}$.
Thirdly, this resonance lies close to the di-$J/\psi$ threshold and may therefore be affected by threshold effects.
Nevertheless, the estimated value of $\mathrm{Br}_{J/\psi J/\psi}$ can serve as a rough consistency check for the properties of this resonance.

To summarize, the lattice data points of $K(s) = \sqrt{s}~(k \cot \delta_0(k))$ exhibit an approximately linear behavior in $s$, while the functional form in Eq.~\eqref{eq:pole-term} yields unstable fit results for the parameters $(g_0, m_0^2, \gamma)$. 
We therefore adopt the linear parametrization in Eq.~\eqref{eq:linear} as the most suitable choice.
A resonance is observed in the $2^{++}$ $S$-wave $J/\psi J/\psi$ scattering, whose parameters are given by
\begin{equation} 
	\begin{aligned}
		&{\rm M420:~}(m_R,\Gamma_R)= (6.543(10),~0.548(34))~{\rm GeV},\\ 
		&{\rm M250:~}(m_R,\Gamma_R)= (6.538(13),~0.537(56))~{\rm GeV}, 
	\end{aligned} 
\end{equation}
along with the effective pole couplings $|c_R| = 4.02(8)~\mathrm{GeV}$ and $3.99(14)~\mathrm{GeV}$ for $m_\pi = 420~\mathrm{MeV}$ and $250~\mathrm{MeV}$, respectively, where the uncertainties are statistical only.  

The ${}^5S_2$ $J/\psi J/\psi$ scattering phase shift $\delta_0(\sqrt{s})$ is calculated explicitly using Eq.~\eqref{eq:phase-th} with different fitted parameters, and its dependence on $\sqrt{s}$ along the real axis is shown in the upper two panels of Fig.~\ref{fig:phase-2pp} (the left panel for M420 and the right panel for M250).
It is seen that $\delta_0(\sqrt{s})$ exhibits a monotonic increase in the resonant energy region $\sqrt{s} \in [6.3,~6.6]~\mathrm{GeV}$, consistent with the presence of a resonance.
Since the observed width $\Gamma_R$ of this resonance is as large as $500$-$600~\mathrm{MeV}$, and its mass $m_R \simeq 6.54~\mathrm{GeV}$ lies close to the di-$J/\psi$ threshold, the energy interval $\sqrt{s} \in [6.2,~6.6]~\mathrm{GeV}$ does not cover the full resonant region. 
As a result, the total variation of the phase shift observed within this energy range is smaller than $\pi$.
On the other hand, one observes that $\delta_0(\sqrt{s})$ evolves from negative to positive values as $\sqrt{s}$ increases, and crosses zero around $\sqrt{s}\simeq 6.45~\mathrm{GeV}$. 
This behavior implies the presence of a zero of the scattering amplitude $t(s)$ at this energy.

We also plot the cross section $\rho^2 |t(s)|^2$ of the $J/\psi J/\psi$ elastic scattering as a function of $\sqrt{s}$ on the real axis in the lower panels of Fig.~\ref{fig:phase-2pp}, which may provide useful theoretical input for experimental analyses of the $J/\psi J/\psi$ invariant mass spectrum.

%%%%%%%%%%%%%%%%%%%%%%%%%%%
\section{Discussion}\label{sectionIV}
%%%%%%%%%%%%%%%%%%%%%%%%%%%
Having presented the numerical results, we make discussions on the physical implications and explore the possible dynamics for dicharmonium scattering which is reflected by the lattice QCD data. 
Since the $\eta_c\eta_c$ and $J/\psi J/\psi$ coupled channel effects are observed to be negligible, we focus on the single-channel scattering. 

\subsection{$\eta_c\eta_c$ scattering}
As shown in Table~\ref{tab:energys} and Table~\ref{tab:energys_2pp}, the $\eta_c\eta_c$ energies in the $(0,2)^{++}$ channels are uniformly higher than the corresponding noninteracting energies (except for the $n^2=4$ energies, which are the highest ones in our analysis and are unreliable owing to the contamination from even higher states and the momentum suppression discussed in Sec.~\ref{secII:NV} and Sec.~\ref{secII:NV-dep}).
In the meaning time, the scattering analysis gives the scattering length $a_0\approx -0.12$~fm for the ${}^1S_0$ $\eta_c\eta_c$ scattering. 
These observations suggest that the $\eta_c \eta_c$ system exhibits a repulsive interaction in both the ${}^1S_0$ and ${}^1D_2$ channels. 
It is intriguing to explore the intrinsic dynamics responsible for this repulsive interaction.

The $\eta_c\eta_c$ correlation functions include the contributions from the `Direct' diagram~(a) and the `quark recombination' diagram~(b) in Fig.~\ref{fig:diagrams}. 
On the hadron level, diagram~(a) reflects the color singlet gluonic interaction between two $\eta_c$ mesons, which can be described either by the Pomeron exchange mechanism~\cite{Gong:2022hgd} or the light meson exchange in the presence of sea quarks.
Similarly, diagram~(b) represents either the charmonium interchange effect or the quark rearrangement (QR) effect. 
In order to analyze the nature of the near-threshold interaction, we consider the correlation function $C_{11}(t)$ from the correlation matrix $C_{\alpha\beta(t)}$ in the $0^{++}$ channel, where the operator $\mathcal{O}_\alpha\in \mathcal{S}_0$ with $\alpha=1$ is the $\eta_c\eta_c$ operator with zero relative momentum (namely, $n^2=0$). 
The correlation $C_{11}(t)$ can be expressed as 
\begin{equation}\label{eq:DQ}
	C_{11}(t)=D(t)+Q(t)\equiv D^{(0)}(t)+D^{(1)}(t)+Q(t)
\end{equation}
where $D(t)$ and $Q(t)$ represent the contribution from the direct diagram (a) and the QR diagram (b), respectively, and $D(t)$ can be split qualitatively into two parts, namely $D^{(0)}(t)$ and $D^{(1)}(t)$, with $D^{(0)}(t)$ being the propagator of two noninteracting $\eta_c$ and $D^{(1)}(t)$ representing the gluon (light hadron) exchange effects. Let $D^{(0)}(t)$ be $\langle \mathcal{O}_{\eta_c}(t)\mathcal{O}_{\eta_c}(0)\rangle^2$, $D^{(1)}(t)$ is calculated by $D^{(1)}(t)=D(t)-D^{(0)}(t)$.
When only $D(t)$ is included in $C_{11}(t)$, the obtained energy is smaller than the $\eta_c\eta_c$ threshold, namely, $E_1<2m_{\eta_c}$. 
This indicates an attractive interaction. 
When both $D(t)$ and $Q(t)$ are included, we find that $E_1>2m_{\eta_c}$. 
This manifests that the repulsive interaction reflected by diagram~(b) overcomes the attractive one result from gluon (or light hadron) exchanges (diagram~(a)). 

%%%%%%%%%%%%%%%%%%%%%%%%%%%
\begin{figure*}[t]
	\centering
	\includegraphics[width=0.45\linewidth]{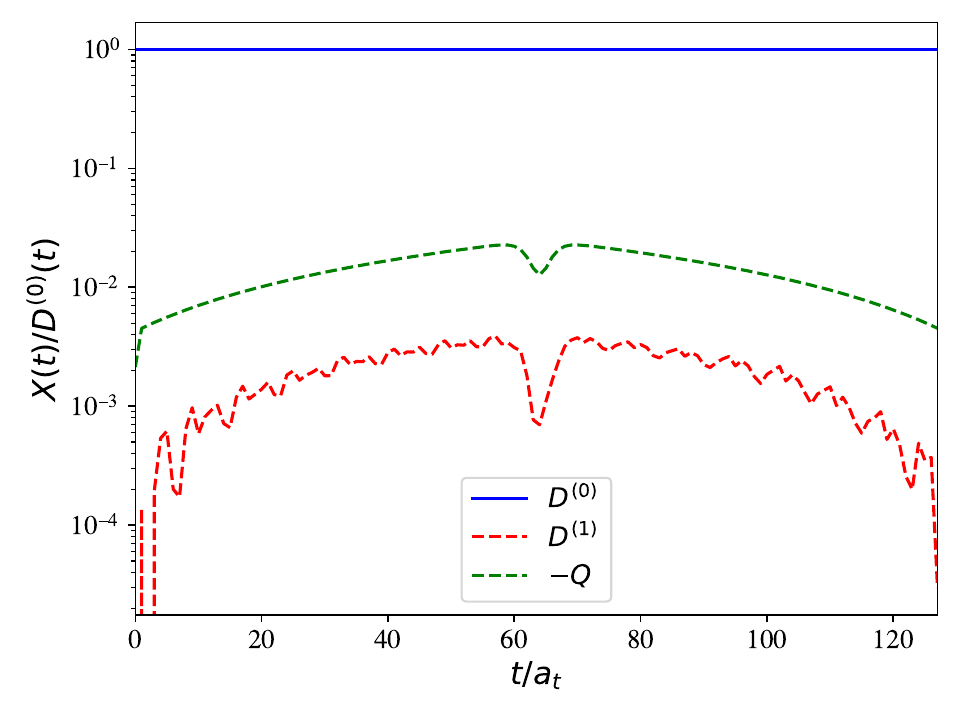}
	\includegraphics[width=0.45\linewidth]{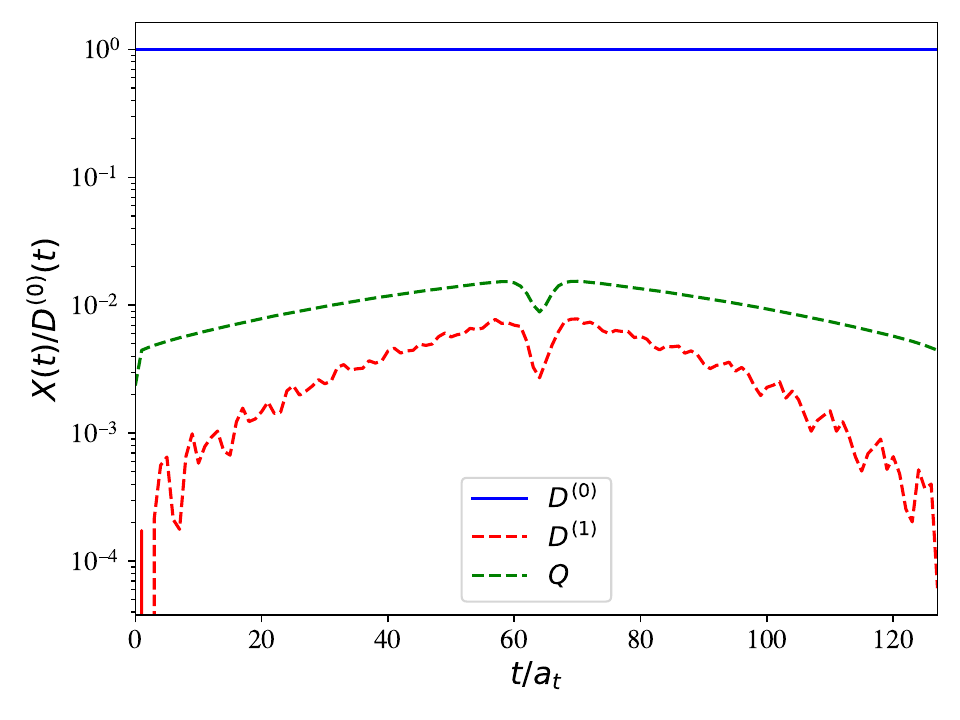}
	\caption{Ratio functions $X(t)/D^{(0)}(t)$ for the ${}^1S_0$ $\eta_c \eta_c$ system~(left) and the ${}^1S_0$ $J/\psi J/\psi$ system~(right).
		Here $X(t)$ refers to $D^{(0)}$, $D^{(1)}$, and $Q(t)$. in Eq.~\eqref{eq:DQ}.}
	\label{fig:ct-DC}
\end{figure*}
%%%%%%%%%%%%%%%%%%%%%%%%%%%

To illustrate this clearly, we introduce the ratio functions,
\begin{equation}
	\frac{D^{(1)}(t)}{D^{(0)}(t)}\approx \epsilon_D e^{-\delta_D t},~~~\frac{Q(t)}{D^{(0)}(t)}\approx \epsilon_Q e^{-\delta_Q t},
\end{equation}
with $\delta_{D,Q}$ being the energy shift from $2m_{\eta_c}$, as shown in the left panel of Fig.~\ref{fig:ct-DC}, where the ratio functions are labeled by $X(t)/D^{(0)}(t)$ with $X(t)$ referring to $D^{(0)}$, $D^{(1)}$, and $Q(t)$.. 
One can easily read out 
\begin{eqnarray}
	\delta_D<0,&&~~~\epsilon_D\sim \mathcal{O}(10^{-3}),\nonumber\\
	\delta_Q<0,&&~~~\epsilon_Q\sim -\mathcal{O}(10^{-2}),
\end{eqnarray}
and the slopes indicate that $a_t|\delta_Q|\sim a_t|\delta_D|\ll 1$.
Following the strategy in Ref.~\cite{Chen:2022vpo}, we have 
\begin{eqnarray}
	E_1&\approx& \frac{1}{a_t}\ln\frac{C_{11}(t)}{C_{11}(t+a_t)}\nonumber\\
	&\approx& 2m_{\eta_c}+\epsilon_D \delta_D e^{\delta_D t}+\epsilon_Q \delta_Q e^{\delta_Q t}\nonumber\\
	&\approx& 2m_{\eta_c}+\epsilon_D \delta_D +\epsilon_Q \delta_Q.
\end{eqnarray}
It is evident from this equation that when diagram~(b) is neglected ($\epsilon_Q=0$), one has $E_1-2m_{\eta_c}=\epsilon_D \delta_D<0$ and therefore the attractive interaction. 
When diagram~(b) is considered, $\epsilon_Q \delta_Q>0$ and $|\epsilon_Q|\sim 10|\epsilon_D|$ result in the totally repulsive interaction ($E_1-2m_{\eta_c}>0$).

\subsection{${}^1S_0$ $J/\psi J/\psi$ scattering}
The ${}^1S_0$ $J/\psi J/\psi$ scattering exhibits an attractive interaction near the threshold. 
We carry out a similar analysis for this case, as the ratio functions are shown in the right panel of Fig.~\ref{fig:ct-DC}. 
The attractive interaction can be partially understood as Yukawa-type, mediated by the exchange of $0^{++}$ charmonium (such as the $\chi_{c0}$ and its excited states). 
However, these contributions are found to be very short-ranged and are highly suppressed due to the large masses of the charmonia. 
On the other hand, the condition $\epsilon_Q > 0$ now holds, providing a key explanation for the dominant mechanism responsible for the attractive interaction in this channel.
The sign difference between $D(t)$ and $Q(t)$ in the $\eta_c\eta_c$ channel arises from the differing numbers of quark loops; diagram~(a) involves two quark loops, whereas diagram~(b) involves only one. 
Since each quark loop contributes a minus sign, this accounts for the overall sign discrepancy.
The same signs of $D(t)$ and $Q(t)$ in the $J/\psi J/\psi$ case arise from the Fierz rearrangement.
For the convenience within the constituent quark picture, we label the quark configurations of the initial state produced by the source operator at $t=0$ to be $\bar{c}_1\Gamma c_2\bar{c}_3\Gamma c_4$. 
Then diagram~(b) describes the QR process from the initial configuration to the final configuration $\bar{c}_1\Gamma c_4 \bar{c}_3\Gamma c_2$. 
For the $\eta_c\eta_c$ system, one has $\Gamma=\gamma_5$. 
The Fierz rearrangement relates the initial and the final configurations (not fermion fields) by
\begin{equation}
	\bar{c}_1\gamma_5 c_2\bar{c}_3\gamma_5 c_4 \to \frac{1}{4}\times \bar{c}_1\gamma_5 c_4 \bar{c}_3\gamma_5 c_2,
\end{equation}
where there is no additional sign. 
In contrast, for the $J/\psi J/\psi$ system, the Fierz transformation gives
\begin{equation}
	\bar{c}_1\gamma_\mu c_2\bar{c}_3\gamma^\mu c_4 \to -\frac{1}{2}\times \bar{c}_1\gamma_\mu c_4\bar{c}_3\gamma^\mu c_2,
\end{equation}
where the minus sign here cancels the additional minus sign in diagram~(a) due to the one more quark loop. 

%%%%%%%%%%%%%%%%%%%%%%%%%%%
\begin{figure}[t]
	\centering
	\includegraphics[width=0.45\linewidth]{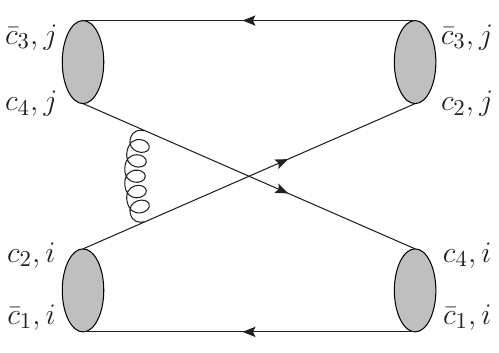}
	\includegraphics[width=0.45\linewidth]{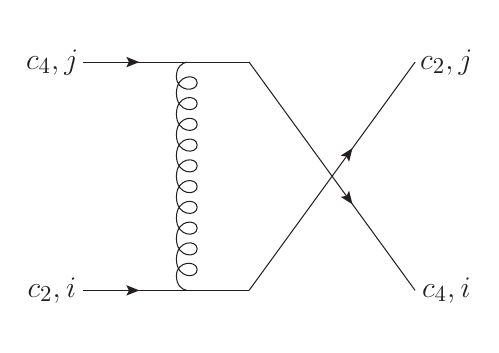}
	\caption{Schematic illustrations of one-gluon exchange between the interchanging (anti)quarks.}
	\label{fig:one-gluon}
\end{figure}
%%%%%%%%%%%%%%%%%%%%%%%%%%%
It seems plausible that the underlying dynamics might be driven by the one-gluon exchange (OGE) between the two interchanging (anti)quarks, as illustrated schematically in Fig.~\ref{fig:one-gluon}. 
For the purpose of a purely qualitative discussion, the initial and final quark pairs $c^{i}c^{j}$ are treated as free particles.
The scattering takes place through the $u$-channel process and the corresponding amplitude of the $(ij\to ij)$ is 
\begin{equation}\label{eq:amp}
	i\mathcal{M}_{ij}^{ij}=\frac{(-)ig_s^2}{(p_i-k_j)^2} t^a_{ji} t^a_{ij} \bar{u}_s(p_i)\gamma^\mu u_{s'}(k_j) \bar{u}_t(p_j)\gamma^\mu u_{t'}(k_i)
\end{equation}
where $g_s$ is the strong coupling constant and the extra minus sign is due to the Dirac statistics in $u$-channel. In the nonrelativistic limit,
this amplitude gives an attractive interaction potential,
\begin{equation}
	V(r)\propto -\frac{g_s^2}{4\pi r} t^a_{ji}t^a_{ij}\times\frac{1}{9}=-\frac{4\alpha_s}{9r},
\end{equation}
where $\alpha_s=g_s^2/(4\pi)$ and the fractor $1/9$ comes from the color wave functions of the four charmonia (note that the $t$-channel gluon exchange gives a repulsive potential between two quarks). 

Figure~\ref{fig:ct-DC} indicates that the QR effect term $Q(t)$ plays a dominant role over the two-gluonic exchange term $D^{(1)}(t)$ in the dicharmonium interaction.
If the OGE mechanism offers a reasonable description of the intercharmonium interaction, then it allows us to make predictions regarding other dicharmonium scattering processes, with the Fierz rearrangement further providing,
\begin{eqnarray}
	\bar{c}_1 c_2\bar{c}_3c_4 &\to& \frac{1}{4}\times \bar{c}_1 c_4 \bar{c}_3 c_2,\nonumber\\
	\bar{c}_1\gamma_5\gamma^\mu c_2\bar{c}_3\gamma_5\gamma_\mu c_4 &\to& -\frac{1}{2}\times \bar{c}_1\gamma_5\gamma^\mu c_4 \bar{c}_3\gamma_5\gamma_\mu c_2,\nonumber\\
	\bar{c}_1\sigma^{\mu\nu} c_2\bar{c}_3\sigma_{\mu\nu} c_4 &\to& -\frac{1}{2}\times \bar{c}_1\sigma^{\mu\nu} c_4 \bar{c}_3\sigma_{\mu\nu} c_2,
\end{eqnarray}
so it is expected that the $\chi_{c0}\chi_{c0}$ interaction should be repulsive, similar to the $\eta_c\eta_c$ case.
In contrast, the $S$-wave $h_c h_c$ (corresponding to $\sigma_{\mu\nu}$) and $\chi_{c1}\chi_{c1}$ interactions are attractive, similar to the $J/\psi J/\psi$ case.  

In principle, such an attractive interaction may give rise to pole singularities near the di-$J/\psi$ threshold, which can correspond to bound states, virtual states, or resonance poles.
As presented in Sec.~\ref{secIII:0pp}, the ${}^1S_0$ $J/\psi J/\psi$ scattering phase suggests the presence of a virtual state below the di-$J/\psi$ threshold, with a binding energy $\Delta E_V = E_V - 2m_{J/\psi}\approx -28(10)~\mathrm{MeV}$ on the M420 ensemble, and $-38(20)~\mathrm{MeV}$ on the M250 ensemble.
However, since the virtual-state pole position $k_V^2$ is located to the left of the lhc point on the M250 ensemble, no definitive conclusion can be drawn regarding the existence of this near-threshold pole at lighter pion masses, where the left-hand cut issue matters.

Experimentally, the $J/\psi J/\psi$ invariant-mass spectra measured by LHCb~\cite{LHCb:2020bwg}, ATLAS~\cite{ATLAS:2023bft}, and CMS~\cite{CMS:2023owd,CMS:2025xwt,CMS:2025fpt} exhibit rapid rises near the $J/\psi J/\psi$ threshold.
A phenomenological study~\cite{Guo:2014iya} argues that a rapid rise near the threshold is generally associated with the presence of a nearby $\mathbf{S}$-matrix pole.
A coupled-channel analysis of the LHCb $J/\psi J/\psi$ spectra suggests the existence of a fully-charmed state, $X(6200)$, near the threshold~\cite{Dong:2020nwy}.
To be specific, the $J/\psi J/\psi - J/\psi \psi(2S)$ coupled-channel fit allows for a virtual state, resonance, or bound state, while the $J/\psi J/\psi - J/\psi \psi(2S) - J/\psi \psi(3770)$ three-channel fit suggests a bound state at $E_B = 6.163_{-32}^{+18}$~GeV, a shallow virtual state at $E_V=6.189_{-10}^{+5}$~GeV, or a bound state within the energy range $[6.159, 6.194]$~GeV.
A similar analysis has been extended to the di-$J/\psi$ spectrum from CMS and ATLAS, confirming the existence of $X(6200)$~\cite{Song:2024ykq}.

It should be noted that the channel selection in the aforementioned phenomenological studies is based on heavy quark spin symmetry (HQSS) and on the assumption that the dynamics are dominated by light-hadron (two-pion or $\sigma$) exchanges.
These model assumptions appear to differ from our observation that quark rearrangement effects play a more significant role in near-threshold $J/\psi J/\psi$ scattering.
However, the parameters of the phenomenological models (including those of the interaction potentials) are entirely fitted from experimental data, and do not necessarily reflect the underlying dynamics assumed in the models.
In this respect, there is no discrepancy between our observations and the phenomenological findings. 
In fact, the $0^{++}$ $J/\psi J/\psi$ scattering amplitude obtained from our calculation leads to a rapid rise of the cross section near the threshold, exhibiting features similar to those observed in the experimental $J/\psi J/\psi$ invariant mass spectra. 
On the other hand, the coupled-channel effects discussed above become significant only when the $J/\psi \psi(2S)$ and $J/\psi \psi(3770)$ channels are open, and are thus relevant to the properties of $X(6900)$ and $X(7100)$.
For $X(6200)$, which lies well below the $J/\psi \psi(2S)$ and $J/\psi \psi(3770)$ thresholds, the coupled-channel effects are expected to be negligible.
Moreover, for the $S$-wave scattering of two vector charmonia, the spin configuration does not affect the dynamics within the HQSS framework. 
Therefore, the quantum numbers of $X(6200)$, whether $0^{++}$ or $2^{++}$, cannot be determined from this model.
In our study, the near-threshold interaction of $J/\psi J/\psi$ is found to be attractive in the $0^{++}$ channel and repulsive in the $2^{++}$ channel, which favors the $0^{++}$ assignment for $X(6200)$, if it exists.

%%%%%%%%%%%%%%%%%%%%%%%
\subsection{${}^5S_2$ $J/\psi J/\psi$ scattering}

%%%%%%%%%%%%%%%%%%%%%%%%%%%
\begin{figure}[t]
	\centering
	\includegraphics[width=0.9\linewidth]{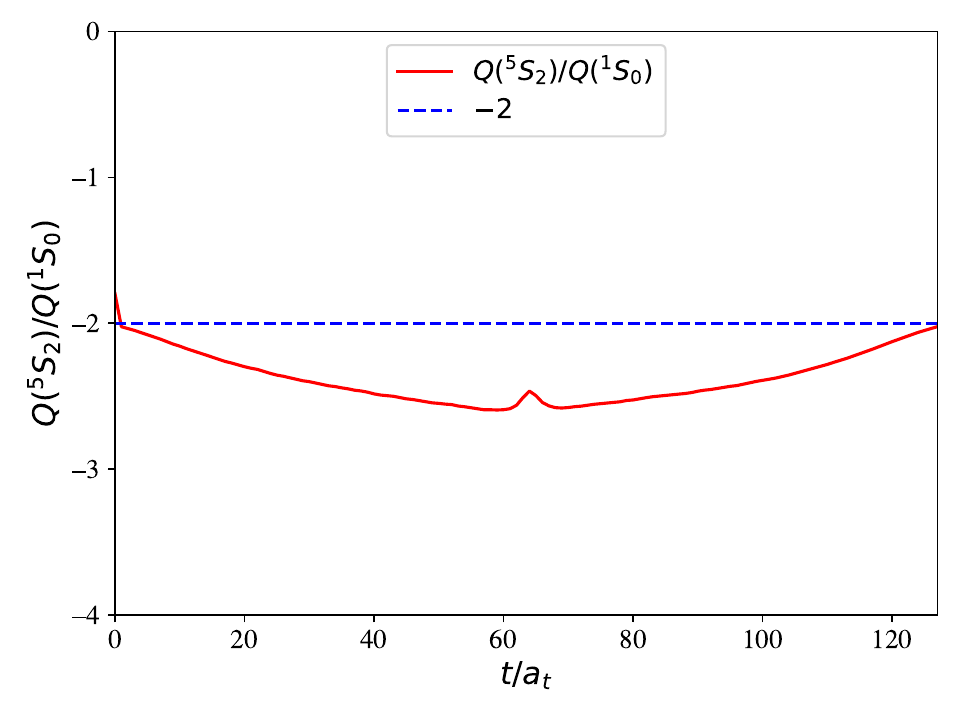}
	\caption{Ratio of the quark rearrangement components of the (${}^5S_2$,~${}^1S_0$) $J/\psi J/\psi$ systems. 
		The blue curve is for the $C(n^2=0,~t)$ case.}
	\label{fig:ct-ratio}
\end{figure}
%%%%%%%%%%%%%%%%%%%%%%%%%%%

We have observed an attractive interaction in the ${}^1S_0$ $J/\psi J/\psi$ channel.
However, in the ${}^5S_2$ channel, there is a repulsive interaction near the di-$J/\psi$ threshold, characterized by a negative scattering length $a_0 \approx -0.16$~fm, as shown in Fig.~\ref{fig:kcot_jpsi} and Table~\ref{tab:energys_2pp}. 
We interpret this difference based on the QR effect.
In the nonrelativistic limit, the bispinor in Eq.~\eqref{eq:amp} can be simplified as 
\begin{equation}
	\bar{u}_s(p_i)\gamma^0 u_{s'}(k_j)\approx 2 m_c \zeta_s^\dagger \zeta_{s'}=2 m_c \delta_{s s'},
\end{equation}
where $m_c$ is the charm quark mass, $\zeta_s$ is the a spinor. 
This means that QR does not change the spin state of the interchanging charm quarks. 
Consequently, the spin configuration of the initial dicharmonium system has a definite relation to that of the final state after the quark rearrangement. 
Let $|SM_S\rangle$ and $|1m\rangle$ be the spin state of the $J/\psi J/\psi$ and $J/\psi$, respectively. 
The spin states $|SM_S\rangle =|00\rangle, |20\rangle, |22\rangle$ are express in terms of $|1m\rangle $ as follows:
\begin{eqnarray}\label{eq:spin}
	|00\rangle&=&\frac{1}{\sqrt{3}}\left(|11\rangle |1-1\rangle+|1-1\rangle |11\rangle-~|10\rangle 10\rangle  \right)\nonumber\\
	|20\rangle&=&\frac{1}{\sqrt{6}}\left(|11\rangle |1-1\rangle+|1-1\rangle |11\rangle+2|10\rangle 10\rangle  \right)\nonumber\\
	|22\rangle&=& |11\rangle |11\rangle
\end{eqnarray}
If the exchanging two (anti)quarks are spin aligned, then the spin configurations of the initial and final states keep the same. 
If spin directions of the two quarks are opposite, then the spin configuration of the final state changes. 
The explicit relations are as follows:
\begin{eqnarray}
	|10\rangle |10\rangle  &\overset{\mathrm{QR}}{\longrightarrow}& \frac{1}{2}\left(|10\rangle |10\rangle+|11\rangle|1-1\rangle+|1-1\rangle|11\rangle\right)\nonumber\\ 
	|11\rangle |1-1\rangle &\overset{\mathrm{QR}}{\longrightarrow}& \frac{1}{2}|10\rangle |10\rangle\nonumber\\
	|1-1\rangle |11\rangle &\overset{\mathrm{QR}}{\longrightarrow}& \frac{1}{2}|10\rangle |10\rangle\nonumber\\
	|11\rangle |11\rangle  &\overset{\mathrm{QR}}{\longrightarrow}& |11\rangle |11\rangle,
\end{eqnarray}
where the factors $1/2$ comes from the spin wave function $|10\rangle=(|\uparrow\downarrow\rangle+|\downarrow\uparrow\rangle)/\sqrt{2}$. 
Let $H_\mathrm{QR}$ be the effective Hamiltonian for the QR effect, then one can easily obtain the relation,
\begin{equation}
	r_\mathrm{QR}=\frac{\langle 2M_S|H_\mathrm{QR}|2M_S\rangle}{\langle 00|H_\mathrm{QR}|00\rangle}\approx -2.
\end{equation}
It provides a reasonable explanation for our observation in Fig.~\ref{fig:ct-ratio}, where the $Q(t)$ component of the $2^{++}$ correlation function and that of $0^{++}$ have a ratio which is approximately $-2$ in the small $t$ region and deviates slowly from $-2$ due to their different time dependence. 
In other words, the QR mechanism is justified here, as it is supported by observations from lattice QCD. 

%%%%%%%%%%%%%%%%%%%%%%%%%%%
\begin{figure}[t]
	\centering
	\includegraphics[width=0.99\linewidth]{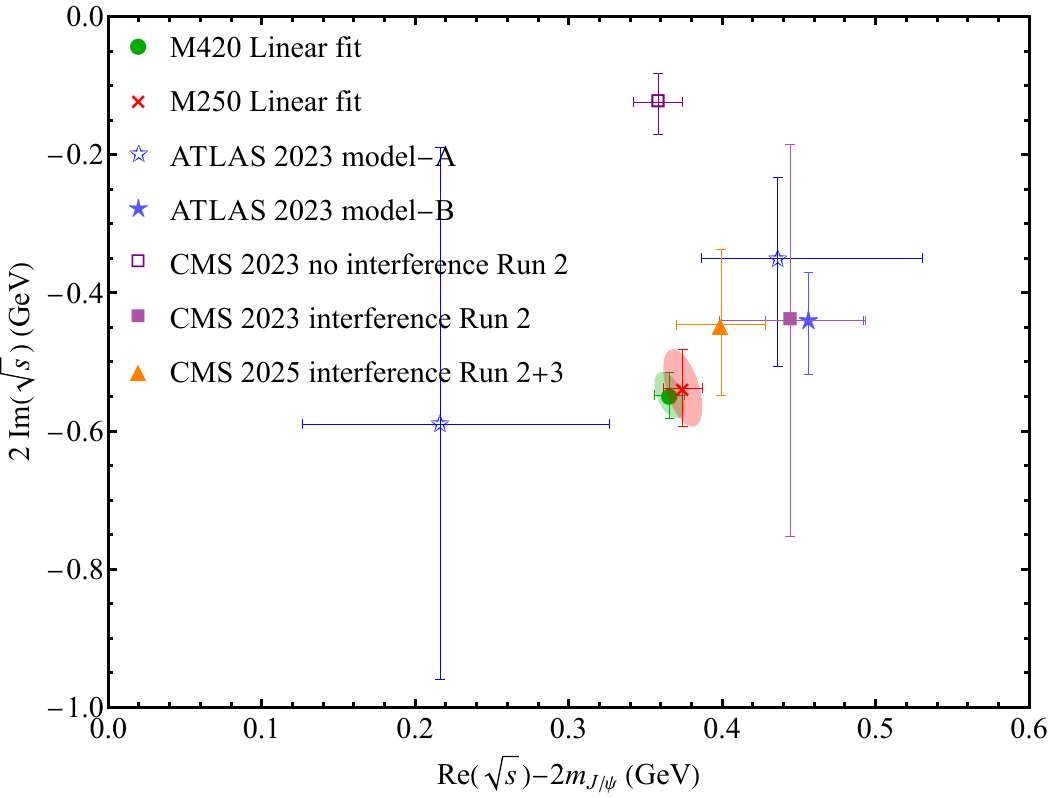}
	\caption{Pole positions in the complex $\sqrt{s}$-plane with linear fit at two pion masses (green dot for M420 and red cross for M250).
		The experimental measurements from ATLAS~\cite{ATLAS:2023bft} and CMS~\cite{CMS:2023owd, CMS:2025fpt, CMS:2025xwt} are represented by blue and purple (orange) points, respectively.}
	\label{fig:pole-exp}
\end{figure}
%%%%%%%%%%%%%%%%%%%%%%%%%%%

Apart from the near-threshold scattering, a wide $2^{++}$ resonance $R_{2^{++}}$ is observed in the ${}^5S_2$ $J/\psi J/\psi$ channel at the center-of-mass energy $E_\mathrm{cm}\approx 6.5- 6.6$~GeV. 
The parameters of $R_{2^{++}}$ are compatible with the $X(6400)$ or $X(6600)$ reported by ATLAS~\cite{ATLAS:2023bft}, the $X(6600)$ reported by CMS~\cite{CMS:2023owd,Wang:2024koq,CMS:2025xwt}. 
Recently, CMS repeats their analysis on a larger data ensemble and confirms the previous results~\cite{CMS:2025xwt}. 
Figure~\ref{fig:pole-exp} shows the comparison of the pole positions of $R_{2^{++}}$ in this study with those of $X(6600)$ (or $X(6400)$) reported by experiments. 
The $\operatorname{Re}(\sqrt{s})$-axis is shifted by $2~m_{J/\psi}$ to facilitate comparison with physical values, as our lattice $m_{J/\psi}$ is 10-15~MeV lower than the experimental value. 

The $J^{PC}$ quantum numbers of these fully-charmed structures remain a focal point of both experimental and theoretical investigations. 
The most probable quantum number assignments are $0^{++}$ and $2^{++}$.
A recent phenomenological study~\cite{Belov:2024qyi} analyzes the yields of fully-charmed states observed by LHCb within the diquark-antidiquark tetraquark framework and concludes that the $X(6900)$ is compatible with a $2^{++}(2S)$ state.
In a recent analysis, CMS~\cite{CMS:2025fpt} claims that the di-$J/\psi$ spectrum is more coherently described by three structures: $X(6600)$, $X(6900)$, and $X(7100)$.
This observation indicates that the three states likely form a family sharing the same $J^{PC}$ quantum number.
CMS has also carried out the first angular distribution analysis of the $4\mu$ decay products of these states, concluding that $2^{++}$ is the most probable quantum number assignment, with the $0^{++}$ excluded at more than 95\% confidence level~\cite{CMS:2025fpt}.
These findings for the $X(6600)$ are supported by our results.

\section{Summary}\label{sectionV}
%%%%%%%%%%%%%%%%%%%%%%%%%%%
We investigate the $0^{++}$ and $2^{++}$ channels of $J/\psi J/\psi$ scattering up to a center-of-mass energy of 6.6 GeV using $N_f = 2$ lattice QCD. 
The analysis is based on lattice ensembles of two different volumes ($12^3 \times 96$ and $16^3 \times 128$) at unphysical pion masses of 250 and 420~MeV, with a single lattice spacing of approximately 0.136~fm. 
We note that relatively small lattice volumes are employed to access higher-energy levels.
The distillation method is adopted to calculate charm quark propagators and the related correlation functions. 
In practice, we do not consider the contribution from charm quark annihilation diagrams, and effectively ignore any coupling to light hadron states in our scattering analysis.

In the numerical technique sector, it is found the LHS smearing built in the distillation method causes strong suppression of the operator coupling to high momentum states, and puts strong a limit to the dicharmonium energies that we can access. 
On the other hand, the high-energy levels suffer from the $N_V$ dependence with $N_V$ being the dimension of the LHS subspace. 
Therefore, on each lattice we use several $N_V$'s to check the reliability of lattice energy levels. 
Finally, we confirm that the center-of-mass energy can be accessed up to 6.6~GeV.

In the $J^{PC}=0^{++}$ channel, the analysis reveals negligible coupled-channel effects between the ${}^1 S_0$ $\eta_c\eta_c$ and ${}^1 S_0$ $J/\psi J/\psi$ channels.
Accordingly, single-channel scattering analyses are conducted separately for each system.
The $0^{++}$ $\eta_c\eta_c$ system exhibits a repulsive interaction, while the $J/\psi J/\psi$ (${}^1 S_0$) system have a attractive interaction. 
The quark rearrangement effect plays a dominant role for the dicharmonium interaction over the direct gluonic exchange between the two scattering charmonia. 
The former can be understood by the one-gluon-exchange mechanism, and the different interactions of $\eta_c\eta_c$ and $J/\psi J/\psi$ is due to the Fierz rearrangement. 
The phase shift of ${}^1S_0$ $J/\psi J/\psi$ scattering hints at the existence of a virtual state lying approximately $30$--$40~\rm{MeV}$ below the di-$J/\psi$ threshold, which is compatible with the $X(6200)$ obtained through the coupled-channel analysis of the $J/\psi J/\psi$ spectrum by LHCb.
Our results support $X(6200)$ to be a $0^{++}$ state if it exists.

In the $J^{PC}=2^{++}$ channel, the ${}^1D_2$ $\eta_c\eta_c$ scattering decouples completely from the ${}^5 S_2$ $J/\psi J/\psi$ scattering. 
We carry out the single channel analysis to the ${}^5 S_2$ $J/\psi J/\psi$ channel. 
In contrast to the ${}^1 S_0$ case, the ${}^5S_2$ $J/\psi J/\psi$ shows repulsive interaction near the threshold which disfavors the existence of a near-threshold state like $X(6200)$. 
The differing characteristics of $J/\psi J/\psi$ interactions in these two channels are further explained by one-gluon-exchange dynamics.
In the nonrelativistic limit, this dynamics predicts that the relative strength of the ${}^5S_2$ channel to the ${}^1S_0$ channel is $-2$, a prediction confirmed by our lattice calculations.

A $2^{++}$ resonance~($R$) is observed in the ${}^5S_2$ $J/\psi J/\psi$ scattering channel with properties of $(m_R, ~\Gamma_R) \approx \big(6540(13), ~ 540(60)\big)$~MeV, consistent with the broad structure $X(6600)$ (or $X(6400)$) reported by ATLAS and CMS. 
It is noted that the angular distribution of the $4\mu$ decay products of these states measured by CMS favors the $2^{++}$ quantum numbers and excludes the $0^{++}$ assignment at more than 95\% confidence level. 

We would like to stress that the calculations are carried out on a single lattice spacing with pion masses heavier than the physical $m_\pi$ and the lattice volumes are relatively small. 
Although its observed that the dynamics for the $\eta_c \eta_c$ and $J/\psi J/\psi$ scattering are dominated by the charm quark rearrangement effects and the light quark mass dependence are unimportant, the relatively long range interaction owing to the light meson exchange should be further explored at lighter quark masses. 
The systematic uncertainties due to the lattice discretization and the finite volume effects have not been well under control, and should be investigated in a more sophisticated study in the future. 
On the other hand, the appearance of the CDD zero the $^5S_2$ $J/\psi J/\psi$ channel is intriguing and needs to be understood theoretically.

%%%%%%%%%%%%%%%%%%%%%%%%%%%
\section*{ACKNOWLEDGMENTS}
%%%%%%%%%%%%%%%%%%%%%%%%%%%
We thank Qiang Zhao and Feng-Kun Guo for insightful discussions. 
This work is supported by the National Key Research and Development Program of China (Grant No.~2025YFB3003602) and the National Natural Science Foundation of China (Grants No.~12293060, No.~12293065, No.~12293061, No.~12205311, and No.~11935017). 
W.~S. and G.~L. are supported by the Chinese Academy of Sciences (Grant No.~YSBR-101), and G.~L. also by the China Postdoctoral Science Foundation (Grant No.~2025M773362). 
We acknowledge the Chroma software~\cite{Edwards:2004sx}, QUDA library~\cite{Clark:2009wm,Babich:2011np}, and PyQUDA package~\cite{jiang2024usequdalatticeqcd}. 
Computations were performed on HPC clusters at the Institute of High Energy Physics (Beijing), China Spallation Neutron Source (Dongguan), and the ORISE computing environment.
\bibliography{reference}

%apsrev4-2.bst 2019-01-14 (MD) hand-edited version of apsrev4-1.bst
%Control: key (0)
%Control: author (8) initials jnrlst
%Control: editor formatted (1) identically to author
%Control: production of article title (0) allowed
%Control: page (0) single
%Control: year (1) truncated
%Control: production of eprint (0) enabled
\begin{thebibliography}{76}%
\makeatletter
\providecommand \@ifxundefined [1]{%
 \@ifx{#1\undefined}
}%
\providecommand \@ifnum [1]{%
 \ifnum #1\expandafter \@firstoftwo
 \else \expandafter \@secondoftwo
 \fi
}%
\providecommand \@ifx [1]{%
 \ifx #1\expandafter \@firstoftwo
 \else \expandafter \@secondoftwo
 \fi
}%
\providecommand \natexlab [1]{#1}%
\providecommand \enquote  [1]{``#1''}%
\providecommand \bibnamefont  [1]{#1}%
\providecommand \bibfnamefont [1]{#1}%
\providecommand \citenamefont [1]{#1}%
\providecommand \href@noop [0]{\@secondoftwo}%
\providecommand \href [0]{\begingroup \@sanitize@url \@href}%
\providecommand \@href[1]{\@@startlink{#1}\@@href}%
\providecommand \@@href[1]{\endgroup#1\@@endlink}%
\providecommand \@sanitize@url [0]{\catcode `\\12\catcode `\$12\catcode
  `\&12\catcode `\#12\catcode `\^12\catcode `\_12\catcode `\%12\relax}%
\providecommand \@@startlink[1]{}%
\providecommand \@@endlink[0]{}%
\providecommand \url  [0]{\begingroup\@sanitize@url \@url }%
\providecommand \@url [1]{\endgroup\@href {#1}{\urlprefix }}%
\providecommand \urlprefix  [0]{URL }%
\providecommand \Eprint [0]{\href }%
\providecommand \doibase [0]{https://doi.org/}%
\providecommand \selectlanguage [0]{\@gobble}%
\providecommand \bibinfo  [0]{\@secondoftwo}%
\providecommand \bibfield  [0]{\@secondoftwo}%
\providecommand \translation [1]{[#1]}%
\providecommand \BibitemOpen [0]{}%
\providecommand \bibitemStop [0]{}%
\providecommand \bibitemNoStop [0]{.\EOS\space}%
\providecommand \EOS [0]{\spacefactor3000\relax}%
\providecommand \BibitemShut  [1]{\csname bibitem#1\endcsname}%
\let\auto@bib@innerbib\@empty
%</preamble>
\bibitem [{\citenamefont {Aaij}\ \emph {et~al.}(2020)\citenamefont {Aaij} \emph
  {et~al.}}]{LHCb:2020bwg}%
  \BibitemOpen
  \bibfield  {author} {\bibinfo {author} {\bibfnamefont {R.}~\bibnamefont
  {Aaij}} \emph {et~al.} (\bibinfo {collaboration} {LHCb}),\ }\bibfield
  {title} {\bibinfo {title} {{Observation of structure in the $J/\psi$ -pair
  mass spectrum}},\ }\href {https://doi.org/10.1016/j.scib.2020.08.032}
  {\bibfield  {journal} {\bibinfo  {journal} {Sci. Bull.}\ }\textbf {\bibinfo
  {volume} {65}},\ \bibinfo {pages} {1983} (\bibinfo {year} {2020})},\ \Eprint
  {https://arxiv.org/abs/2006.16957} {arXiv:2006.16957 [hep-ex]} \BibitemShut
  {NoStop}%
\bibitem [{\citenamefont {Aad}\ \emph {et~al.}(2023)\citenamefont {Aad} \emph
  {et~al.}}]{ATLAS:2023bft}%
  \BibitemOpen
  \bibfield  {author} {\bibinfo {author} {\bibfnamefont {G.}~\bibnamefont
  {Aad}} \emph {et~al.} (\bibinfo {collaboration} {ATLAS}),\ }\bibfield
  {title} {\bibinfo {title} {{Observation of an Excess of Dicharmonium Events
  in the Four-Muon Final State with the ATLAS Detector}},\ }\href
  {https://doi.org/10.1103/PhysRevLett.131.151902} {\bibfield  {journal}
  {\bibinfo  {journal} {Phys. Rev. Lett.}\ }\textbf {\bibinfo {volume} {131}},\
  \bibinfo {pages} {151902} (\bibinfo {year} {2023})},\ \Eprint
  {https://arxiv.org/abs/2304.08962} {arXiv:2304.08962 [hep-ex]} \BibitemShut
  {NoStop}%
\bibitem [{\citenamefont {Hayrapetyan}\ \emph {et~al.}(2024)\citenamefont
  {Hayrapetyan} \emph {et~al.}}]{CMS:2023owd}%
  \BibitemOpen
  \bibfield  {author} {\bibinfo {author} {\bibfnamefont {A.}~\bibnamefont
  {Hayrapetyan}} \emph {et~al.} (\bibinfo {collaboration} {CMS}),\ }\bibfield
  {title} {\bibinfo {title} {{New Structures in the $J/\psi$ $J/\psi$ Mass
  Spectrum in Proton-Proton Collisions at $s=13$ TeV}},\ }\href
  {https://doi.org/10.1103/PhysRevLett.132.111901} {\bibfield  {journal}
  {\bibinfo  {journal} {Phys. Rev. Lett.}\ }\textbf {\bibinfo {volume} {132}},\
  \bibinfo {pages} {111901} (\bibinfo {year} {2024})},\ \Eprint
  {https://arxiv.org/abs/2306.07164} {arXiv:2306.07164 [hep-ex]} \BibitemShut
  {NoStop}%
\bibitem [{\citenamefont {Wang}\ and\ \citenamefont {Yi}(2024)}]{Wang:2024koq}%
  \BibitemOpen
  \bibfield  {author} {\bibinfo {author} {\bibfnamefont {X.}~\bibnamefont
  {Wang}}\ and\ \bibinfo {author} {\bibfnamefont {K.}~\bibnamefont {Yi}},\
  }\bibfield  {title} {\bibinfo {title} {{New Structures in the $J/\psi$
  $J/\psi$ Mass Spectrum at CMS}},\ }in\ \href@noop {} {\emph {\bibinfo
  {booktitle} {{22nd Conference on Flavor Physics and CP Violation}}}}\
  (\bibinfo {year} {2024})\ \Eprint {https://arxiv.org/abs/2408.04722}
  {arXiv:2408.04722 [hep-ex]} \BibitemShut {NoStop}%
\bibitem [{\citenamefont {Hayrapetyan}\ \emph
  {et~al.}(2025{\natexlab{a}})\citenamefont {Hayrapetyan} \emph
  {et~al.}}]{CMS:2025xwt}%
  \BibitemOpen
  \bibfield  {author} {\bibinfo {author} {\bibfnamefont {A.}~\bibnamefont
  {Hayrapetyan}} \emph {et~al.} (\bibinfo {collaboration} {CMS}),\ }\href@noop
  {} {\bibinfo {title} {{Observation of a family of all-charm tetraquark
  candidates at the LHC, report number: CMS-PAS-BPH-24-003}}} (\bibinfo {year}
  {2025}{\natexlab{a}})\BibitemShut {NoStop}%
\bibitem [{\citenamefont {Hayrapetyan}\ \emph
  {et~al.}(2025{\natexlab{b}})\citenamefont {Hayrapetyan} \emph
  {et~al.}}]{CMS:2025fpt}%
  \BibitemOpen
  \bibfield  {author} {\bibinfo {author} {\bibfnamefont {A.}~\bibnamefont
  {Hayrapetyan}} \emph {et~al.} (\bibinfo {collaboration} {CMS}),\ }\bibfield
  {title} {\bibinfo {title} {{Determination of the spin and parity of all-charm
  tetraquarks}},\ }\href {https://doi.org/10.1038/s41586-025-09711-7}
  {\bibfield  {journal} {\bibinfo  {journal} {Nature}\ }\textbf {\bibinfo
  {volume} {648}},\ \bibinfo {pages} {58} (\bibinfo {year}
  {2025}{\natexlab{b}})},\ \Eprint {https://arxiv.org/abs/2506.07944}
  {arXiv:2506.07944 [hep-ex]} \BibitemShut {NoStop}%
\bibitem [{\citenamefont {Guo}\ and\ \citenamefont
  {Oller}(2021)}]{Guo:2020pvt}%
  \BibitemOpen
  \bibfield  {author} {\bibinfo {author} {\bibfnamefont {Z.-H.}\ \bibnamefont
  {Guo}}\ and\ \bibinfo {author} {\bibfnamefont {J.~A.}\ \bibnamefont
  {Oller}},\ }\bibfield  {title} {\bibinfo {title} {{Insights into the inner
  structures of the fully charmed tetraquark state $X(6900)$}},\ }\href
  {https://doi.org/10.1103/PhysRevD.103.034024} {\bibfield  {journal} {\bibinfo
   {journal} {Phys. Rev. D}\ }\textbf {\bibinfo {volume} {103}},\ \bibinfo
  {pages} {034024} (\bibinfo {year} {2021})},\ \Eprint
  {https://arxiv.org/abs/2011.00978} {arXiv:2011.00978 [hep-ph]} \BibitemShut
  {NoStop}%
\bibitem [{\citenamefont {Liu}\ \emph {et~al.}(2024)\citenamefont {Liu},
  \citenamefont {Liu}, \citenamefont {Zhong},\ and\ \citenamefont
  {Zhao}}]{liu:2020eha}%
  \BibitemOpen
  \bibfield  {author} {\bibinfo {author} {\bibfnamefont {M.-S.}\ \bibnamefont
  {Liu}}, \bibinfo {author} {\bibfnamefont {F.-X.}\ \bibnamefont {Liu}},
  \bibinfo {author} {\bibfnamefont {X.-H.}\ \bibnamefont {Zhong}},\ and\
  \bibinfo {author} {\bibfnamefont {Q.}~\bibnamefont {Zhao}},\ }\bibfield
  {title} {\bibinfo {title} {{Fully heavy tetraquark states and their evidences
  in LHC observations}},\ }\href {https://doi.org/10.1103/PhysRevD.109.076017}
  {\bibfield  {journal} {\bibinfo  {journal} {Phys. Rev. D}\ }\textbf {\bibinfo
  {volume} {109}},\ \bibinfo {pages} {076017} (\bibinfo {year} {2024})},\
  \Eprint {https://arxiv.org/abs/2006.11952} {arXiv:2006.11952 [hep-ph]}
  \BibitemShut {NoStop}%
\bibitem [{\citenamefont {Chen}\ \emph {et~al.}(2020)\citenamefont {Chen},
  \citenamefont {Chen}, \citenamefont {Liu},\ and\ \citenamefont
  {Zhu}}]{Chen:2020xwe}%
  \BibitemOpen
  \bibfield  {author} {\bibinfo {author} {\bibfnamefont {H.-X.}\ \bibnamefont
  {Chen}}, \bibinfo {author} {\bibfnamefont {W.}~\bibnamefont {Chen}}, \bibinfo
  {author} {\bibfnamefont {X.}~\bibnamefont {Liu}},\ and\ \bibinfo {author}
  {\bibfnamefont {S.-L.}\ \bibnamefont {Zhu}},\ }\bibfield  {title} {\bibinfo
  {title} {{Strong decays of fully-charm tetraquarks into di-charmonia}},\
  }\href {https://doi.org/10.1016/j.scib.2020.08.038} {\bibfield  {journal}
  {\bibinfo  {journal} {Sci. Bull.}\ }\textbf {\bibinfo {volume} {65}},\
  \bibinfo {pages} {1994} (\bibinfo {year} {2020})},\ \Eprint
  {https://arxiv.org/abs/2006.16027} {arXiv:2006.16027 [hep-ph]} \BibitemShut
  {NoStop}%
\bibitem [{\citenamefont {Maiani}(2020)}]{Maiani:2020pur}%
  \BibitemOpen
  \bibfield  {author} {\bibinfo {author} {\bibfnamefont {L.}~\bibnamefont
  {Maiani}},\ }\bibfield  {title} {\bibinfo {title} {{$J/\psi$-pair resonance
  by LHCb: a new revolution?}},\ }\href
  {https://doi.org/10.1016/j.scib.2020.08.019} {\bibfield  {journal} {\bibinfo
  {journal} {Sci. Bull.}\ }\textbf {\bibinfo {volume} {65}},\ \bibinfo {pages}
  {1949} (\bibinfo {year} {2020})},\ \Eprint {https://arxiv.org/abs/2008.01637}
  {arXiv:2008.01637 [hep-ph]} \BibitemShut {NoStop}%
\bibitem [{\citenamefont {Dong}\ \emph
  {et~al.}(2021{\natexlab{a}})\citenamefont {Dong}, \citenamefont {Baru},
  \citenamefont {Guo}, \citenamefont {Hanhart},\ and\ \citenamefont
  {Nefediev}}]{Dong:2020nwy}%
  \BibitemOpen
  \bibfield  {author} {\bibinfo {author} {\bibfnamefont {X.-K.}\ \bibnamefont
  {Dong}}, \bibinfo {author} {\bibfnamefont {V.}~\bibnamefont {Baru}}, \bibinfo
  {author} {\bibfnamefont {F.-K.}\ \bibnamefont {Guo}}, \bibinfo {author}
  {\bibfnamefont {C.}~\bibnamefont {Hanhart}},\ and\ \bibinfo {author}
  {\bibfnamefont {A.}~\bibnamefont {Nefediev}},\ }\bibfield  {title} {\bibinfo
  {title} {{Coupled-Channel Interpretation of the LHCb
  Double-~$J/\psi$~Spectrum and Hints of a New State Near the~ $J/\psi
  J/\psi$~~Threshold}},\ }\href
  {https://doi.org/10.1103/PhysRevLett.127.119901} {\bibfield  {journal}
  {\bibinfo  {journal} {Phys. Rev. Lett.}\ }\textbf {\bibinfo {volume} {126}},\
  \bibinfo {pages} {132001} (\bibinfo {year} {2021}{\natexlab{a}})},\ \bibinfo
  {note} {[Erratum: Phys.Rev.Lett. 127, 119901 (2021)]},\ \Eprint
  {https://arxiv.org/abs/2009.07795} {arXiv:2009.07795 [hep-ph]} \BibitemShut
  {NoStop}%
\bibitem [{\citenamefont {Chao}\ and\ \citenamefont
  {Zhu}(2020)}]{Chao:2020dml}%
  \BibitemOpen
  \bibfield  {author} {\bibinfo {author} {\bibfnamefont {K.-T.}\ \bibnamefont
  {Chao}}\ and\ \bibinfo {author} {\bibfnamefont {S.-L.}\ \bibnamefont {Zhu}},\
  }\bibfield  {title} {\bibinfo {title} {{The possible tetraquark states $cc
  \bar c \bar c$ observed by the LHCb experiment}},\ }\href
  {https://doi.org/10.1016/j.scib.2020.08.031} {\bibfield  {journal} {\bibinfo
  {journal} {Sci. Bull.}\ }\textbf {\bibinfo {volume} {65}},\ \bibinfo {pages}
  {1952} (\bibinfo {year} {2020})},\ \Eprint {https://arxiv.org/abs/2008.07670}
  {arXiv:2008.07670 [hep-ph]} \BibitemShut {NoStop}%
\bibitem [{\citenamefont {Zhu}(2021)}]{Zhu:2020xni}%
  \BibitemOpen
  \bibfield  {author} {\bibinfo {author} {\bibfnamefont {R.}~\bibnamefont
  {Zhu}},\ }\bibfield  {title} {\bibinfo {title} {{Fully-heavy tetraquark
  spectra and production at hadron colliders}},\ }\href
  {https://doi.org/10.1016/j.nuclphysb.2021.115393} {\bibfield  {journal}
  {\bibinfo  {journal} {Nucl. Phys. B}\ }\textbf {\bibinfo {volume} {966}},\
  \bibinfo {pages} {115393} (\bibinfo {year} {2021})},\ \Eprint
  {https://arxiv.org/abs/2010.09082} {arXiv:2010.09082 [hep-ph]} \BibitemShut
  {NoStop}%
\bibitem [{\citenamefont {Gong}\ \emph
  {et~al.}(2022{\natexlab{a}})\citenamefont {Gong}, \citenamefont {Du},
  \citenamefont {Zhao}, \citenamefont {Zhong},\ and\ \citenamefont
  {Zhou}}]{Gong:2020bmg}%
  \BibitemOpen
  \bibfield  {author} {\bibinfo {author} {\bibfnamefont {C.}~\bibnamefont
  {Gong}}, \bibinfo {author} {\bibfnamefont {M.-C.}\ \bibnamefont {Du}},
  \bibinfo {author} {\bibfnamefont {Q.}~\bibnamefont {Zhao}}, \bibinfo {author}
  {\bibfnamefont {X.-H.}\ \bibnamefont {Zhong}},\ and\ \bibinfo {author}
  {\bibfnamefont {B.}~\bibnamefont {Zhou}},\ }\bibfield  {title} {\bibinfo
  {title} {{Nature of X(6900) and its production mechanism at LHCb}},\ }\href
  {https://doi.org/10.1016/j.physletb.2021.136794} {\bibfield  {journal}
  {\bibinfo  {journal} {Phys. Lett. B}\ }\textbf {\bibinfo {volume} {824}},\
  \bibinfo {pages} {136794} (\bibinfo {year} {2022}{\natexlab{a}})},\ \Eprint
  {https://arxiv.org/abs/2011.11374} {arXiv:2011.11374 [hep-ph]} \BibitemShut
  {NoStop}%
\bibitem [{\citenamefont {Liang}\ \emph {et~al.}(2021)\citenamefont {Liang},
  \citenamefont {Wu},\ and\ \citenamefont {Yao}}]{Liang:2021fzr}%
  \BibitemOpen
  \bibfield  {author} {\bibinfo {author} {\bibfnamefont {Z.-R.}\ \bibnamefont
  {Liang}}, \bibinfo {author} {\bibfnamefont {X.-Y.}\ \bibnamefont {Wu}},\ and\
  \bibinfo {author} {\bibfnamefont {D.-L.}\ \bibnamefont {Yao}},\ }\bibfield
  {title} {\bibinfo {title} {{Hunting for states in the recent LHCb
  di-J/\ensuremath{\psi} invariant mass spectrum}},\ }\href
  {https://doi.org/10.1103/PhysRevD.104.034034} {\bibfield  {journal} {\bibinfo
   {journal} {Phys. Rev. D}\ }\textbf {\bibinfo {volume} {104}},\ \bibinfo
  {pages} {034034} (\bibinfo {year} {2021})},\ \Eprint
  {https://arxiv.org/abs/2104.08589} {arXiv:2104.08589 [hep-ph]} \BibitemShut
  {NoStop}%
\bibitem [{\citenamefont {Liu}\ \emph {et~al.}(2021)\citenamefont {Liu},
  \citenamefont {Liu}, \citenamefont {Zhong},\ and\ \citenamefont
  {Zhao}}]{Liu:2021rtn}%
  \BibitemOpen
  \bibfield  {author} {\bibinfo {author} {\bibfnamefont {F.-X.}\ \bibnamefont
  {Liu}}, \bibinfo {author} {\bibfnamefont {M.-S.}\ \bibnamefont {Liu}},
  \bibinfo {author} {\bibfnamefont {X.-H.}\ \bibnamefont {Zhong}},\ and\
  \bibinfo {author} {\bibfnamefont {Q.}~\bibnamefont {Zhao}},\ }\bibfield
  {title} {\bibinfo {title} {{Higher mass spectra of the fully-charmed and
  fully-bottom tetraquarks}},\ }\href
  {https://doi.org/10.1103/PhysRevD.104.116029} {\bibfield  {journal} {\bibinfo
   {journal} {Phys. Rev. D}\ }\textbf {\bibinfo {volume} {104}},\ \bibinfo
  {pages} {116029} (\bibinfo {year} {2021})},\ \Eprint
  {https://arxiv.org/abs/2110.09052} {arXiv:2110.09052 [hep-ph]} \BibitemShut
  {NoStop}%
\bibitem [{\citenamefont {Dong}\ \emph
  {et~al.}(2021{\natexlab{b}})\citenamefont {Dong}, \citenamefont {Baru},
  \citenamefont {Guo}, \citenamefont {Hanhart}, \citenamefont {Nefediev},\ and\
  \citenamefont {Zou}}]{Dong:2021lkh}%
  \BibitemOpen
  \bibfield  {author} {\bibinfo {author} {\bibfnamefont {X.-K.}\ \bibnamefont
  {Dong}}, \bibinfo {author} {\bibfnamefont {V.}~\bibnamefont {Baru}}, \bibinfo
  {author} {\bibfnamefont {F.-K.}\ \bibnamefont {Guo}}, \bibinfo {author}
  {\bibfnamefont {C.}~\bibnamefont {Hanhart}}, \bibinfo {author} {\bibfnamefont
  {A.}~\bibnamefont {Nefediev}},\ and\ \bibinfo {author} {\bibfnamefont
  {B.-S.}\ \bibnamefont {Zou}},\ }\bibfield  {title} {\bibinfo {title} {{Is the
  existence of a J/\ensuremath{\psi}J/\ensuremath{\psi} bound state
  plausible?}},\ }\href {https://doi.org/10.1016/j.scib.2021.09.009} {\bibfield
   {journal} {\bibinfo  {journal} {Sci. Bull.}\ }\textbf {\bibinfo {volume}
  {66}},\ \bibinfo {pages} {2462} (\bibinfo {year} {2021}{\natexlab{b}})},\
  \Eprint {https://arxiv.org/abs/2107.03946} {arXiv:2107.03946 [hep-ph]}
  \BibitemShut {NoStop}%
\bibitem [{\citenamefont {Niu}\ \emph {et~al.}(2023)\citenamefont {Niu},
  \citenamefont {Zhang}, \citenamefont {Wang},\ and\ \citenamefont
  {Du}}]{Niu:2022jqp}%
  \BibitemOpen
  \bibfield  {author} {\bibinfo {author} {\bibfnamefont {P.}~\bibnamefont
  {Niu}}, \bibinfo {author} {\bibfnamefont {Z.}~\bibnamefont {Zhang}}, \bibinfo
  {author} {\bibfnamefont {Q.}~\bibnamefont {Wang}},\ and\ \bibinfo {author}
  {\bibfnamefont {M.-L.}\ \bibnamefont {Du}},\ }\bibfield  {title} {\bibinfo
  {title} {{The third peak structure in the double $J/\psi$ spectrum}},\ }\href
  {https://doi.org/10.1016/j.scib.2023.03.025} {\bibfield  {journal} {\bibinfo
  {journal} {Sci. Bull.}\ }\textbf {\bibinfo {volume} {68}},\ \bibinfo {pages}
  {800} (\bibinfo {year} {2023})},\ \Eprint {https://arxiv.org/abs/2212.06535}
  {arXiv:2212.06535 [hep-ph]} \BibitemShut {NoStop}%
\bibitem [{\citenamefont {Dong}\ and\ \citenamefont
  {Wang}(2023)}]{Dong:2022sef}%
  \BibitemOpen
  \bibfield  {author} {\bibinfo {author} {\bibfnamefont {W.-C.}\ \bibnamefont
  {Dong}}\ and\ \bibinfo {author} {\bibfnamefont {Z.-G.}\ \bibnamefont
  {Wang}},\ }\bibfield  {title} {\bibinfo {title} {{Going in quest of potential
  tetraquark interpretations for the newly observed
  T\ensuremath{\psi}\ensuremath{\psi} states in light of the
  diquark-antidiquark scenarios}},\ }\href
  {https://doi.org/10.1103/PhysRevD.107.074010} {\bibfield  {journal} {\bibinfo
   {journal} {Phys. Rev. D}\ }\textbf {\bibinfo {volume} {107}},\ \bibinfo
  {pages} {074010} (\bibinfo {year} {2023})},\ \Eprint
  {https://arxiv.org/abs/2211.11989} {arXiv:2211.11989 [hep-ph]} \BibitemShut
  {NoStop}%
\bibitem [{\citenamefont {Kuang}\ \emph {et~al.}(2023)\citenamefont {Kuang},
  \citenamefont {Zhou}, \citenamefont {Guo}, \citenamefont {Yang},\ and\
  \citenamefont {Dai}}]{Kuang:2023vac}%
  \BibitemOpen
  \bibfield  {author} {\bibinfo {author} {\bibfnamefont {S.-Q.}\ \bibnamefont
  {Kuang}}, \bibinfo {author} {\bibfnamefont {Q.}~\bibnamefont {Zhou}},
  \bibinfo {author} {\bibfnamefont {D.}~\bibnamefont {Guo}}, \bibinfo {author}
  {\bibfnamefont {Q.-H.}\ \bibnamefont {Yang}},\ and\ \bibinfo {author}
  {\bibfnamefont {L.-Y.}\ \bibnamefont {Dai}},\ }\bibfield  {title} {\bibinfo
  {title} {{Study of X(6900) with unitarized coupled channel scattering
  amplitudes}},\ }\href {https://doi.org/10.1140/epjc/s10052-023-11473-3}
  {\bibfield  {journal} {\bibinfo  {journal} {Eur. Phys. J. C}\ }\textbf
  {\bibinfo {volume} {83}},\ \bibinfo {pages} {383} (\bibinfo {year} {2023})},\
  \Eprint {https://arxiv.org/abs/2302.03968} {arXiv:2302.03968 [hep-ph]}
  \BibitemShut {NoStop}%
\bibitem [{\citenamefont {Huang}\ \emph {et~al.}(2024)\citenamefont {Huang},
  \citenamefont {Chen}, \citenamefont {He},\ and\ \citenamefont
  {Liu}}]{Huang:2024jin}%
  \BibitemOpen
  \bibfield  {author} {\bibinfo {author} {\bibfnamefont {Q.}~\bibnamefont
  {Huang}}, \bibinfo {author} {\bibfnamefont {R.}~\bibnamefont {Chen}},
  \bibinfo {author} {\bibfnamefont {J.}~\bibnamefont {He}},\ and\ \bibinfo
  {author} {\bibfnamefont {X.}~\bibnamefont {Liu}},\ }\href@noop {} {\bibinfo
  {title} {{Discovering a Novel Dynamics Mechanism for Charmonium Scattering}}}
  (\bibinfo {year} {2024}),\ \Eprint {https://arxiv.org/abs/2407.16316}
  {arXiv:2407.16316 [hep-ph]} \BibitemShut {NoStop}%
\bibitem [{\citenamefont {Wu}\ \emph {et~al.}(2024)\citenamefont {Wu},
  \citenamefont {Chen}, \citenamefont {Meng},\ and\ \citenamefont
  {Zhu}}]{Wu:2024euj}%
  \BibitemOpen
  \bibfield  {author} {\bibinfo {author} {\bibfnamefont {W.-L.}\ \bibnamefont
  {Wu}}, \bibinfo {author} {\bibfnamefont {Y.-K.}\ \bibnamefont {Chen}},
  \bibinfo {author} {\bibfnamefont {L.}~\bibnamefont {Meng}},\ and\ \bibinfo
  {author} {\bibfnamefont {S.-L.}\ \bibnamefont {Zhu}},\ }\bibfield  {title}
  {\bibinfo {title} {{Benchmark calculations of fully heavy compact and
  molecular tetraquark states}},\ }\href
  {https://doi.org/10.1103/PhysRevD.109.054034} {\bibfield  {journal} {\bibinfo
   {journal} {Phys. Rev. D}\ }\textbf {\bibinfo {volume} {109}},\ \bibinfo
  {pages} {054034} (\bibinfo {year} {2024})},\ \Eprint
  {https://arxiv.org/abs/2401.14899} {arXiv:2401.14899 [hep-ph]} \BibitemShut
  {NoStop}%
\bibitem [{\citenamefont {Zhang}\ and\ \citenamefont
  {Guo}(2024)}]{Zhang:2024qkg}%
  \BibitemOpen
  \bibfield  {author} {\bibinfo {author} {\bibfnamefont {Z.-H.}\ \bibnamefont
  {Zhang}}\ and\ \bibinfo {author} {\bibfnamefont {F.-K.}\ \bibnamefont
  {Guo}},\ }\href@noop {} {\bibinfo {title} {{Classification of Coupled-Channel
  Near-Threshold Structures}}} (\bibinfo {year} {2024}),\ \Eprint
  {https://arxiv.org/abs/2407.10620} {arXiv:2407.10620 [hep-ph]} \BibitemShut
  {NoStop}%
\bibitem [{\citenamefont {Chen}\ \emph {et~al.}(2023)\citenamefont {Chen},
  \citenamefont {Chen}, \citenamefont {Liu}, \citenamefont {Liu},\ and\
  \citenamefont {Zhu}}]{Chen:2022asf}%
  \BibitemOpen
  \bibfield  {author} {\bibinfo {author} {\bibfnamefont {H.-X.}\ \bibnamefont
  {Chen}}, \bibinfo {author} {\bibfnamefont {W.}~\bibnamefont {Chen}}, \bibinfo
  {author} {\bibfnamefont {X.}~\bibnamefont {Liu}}, \bibinfo {author}
  {\bibfnamefont {Y.-R.}\ \bibnamefont {Liu}},\ and\ \bibinfo {author}
  {\bibfnamefont {S.-L.}\ \bibnamefont {Zhu}},\ }\bibfield  {title} {\bibinfo
  {title} {{An updated review of the new hadron states}},\ }\href
  {https://doi.org/10.1088/1361-6633/aca3b6} {\bibfield  {journal} {\bibinfo
  {journal} {Rept. Prog. Phys.}\ }\textbf {\bibinfo {volume} {86}},\ \bibinfo
  {pages} {026201} (\bibinfo {year} {2023})},\ \Eprint
  {https://arxiv.org/abs/2204.02649} {arXiv:2204.02649 [hep-ph]} \BibitemShut
  {NoStop}%
\bibitem [{\citenamefont {Iwasaki}(1975)}]{Iwasaki:1975pv}%
  \BibitemOpen
  \bibfield  {author} {\bibinfo {author} {\bibfnamefont {Y.}~\bibnamefont
  {Iwasaki}},\ }\bibfield  {title} {\bibinfo {title} {{A Possible Model for New
  Resonances-Exotics and Hidden Charm}},\ }\href
  {https://doi.org/10.1143/PTP.54.492} {\bibfield  {journal} {\bibinfo
  {journal} {Prog. Theor. Phys.}\ }\textbf {\bibinfo {volume} {54}},\ \bibinfo
  {pages} {492} (\bibinfo {year} {1975})}\BibitemShut {NoStop}%
\bibitem [{\citenamefont {Chao}(1981)}]{Chao:1980dv}%
  \BibitemOpen
  \bibfield  {author} {\bibinfo {author} {\bibfnamefont {K.-T.}\ \bibnamefont
  {Chao}},\ }\bibfield  {title} {\bibinfo {title} {{The (cc) - ($\bar{cc}$)
  (Diquark - Anti-Diquark) States in $e^+ e^-$ Annihilation}},\ }\href
  {https://doi.org/10.1007/BF01431564} {\bibfield  {journal} {\bibinfo
  {journal} {Z. Phys. C}\ }\textbf {\bibinfo {volume} {7}},\ \bibinfo {pages}
  {317} (\bibinfo {year} {1981})}\BibitemShut {NoStop}%
\bibitem [{\citenamefont {Ader}\ \emph {et~al.}(1982)\citenamefont {Ader},
  \citenamefont {Richard},\ and\ \citenamefont {Taxil}}]{Ader:1981db}%
  \BibitemOpen
  \bibfield  {author} {\bibinfo {author} {\bibfnamefont {J.~P.}\ \bibnamefont
  {Ader}}, \bibinfo {author} {\bibfnamefont {J.~M.}\ \bibnamefont {Richard}},\
  and\ \bibinfo {author} {\bibfnamefont {P.}~\bibnamefont {Taxil}},\ }\bibfield
   {title} {\bibinfo {title} {{DO NARROW HEAVY MULTI - QUARK STATES EXIST?}},\
  }\href {https://doi.org/10.1103/PhysRevD.25.2370} {\bibfield  {journal}
  {\bibinfo  {journal} {Phys. Rev. D}\ }\textbf {\bibinfo {volume} {25}},\
  \bibinfo {pages} {2370} (\bibinfo {year} {1982})}\BibitemShut {NoStop}%
\bibitem [{\citenamefont {Li}\ and\ \citenamefont {Liu}(1984)}]{Li:1983ru}%
  \BibitemOpen
  \bibfield  {author} {\bibinfo {author} {\bibfnamefont {B.-A.}\ \bibnamefont
  {Li}}\ and\ \bibinfo {author} {\bibfnamefont {K.-F.}\ \bibnamefont {Liu}},\
  }\bibfield  {title} {\bibinfo {title} {{$J/\psi$ Pair Production in Hadronic
  Collisions}},\ }\href {https://doi.org/10.1103/PhysRevD.29.426} {\bibfield
  {journal} {\bibinfo  {journal} {Phys. Rev. D}\ }\textbf {\bibinfo {volume}
  {29}},\ \bibinfo {pages} {426} (\bibinfo {year} {1984})}\BibitemShut
  {NoStop}%
\bibitem [{\citenamefont {Heller}\ and\ \citenamefont
  {Tjon}(1985)}]{Heller:1985cb}%
  \BibitemOpen
  \bibfield  {author} {\bibinfo {author} {\bibfnamefont {L.}~\bibnamefont
  {Heller}}\ and\ \bibinfo {author} {\bibfnamefont {J.~A.}\ \bibnamefont
  {Tjon}},\ }\bibfield  {title} {\bibinfo {title} {{On Bound States of Heavy
  $Q^2 \bar{Q}^2$ Systems}},\ }\href {https://doi.org/10.1103/PhysRevD.32.755}
  {\bibfield  {journal} {\bibinfo  {journal} {Phys. Rev. D}\ }\textbf {\bibinfo
  {volume} {32}},\ \bibinfo {pages} {755} (\bibinfo {year} {1985})}\BibitemShut
  {NoStop}%
\bibitem [{\citenamefont {Badalian}\ \emph {et~al.}(1987)\citenamefont
  {Badalian}, \citenamefont {Ioffe},\ and\ \citenamefont
  {Smilga}}]{Badalian:1985es}%
  \BibitemOpen
  \bibfield  {author} {\bibinfo {author} {\bibfnamefont {A.~M.}\ \bibnamefont
  {Badalian}}, \bibinfo {author} {\bibfnamefont {B.~L.}\ \bibnamefont
  {Ioffe}},\ and\ \bibinfo {author} {\bibfnamefont {A.~V.}\ \bibnamefont
  {Smilga}},\ }\bibfield  {title} {\bibinfo {title} {{FOUR QUARK STATES IN THE
  HEAVY QUARK SYSTEM}},\ }\href {https://doi.org/10.1016/0550-3213(87)90248-3}
  {\bibfield  {journal} {\bibinfo  {journal} {Nucl. Phys. B}\ }\textbf
  {\bibinfo {volume} {281}},\ \bibinfo {pages} {85} (\bibinfo {year}
  {1987})}\BibitemShut {NoStop}%
\bibitem [{\citenamefont {Brambilla}\ \emph {et~al.}(2016)\citenamefont
  {Brambilla}, \citenamefont {Krein}, \citenamefont {Tarr\'us~Castell\`a},\
  and\ \citenamefont {Vairo}}]{Brambilla:2015rqa}%
  \BibitemOpen
  \bibfield  {author} {\bibinfo {author} {\bibfnamefont {N.}~\bibnamefont
  {Brambilla}}, \bibinfo {author} {\bibfnamefont {G.~a.}\ \bibnamefont
  {Krein}}, \bibinfo {author} {\bibfnamefont {J.}~\bibnamefont
  {Tarr\'us~Castell\`a}},\ and\ \bibinfo {author} {\bibfnamefont
  {A.}~\bibnamefont {Vairo}},\ }\bibfield  {title} {\bibinfo {title}
  {{Long-range properties of $1S$ bottomonium states}},\ }\href
  {https://doi.org/10.1103/PhysRevD.93.054002} {\bibfield  {journal} {\bibinfo
  {journal} {Phys. Rev. D}\ }\textbf {\bibinfo {volume} {93}},\ \bibinfo
  {pages} {054002} (\bibinfo {year} {2016})},\ \Eprint
  {https://arxiv.org/abs/1510.05895} {arXiv:1510.05895 [hep-ph]} \BibitemShut
  {NoStop}%
\bibitem [{\citenamefont {Chen}\ \emph {et~al.}(2017)\citenamefont {Chen},
  \citenamefont {Chen}, \citenamefont {Liu}, \citenamefont {Steele},\ and\
  \citenamefont {Zhu}}]{Chen:2016jxd}%
  \BibitemOpen
  \bibfield  {author} {\bibinfo {author} {\bibfnamefont {W.}~\bibnamefont
  {Chen}}, \bibinfo {author} {\bibfnamefont {H.-X.}\ \bibnamefont {Chen}},
  \bibinfo {author} {\bibfnamefont {X.}~\bibnamefont {Liu}}, \bibinfo {author}
  {\bibfnamefont {T.~G.}\ \bibnamefont {Steele}},\ and\ \bibinfo {author}
  {\bibfnamefont {S.-L.}\ \bibnamefont {Zhu}},\ }\bibfield  {title} {\bibinfo
  {title} {{Hunting for exotic doubly hidden-charm/bottom tetraquark states}},\
  }\href {https://doi.org/10.1016/j.physletb.2017.08.034} {\bibfield  {journal}
  {\bibinfo  {journal} {Phys. Lett. B}\ }\textbf {\bibinfo {volume} {773}},\
  \bibinfo {pages} {247} (\bibinfo {year} {2017})},\ \Eprint
  {https://arxiv.org/abs/1605.01647} {arXiv:1605.01647 [hep-ph]} \BibitemShut
  {NoStop}%
\bibitem [{\citenamefont {Wang}(2017)}]{Wang:2017jtz}%
  \BibitemOpen
  \bibfield  {author} {\bibinfo {author} {\bibfnamefont {Z.-G.}\ \bibnamefont
  {Wang}},\ }\bibfield  {title} {\bibinfo {title} {{Analysis of the
  $QQ\bar{Q}\bar{Q}$ tetraquark states with QCD sum rules}},\ }\href
  {https://doi.org/10.1140/epjc/s10052-017-4997-0} {\bibfield  {journal}
  {\bibinfo  {journal} {Eur. Phys. J. C}\ }\textbf {\bibinfo {volume} {77}},\
  \bibinfo {pages} {432} (\bibinfo {year} {2017})},\ \Eprint
  {https://arxiv.org/abs/1701.04285} {arXiv:1701.04285 [hep-ph]} \BibitemShut
  {NoStop}%
\bibitem [{\citenamefont {Wang}\ and\ \citenamefont {Di}(2019)}]{Wang:2018poa}%
  \BibitemOpen
  \bibfield  {author} {\bibinfo {author} {\bibfnamefont {Z.-G.}\ \bibnamefont
  {Wang}}\ and\ \bibinfo {author} {\bibfnamefont {Z.-Y.}\ \bibnamefont {Di}},\
  }\bibfield  {title} {\bibinfo {title} {{Analysis of the vector and
  axialvector $QQ\bar{Q}\bar{Q}$ tetraquark states with QCD sum rules}},\
  }\href {https://doi.org/10.5506/APhysPolB.50.1335} {\bibfield  {journal}
  {\bibinfo  {journal} {Acta Phys. Polon. B}\ }\textbf {\bibinfo {volume}
  {50}},\ \bibinfo {pages} {1335} (\bibinfo {year} {2019})},\ \Eprint
  {https://arxiv.org/abs/1807.08520} {arXiv:1807.08520 [hep-ph]} \BibitemShut
  {NoStop}%
\bibitem [{\citenamefont {Wang}(2022)}]{Wang:2022xja}%
  \BibitemOpen
  \bibfield  {author} {\bibinfo {author} {\bibfnamefont {Z.-G.}\ \bibnamefont
  {Wang}},\ }\bibfield  {title} {\bibinfo {title} {{Analysis of the X(6600),
  X(6900), X(7300) and related tetraquark states with the QCD sum rules}},\
  }\href {https://doi.org/10.1016/j.nuclphysb.2022.115983} {\bibfield
  {journal} {\bibinfo  {journal} {Nucl. Phys. B}\ }\textbf {\bibinfo {volume}
  {985}},\ \bibinfo {pages} {115983} (\bibinfo {year} {2022})},\ \Eprint
  {https://arxiv.org/abs/2207.08059} {arXiv:2207.08059 [hep-ph]} \BibitemShut
  {NoStop}%
\bibitem [{\citenamefont {Belov}\ \emph {et~al.}(2025)\citenamefont {Belov},
  \citenamefont {Giachino},\ and\ \citenamefont {Santopinto}}]{Belov:2024qyi}%
  \BibitemOpen
  \bibfield  {author} {\bibinfo {author} {\bibfnamefont {I.}~\bibnamefont
  {Belov}}, \bibinfo {author} {\bibfnamefont {A.}~\bibnamefont {Giachino}},\
  and\ \bibinfo {author} {\bibfnamefont {E.}~\bibnamefont {Santopinto}},\
  }\bibfield  {title} {\bibinfo {title} {{Fully charmed tetraquark production
  at the LHC experiments}},\ }\href {https://doi.org/10.1007/JHEP01(2025)093}
  {\bibfield  {journal} {\bibinfo  {journal} {JHEP}\ }\textbf {\bibinfo
  {volume} {01}},\ \bibinfo {pages} {093}},\ \Eprint
  {https://arxiv.org/abs/2409.12070} {arXiv:2409.12070 [hep-ph]} \BibitemShut
  {NoStop}%
\bibitem [{\citenamefont {L\"uscher}(1986)}]{Luscher:1986pf}%
  \BibitemOpen
  \bibfield  {author} {\bibinfo {author} {\bibfnamefont {M.}~\bibnamefont
  {L\"uscher}},\ }\bibfield  {title} {\bibinfo {title} {{Volume dependence of
  the energy spectrum in massive quantum field theories. II. Scattering
  states}},\ }\href {https://doi.org/10.1007/BF01211097} {\bibfield  {journal}
  {\bibinfo  {journal} {Commun. Math. Phys.}\ }\textbf {\bibinfo {volume}
  {105}},\ \bibinfo {pages} {153} (\bibinfo {year} {1986})}\BibitemShut
  {NoStop}%
\bibitem [{\citenamefont {L\"uscher}(1991{\natexlab{a}})}]{Luscher:1990ux}%
  \BibitemOpen
  \bibfield  {author} {\bibinfo {author} {\bibfnamefont {M.}~\bibnamefont
  {L\"uscher}},\ }\bibfield  {title} {\bibinfo {title} {{Two-particle states on
  a torus and their relation to the scattering matrix}},\ }\href
  {https://doi.org/10.1016/0550-3213(91)90366-6} {\bibfield  {journal}
  {\bibinfo  {journal} {Nucl. Phys. B}\ }\textbf {\bibinfo {volume} {354}},\
  \bibinfo {pages} {531} (\bibinfo {year} {1991}{\natexlab{a}})}\BibitemShut
  {NoStop}%
\bibitem [{\citenamefont {L\"uscher}(1991{\natexlab{b}})}]{Luscher:1991cf}%
  \BibitemOpen
  \bibfield  {author} {\bibinfo {author} {\bibfnamefont {M.}~\bibnamefont
  {L\"uscher}},\ }\bibfield  {title} {\bibinfo {title} {{Signatures of unstable
  particles in finite volume}},\ }\href
  {https://doi.org/10.1016/0550-3213(91)90584-K} {\bibfield  {journal}
  {\bibinfo  {journal} {Nucl. Phys. B}\ }\textbf {\bibinfo {volume} {364}},\
  \bibinfo {pages} {237} (\bibinfo {year} {1991}{\natexlab{b}})}\BibitemShut
  {NoStop}%
\bibitem [{\citenamefont {Peardon}\ \emph
  {et~al.}(2009{\natexlab{a}})\citenamefont {Peardon}, \citenamefont {Bulava},
  \citenamefont {Foley}, \citenamefont {Morningstar}, \citenamefont {Dudek},
  \citenamefont {Edwards}, \citenamefont {Joo}, \citenamefont {Lin},
  \citenamefont {Richards},\ and\ \citenamefont {Juge}}]{Peardon:2009gh}%
  \BibitemOpen
  \bibfield  {author} {\bibinfo {author} {\bibfnamefont {M.}~\bibnamefont
  {Peardon}}, \bibinfo {author} {\bibfnamefont {J.}~\bibnamefont {Bulava}},
  \bibinfo {author} {\bibfnamefont {J.}~\bibnamefont {Foley}}, \bibinfo
  {author} {\bibfnamefont {C.}~\bibnamefont {Morningstar}}, \bibinfo {author}
  {\bibfnamefont {J.}~\bibnamefont {Dudek}}, \bibinfo {author} {\bibfnamefont
  {R.~G.}\ \bibnamefont {Edwards}}, \bibinfo {author} {\bibfnamefont
  {B.}~\bibnamefont {Joo}}, \bibinfo {author} {\bibfnamefont {H.-W.}\
  \bibnamefont {Lin}}, \bibinfo {author} {\bibfnamefont {D.~G.}\ \bibnamefont
  {Richards}},\ and\ \bibinfo {author} {\bibfnamefont {K.~J.}\ \bibnamefont
  {Juge}} (\bibinfo {collaboration} {Hadron Spectrum}),\ }\bibfield  {title}
  {\bibinfo {title} {{Novel quark-field creation operator construction for
  hadronic physics in lattice QCD}},\ }\href
  {https://doi.org/10.1103/PhysRevD.80.054506} {\bibfield  {journal} {\bibinfo
  {journal} {Phys. Rev. D}\ }\textbf {\bibinfo {volume} {80}},\ \bibinfo
  {pages} {054506} (\bibinfo {year} {2009}{\natexlab{a}})},\ \Eprint
  {https://arxiv.org/abs/0905.2160} {arXiv:0905.2160 [hep-lat]} \BibitemShut
  {NoStop}%
\bibitem [{\citenamefont {Morningstar}\ and\ \citenamefont
  {Peardon}(1997)}]{Morningstar:1997ff}%
  \BibitemOpen
  \bibfield  {author} {\bibinfo {author} {\bibfnamefont {C.~J.}\ \bibnamefont
  {Morningstar}}\ and\ \bibinfo {author} {\bibfnamefont {M.~J.}\ \bibnamefont
  {Peardon}},\ }\bibfield  {title} {\bibinfo {title} {{Efficient glueball
  simulations on anisotropic lattices}},\ }\href
  {https://doi.org/10.1103/PhysRevD.56.4043} {\bibfield  {journal} {\bibinfo
  {journal} {Phys. Rev. D}\ }\textbf {\bibinfo {volume} {56}},\ \bibinfo
  {pages} {4043} (\bibinfo {year} {1997})},\ \Eprint
  {https://arxiv.org/abs/hep-lat/9704011} {arXiv:hep-lat/9704011} \BibitemShut
  {NoStop}%
\bibitem [{\citenamefont {Chen}\ \emph {et~al.}(2006)\citenamefont {Chen} \emph
  {et~al.}}]{Chen:2005mg}%
  \BibitemOpen
  \bibfield  {author} {\bibinfo {author} {\bibfnamefont {Y.}~\bibnamefont
  {Chen}} \emph {et~al.},\ }\bibfield  {title} {\bibinfo {title} {{Glueball
  spectrum and matrix elements on anisotropic lattices}},\ }\href
  {https://doi.org/10.1103/PhysRevD.73.014516} {\bibfield  {journal} {\bibinfo
  {journal} {Phys. Rev. D}\ }\textbf {\bibinfo {volume} {73}},\ \bibinfo
  {pages} {014516} (\bibinfo {year} {2006})},\ \Eprint
  {https://arxiv.org/abs/hep-lat/0510074} {arXiv:hep-lat/0510074} \BibitemShut
  {NoStop}%
\bibitem [{\citenamefont {Zhang}\ and\ \citenamefont
  {Liu}(2001)}]{Zhang:2001in}%
  \BibitemOpen
  \bibfield  {author} {\bibinfo {author} {\bibfnamefont {J.-h.}\ \bibnamefont
  {Zhang}}\ and\ \bibinfo {author} {\bibfnamefont {C.}~\bibnamefont {Liu}},\
  }\bibfield  {title} {\bibinfo {title} {{Tuning the tadpole improved clover
  Wilson action on coarse anisotropic lattices}},\ }\href
  {https://doi.org/10.1142/S0217732301005096} {\bibfield  {journal} {\bibinfo
  {journal} {Mod. Phys. Lett. A}\ }\textbf {\bibinfo {volume} {16}},\ \bibinfo
  {pages} {1841} (\bibinfo {year} {2001})},\ \Eprint
  {https://arxiv.org/abs/hep-lat/0107005} {arXiv:hep-lat/0107005} \BibitemShut
  {NoStop}%
\bibitem [{\citenamefont {Su}\ \emph {et~al.}(2006)\citenamefont {Su},
  \citenamefont {Liu}, \citenamefont {Li},\ and\ \citenamefont
  {Liu}}]{Su:2004sc}%
  \BibitemOpen
  \bibfield  {author} {\bibinfo {author} {\bibfnamefont {S.-q.}\ \bibnamefont
  {Su}}, \bibinfo {author} {\bibfnamefont {L.-m.}\ \bibnamefont {Liu}},
  \bibinfo {author} {\bibfnamefont {X.}~\bibnamefont {Li}},\ and\ \bibinfo
  {author} {\bibfnamefont {C.}~\bibnamefont {Liu}},\ }\bibfield  {title}
  {\bibinfo {title} {{A Numerical study of improved quark actions on
  anisotropic lattices}},\ }\href {https://doi.org/10.1142/S0217751X06024967}
  {\bibfield  {journal} {\bibinfo  {journal} {Int. J. Mod. Phys. A}\ }\textbf
  {\bibinfo {volume} {21}},\ \bibinfo {pages} {1015} (\bibinfo {year}
  {2006})},\ \Eprint {https://arxiv.org/abs/hep-lat/0412034}
  {arXiv:hep-lat/0412034} \BibitemShut {NoStop}%
\bibitem [{\citenamefont {Meng}\ \emph {et~al.}(2009)\citenamefont {Meng} \emph
  {et~al.}}]{CLQCD:2009nvn}%
  \BibitemOpen
  \bibfield  {author} {\bibinfo {author} {\bibfnamefont {G.-Z.}\ \bibnamefont
  {Meng}} \emph {et~al.} (\bibinfo {collaboration} {CLQCD}),\ }\bibfield
  {title} {\bibinfo {title} {{Low-energy $D^{*+} \bar{D}^0_1$ scattering and
  the resonancelike structure $Z^+$ (4430)}},\ }\href
  {https://doi.org/10.1103/PhysRevD.80.034503} {\bibfield  {journal} {\bibinfo
  {journal} {Phys. Rev. D}\ }\textbf {\bibinfo {volume} {80}},\ \bibinfo
  {pages} {034503} (\bibinfo {year} {2009})},\ \Eprint
  {https://arxiv.org/abs/0905.0752} {arXiv:0905.0752 [hep-lat]} \BibitemShut
  {NoStop}%
\bibitem [{\citenamefont {Li}\ \emph {et~al.}(2024)\citenamefont {Li},
  \citenamefont {Shi}, \citenamefont {Chen}, \citenamefont {Gong},
  \citenamefont {Liang}, \citenamefont {Liu},\ and\ \citenamefont
  {Sun}}]{Li:2024pfg}%
  \BibitemOpen
  \bibfield  {author} {\bibinfo {author} {\bibfnamefont {H.}~\bibnamefont
  {Li}}, \bibinfo {author} {\bibfnamefont {C.}~\bibnamefont {Shi}}, \bibinfo
  {author} {\bibfnamefont {Y.}~\bibnamefont {Chen}}, \bibinfo {author}
  {\bibfnamefont {M.}~\bibnamefont {Gong}}, \bibinfo {author} {\bibfnamefont
  {J.}~\bibnamefont {Liang}}, \bibinfo {author} {\bibfnamefont
  {Z.}~\bibnamefont {Liu}},\ and\ \bibinfo {author} {\bibfnamefont
  {W.}~\bibnamefont {Sun}},\ }\href@noop {} {\bibinfo {title} {{$X(3872)$
  Relevant $D\bar{D}^*$ Scattering in $N_f=2$ Lattice QCD}}} (\bibinfo {year}
  {2024}),\ \Eprint {https://arxiv.org/abs/2402.14541} {arXiv:2402.14541
  [hep-lat]} \BibitemShut {NoStop}%
\bibitem [{\citenamefont {L\"uscher}(2010)}]{Luscher:2010iy}%
  \BibitemOpen
  \bibfield  {author} {\bibinfo {author} {\bibfnamefont {M.}~\bibnamefont
  {L\"uscher}},\ }\bibfield  {title} {\bibinfo {title} {{Properties and uses of
  the Wilson flow in lattice QCD}},\ }\href
  {https://doi.org/10.1007/JHEP08(2010)071} {\bibfield  {journal} {\bibinfo
  {journal} {JHEP}\ }\textbf {\bibinfo {volume} {08}},\ \bibinfo {pages}
  {071}},\ \bibinfo {note} {[Erratum: JHEP 03, 092 (2014)]},\ \Eprint
  {https://arxiv.org/abs/1006.4518} {arXiv:1006.4518 [hep-lat]} \BibitemShut
  {NoStop}%
\bibitem [{\citenamefont {Bors\'anyi}\ \emph {et~al.}(2012)\citenamefont
  {Bors\'anyi}, \citenamefont {D\"urr}, \citenamefont {Fodor}, \citenamefont
  {Hoelbling}, \citenamefont {Katz}, \citenamefont {Krieg}, \citenamefont
  {Kurth}, \citenamefont {Lellouch}, \citenamefont {Lippert},\ and\
  \citenamefont {McNeile}}]{BMW:2012hcm}%
  \BibitemOpen
  \bibfield  {author} {\bibinfo {author} {\bibfnamefont {S.}~\bibnamefont
  {Bors\'anyi}}, \bibinfo {author} {\bibfnamefont {S.}~\bibnamefont {D\"urr}},
  \bibinfo {author} {\bibfnamefont {Z.}~\bibnamefont {Fodor}}, \bibinfo
  {author} {\bibfnamefont {C.}~\bibnamefont {Hoelbling}}, \bibinfo {author}
  {\bibfnamefont {S.~D.}\ \bibnamefont {Katz}}, \bibinfo {author}
  {\bibfnamefont {S.}~\bibnamefont {Krieg}}, \bibinfo {author} {\bibfnamefont
  {T.}~\bibnamefont {Kurth}}, \bibinfo {author} {\bibfnamefont
  {L.}~\bibnamefont {Lellouch}}, \bibinfo {author} {\bibfnamefont
  {T.}~\bibnamefont {Lippert}},\ and\ \bibinfo {author} {\bibfnamefont
  {C.}~\bibnamefont {McNeile}} (\bibinfo {collaboration} {BMW}),\ }\bibfield
  {title} {\bibinfo {title} {{High-precision scale setting in lattice QCD}},\
  }\href {https://doi.org/10.1007/JHEP09(2012)010} {\bibfield  {journal}
  {\bibinfo  {journal} {JHEP}\ }\textbf {\bibinfo {volume} {09}},\ \bibinfo
  {pages} {010}},\ \Eprint {https://arxiv.org/abs/1203.4469} {arXiv:1203.4469
  [hep-lat]} \BibitemShut {NoStop}%
\bibitem [{\citenamefont {Bali}\ \emph {et~al.}(2016)\citenamefont {Bali},
  \citenamefont {Lang}, \citenamefont {Musch},\ and\ \citenamefont
  {Sch{\"a}fer}}]{Bali:2016lva}%
  \BibitemOpen
  \bibfield  {author} {\bibinfo {author} {\bibfnamefont {G.~S.}\ \bibnamefont
  {Bali}}, \bibinfo {author} {\bibfnamefont {B.}~\bibnamefont {Lang}}, \bibinfo
  {author} {\bibfnamefont {B.~U.}\ \bibnamefont {Musch}},\ and\ \bibinfo
  {author} {\bibfnamefont {A.}~\bibnamefont {Sch{\"a}fer}},\ }\bibfield
  {title} {\bibinfo {title} {{Novel quark smearing for hadrons with high
  momenta in lattice QCD}},\ }\href
  {https://doi.org/10.1103/PhysRevD.93.094515} {\bibfield  {journal} {\bibinfo
  {journal} {Phys. Rev. D}\ }\textbf {\bibinfo {volume} {93}},\ \bibinfo
  {pages} {094515} (\bibinfo {year} {2016})},\ \Eprint
  {https://arxiv.org/abs/1602.05525} {arXiv:1602.05525 [hep-lat]} \BibitemShut
  {NoStop}%
\bibitem [{\citenamefont {Wilson}\ \emph {et~al.}(2024)\citenamefont {Wilson},
  \citenamefont {Thomas}, \citenamefont {Dudek},\ and\ \citenamefont
  {Edwards}}]{Wilson:2023anv}%
  \BibitemOpen
  \bibfield  {author} {\bibinfo {author} {\bibfnamefont {D.~J.}\ \bibnamefont
  {Wilson}}, \bibinfo {author} {\bibfnamefont {C.~E.}\ \bibnamefont {Thomas}},
  \bibinfo {author} {\bibfnamefont {J.~J.}\ \bibnamefont {Dudek}},\ and\
  \bibinfo {author} {\bibfnamefont {R.~G.}\ \bibnamefont {Edwards}} (\bibinfo
  {collaboration} {Hadron Spectrum}),\ }\bibfield  {title} {\bibinfo {title}
  {{Charmonium {\ensuremath{\chi}}c0 and {\ensuremath{\chi}}c2 resonances in
  coupled-channel scattering from lattice QCD}},\ }\href
  {https://doi.org/10.1103/PhysRevD.109.114503} {\bibfield  {journal} {\bibinfo
   {journal} {Phys. Rev. D}\ }\textbf {\bibinfo {volume} {109}},\ \bibinfo
  {pages} {114503} (\bibinfo {year} {2024})},\ \Eprint
  {https://arxiv.org/abs/2309.14071} {arXiv:2309.14071 [hep-lat]} \BibitemShut
  {NoStop}%
\bibitem [{\citenamefont {Cheung}\ \emph {et~al.}(2017)\citenamefont {Cheung},
  \citenamefont {Thomas}, \citenamefont {Dudek},\ and\ \citenamefont
  {Edwards}}]{Cheung:2017tnt}%
  \BibitemOpen
  \bibfield  {author} {\bibinfo {author} {\bibfnamefont {G.~K.~C.}\
  \bibnamefont {Cheung}}, \bibinfo {author} {\bibfnamefont {C.~E.}\
  \bibnamefont {Thomas}}, \bibinfo {author} {\bibfnamefont {J.~J.}\
  \bibnamefont {Dudek}},\ and\ \bibinfo {author} {\bibfnamefont {R.~G.}\
  \bibnamefont {Edwards}} (\bibinfo {collaboration} {Hadron Spectrum}),\
  }\bibfield  {title} {\bibinfo {title} {{Tetraquark operators in lattice QCD
  and exotic flavour states in the charm sector}},\ }\href
  {https://doi.org/10.1007/JHEP11(2017)033} {\bibfield  {journal} {\bibinfo
  {journal} {JHEP}\ }\textbf {\bibinfo {volume} {11}},\ \bibinfo {pages}
  {033}},\ \Eprint {https://arxiv.org/abs/1709.01417} {arXiv:1709.01417
  [hep-lat]} \BibitemShut {NoStop}%
\bibitem [{\citenamefont {Peardon}\ \emph
  {et~al.}(2009{\natexlab{b}})\citenamefont {Peardon}, \citenamefont {Bulava},
  \citenamefont {Foley}, \citenamefont {Morningstar}, \citenamefont {Dudek},
  \citenamefont {Edwards}, \citenamefont {Joo}, \citenamefont {Lin},
  \citenamefont {Richards},\ and\ \citenamefont
  {Juge}}]{HadronSpectrum:2009krc}%
  \BibitemOpen
  \bibfield  {author} {\bibinfo {author} {\bibfnamefont {M.}~\bibnamefont
  {Peardon}}, \bibinfo {author} {\bibfnamefont {J.}~\bibnamefont {Bulava}},
  \bibinfo {author} {\bibfnamefont {J.}~\bibnamefont {Foley}}, \bibinfo
  {author} {\bibfnamefont {C.}~\bibnamefont {Morningstar}}, \bibinfo {author}
  {\bibfnamefont {J.}~\bibnamefont {Dudek}}, \bibinfo {author} {\bibfnamefont
  {R.~G.}\ \bibnamefont {Edwards}}, \bibinfo {author} {\bibfnamefont
  {B.}~\bibnamefont {Joo}}, \bibinfo {author} {\bibfnamefont {H.-W.}\
  \bibnamefont {Lin}}, \bibinfo {author} {\bibfnamefont {D.~G.}\ \bibnamefont
  {Richards}},\ and\ \bibinfo {author} {\bibfnamefont {K.~J.}\ \bibnamefont
  {Juge}} (\bibinfo {collaboration} {Hadron Spectrum}),\ }\bibfield  {title}
  {\bibinfo {title} {{A Novel quark-field creation operator construction for
  hadronic physics in lattice QCD}},\ }\href
  {https://doi.org/10.1103/PhysRevD.80.054506} {\bibfield  {journal} {\bibinfo
  {journal} {Phys. Rev. D}\ }\textbf {\bibinfo {volume} {80}},\ \bibinfo
  {pages} {054506} (\bibinfo {year} {2009}{\natexlab{b}})},\ \Eprint
  {https://arxiv.org/abs/0905.2160} {arXiv:0905.2160 [hep-lat]} \BibitemShut
  {NoStop}%
\bibitem [{\citenamefont {Zhang}\ \emph {et~al.}(2022)\citenamefont {Zhang},
  \citenamefont {Sun}, \citenamefont {Chen}, \citenamefont {Chen},
  \citenamefont {Gong}, \citenamefont {Jiang},\ and\ \citenamefont
  {Liu}}]{ZhangRenQiang:2021gnn}%
  \BibitemOpen
  \bibfield  {author} {\bibinfo {author} {\bibfnamefont {R.}~\bibnamefont
  {Zhang}}, \bibinfo {author} {\bibfnamefont {W.}~\bibnamefont {Sun}}, \bibinfo
  {author} {\bibfnamefont {F.}~\bibnamefont {Chen}}, \bibinfo {author}
  {\bibfnamefont {Y.}~\bibnamefont {Chen}}, \bibinfo {author} {\bibfnamefont
  {M.}~\bibnamefont {Gong}}, \bibinfo {author} {\bibfnamefont {X.}~\bibnamefont
  {Jiang}},\ and\ \bibinfo {author} {\bibfnamefont {Z.}~\bibnamefont {Liu}},\
  }\bibfield  {title} {\bibinfo {title} {{Annihilation diagram contribution to
  charmonium masses *}},\ }\href {https://doi.org/10.1088/1674-1137/ac3d8c}
  {\bibfield  {journal} {\bibinfo  {journal} {Chin. Phys. C}\ }\textbf
  {\bibinfo {volume} {46}},\ \bibinfo {pages} {043102} (\bibinfo {year}
  {2022})},\ \Eprint {https://arxiv.org/abs/2110.01755} {arXiv:2110.01755
  [hep-lat]} \BibitemShut {NoStop}%
\bibitem [{\citenamefont {Umeda}(2007)}]{Umeda:2007hy}%
  \BibitemOpen
  \bibfield  {author} {\bibinfo {author} {\bibfnamefont {T.}~\bibnamefont
  {Umeda}},\ }\bibfield  {title} {\bibinfo {title} {{A Constant contribution in
  meson correlators at finite temperature}},\ }\href
  {https://doi.org/10.1103/PhysRevD.75.094502} {\bibfield  {journal} {\bibinfo
  {journal} {Phys. Rev. D}\ }\textbf {\bibinfo {volume} {75}},\ \bibinfo
  {pages} {094502} (\bibinfo {year} {2007})},\ \Eprint
  {https://arxiv.org/abs/hep-lat/0701005} {arXiv:hep-lat/0701005} \BibitemShut
  {NoStop}%
\bibitem [{\citenamefont {Feng}\ \emph {et~al.}(2010)\citenamefont {Feng},
  \citenamefont {Jansen},\ and\ \citenamefont {Renner}}]{Feng:2009ij}%
  \BibitemOpen
  \bibfield  {author} {\bibinfo {author} {\bibfnamefont {X.}~\bibnamefont
  {Feng}}, \bibinfo {author} {\bibfnamefont {K.}~\bibnamefont {Jansen}},\ and\
  \bibinfo {author} {\bibfnamefont {D.~B.}\ \bibnamefont {Renner}},\ }\href
  {https://doi.org/10.1016/j.physletb.2010.01.018} {\bibfield  {journal}
  {\bibinfo  {journal} {Phys. Lett. B}\ }\textbf {\bibinfo {volume} {684}},\
  \bibinfo {pages} {268} (\bibinfo {year} {2010})},\ \Eprint
  {https://arxiv.org/abs/0909.3255} {arXiv:0909.3255 [hep-lat]} \BibitemShut
  {NoStop}%
\bibitem [{\citenamefont {Dudek}\ \emph {et~al.}(2012)\citenamefont {Dudek},
  \citenamefont {Edwards},\ and\ \citenamefont {Thomas}}]{Dudek:2012gj}%
  \BibitemOpen
  \bibfield  {author} {\bibinfo {author} {\bibfnamefont {J.~J.}\ \bibnamefont
  {Dudek}}, \bibinfo {author} {\bibfnamefont {R.~G.}\ \bibnamefont {Edwards}},\
  and\ \bibinfo {author} {\bibfnamefont {C.~E.}\ \bibnamefont {Thomas}},\
  }\bibfield  {title} {\bibinfo {title} {{S and D-wave phase shifts in
  isospin-2 pi pi scattering from lattice QCD}},\ }\href
  {https://doi.org/10.1103/PhysRevD.86.034031} {\bibfield  {journal} {\bibinfo
  {journal} {Phys. Rev. D}\ }\textbf {\bibinfo {volume} {86}},\ \bibinfo
  {pages} {034031} (\bibinfo {year} {2012})},\ \Eprint
  {https://arxiv.org/abs/1203.6041} {arXiv:1203.6041 [hep-ph]} \BibitemShut
  {NoStop}%
\bibitem [{\citenamefont {Meng}\ \emph {et~al.}(2025)\citenamefont {Meng},
  \citenamefont {Liu}, \citenamefont {Tuo}, \citenamefont {Yan},\ and\
  \citenamefont {Zhang}}]{Meng:2024czd}%
  \BibitemOpen
  \bibfield  {author} {\bibinfo {author} {\bibfnamefont {Y.}~\bibnamefont
  {Meng}}, \bibinfo {author} {\bibfnamefont {C.}~\bibnamefont {Liu}}, \bibinfo
  {author} {\bibfnamefont {X.-Y.}\ \bibnamefont {Tuo}}, \bibinfo {author}
  {\bibfnamefont {H.}~\bibnamefont {Yan}},\ and\ \bibinfo {author}
  {\bibfnamefont {Z.}~\bibnamefont {Zhang}},\ }\bibfield  {title} {\bibinfo
  {title} {{Lattice calculation of the $\eta _c\eta _c$ and $J/\psi J/\psi $
  s-wave scattering length}},\ }\href
  {https://doi.org/10.1140/epjc/s10052-025-14192-z} {\bibfield  {journal}
  {\bibinfo  {journal} {Eur. Phys. J. C}\ }\textbf {\bibinfo {volume} {85}},\
  \bibinfo {pages} {458} (\bibinfo {year} {2025})},\ \Eprint
  {https://arxiv.org/abs/2411.11533} {arXiv:2411.11533 [hep-lat]} \BibitemShut
  {NoStop}%
\bibitem [{\citenamefont {Du}\ \emph {et~al.}(2023)\citenamefont {Du},
  \citenamefont {Filin}, \citenamefont {Baru}, \citenamefont {Dong},
  \citenamefont {Epelbaum}, \citenamefont {Guo}, \citenamefont {Hanhart},
  \citenamefont {Nefediev}, \citenamefont {Nieves},\ and\ \citenamefont
  {Wang}}]{Du:2023hlu}%
  \BibitemOpen
  \bibfield  {author} {\bibinfo {author} {\bibfnamefont {M.-L.}\ \bibnamefont
  {Du}}, \bibinfo {author} {\bibfnamefont {A.}~\bibnamefont {Filin}}, \bibinfo
  {author} {\bibfnamefont {V.}~\bibnamefont {Baru}}, \bibinfo {author}
  {\bibfnamefont {X.-K.}\ \bibnamefont {Dong}}, \bibinfo {author}
  {\bibfnamefont {E.}~\bibnamefont {Epelbaum}}, \bibinfo {author}
  {\bibfnamefont {F.-K.}\ \bibnamefont {Guo}}, \bibinfo {author} {\bibfnamefont
  {C.}~\bibnamefont {Hanhart}}, \bibinfo {author} {\bibfnamefont
  {A.}~\bibnamefont {Nefediev}}, \bibinfo {author} {\bibfnamefont
  {J.}~\bibnamefont {Nieves}},\ and\ \bibinfo {author} {\bibfnamefont
  {Q.}~\bibnamefont {Wang}},\ }\bibfield  {title} {\bibinfo {title} {{Role of
  Left-Hand Cut Contributions on Pole Extractions from Lattice Data: Case Study
  for Tcc(3875)+}},\ }\href {https://doi.org/10.1103/PhysRevLett.131.131903}
  {\bibfield  {journal} {\bibinfo  {journal} {Phys. Rev. Lett.}\ }\textbf
  {\bibinfo {volume} {131}},\ \bibinfo {pages} {131903} (\bibinfo {year}
  {2023})},\ \Eprint {https://arxiv.org/abs/2303.09441} {arXiv:2303.09441
  [hep-ph]} \BibitemShut {NoStop}%
\bibitem [{\citenamefont {Meng}\ \emph {et~al.}(2024)\citenamefont {Meng},
  \citenamefont {Baru}, \citenamefont {Epelbaum}, \citenamefont {Filin},\ and\
  \citenamefont {Gasparyan}}]{Meng:2023bmz}%
  \BibitemOpen
  \bibfield  {author} {\bibinfo {author} {\bibfnamefont {L.}~\bibnamefont
  {Meng}}, \bibinfo {author} {\bibfnamefont {V.}~\bibnamefont {Baru}}, \bibinfo
  {author} {\bibfnamefont {E.}~\bibnamefont {Epelbaum}}, \bibinfo {author}
  {\bibfnamefont {A.~A.}\ \bibnamefont {Filin}},\ and\ \bibinfo {author}
  {\bibfnamefont {A.~M.}\ \bibnamefont {Gasparyan}},\ }\bibfield  {title}
  {\bibinfo {title} {{Solving the left-hand cut problem in lattice QCD:
  Tcc(3875)+ from finite volume energy levels}},\ }\href
  {https://doi.org/10.1103/PhysRevD.109.L071506} {\bibfield  {journal}
  {\bibinfo  {journal} {Phys. Rev. D}\ }\textbf {\bibinfo {volume} {109}},\
  \bibinfo {pages} {L071506} (\bibinfo {year} {2024})},\ \Eprint
  {https://arxiv.org/abs/2312.01930} {arXiv:2312.01930 [hep-lat]} \BibitemShut
  {NoStop}%
\bibitem [{\citenamefont {Raposo}\ and\ \citenamefont
  {Hansen}(2024)}]{Raposo:2023oru}%
  \BibitemOpen
  \bibfield  {author} {\bibinfo {author} {\bibfnamefont {A.~B.}\ \bibnamefont
  {Raposo}}\ and\ \bibinfo {author} {\bibfnamefont {M.~T.}\ \bibnamefont
  {Hansen}},\ }\bibfield  {title} {\bibinfo {title} {{Finite-volume scattering
  on the left-hand cut}},\ }\href {https://doi.org/10.1007/JHEP08(2024)075}
  {\bibfield  {journal} {\bibinfo  {journal} {JHEP}\ }\textbf {\bibinfo
  {volume} {08}},\ \bibinfo {pages} {075}},\ \Eprint
  {https://arxiv.org/abs/2311.18793} {arXiv:2311.18793 [hep-lat]} \BibitemShut
  {NoStop}%
\bibitem [{\citenamefont {Iritani}\ \emph {et~al.}(2017)\citenamefont
  {Iritani}, \citenamefont {Aoki}, \citenamefont {Doi}, \citenamefont
  {Hatsuda}, \citenamefont {Ikeda}, \citenamefont {Inoue}, \citenamefont
  {Ishii}, \citenamefont {Nemura},\ and\ \citenamefont
  {Sasaki}}]{Iritani:2017rlk}%
  \BibitemOpen
  \bibfield  {author} {\bibinfo {author} {\bibfnamefont {T.}~\bibnamefont
  {Iritani}}, \bibinfo {author} {\bibfnamefont {S.}~\bibnamefont {Aoki}},
  \bibinfo {author} {\bibfnamefont {T.}~\bibnamefont {Doi}}, \bibinfo {author}
  {\bibfnamefont {T.}~\bibnamefont {Hatsuda}}, \bibinfo {author} {\bibfnamefont
  {Y.}~\bibnamefont {Ikeda}}, \bibinfo {author} {\bibfnamefont
  {T.}~\bibnamefont {Inoue}}, \bibinfo {author} {\bibfnamefont
  {N.}~\bibnamefont {Ishii}}, \bibinfo {author} {\bibfnamefont
  {H.}~\bibnamefont {Nemura}},\ and\ \bibinfo {author} {\bibfnamefont
  {K.}~\bibnamefont {Sasaki}},\ }\bibfield  {title} {\bibinfo {title} {{Are two
  nucleons bound in lattice QCD for heavy quark masses? Consistency check with
  L\"uscher\textquoteright{}s finite volume formula}},\ }\href
  {https://doi.org/10.1103/PhysRevD.96.034521} {\bibfield  {journal} {\bibinfo
  {journal} {Phys. Rev. D}\ }\textbf {\bibinfo {volume} {96}},\ \bibinfo
  {pages} {034521} (\bibinfo {year} {2017})},\ \Eprint
  {https://arxiv.org/abs/1703.07210} {arXiv:1703.07210 [hep-lat]} \BibitemShut
  {NoStop}%
\bibitem [{\citenamefont {Castillejo}\ \emph {et~al.}(1956)\citenamefont
  {Castillejo}, \citenamefont {Dalitz},\ and\ \citenamefont
  {Dyson}}]{Castillejo:1955ed}%
  \BibitemOpen
  \bibfield  {author} {\bibinfo {author} {\bibfnamefont {L.}~\bibnamefont
  {Castillejo}}, \bibinfo {author} {\bibfnamefont {R.~H.}\ \bibnamefont
  {Dalitz}},\ and\ \bibinfo {author} {\bibfnamefont {F.~J.}\ \bibnamefont
  {Dyson}},\ }\bibfield  {title} {\bibinfo {title} {{Low's scattering equation
  for the charged and neutral scalar theories}},\ }\href
  {https://doi.org/10.1103/PhysRev.101.453} {\bibfield  {journal} {\bibinfo
  {journal} {Phys. Rev.}\ }\textbf {\bibinfo {volume} {101}},\ \bibinfo {pages}
  {453} (\bibinfo {year} {1956})}\BibitemShut {NoStop}%
\bibitem [{\citenamefont {Oller}(2024)}]{Oller:2024lrk}%
  \BibitemOpen
  \bibfield  {author} {\bibinfo {author} {\bibfnamefont {J.~A.}\ \bibnamefont
  {Oller}},\ }\href@noop {} {\bibinfo {title} {{Lectures on scattering theory
  in partial-wave amplitudes}}} (\bibinfo {year} {2024}),\ \Eprint
  {https://arxiv.org/abs/2409.16790} {arXiv:2409.16790 [hep-ph]} \BibitemShut
  {NoStop}%
\bibitem [{\citenamefont {Dyson}(1957)}]{Dyson:1957rgq}%
  \BibitemOpen
  \bibfield  {author} {\bibinfo {author} {\bibfnamefont {F.~J.}\ \bibnamefont
  {Dyson}},\ }\bibfield  {title} {\bibinfo {title} {{Meaning of the solutions
  of Low's scattering equation}},\ }\href
  {https://doi.org/10.1103/physrev.106.157} {\bibfield  {journal} {\bibinfo
  {journal} {Phys. Rev.}\ }\textbf {\bibinfo {volume} {106}},\ \bibinfo {pages}
  {157} (\bibinfo {year} {1957})}\BibitemShut {NoStop}%
\bibitem [{\citenamefont {Krivoruchenko}(2010)}]{Krivoruchenko:2010ft}%
  \BibitemOpen
  \bibfield  {author} {\bibinfo {author} {\bibfnamefont {M.~I.}\ \bibnamefont
  {Krivoruchenko}},\ }\bibfield  {title} {\bibinfo {title} {{Remarks on the
  origin of Castillejo-Dalitz-Dyson poles}},\ }\href
  {https://doi.org/10.1103/PhysRevC.82.018201} {\bibfield  {journal} {\bibinfo
  {journal} {Phys. Rev. C}\ }\textbf {\bibinfo {volume} {82}},\ \bibinfo
  {pages} {018201} (\bibinfo {year} {2010})},\ \Eprint
  {https://arxiv.org/abs/1001.1659} {arXiv:1001.1659 [nucl-th]} \BibitemShut
  {NoStop}%
\bibitem [{\citenamefont {Li}\ \emph {et~al.}(2022)\citenamefont {Li},
  \citenamefont {Guo}, \citenamefont {Pang},\ and\ \citenamefont
  {Wu}}]{Li:2021cue}%
  \BibitemOpen
  \bibfield  {author} {\bibinfo {author} {\bibfnamefont {Y.}~\bibnamefont
  {Li}}, \bibinfo {author} {\bibfnamefont {F.-K.}\ \bibnamefont {Guo}},
  \bibinfo {author} {\bibfnamefont {J.-Y.}\ \bibnamefont {Pang}},\ and\
  \bibinfo {author} {\bibfnamefont {J.-J.}\ \bibnamefont {Wu}},\ }\bibfield
  {title} {\bibinfo {title} {{Generalization of Weinberg\textquoteright{}s
  compositeness relations}},\ }\href
  {https://doi.org/10.1103/PhysRevD.105.L071502} {\bibfield  {journal}
  {\bibinfo  {journal} {Phys. Rev. D}\ }\textbf {\bibinfo {volume} {105}},\
  \bibinfo {pages} {L071502} (\bibinfo {year} {2022})},\ \Eprint
  {https://arxiv.org/abs/2110.02766} {arXiv:2110.02766 [hep-ph]} \BibitemShut
  {NoStop}%
\bibitem [{\citenamefont {L\"u}\ \emph {et~al.}(2020)\citenamefont {L\"u},
  \citenamefont {Chen},\ and\ \citenamefont {Dong}}]{Lu:2020cns}%
  \BibitemOpen
  \bibfield  {author} {\bibinfo {author} {\bibfnamefont {Q.-F.}\ \bibnamefont
  {L\"u}}, \bibinfo {author} {\bibfnamefont {D.-Y.}\ \bibnamefont {Chen}},\
  and\ \bibinfo {author} {\bibfnamefont {Y.-B.}\ \bibnamefont {Dong}},\
  }\bibfield  {title} {\bibinfo {title} {{Masses of fully heavy tetraquarks $QQ
  {\bar{Q}} {\bar{Q}}$ in an extended relativized quark model}},\ }\href
  {https://doi.org/10.1140/epjc/s10052-020-08454-1} {\bibfield  {journal}
  {\bibinfo  {journal} {Eur. Phys. J. C}\ }\textbf {\bibinfo {volume} {80}},\
  \bibinfo {pages} {871} (\bibinfo {year} {2020})},\ \Eprint
  {https://arxiv.org/abs/2006.14445} {arXiv:2006.14445 [hep-ph]} \BibitemShut
  {NoStop}%
\bibitem [{\citenamefont {Wang}\ \emph {et~al.}(2021)\citenamefont {Wang},
  \citenamefont {Meng}, \citenamefont {Oka},\ and\ \citenamefont
  {Zhu}}]{Wang:2021kfv}%
  \BibitemOpen
  \bibfield  {author} {\bibinfo {author} {\bibfnamefont {G.-J.}\ \bibnamefont
  {Wang}}, \bibinfo {author} {\bibfnamefont {L.}~\bibnamefont {Meng}}, \bibinfo
  {author} {\bibfnamefont {M.}~\bibnamefont {Oka}},\ and\ \bibinfo {author}
  {\bibfnamefont {S.-L.}\ \bibnamefont {Zhu}},\ }\bibfield  {title} {\bibinfo
  {title} {{Higher fully charmed tetraquarks: Radial excitations and P-wave
  states}},\ }\href {https://doi.org/10.1103/PhysRevD.104.036016} {\bibfield
  {journal} {\bibinfo  {journal} {Phys. Rev. D}\ }\textbf {\bibinfo {volume}
  {104}},\ \bibinfo {pages} {036016} (\bibinfo {year} {2021})},\ \Eprint
  {https://arxiv.org/abs/2105.13109} {arXiv:2105.13109 [hep-ph]} \BibitemShut
  {NoStop}%
\bibitem [{\citenamefont {Gong}\ \emph
  {et~al.}(2022{\natexlab{b}})\citenamefont {Gong}, \citenamefont {Du},\ and\
  \citenamefont {Zhao}}]{Gong:2022hgd}%
  \BibitemOpen
  \bibfield  {author} {\bibinfo {author} {\bibfnamefont {C.}~\bibnamefont
  {Gong}}, \bibinfo {author} {\bibfnamefont {M.-C.}\ \bibnamefont {Du}},\ and\
  \bibinfo {author} {\bibfnamefont {Q.}~\bibnamefont {Zhao}},\ }\bibfield
  {title} {\bibinfo {title} {{Pseudoscalar charmonium pair interactions via the
  Pomeron exchange mechanism}},\ }\href
  {https://doi.org/10.1103/PhysRevD.106.054011} {\bibfield  {journal} {\bibinfo
   {journal} {Phys. Rev. D}\ }\textbf {\bibinfo {volume} {106}},\ \bibinfo
  {pages} {054011} (\bibinfo {year} {2022}{\natexlab{b}})},\ \Eprint
  {https://arxiv.org/abs/2206.13867} {arXiv:2206.13867 [hep-ph]} \BibitemShut
  {NoStop}%
\bibitem [{\citenamefont {Chen}\ \emph {et~al.}(2022)\citenamefont {Chen},
  \citenamefont {Shi}, \citenamefont {Chen}, \citenamefont {Gong},
  \citenamefont {Liu}, \citenamefont {Sun},\ and\ \citenamefont
  {Zhang}}]{Chen:2022vpo}%
  \BibitemOpen
  \bibfield  {author} {\bibinfo {author} {\bibfnamefont {S.}~\bibnamefont
  {Chen}}, \bibinfo {author} {\bibfnamefont {C.}~\bibnamefont {Shi}}, \bibinfo
  {author} {\bibfnamefont {Y.}~\bibnamefont {Chen}}, \bibinfo {author}
  {\bibfnamefont {M.}~\bibnamefont {Gong}}, \bibinfo {author} {\bibfnamefont
  {Z.}~\bibnamefont {Liu}}, \bibinfo {author} {\bibfnamefont {W.}~\bibnamefont
  {Sun}},\ and\ \bibinfo {author} {\bibfnamefont {R.}~\bibnamefont {Zhang}},\
  }\bibfield  {title} {\bibinfo {title} {{Tcc+(3875) relevant
  DD\textasteriskcentered{} scattering from Nf=2 lattice QCD}},\ }\href
  {https://doi.org/10.1016/j.physletb.2022.137391} {\bibfield  {journal}
  {\bibinfo  {journal} {Phys. Lett. B}\ }\textbf {\bibinfo {volume} {833}},\
  \bibinfo {pages} {137391} (\bibinfo {year} {2022})},\ \Eprint
  {https://arxiv.org/abs/2206.06185} {arXiv:2206.06185 [hep-lat]} \BibitemShut
  {NoStop}%
\bibitem [{\citenamefont {Guo}\ \emph {et~al.}(2015)\citenamefont {Guo},
  \citenamefont {Hanhart}, \citenamefont {Wang},\ and\ \citenamefont
  {Zhao}}]{Guo:2014iya}%
  \BibitemOpen
  \bibfield  {author} {\bibinfo {author} {\bibfnamefont {F.-K.}\ \bibnamefont
  {Guo}}, \bibinfo {author} {\bibfnamefont {C.}~\bibnamefont {Hanhart}},
  \bibinfo {author} {\bibfnamefont {Q.}~\bibnamefont {Wang}},\ and\ \bibinfo
  {author} {\bibfnamefont {Q.}~\bibnamefont {Zhao}},\ }\bibfield  {title}
  {\bibinfo {title} {{Could the near-threshold $XYZ$ states be simply kinematic
  effects?}},\ }\href {https://doi.org/10.1103/PhysRevD.91.051504} {\bibfield
  {journal} {\bibinfo  {journal} {Phys. Rev. D}\ }\textbf {\bibinfo {volume}
  {91}},\ \bibinfo {pages} {051504} (\bibinfo {year} {2015})},\ \Eprint
  {https://arxiv.org/abs/1411.5584} {arXiv:1411.5584 [hep-ph]} \BibitemShut
  {NoStop}%
\bibitem [{\citenamefont {Song}\ \emph {et~al.}(2025)\citenamefont {Song},
  \citenamefont {Zhang}, \citenamefont {Baru}, \citenamefont {Guo},
  \citenamefont {Hanhart},\ and\ \citenamefont {Nefediev}}]{Song:2024ykq}%
  \BibitemOpen
  \bibfield  {author} {\bibinfo {author} {\bibfnamefont {Y.-L.}\ \bibnamefont
  {Song}}, \bibinfo {author} {\bibfnamefont {Y.}~\bibnamefont {Zhang}},
  \bibinfo {author} {\bibfnamefont {V.}~\bibnamefont {Baru}}, \bibinfo {author}
  {\bibfnamefont {F.-K.}\ \bibnamefont {Guo}}, \bibinfo {author} {\bibfnamefont
  {C.}~\bibnamefont {Hanhart}},\ and\ \bibinfo {author} {\bibfnamefont
  {A.}~\bibnamefont {Nefediev}},\ }\bibfield  {title} {\bibinfo {title}
  {{Toward a precision determination of the X(6200) parameters from data}},\
  }\href {https://doi.org/10.1103/PhysRevD.111.034038} {\bibfield  {journal}
  {\bibinfo  {journal} {Phys. Rev. D}\ }\textbf {\bibinfo {volume} {111}},\
  \bibinfo {pages} {034038} (\bibinfo {year} {2025})},\ \Eprint
  {https://arxiv.org/abs/2411.12062} {arXiv:2411.12062 [hep-ph]} \BibitemShut
  {NoStop}%
\bibitem [{\citenamefont {Edwards}\ and\ \citenamefont
  {Joo}(2005)}]{Edwards:2004sx}%
  \BibitemOpen
  \bibfield  {author} {\bibinfo {author} {\bibfnamefont {R.~G.}\ \bibnamefont
  {Edwards}}\ and\ \bibinfo {author} {\bibfnamefont {B.}~\bibnamefont {Joo}}
  (\bibinfo {collaboration} {SciDAC, LHPC, UKQCD}),\ }\bibfield  {title}
  {\bibinfo {title} {{The Chroma software system for lattice QCD}},\ }\href
  {https://doi.org/10.1016/j.nuclphysbps.2004.11.254} {\bibfield  {journal}
  {\bibinfo  {journal} {Nucl. Phys. B Proc. Suppl.}\ }\textbf {\bibinfo
  {volume} {140}},\ \bibinfo {pages} {832} (\bibinfo {year} {2005})},\ \Eprint
  {https://arxiv.org/abs/hep-lat/0409003} {arXiv:hep-lat/0409003} \BibitemShut
  {NoStop}%
\bibitem [{\citenamefont {Clark}\ \emph {et~al.}(2010)\citenamefont {Clark},
  \citenamefont {Babich}, \citenamefont {Barros}, \citenamefont {Brower},\ and\
  \citenamefont {Rebbi}}]{Clark:2009wm}%
  \BibitemOpen
  \bibfield  {author} {\bibinfo {author} {\bibfnamefont {M.~A.}\ \bibnamefont
  {Clark}}, \bibinfo {author} {\bibfnamefont {R.}~\bibnamefont {Babich}},
  \bibinfo {author} {\bibfnamefont {K.}~\bibnamefont {Barros}}, \bibinfo
  {author} {\bibfnamefont {R.~C.}\ \bibnamefont {Brower}},\ and\ \bibinfo
  {author} {\bibfnamefont {C.}~\bibnamefont {Rebbi}},\ }\bibfield  {title}
  {\bibinfo {title} {{Solving Lattice QCD systems of equations using mixed
  precision solvers on GPUs}},\ }\href
  {https://doi.org/10.1016/j.cpc.2010.05.002} {\bibfield  {journal} {\bibinfo
  {journal} {Comput. Phys. Commun.}\ }\textbf {\bibinfo {volume} {181}},\
  \bibinfo {pages} {1517} (\bibinfo {year} {2010})},\ \Eprint
  {https://arxiv.org/abs/0911.3191} {arXiv:0911.3191 [hep-lat]} \BibitemShut
  {NoStop}%
\bibitem [{\citenamefont {Babich}\ \emph {et~al.}(2011)\citenamefont {Babich},
  \citenamefont {Clark}, \citenamefont {Joo}, \citenamefont {Shi},
  \citenamefont {Brower},\ and\ \citenamefont {Gottlieb}}]{Babich:2011np}%
  \BibitemOpen
  \bibfield  {author} {\bibinfo {author} {\bibfnamefont {R.}~\bibnamefont
  {Babich}}, \bibinfo {author} {\bibfnamefont {M.~A.}\ \bibnamefont {Clark}},
  \bibinfo {author} {\bibfnamefont {B.}~\bibnamefont {Joo}}, \bibinfo {author}
  {\bibfnamefont {G.}~\bibnamefont {Shi}}, \bibinfo {author} {\bibfnamefont
  {R.~C.}\ \bibnamefont {Brower}},\ and\ \bibinfo {author} {\bibfnamefont
  {S.}~\bibnamefont {Gottlieb}} (\bibinfo {collaboration} {QUDA}),\ }\bibfield
  {title} {\bibinfo {title} {{Scaling lattice QCD beyond 100 GPUs}},\ }in\
  \href {https://doi.org/10.1145/2063384.2063478} {\emph {\bibinfo {booktitle}
  {{International Conference for High Performance Computing, Networking,
  Storage and Analysis}}}}\ (\bibinfo {year} {2011})\ \Eprint
  {https://arxiv.org/abs/1109.2935} {arXiv:1109.2935 [hep-lat]} \BibitemShut
  {NoStop}%
\bibitem [{\citenamefont {Jiang}\ \emph {et~al.}(2024)\citenamefont {Jiang},
  \citenamefont {Shi}, \citenamefont {Chen}, \citenamefont {Gong},\ and\
  \citenamefont {Yang}}]{jiang2024usequdalatticeqcd}%
  \BibitemOpen
  \bibfield  {author} {\bibinfo {author} {\bibfnamefont {X.}~\bibnamefont
  {Jiang}}, \bibinfo {author} {\bibfnamefont {C.}~\bibnamefont {Shi}}, \bibinfo
  {author} {\bibfnamefont {Y.}~\bibnamefont {Chen}}, \bibinfo {author}
  {\bibfnamefont {M.}~\bibnamefont {Gong}},\ and\ \bibinfo {author}
  {\bibfnamefont {Y.-B.}\ \bibnamefont {Yang}},\ }\href
  {https://arxiv.org/abs/2411.08461} {\bibinfo {title} {Use quda for lattice
  qcd calculation with python}} (\bibinfo {year} {2024}),\ \Eprint
  {https://arxiv.org/abs/2411.08461} {arXiv:2411.08461 [hep-lat]} \BibitemShut
  {NoStop}%
\end{thebibliography}%

\appendix
%%%%%%%%%%%%%%%%%%%%%%%%%%%
\section{Two-particle operators}\label{appendix:A}

The two-particle operators in the $S$-wave channel, defined in Eq.~\eqref{eq:operator-A1}, include spatial rotation operations $R$ acting on the lattice momenta.
For the $D$-wave channel, additional coefficients $c_R^{E(1)}$, as defined in Eq.~\eqref{eq:operator-E1}, are introduced.
For clarity, their explicit forms are summarized in the following Tables.
\renewcommand{\arraystretch}{1.5}
\begin{widetext}
	\begin{equation*}
		\begin{array}{cc}
			\hline\hline
			\mathcal O^{A_1}_{\eta_c\eta_c}(n^2)&S\text{-wave~Operators} \\ 
			\hline
			n^2=0  &[\bar{c}\gamma^5 c](0,0,0)\times[\bar{c}\gamma^5 c](0,0,0)   \\ 
			\hline
			n^2=1  &[\bar{c}\gamma^5 c](-1,0,0)\times[\bar{c}\gamma^5 c](1,0,0)   
			+[\bar{c}\gamma^5 c](0,-1,0)\times[\bar{c}\gamma^5 c](0,1,0)   
			+[\bar{c}\gamma^5 c](0,0,-1)\times[\bar{c}\gamma^5 c](0,0,1)   \\
			&+[\bar{c}\gamma^5 c](0,0,1)\times[\bar{c}\gamma^5 c](0,0,-1)    
			+[\bar{c}\gamma^5 c](0,1,0)\times[\bar{c}\gamma^5 c](0,-1,0)   
			+[\bar{c}\gamma^5 c](1,0,0)\times[\bar{c}\gamma^5 c](-1,0,0)   \\
			\hline
			n^2=2  &[\bar{c}\gamma^5 c](-1,-1,0)\times[\bar{c}\gamma^5 c](1,1,0)   
			+[\bar{c}\gamma^5 c](-1,0,-1)\times[\bar{c}\gamma^5 c](1,0,1)   
			+[\bar{c}\gamma^5 c](-1, 0, 1)\times[\bar{c}\gamma^5 c](1, 0, -1)   \\
			&+[\bar{c}\gamma^5 c](-1, 1, 0)\times[\bar{c}\gamma^5 c](1, -1, 0)  
			+[\bar{c}\gamma^5 c](0, -1, -1)\times[\bar{c}\gamma^5 c](0, 1, 1)   
			+[\bar{c}\gamma^5 c](0, -1, 1)\times[\bar{c}\gamma^5 c](0, 1, -1)   \\
			&+[\bar{c}\gamma^5 c](0, 1, -1)\times[\bar{c}\gamma^5 c](0, -1, 1) 
			+[\bar{c}\gamma^5 c](0, 1, 1)\times[\bar{c}\gamma^5 c](0, -1, -1)   
			+[\bar{c}\gamma^5 c](1, -1, 0)\times[\bar{c}\gamma^5 c](-1, 1, 0)   \\
			&+[\bar{c}\gamma^5 c](1, 0, -1)\times[\bar{c}\gamma^5 c](-1, 0, 1) 
			+[\bar{c}\gamma^5 c](1, 0, 1)\times[\bar{c}\gamma^5 c](-1, 0, -1)   
			+[\bar{c}\gamma^5 c](1, 1, 0)\times[\bar{c}\gamma^5 c](-1, -1, 0)   \\
			\hline
			n^2=3  &[\bar{c}\gamma^5 c](-1,-1,-1)\times[\bar{c}\gamma^5 c](1,1,1)   
			+[\bar{c}\gamma^5 c](-1,-1,1)\times[\bar{c}\gamma^5 c](1,1,-1)   
			+[\bar{c}\gamma^5 c](-1, 1, -1)\times[\bar{c}\gamma^5 c](1, -1, 1)   \\
			&+[\bar{c}\gamma^5 c](-1, 1, 1)\times[\bar{c}\gamma^5 c](1, -1, -1)  
			+[\bar{c}\gamma^5 c](1, -1, -1)\times[\bar{c}\gamma^5 c](-1, 1, 1)   
			+[\bar{c}\gamma^5 c](1, -1, 1)\times[\bar{c}\gamma^5 c](-1, 1, -1)   \\
			&+[\bar{c}\gamma^5 c](1, 1, -1)\times[\bar{c}\gamma^5 c](-1, -1, 1) 
			+[\bar{c}\gamma^5 c](1, 1, 1)\times[\bar{c}\gamma^5 c](-1, -1, -1)    \\
			\hline
			n^2=4  &[\bar{c}\gamma^5 c](-2,0,0)\times[\bar{c}\gamma^5 c](2,0,0)   
			+[\bar{c}\gamma^5 c](0,-2,0)\times[\bar{c}\gamma^5 c](0,2,0)   
			+[\bar{c}\gamma^5 c](0,0,-2)\times[\bar{c}\gamma^5 c](0,0,2)   \\
			&+[\bar{c}\gamma^5 c](0,0,2)\times[\bar{c}\gamma^5 c](0,0,-2)    
			+[\bar{c}\gamma^5 c](0,2,0)\times[\bar{c}\gamma^5 c](0,-2,0)   
			+[\bar{c}\gamma^5 c](2,0,0)\times[\bar{c}\gamma^5 c](-2,0,0)   \\
			\hline\hline
		\end{array}
	\end{equation*}
	
	\begin{equation*}
		\begin{array}{cc}
			\hline\hline
			\mathcal O^{E(1)}_{\eta_c\eta_c}(n^2)&D\text{-wave~Operators} \\ 
			\hline
			n^2=0  & 0\\
			\hline
			n^2=1  &-[\bar{c}\gamma^5 c](-1,0,0)\times[\bar{c}\gamma^5 c](1,0,0)   
			-[\bar{c}\gamma^5 c](0,-1,0)\times[\bar{c}\gamma^5 c](0,1,0)   
			+2[\bar{c}\gamma^5 c](0,0,-1)\times[\bar{c}\gamma^5 c](0,0,1)   \\
			&+2[\bar{c}\gamma^5 c](0,0,1)\times[\bar{c}\gamma^5 c](0,0,-1)    
			-[\bar{c}\gamma^5 c](0,1,0)\times[\bar{c}\gamma^5 c](0,-1,0)   
			-[\bar{c}\gamma^5 c](1,0,0)\times[\bar{c}\gamma^5 c](-1,0,0)   \\
			\hline
			n^2=2  &-[\bar{c}\gamma^5 c](-1,-1,0)\times[\bar{c}\gamma^5 c](1,1,0)   
			+\frac{1}{2}[\bar{c}\gamma^5 c](-1,0,-1)\times[\bar{c}\gamma^5 c](1,0,1)   
			+\frac{1}{2}[\bar{c}\gamma^5 c](-1, 0, 1)\times[\bar{c}\gamma^5 c](1, 0, -1) \\
			&-[\bar{c}\gamma^5 c](-1, 1, 0)\times[\bar{c}\gamma^5 c](1, -1, 0)  
			+\frac{1}{2}[\bar{c}\gamma^5 c](0, -1, -1)\times[\bar{c}\gamma^5 c](0, 1, 1)   
			+\frac{1}{2}[\bar{c}\gamma^5 c](0, -1, 1)\times[\bar{c}\gamma^5 c](0, 1, -1) \\
			&+\frac{1}{2}[\bar{c}\gamma^5 c](0, 1, -1)\times[\bar{c}\gamma^5 c](0, -1, 1) 
			+\frac{1}{2}[\bar{c}\gamma^5 c](0, 1, 1)\times[\bar{c}\gamma^5 c](0, -1, -1)   
			-[\bar{c}\gamma^5 c](1, -1, 0)\times[\bar{c}\gamma^5 c](-1, 1, 0)   \\
			&+\frac{1}{2}[\bar{c}\gamma^5 c](1, 0, -1)\times[\bar{c}\gamma^5 c](-1, 0, 1) 
			+\frac{1}{2}[\bar{c}\gamma^5 c](1, 0, 1)\times[\bar{c}\gamma^5 c](-1, 0, -1)   
			-[\bar{c}\gamma^5 c](1, 1, 0)\times[\bar{c}\gamma^5 c](-1, -1, 0)   \\
			\hline
			n^2=3  &0               \\
			\hline
			n^2=4  &-[\bar{c}\gamma^5 c](-2,0,0)\times[\bar{c}\gamma^5 c](2,0,0)   
			-[\bar{c}\gamma^5 c](0,-2,0)\times[\bar{c}\gamma^5 c](0,2,0)   
			+2[\bar{c}\gamma^5 c](0,0,-2)\times[\bar{c}\gamma^5 c](0,0,2)   \\
			&+2[\bar{c}\gamma^5 c](0,0,2)\times[\bar{c}\gamma^5 c](0,0,-2)    
			-[\bar{c}\gamma^5 c](0,2,0)\times[\bar{c}\gamma^5 c](0,-2,0)   
			-[\bar{c}\gamma^5 c](2,0,0)\times[\bar{c}\gamma^5 c](-2,0,0)   \\
			\hline\hline
		\end{array}
	\end{equation*}
\end{widetext}

%%%%%%%%%%%%%%%%%%%%%%%%%%%
\section{Energy-level determination}\label{appendix:B}

%%%%%%%%%%%%%%%%%%%%%%%%%%%
\begin{figure*}[h]
	\centering
	\includegraphics[width=0.45\linewidth]{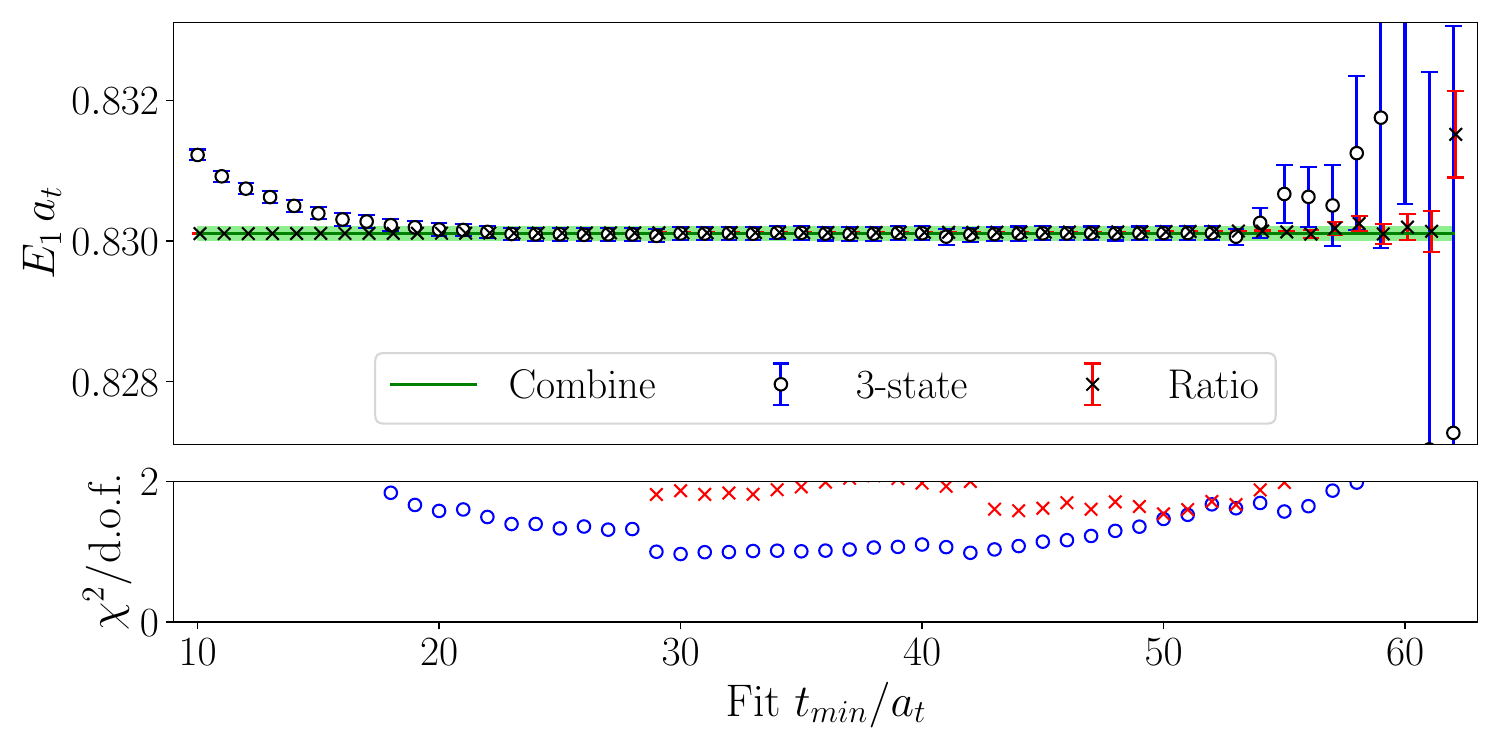}
	\includegraphics[width=0.45\linewidth]{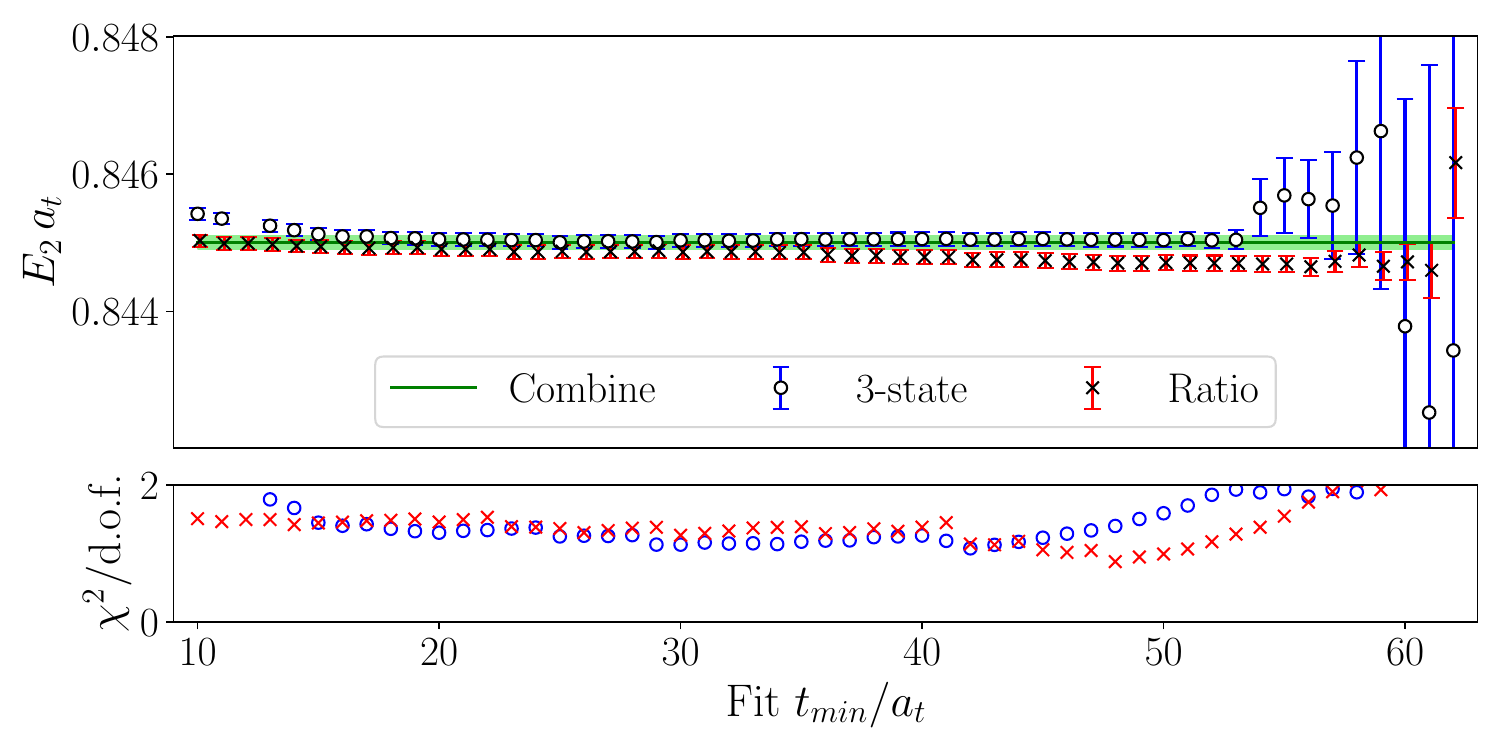}\\
	\includegraphics[width=0.45\linewidth]{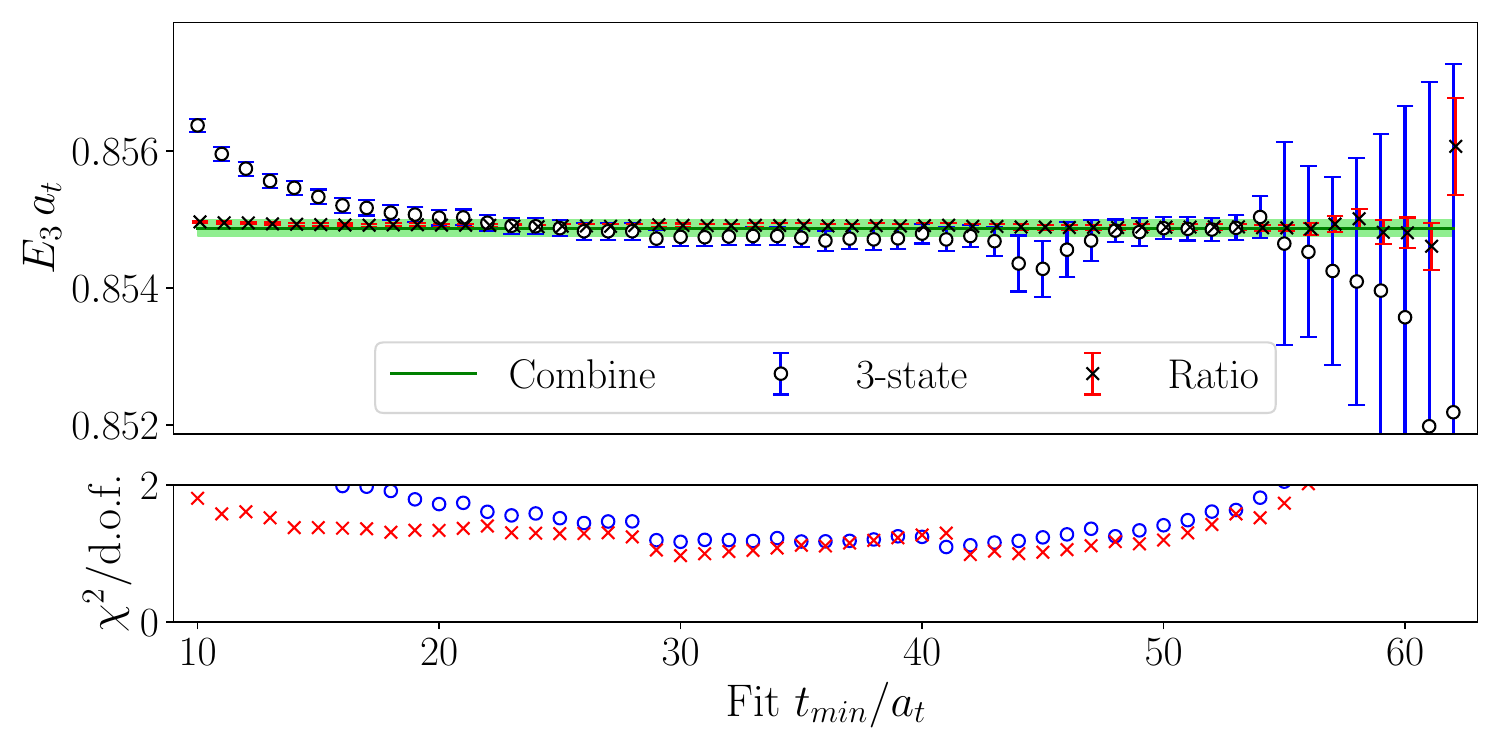}
	\includegraphics[width=0.45\linewidth]{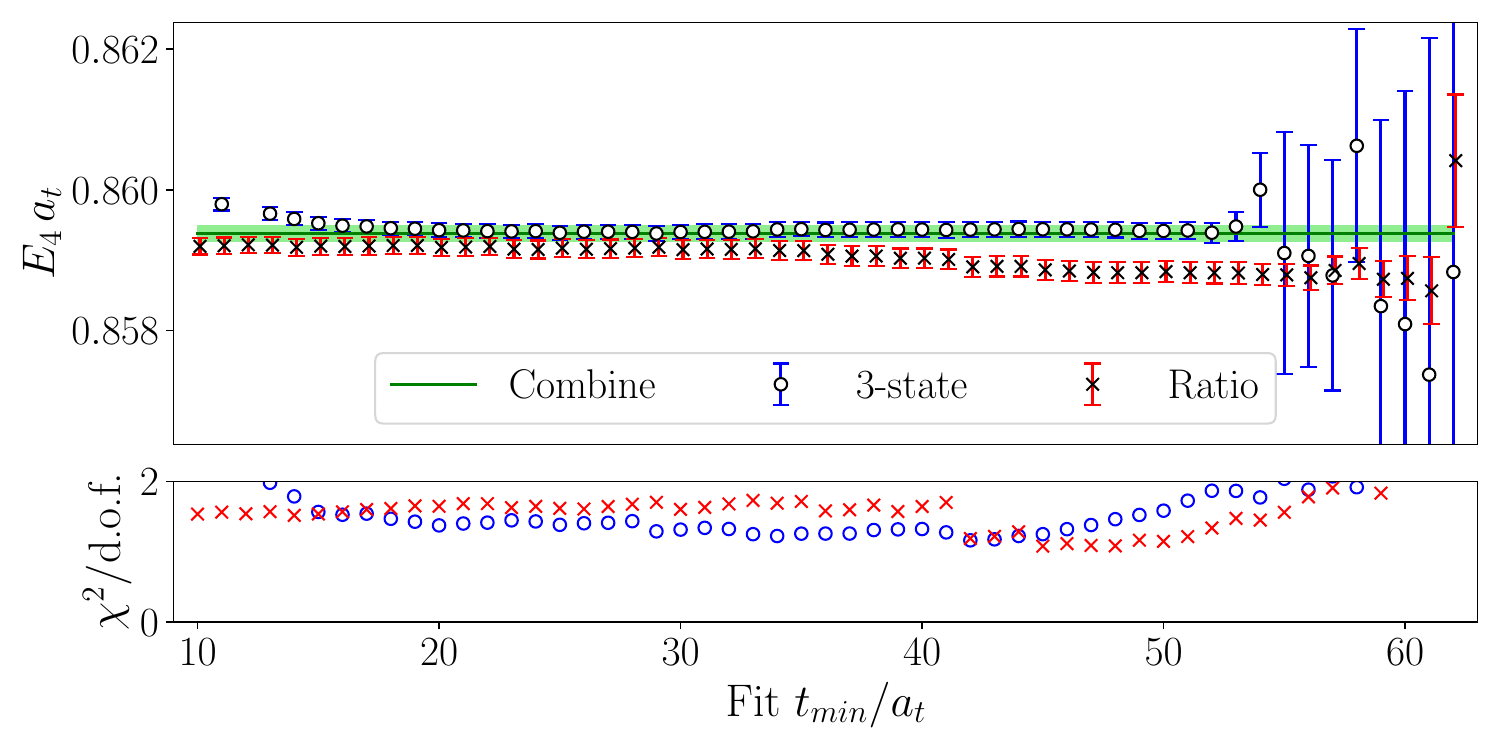}\\
	\includegraphics[width=0.45\linewidth]{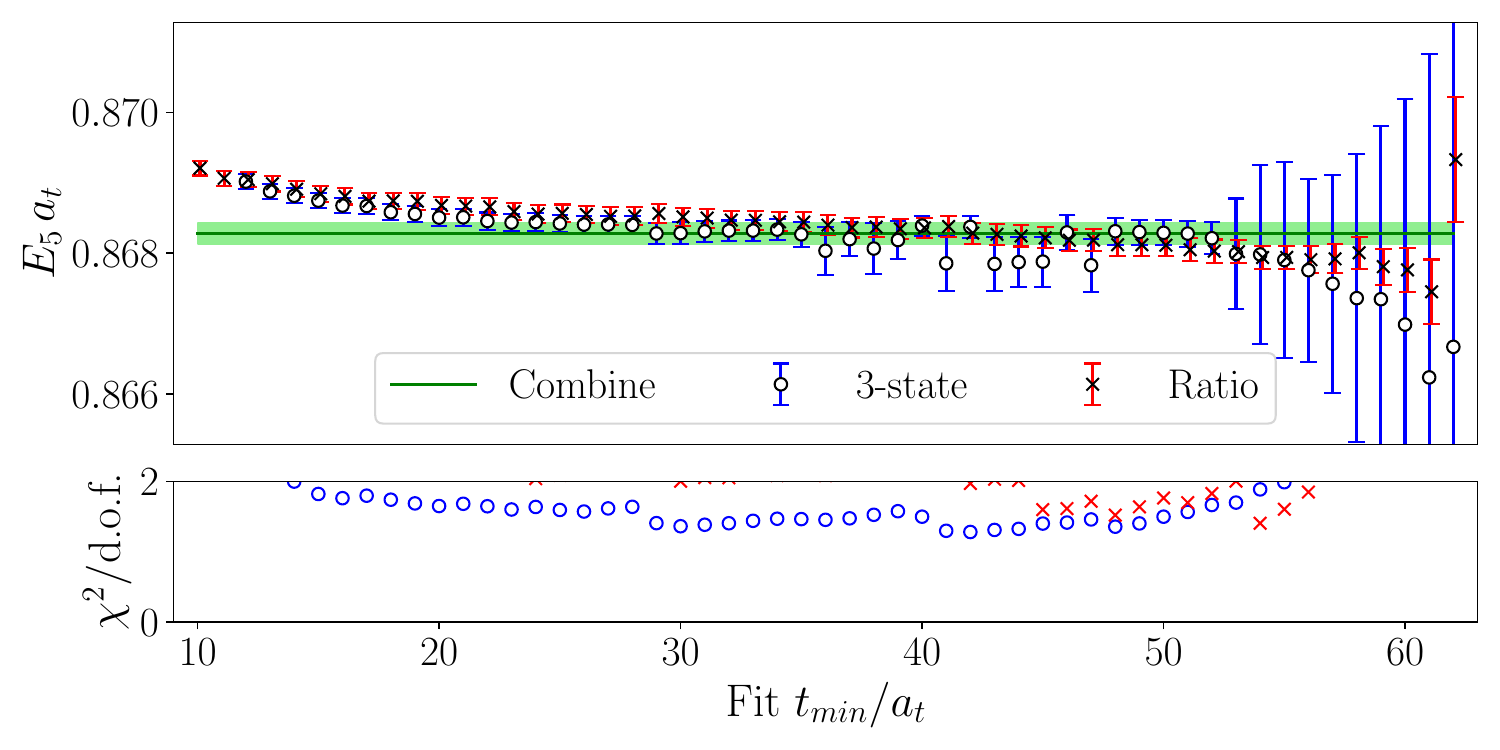}
	\includegraphics[width=0.45\linewidth]{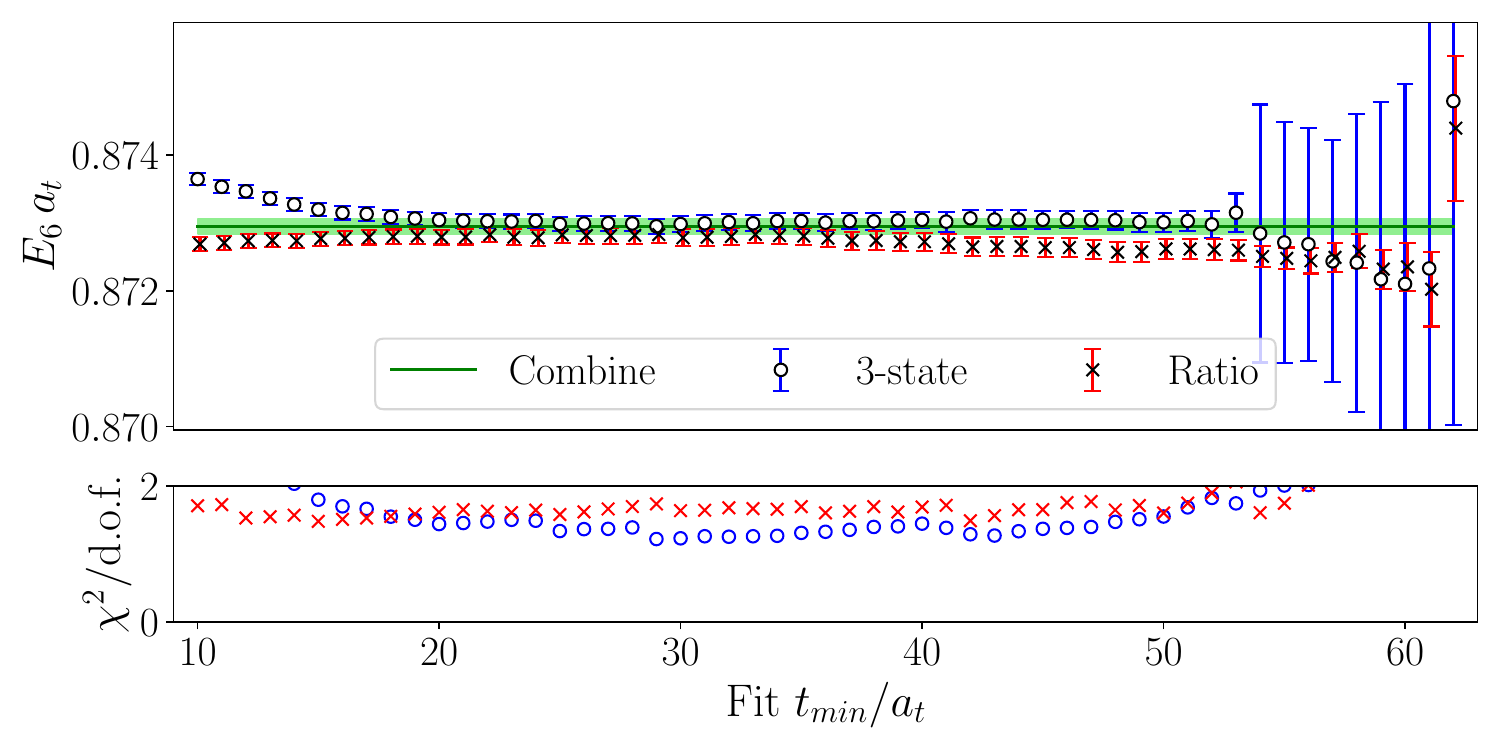}\\
	\includegraphics[width=0.45\linewidth]{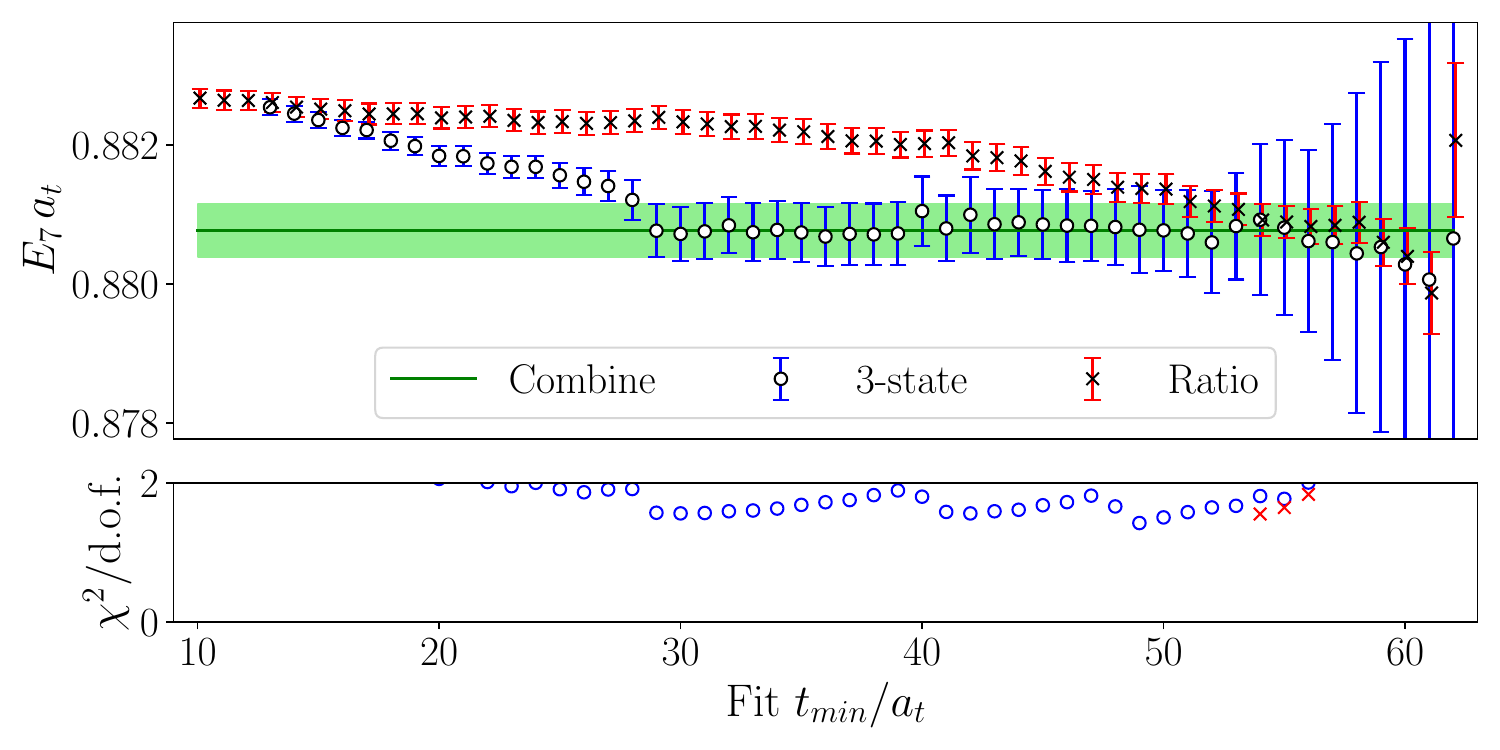}
	\includegraphics[width=0.45\linewidth]{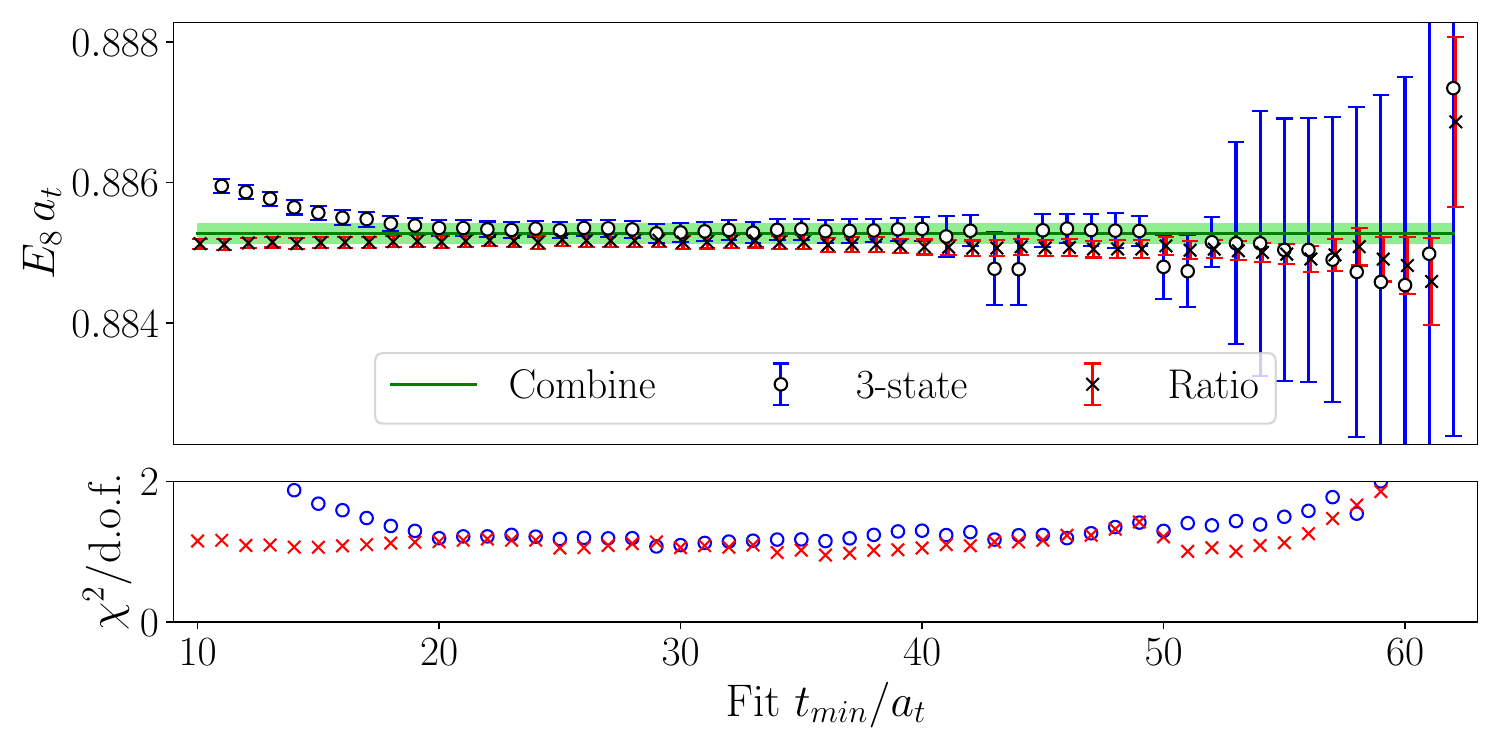}\\
	\includegraphics[width=0.45\linewidth]{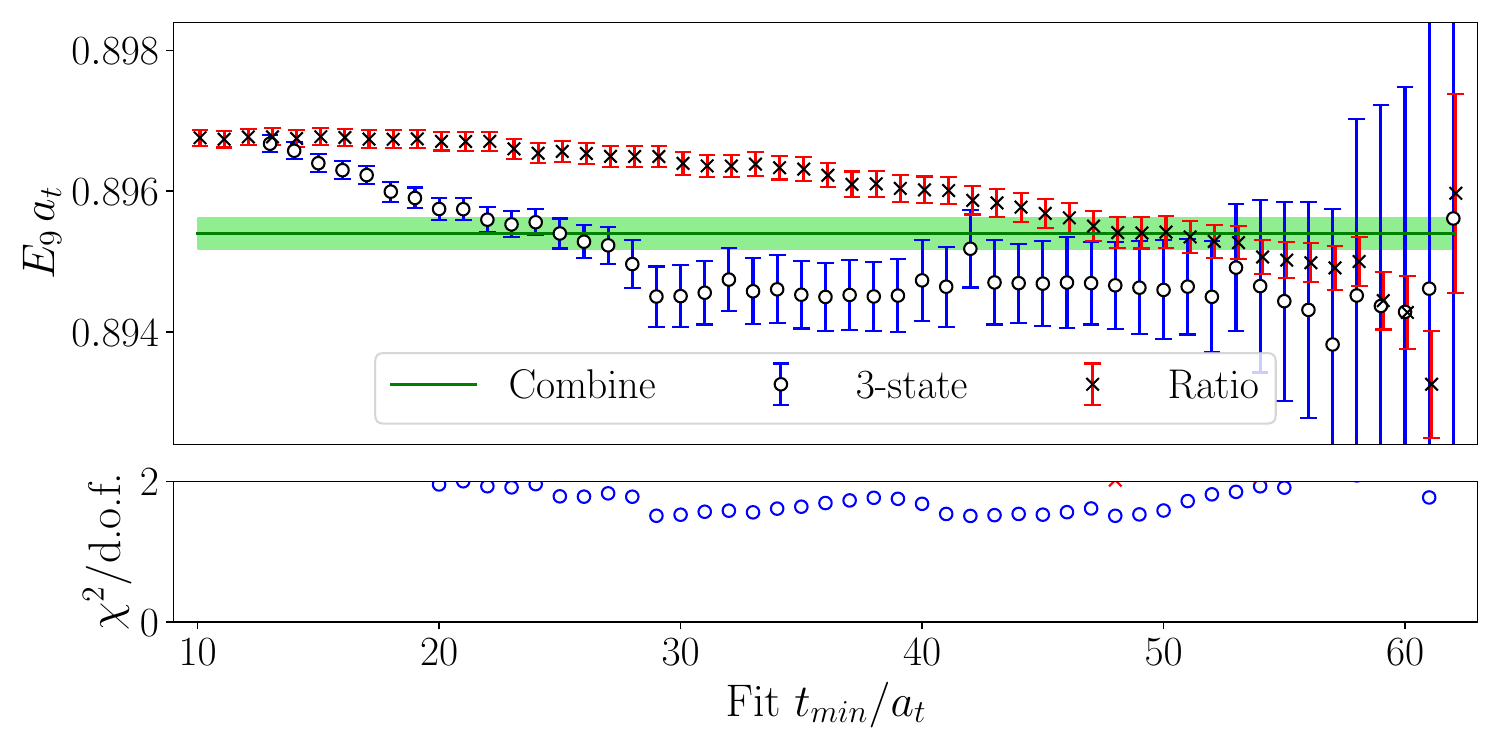}
	\includegraphics[width=0.45\linewidth]{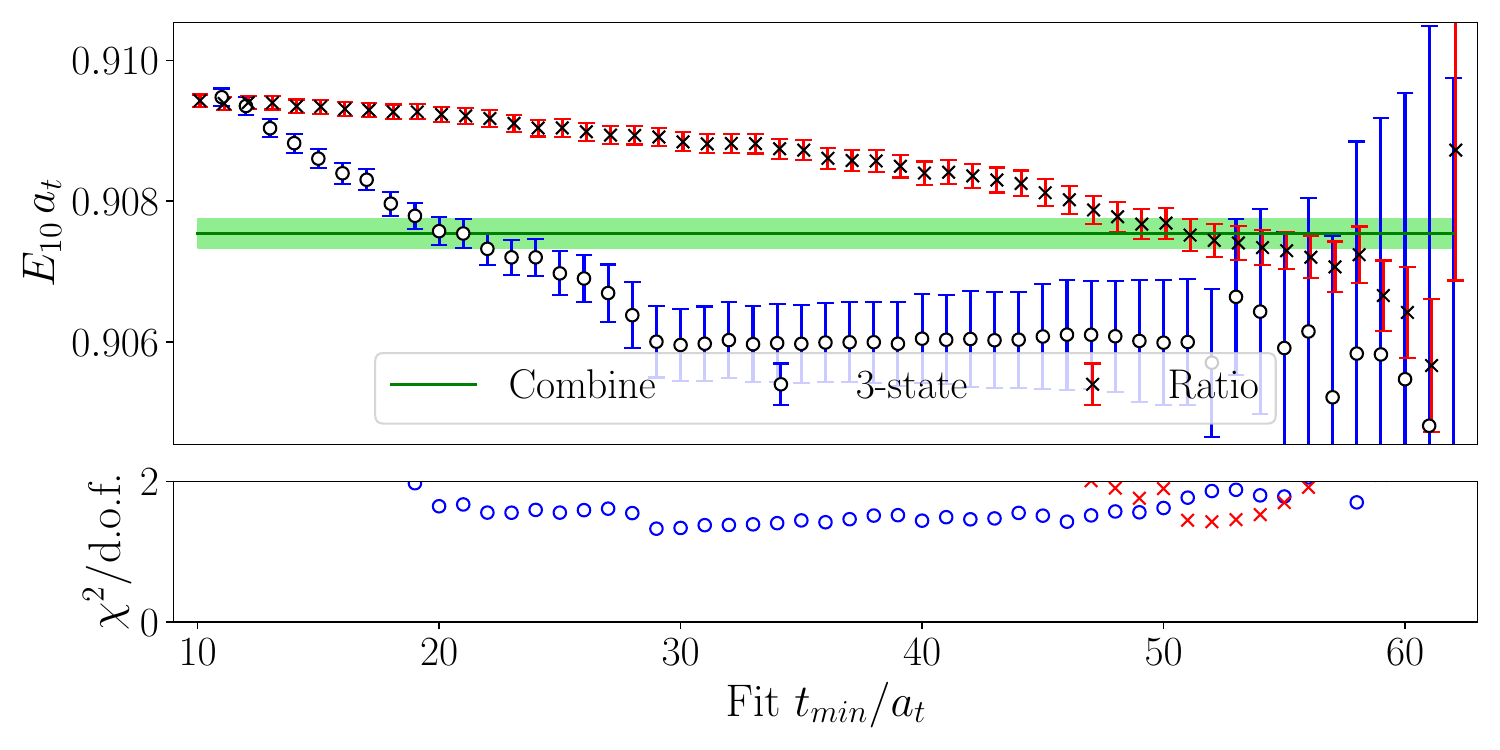}
	\caption{Stability of the two fitting methods on the L16M420 ensemble with $N_V=120$ is demonstrated by varying the minimum fitting time $t_{\rm min}$ in the $0^{++}$ system.}
	\label{fig:L16-fit-tmin}
\end{figure*}
%%%%%%%%%%%%%%%%%%%%%%%%%%%

%%%%%%%%%%%%%%%%%%%%%%%%%%%
\begin{figure*}[h]
	\centering
	\includegraphics[width=0.45\linewidth]{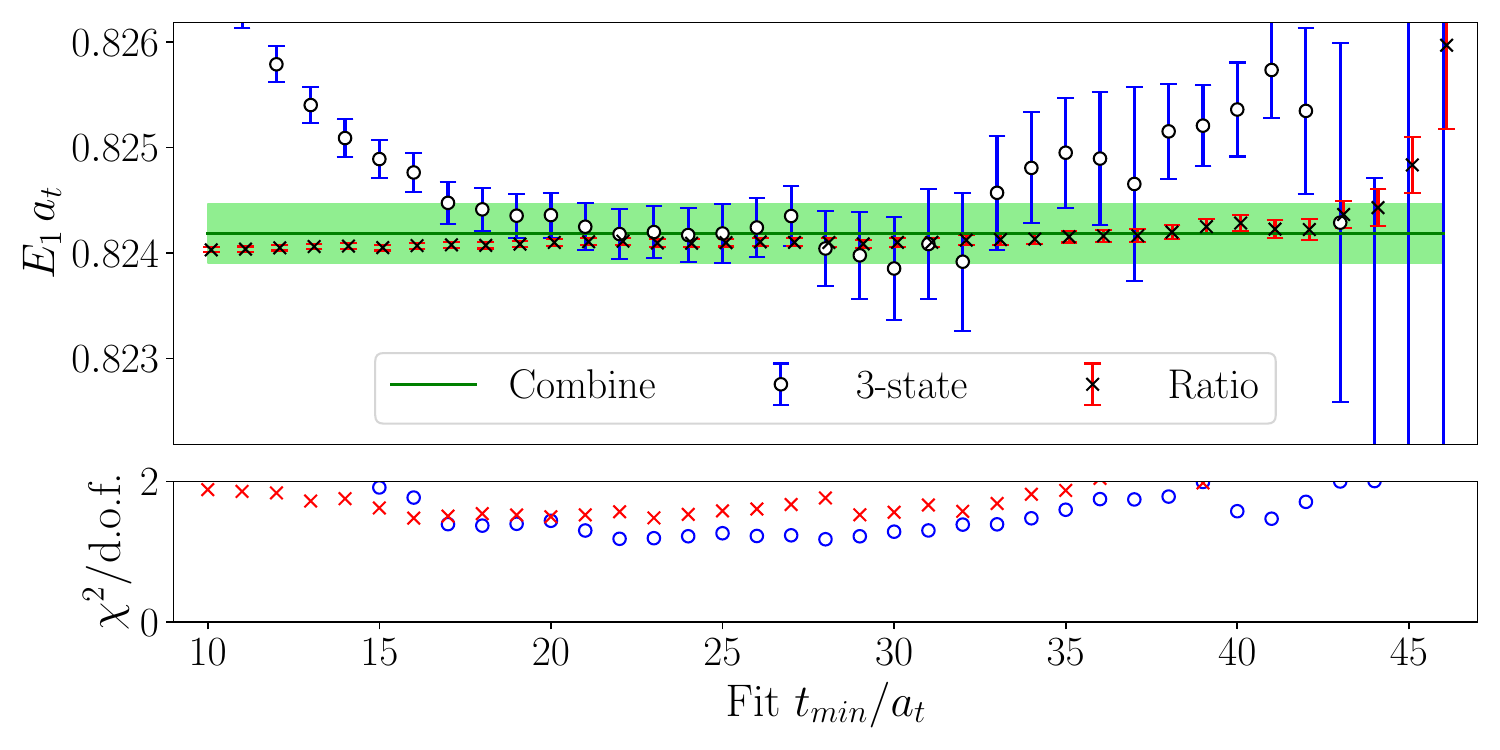}
	\includegraphics[width=0.45\linewidth]{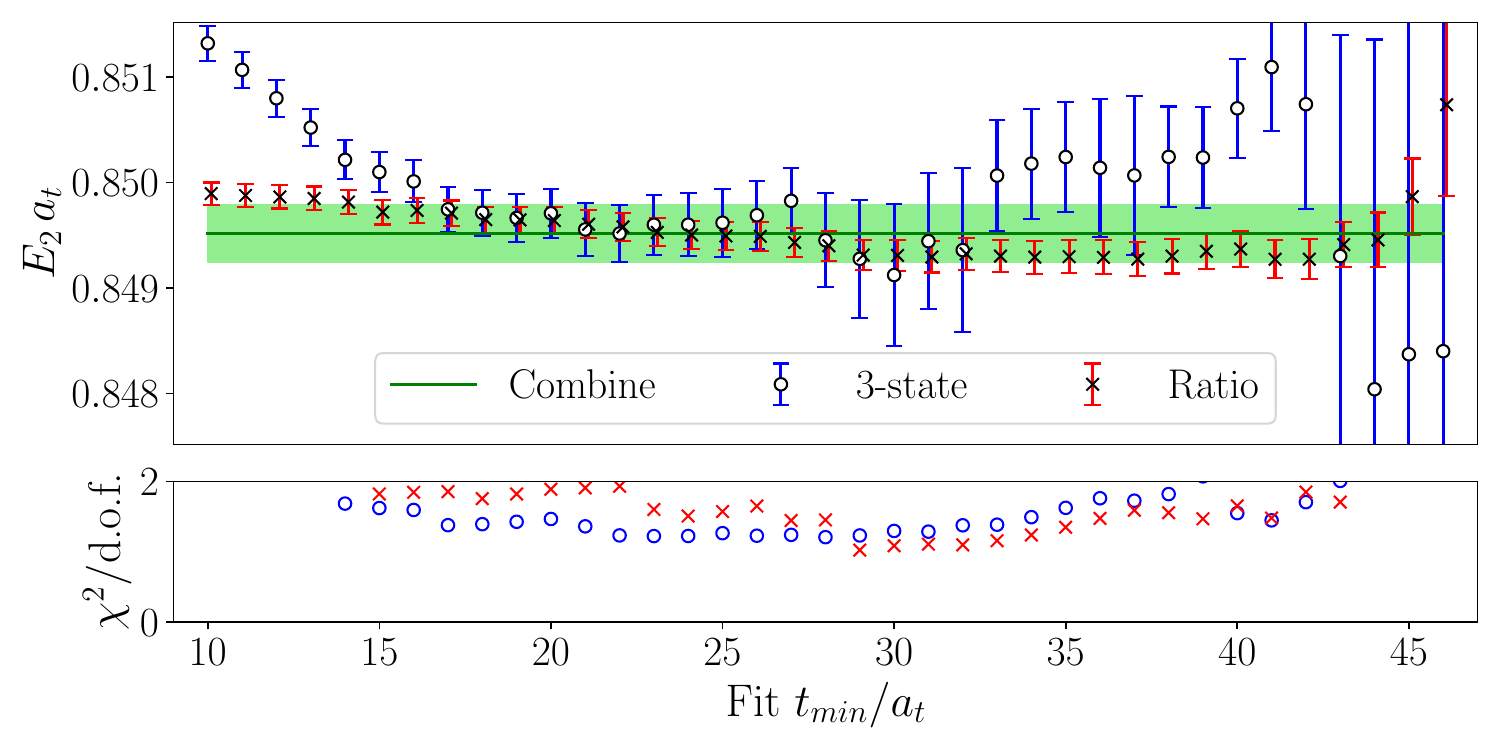}\\
	\includegraphics[width=0.45\linewidth]{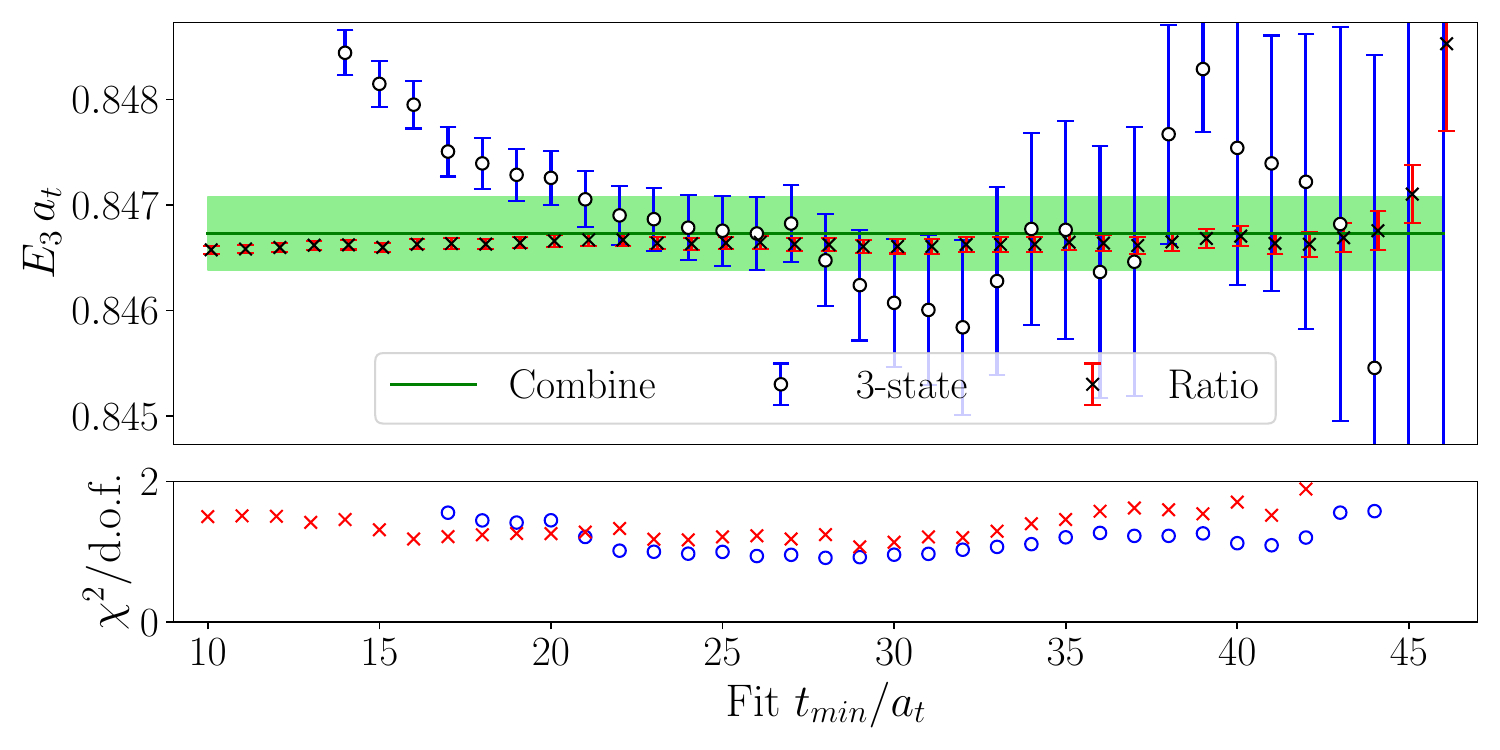}
	\includegraphics[width=0.45\linewidth]{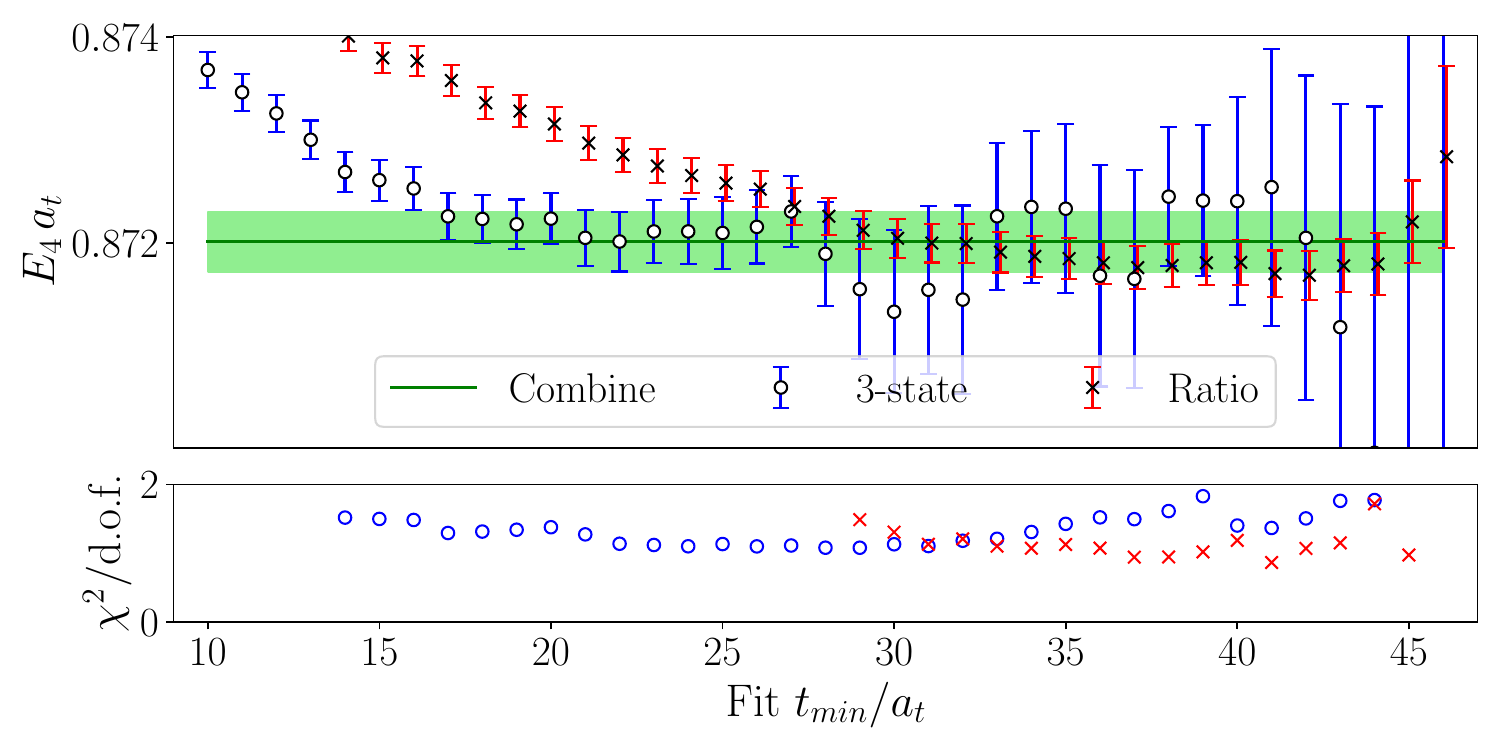}\\
	\includegraphics[width=0.45\linewidth]{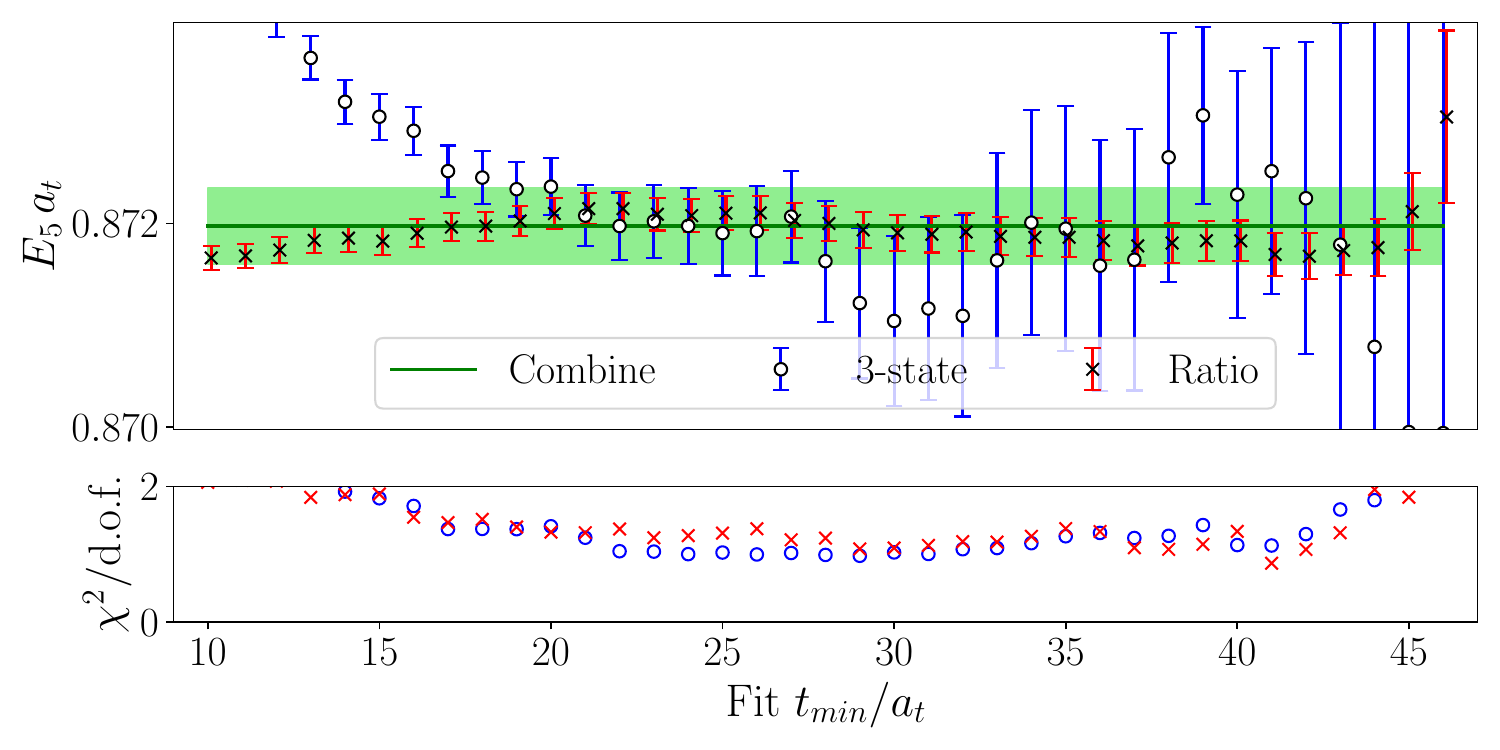}
	\includegraphics[width=0.45\linewidth]{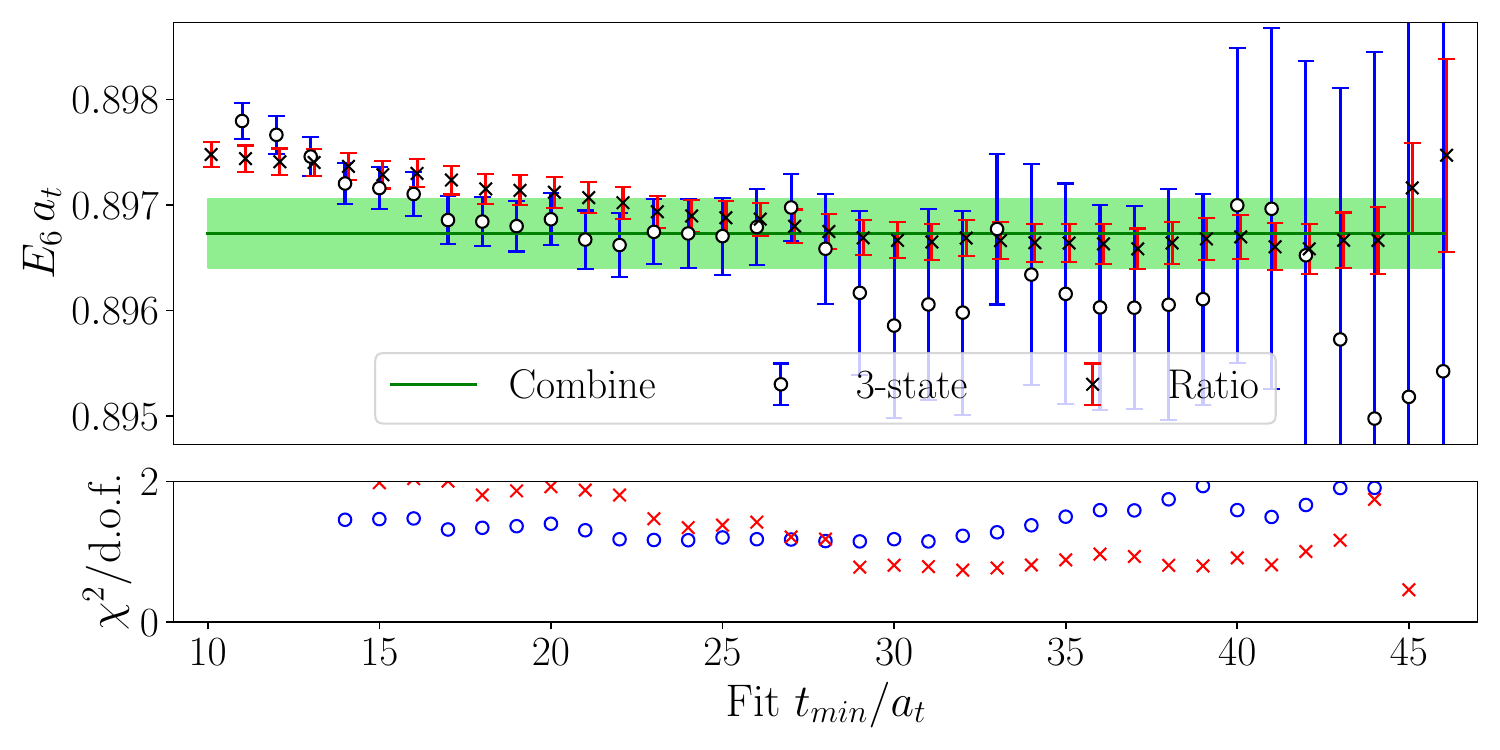}\\
	\includegraphics[width=0.45\linewidth]{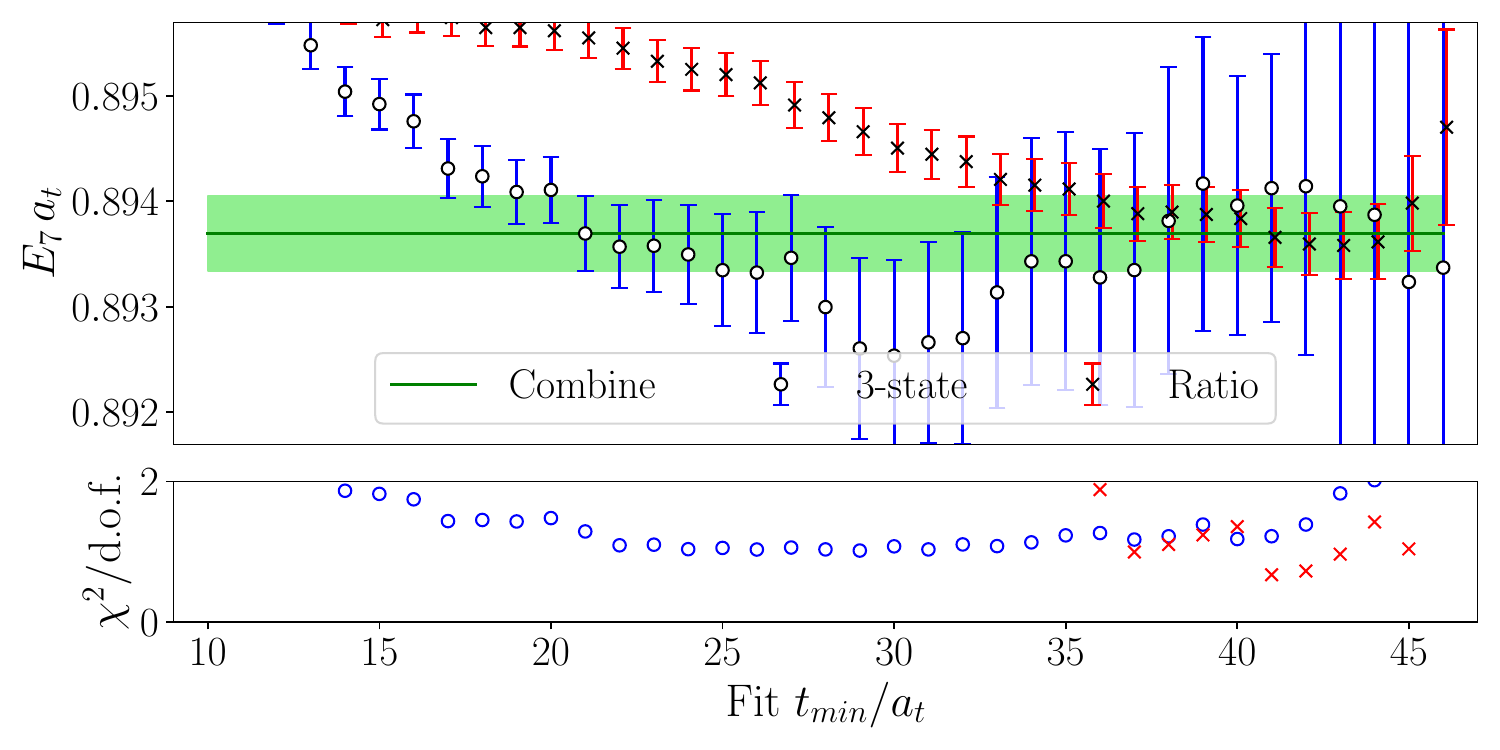}
	\includegraphics[width=0.45\linewidth]{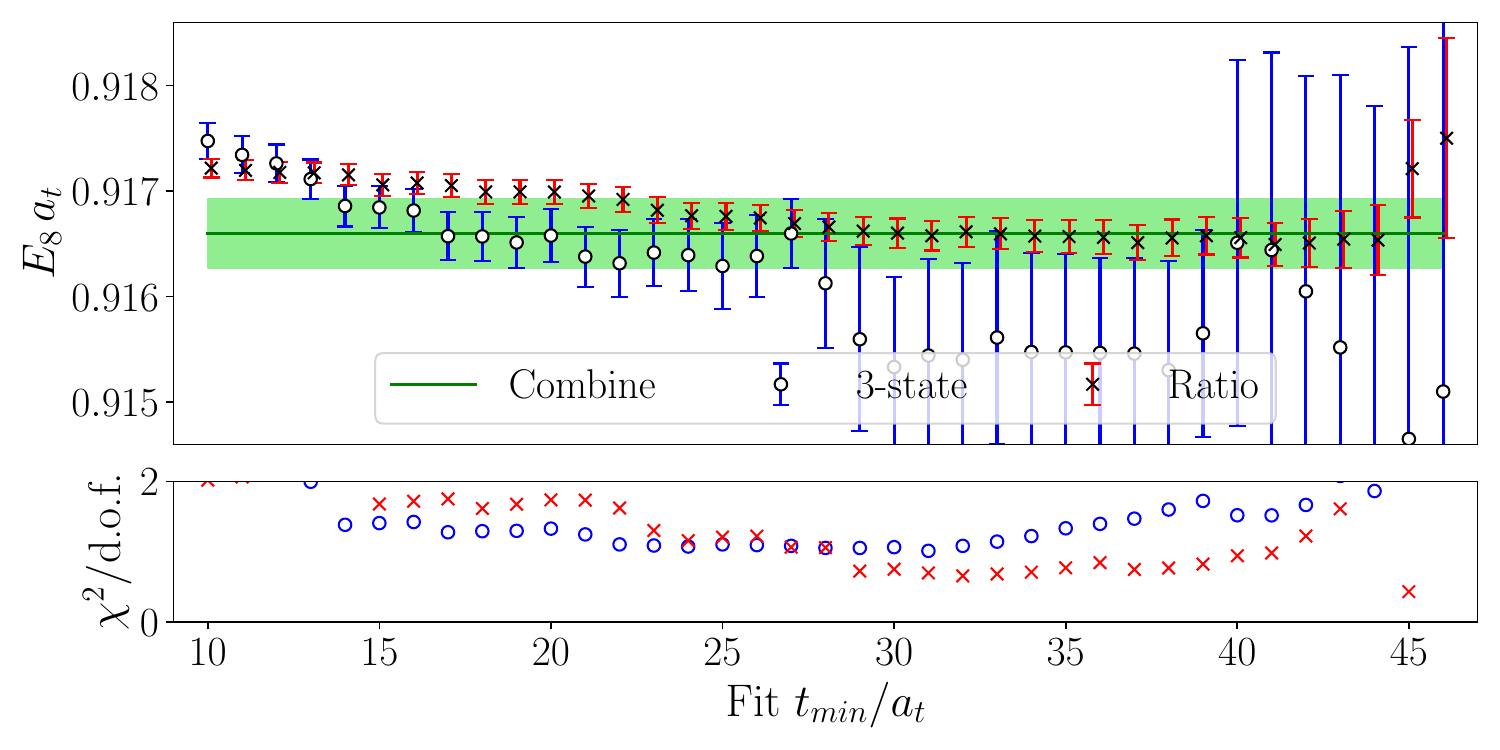}\\
	\includegraphics[width=0.45\linewidth]{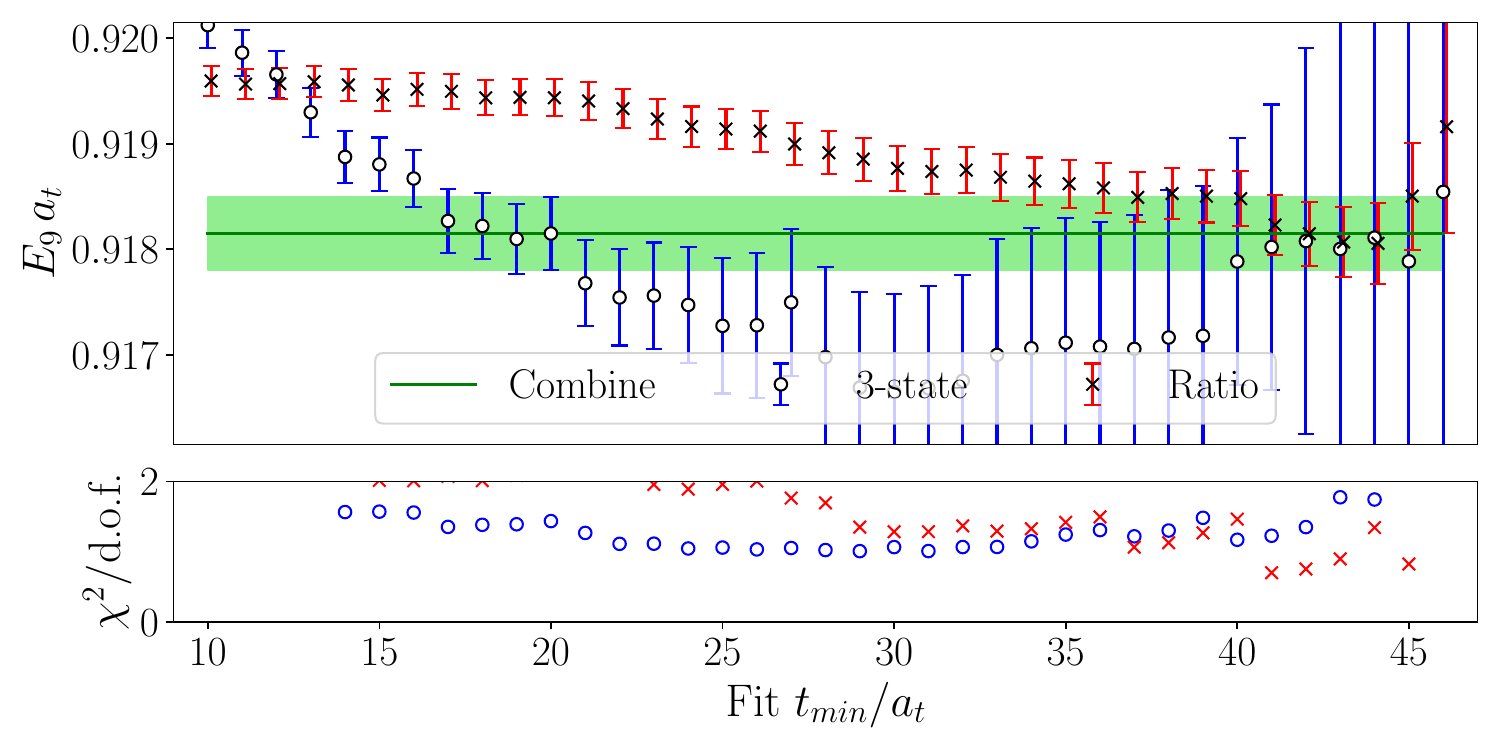}
	\includegraphics[width=0.45\linewidth]{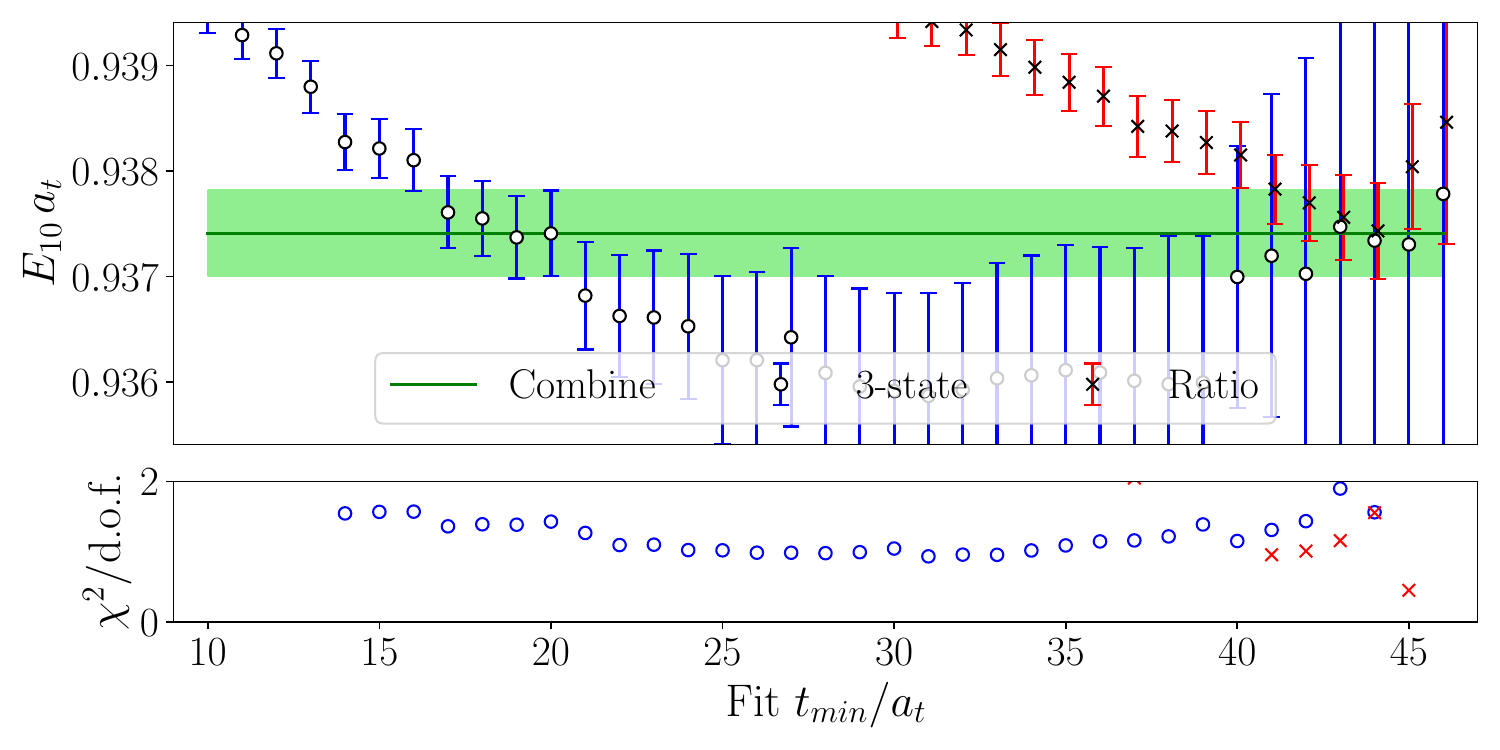}
	\caption{Stability of the two fitting methods on the L12M250 ensemble with $N_V=170$ is demonstrated by varying the minimum fitting time $t_{\rm min}$ in the $0^{++}$ system.}
	\label{fig:L12-pi-fit-tmin}
\end{figure*}
%%%%%%%%%%%%%%%%%%%%%%%%%%%

%%%%%%%%%%%%%%%%%%%%%%%%%%%
\begin{figure*}[h]
	\centering
	\includegraphics[width=0.45\linewidth]{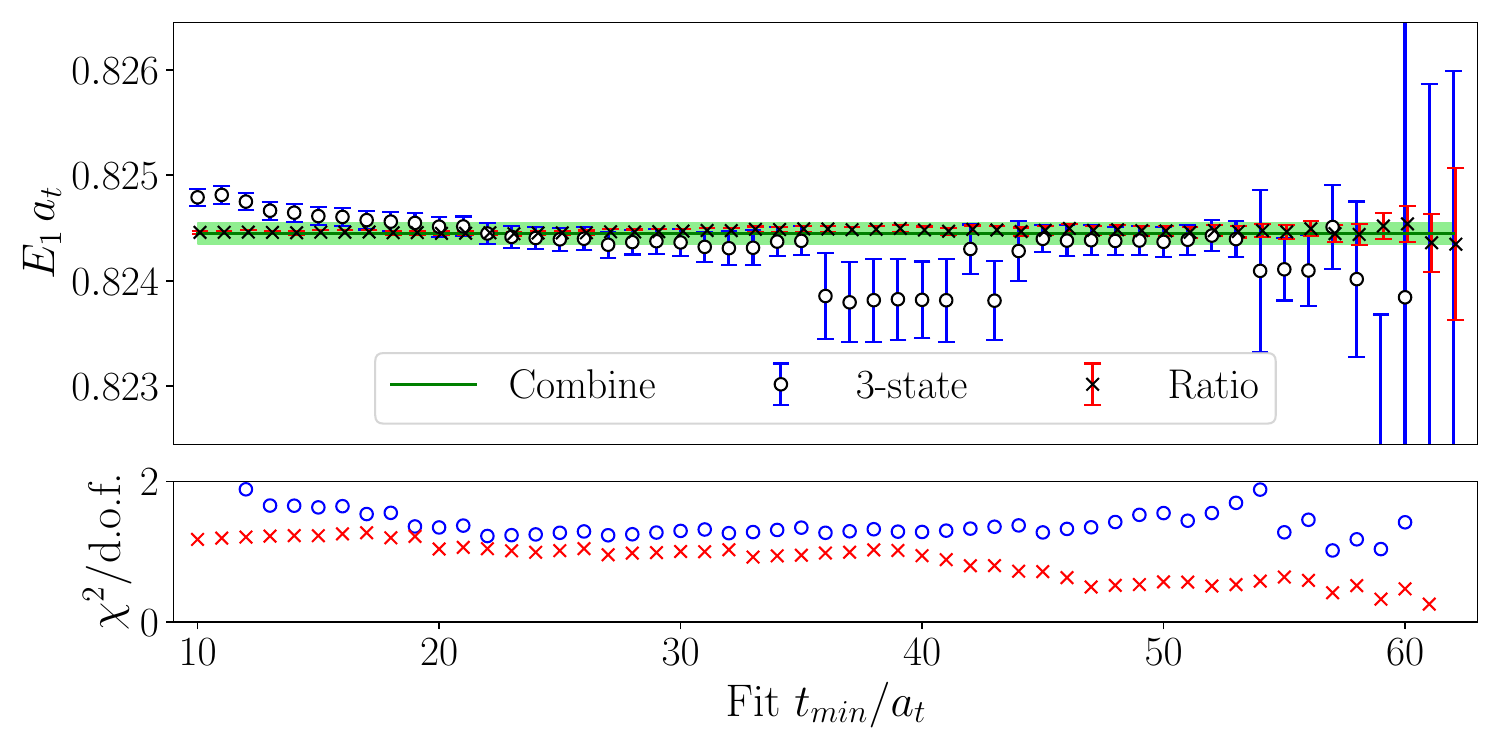}
	\includegraphics[width=0.45\linewidth]{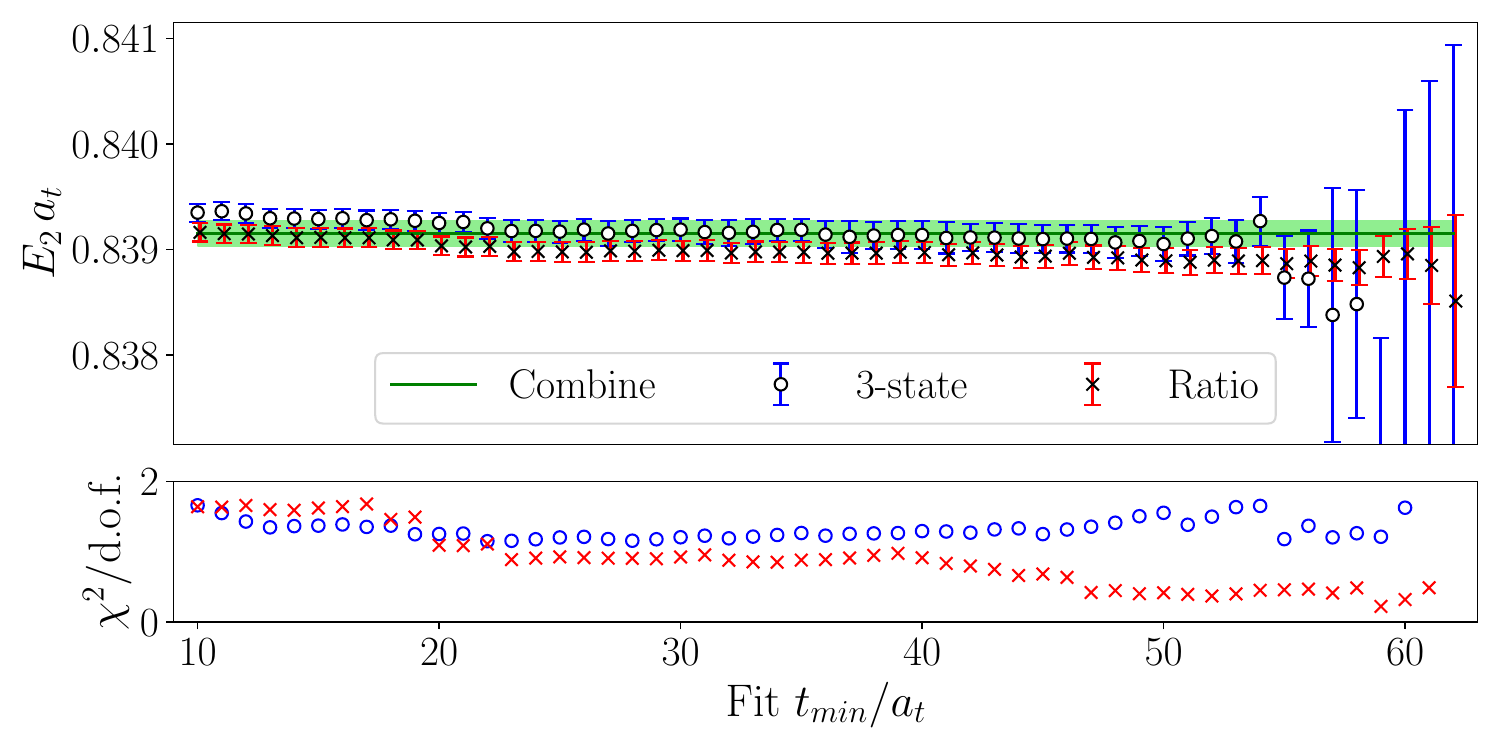}\\
	\includegraphics[width=0.45\linewidth]{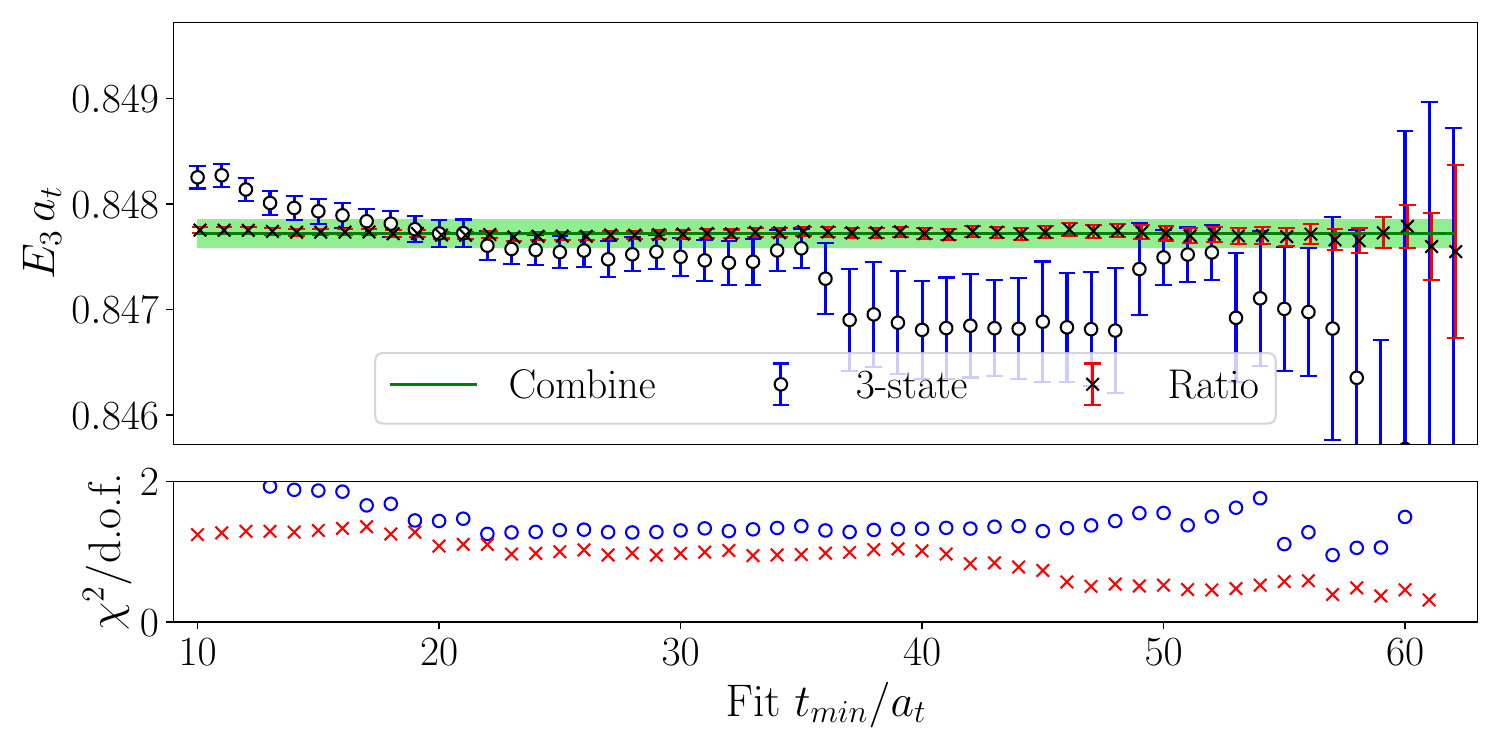}
	\includegraphics[width=0.45\linewidth]{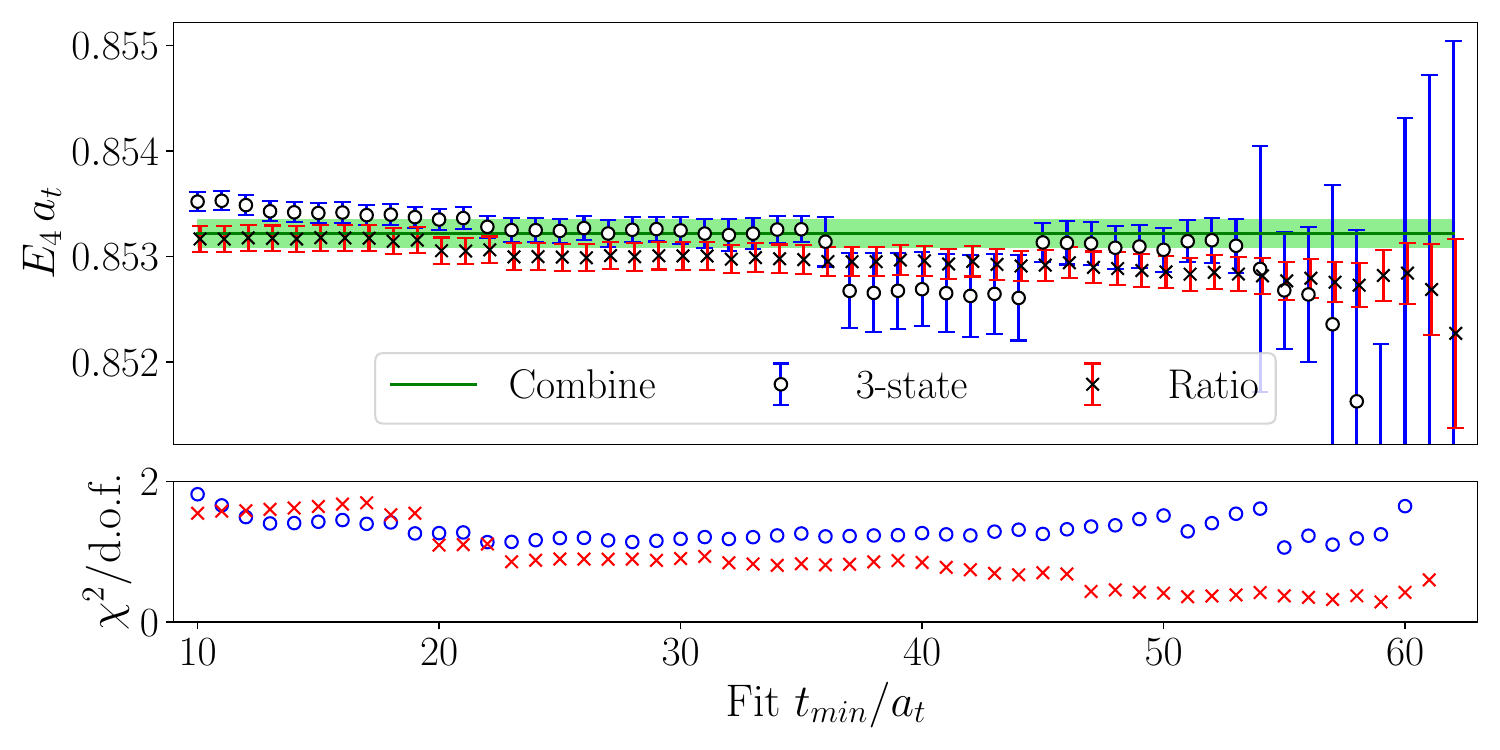}\\
	\includegraphics[width=0.45\linewidth]{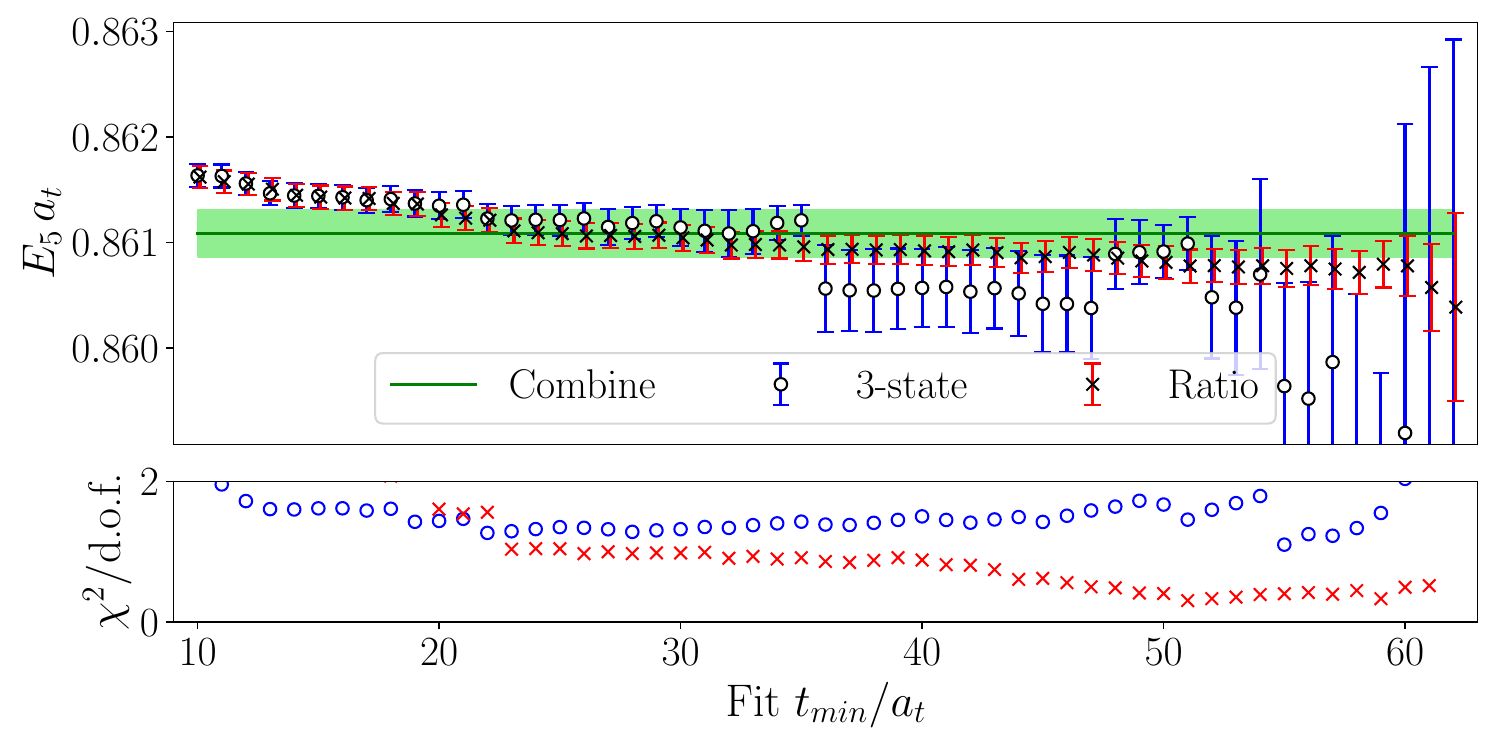}
	\includegraphics[width=0.45\linewidth]{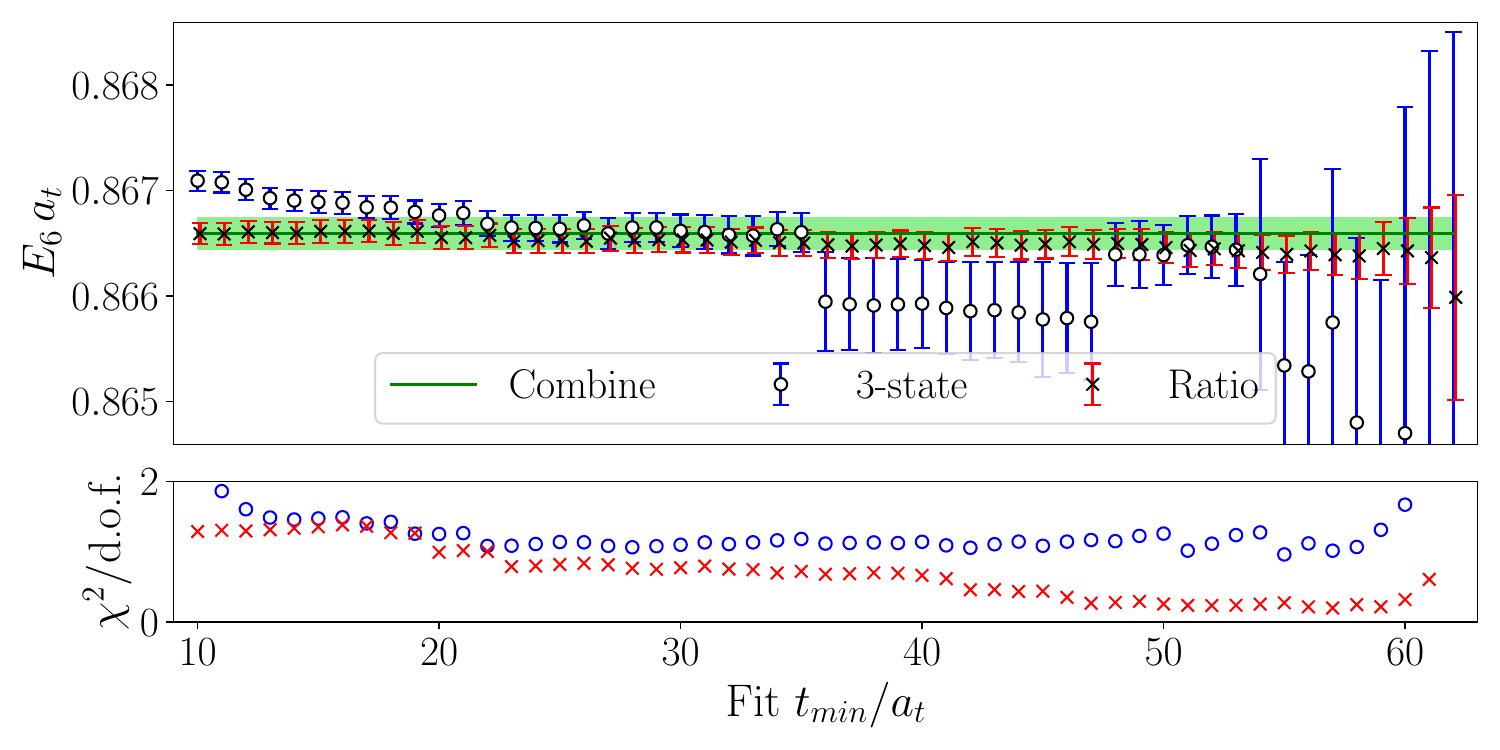}\\
	\includegraphics[width=0.45\linewidth]{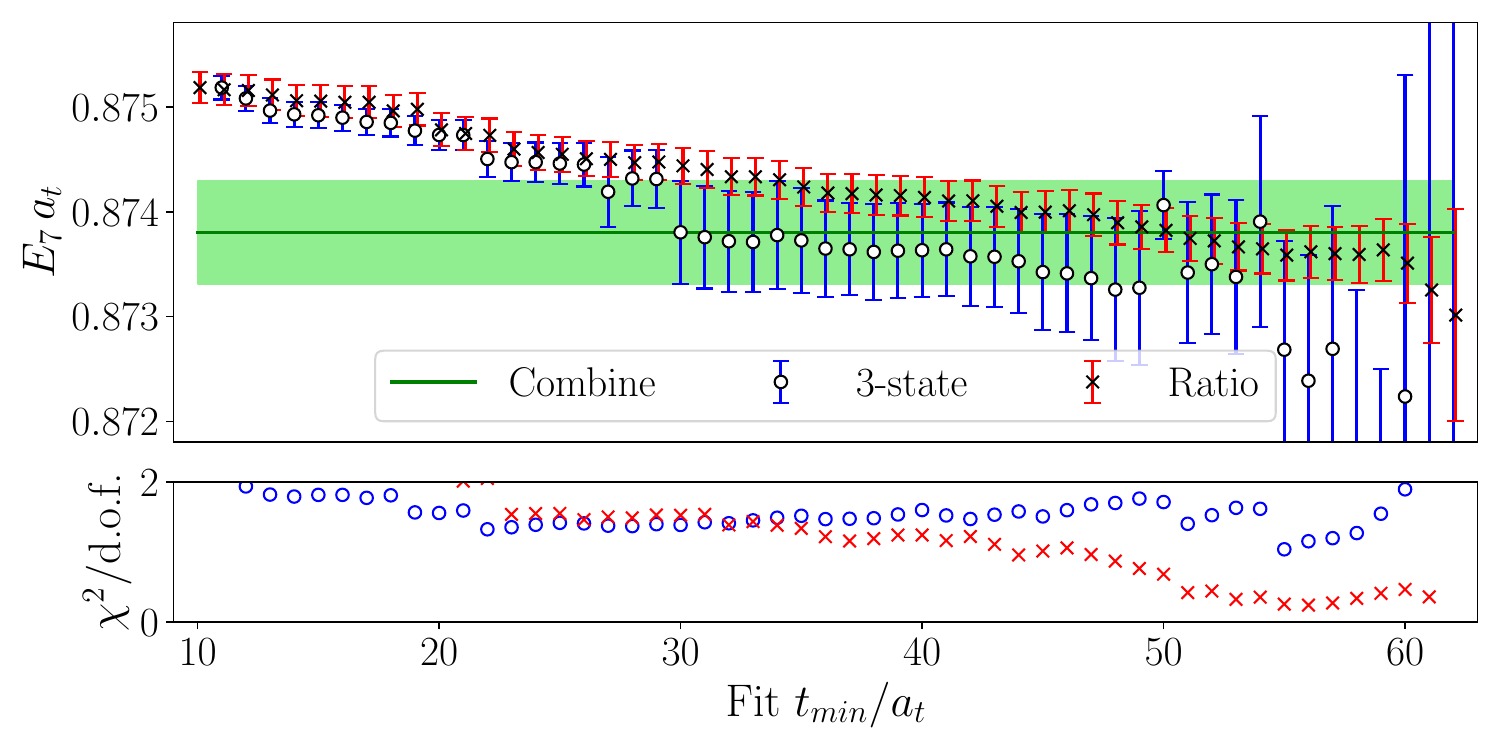}
	\includegraphics[width=0.45\linewidth]{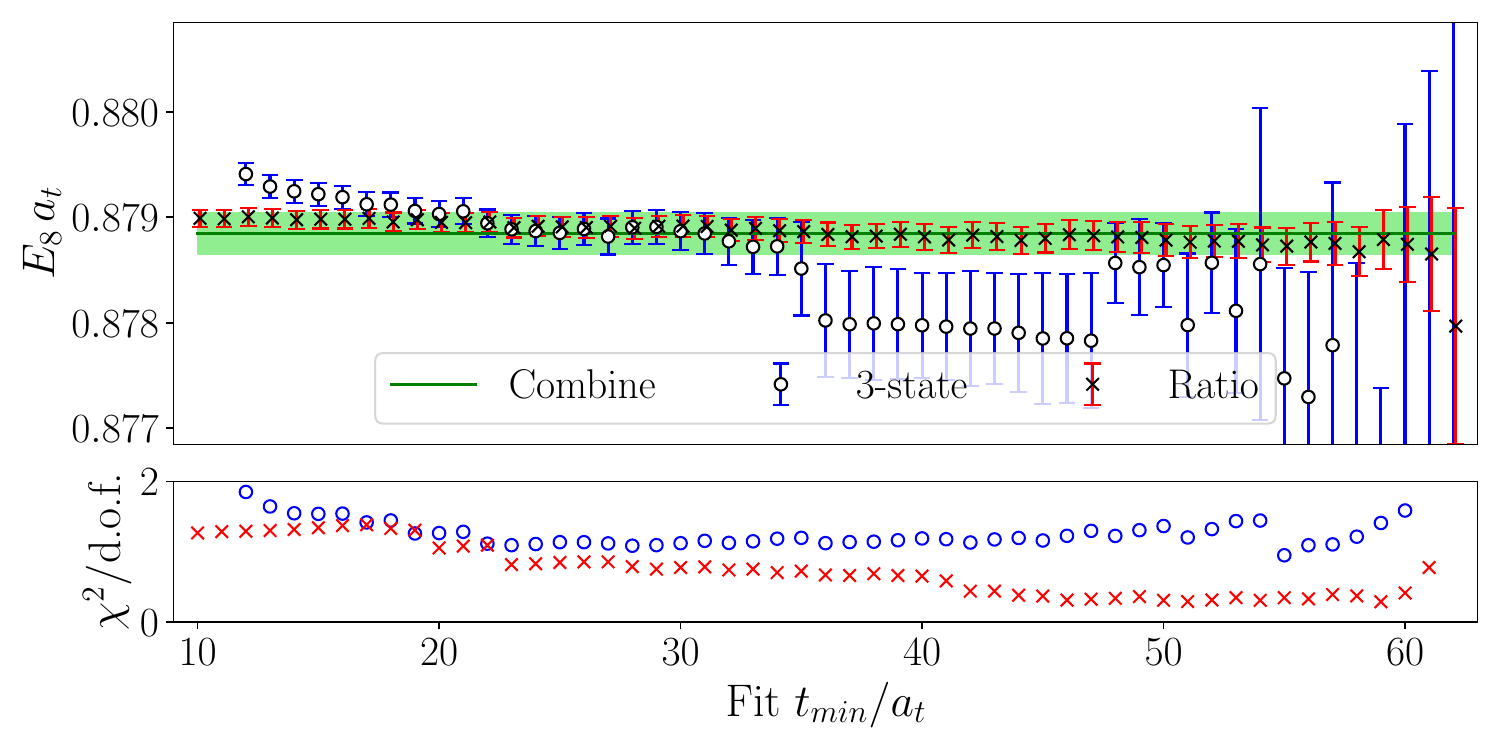}\\
	\includegraphics[width=0.45\linewidth]{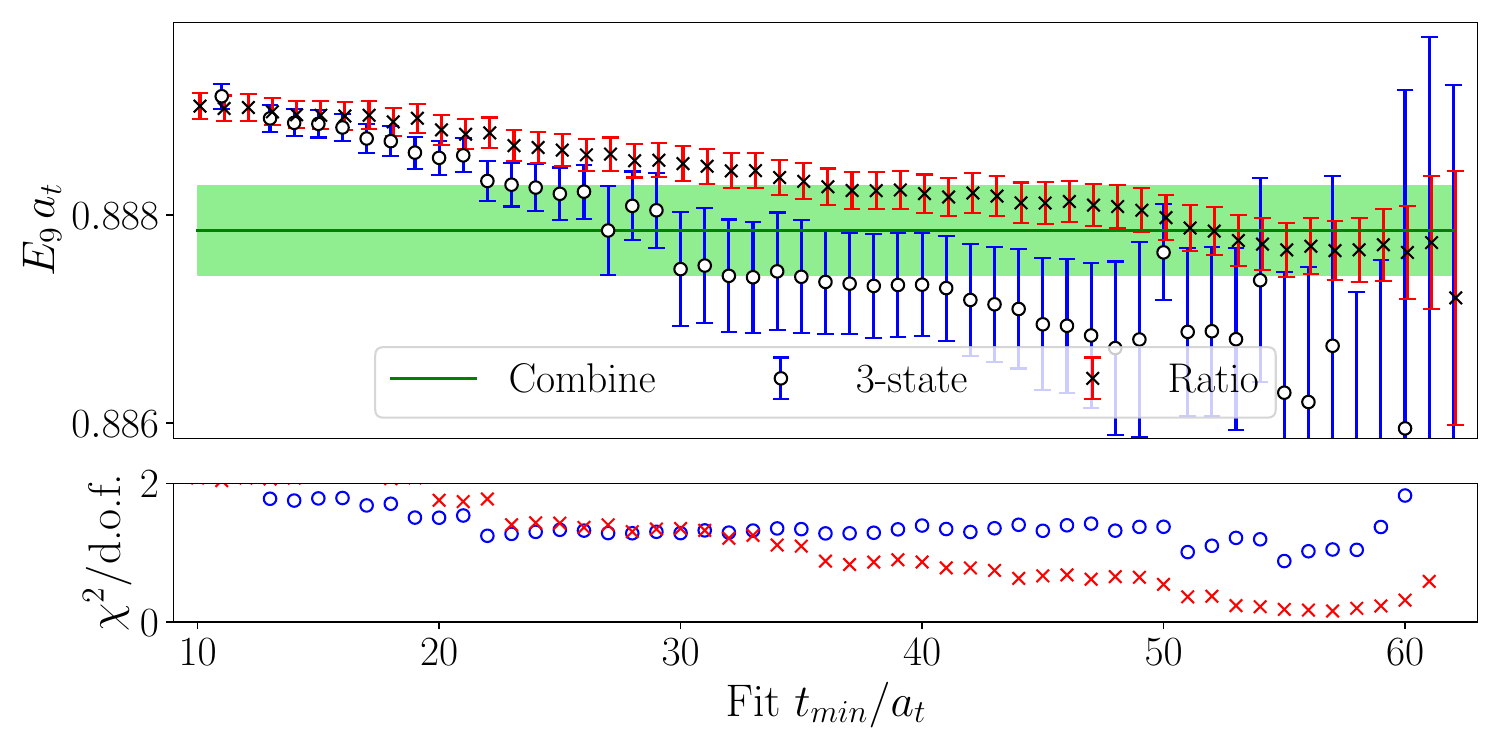}
	\includegraphics[width=0.45\linewidth]{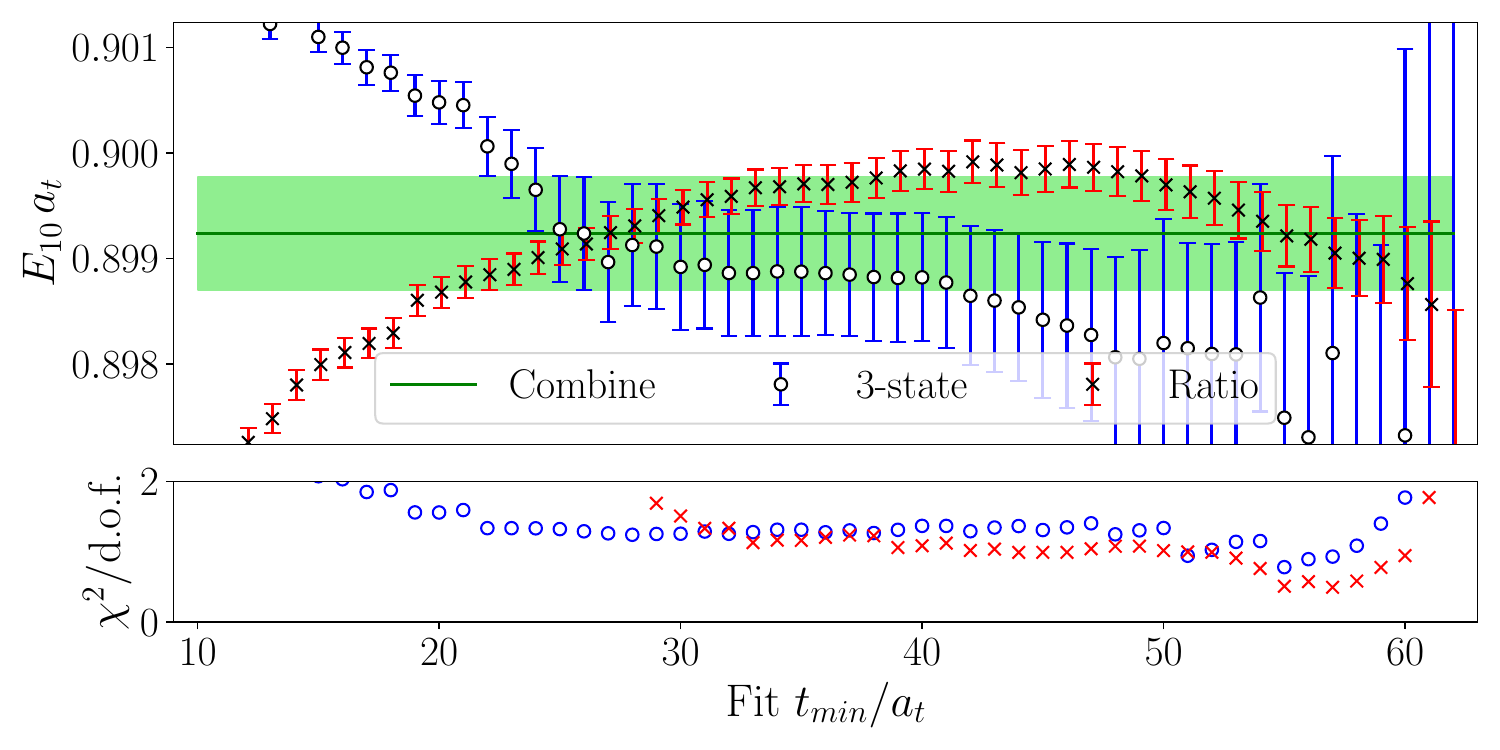}
	\caption{Stability of the two fitting methods on the L16M250 ensemble with $N_V=170$ is demonstrated by varying the minimum fitting time $t_{\rm min}$ in the $0^{++}$ system.}
	\label{fig:L16-pi-fit-tmin}
\end{figure*}
%%%%%%%%%%%%%%%%%%%%%%%%%%%

%%%%%%%%%%%%%%%%%%%%%%%%%%%
\begin{figure*}[h]
	\centering
	\includegraphics[width=0.45\linewidth]{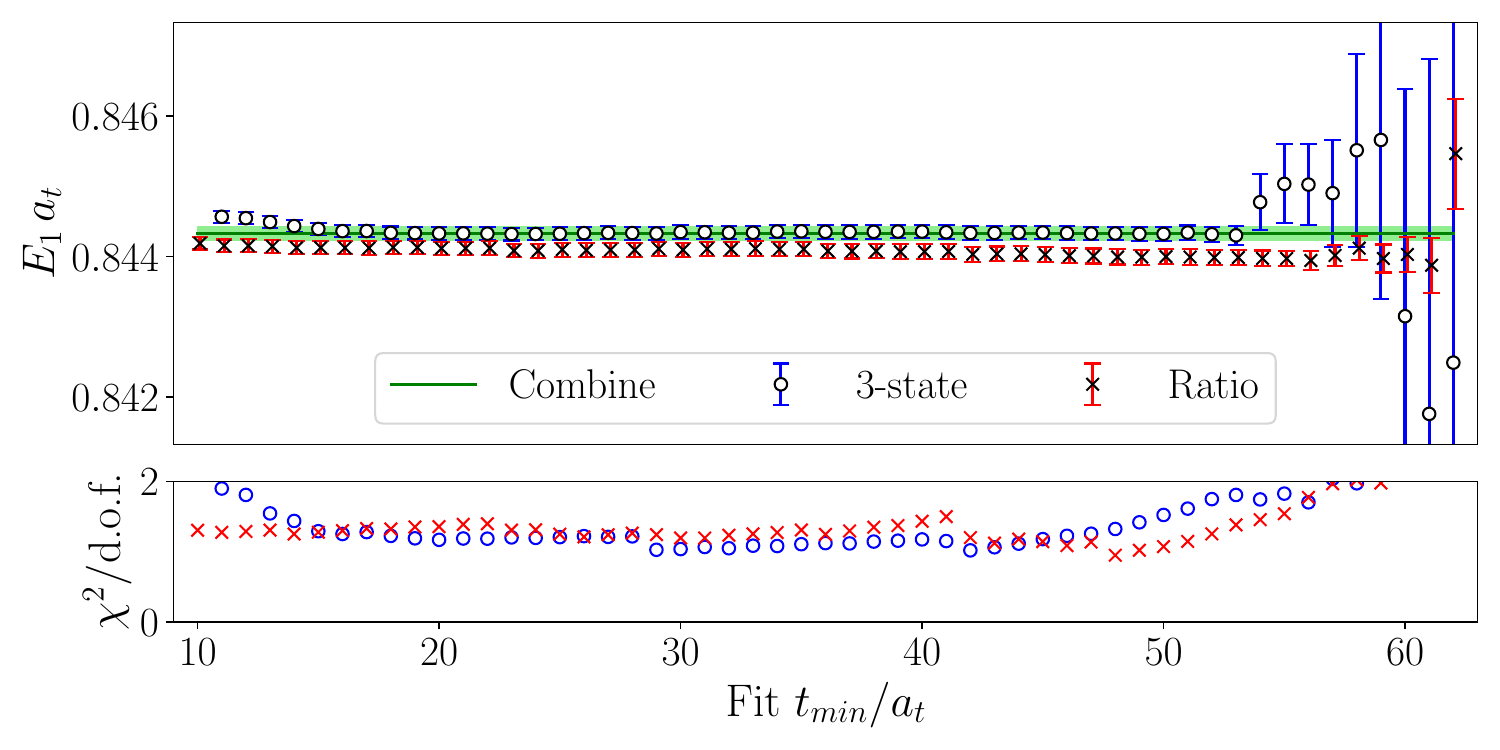}
	\includegraphics[width=0.45\linewidth]{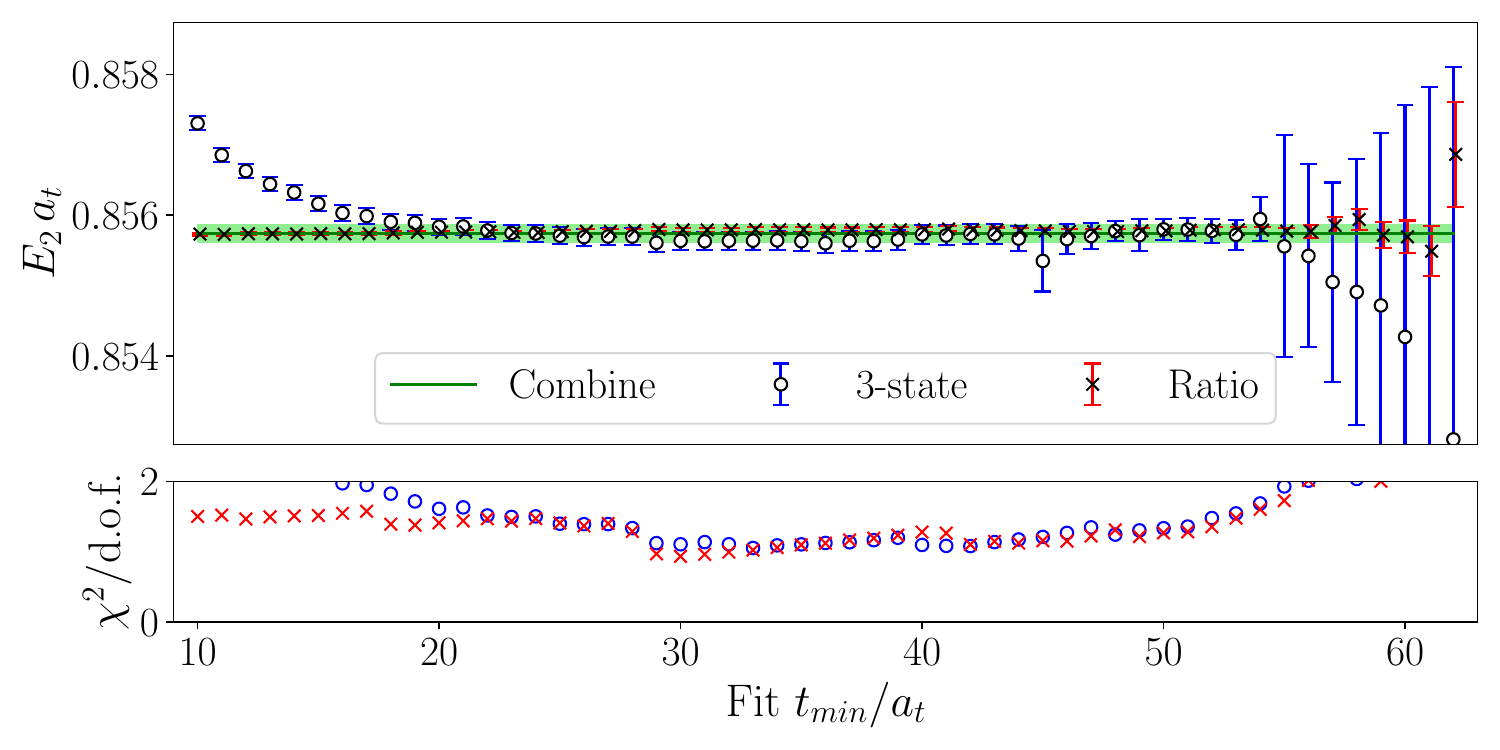}\\
	\includegraphics[width=0.45\linewidth]{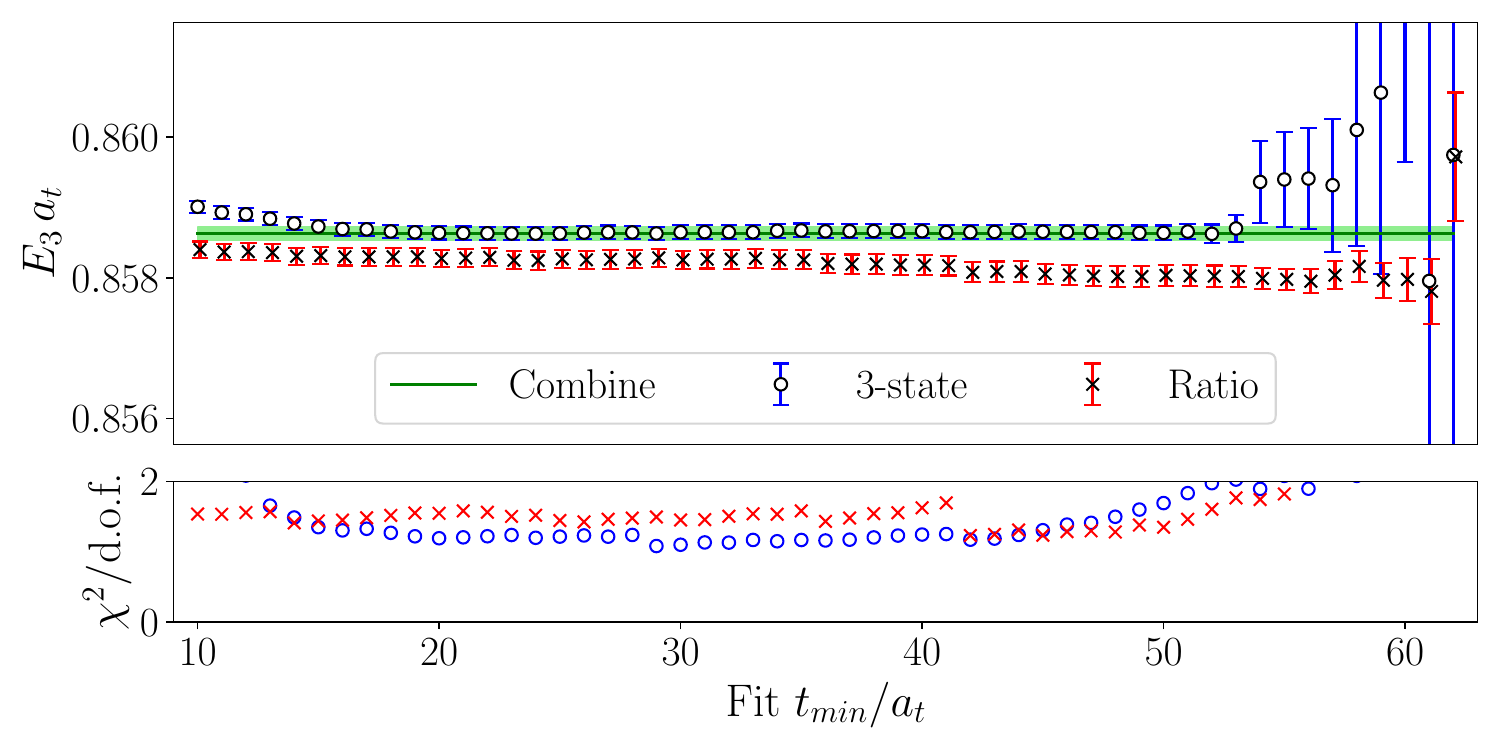}
	\includegraphics[width=0.45\linewidth]{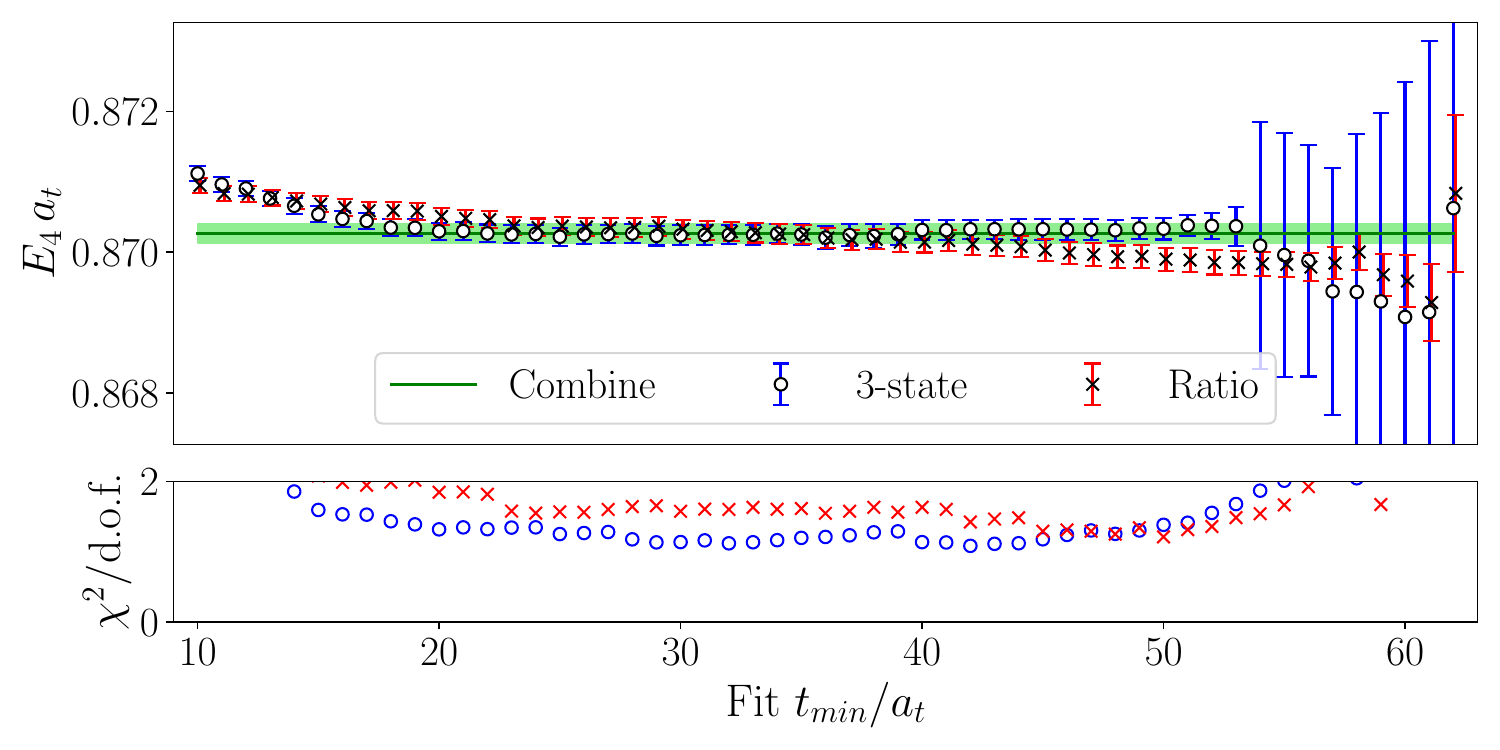}\\
	\includegraphics[width=0.45\linewidth]{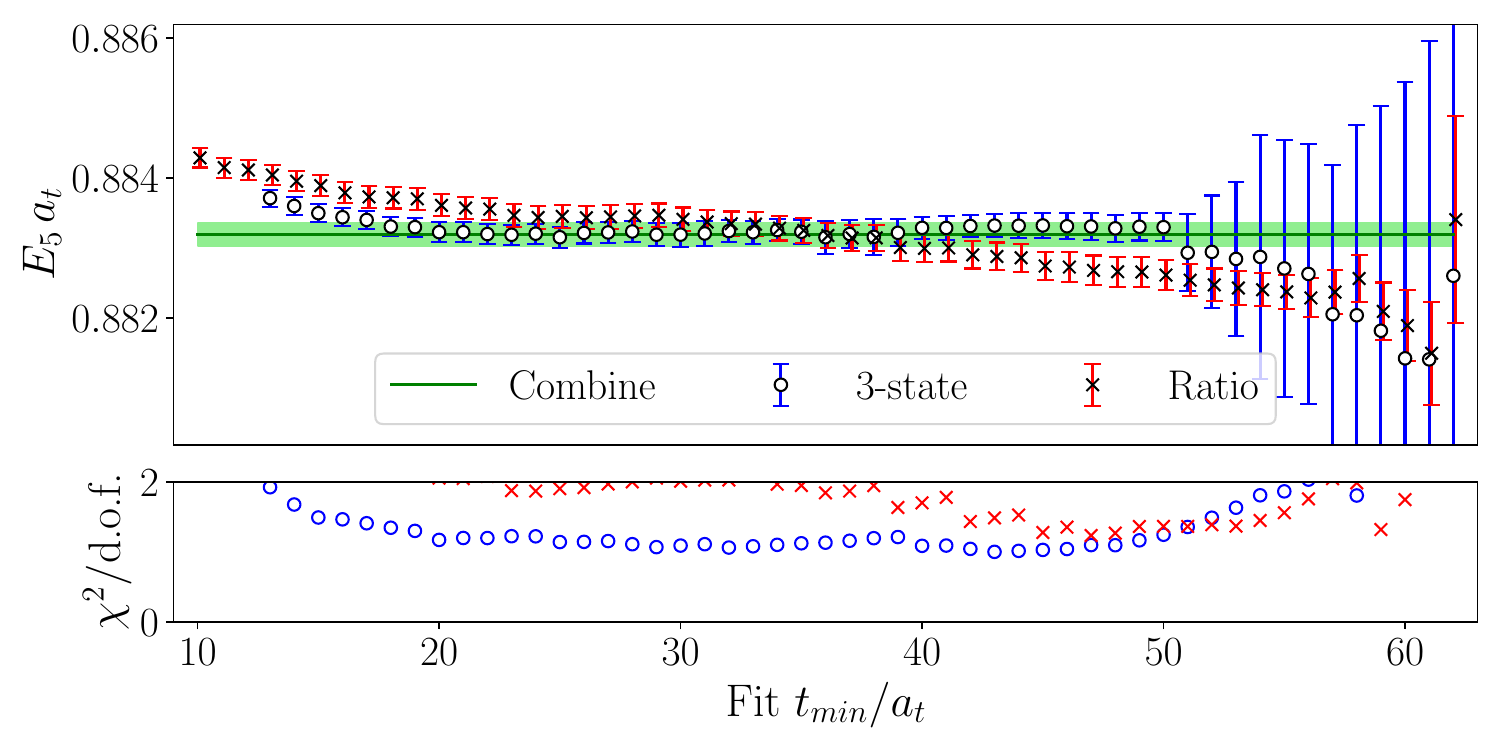}
	\includegraphics[width=0.45\linewidth]{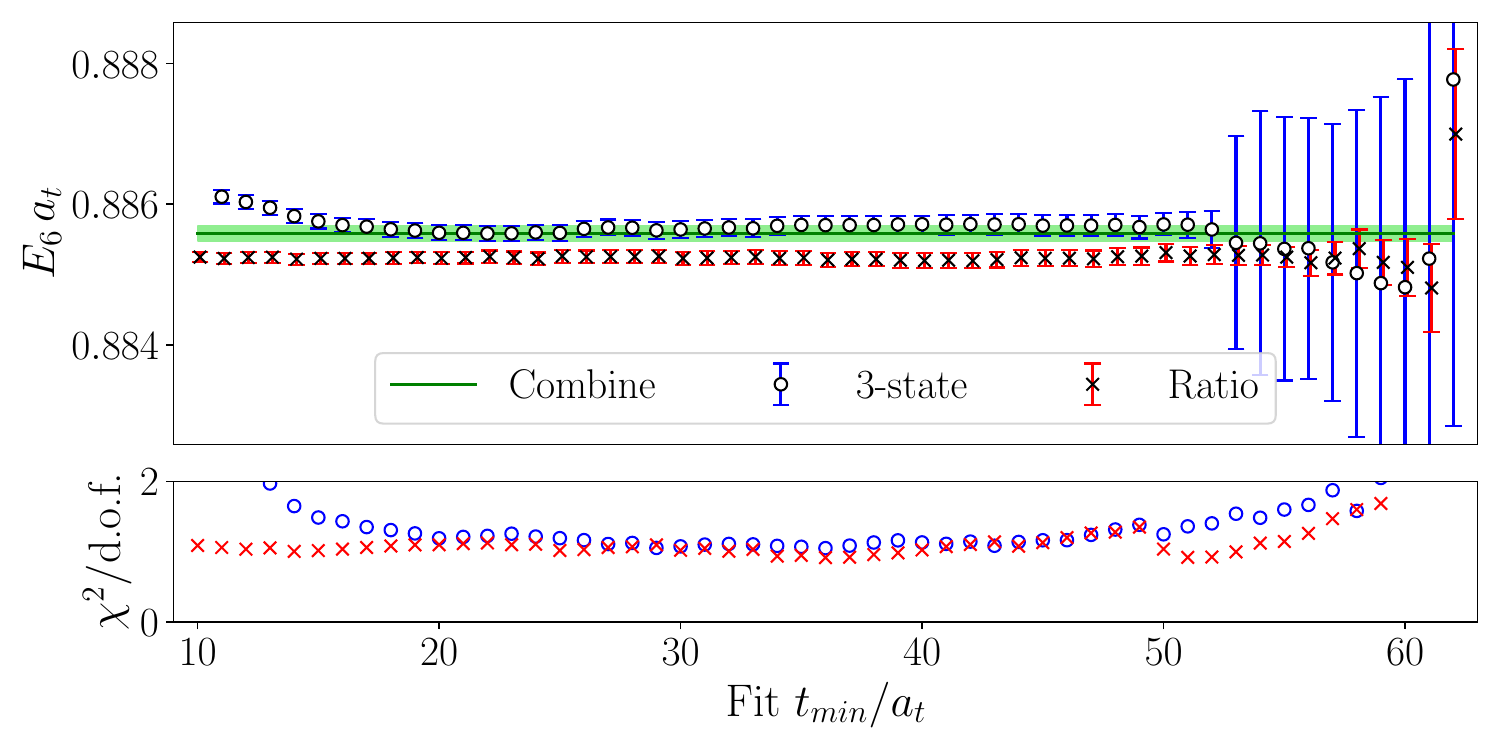}\\
	\includegraphics[width=0.45\linewidth]{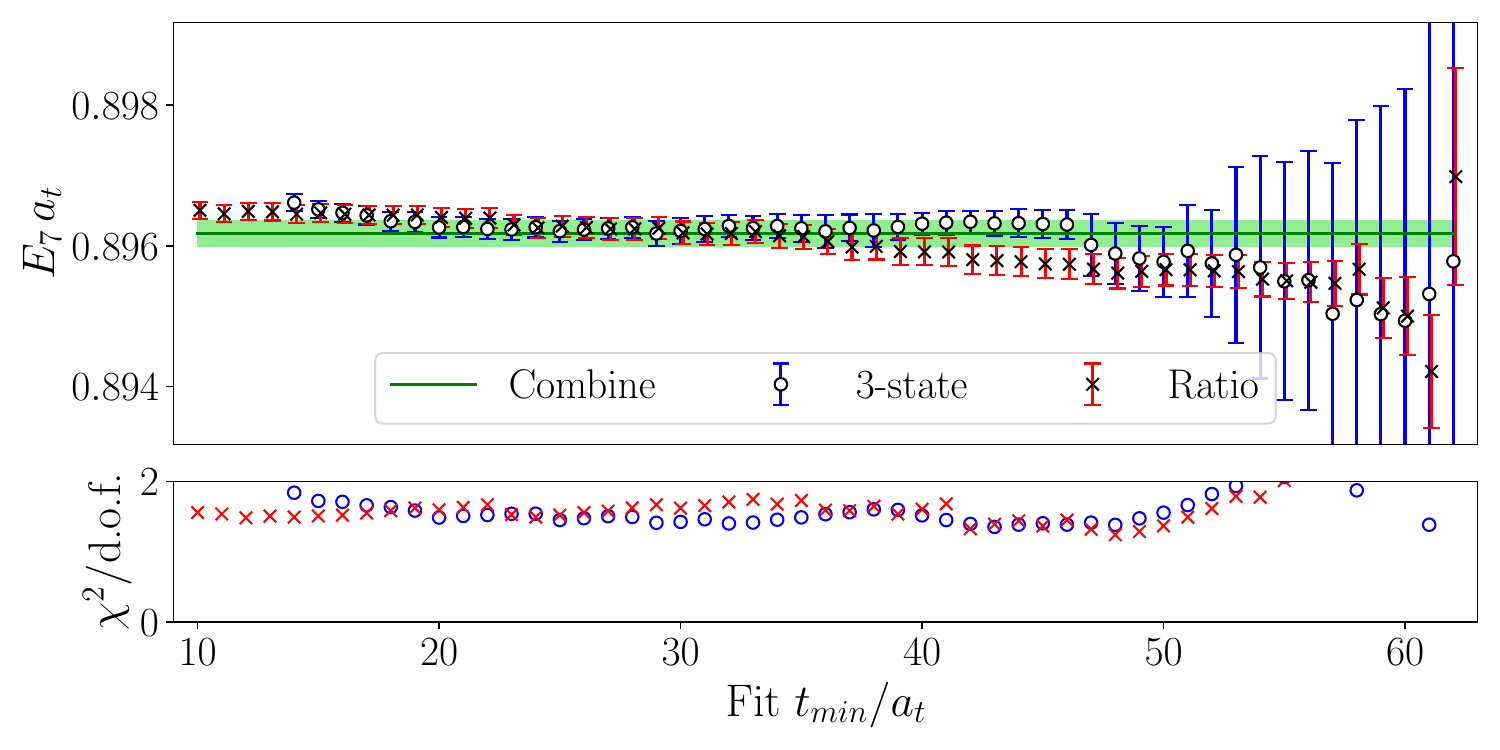}
	\includegraphics[width=0.45\linewidth]{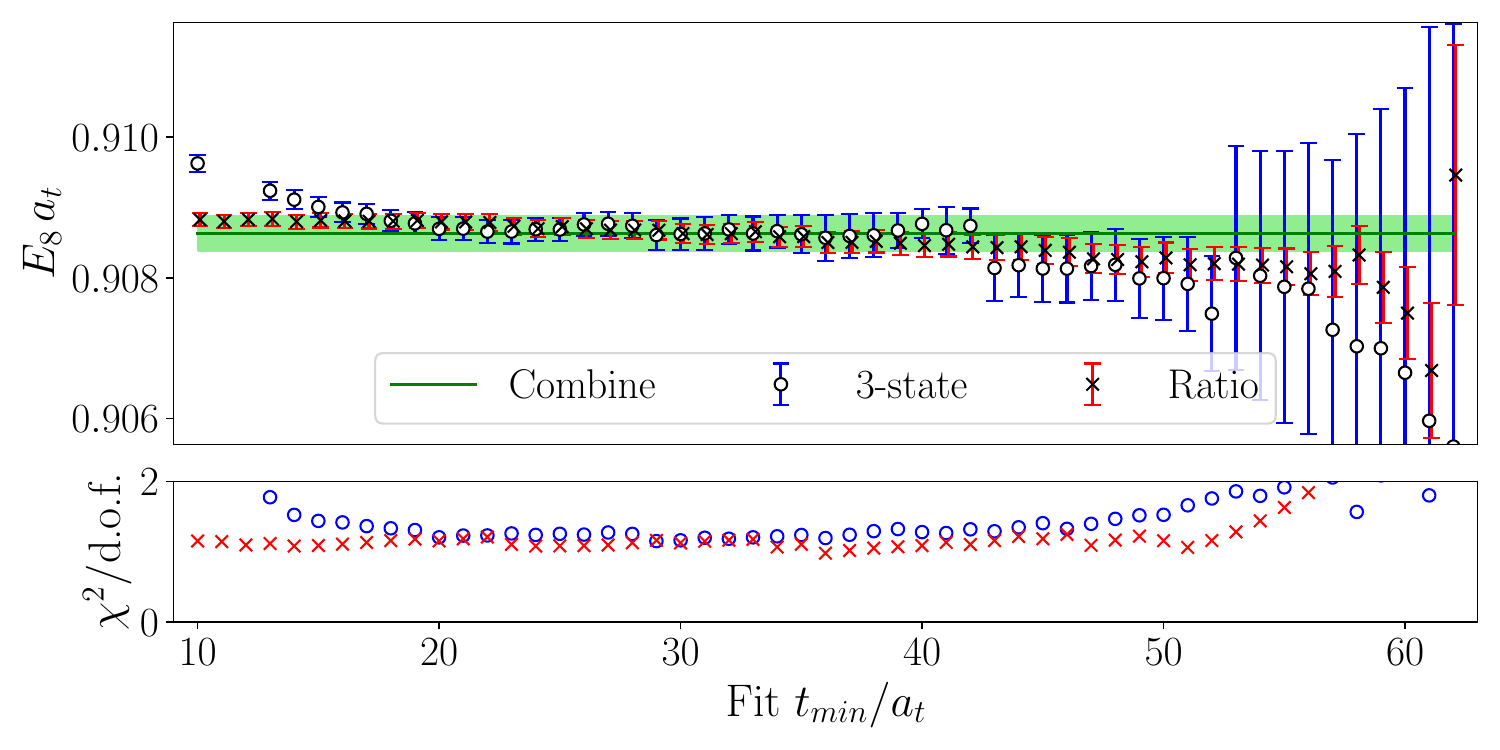}
	\caption{Stability of the two fitting methods on the L16M420 ensemble with $N_V=120$ is demonstrated by varying the minimum fitting time $t_{\rm min}$ in the $2^{++}$ system.}
	\label{fig:L16-2pp-fit-tmin}
\end{figure*}
%%%%%%%%%%%%%%%%%%%%%%%%%%%

%%%%%%%%%%%%%%%%%%%%%%%%%%%
\begin{figure*}[h]
	\centering
	\includegraphics[width=0.45\linewidth]{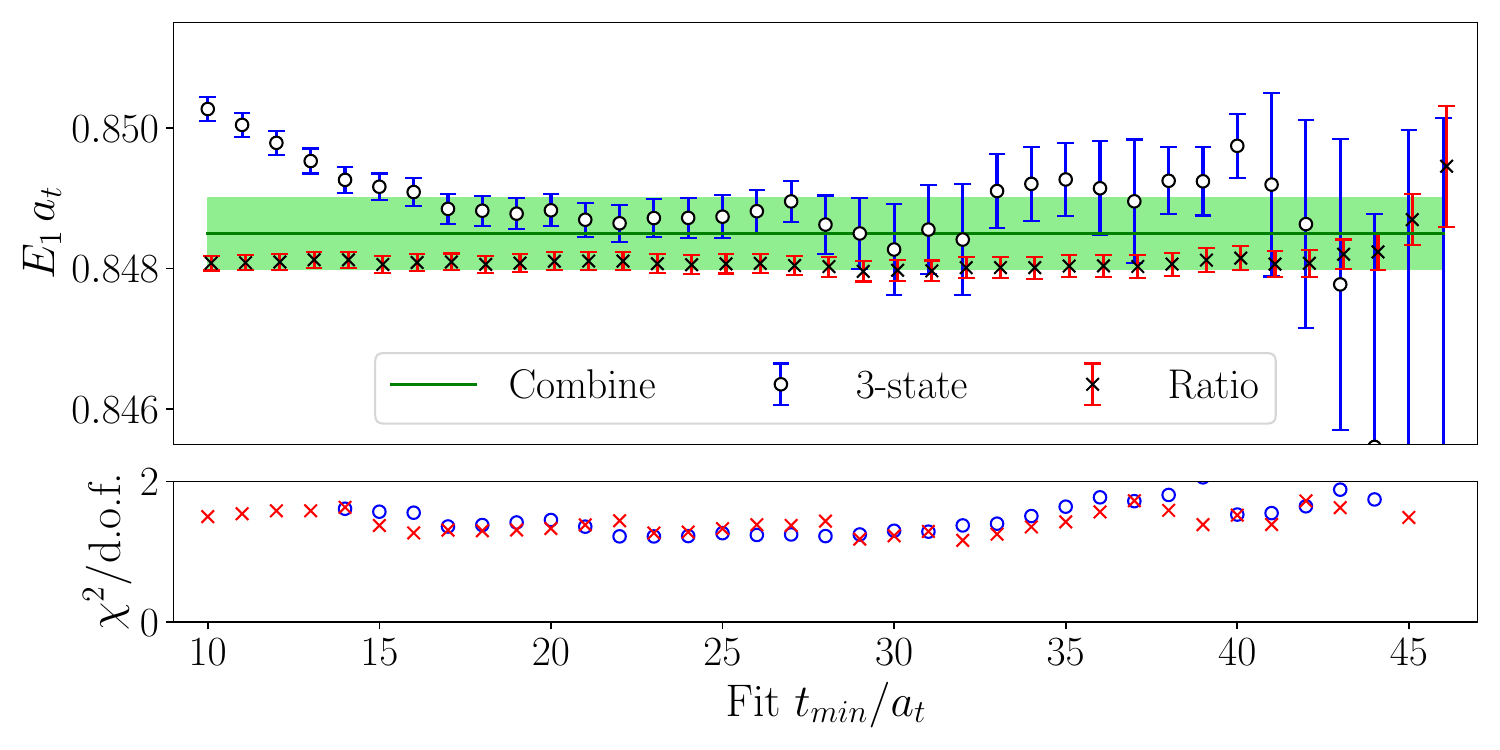}
	\includegraphics[width=0.45\linewidth]{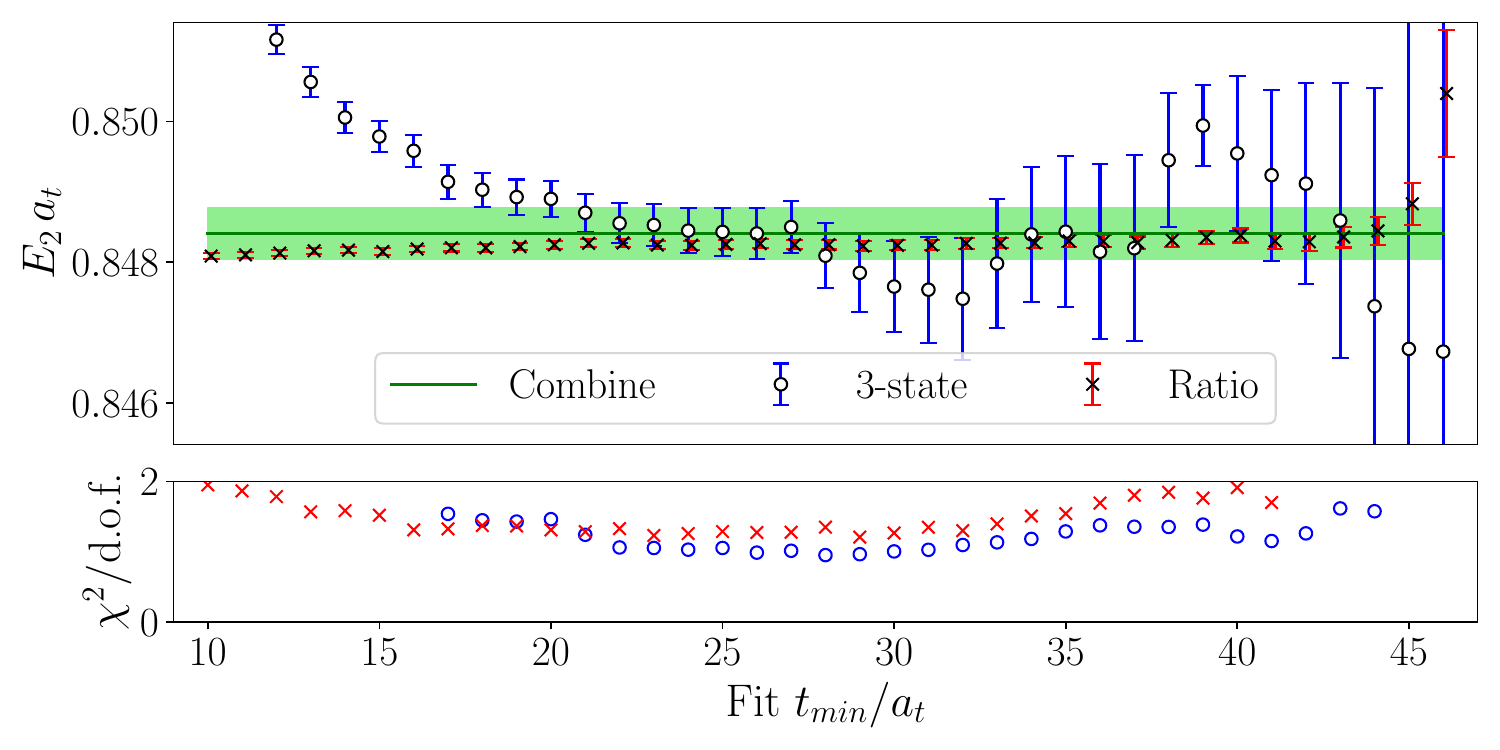}\\
	\includegraphics[width=0.45\linewidth]{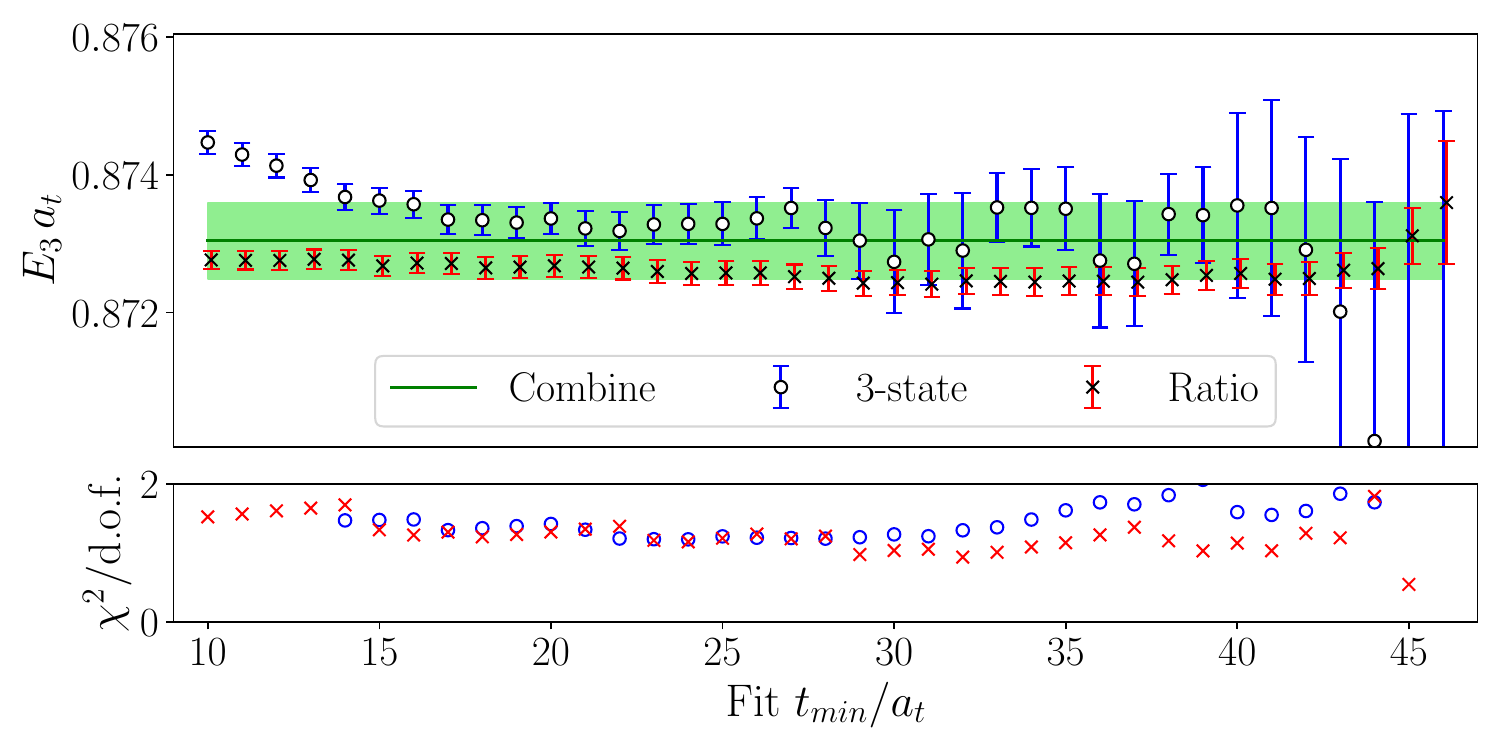}
	\includegraphics[width=0.45\linewidth]{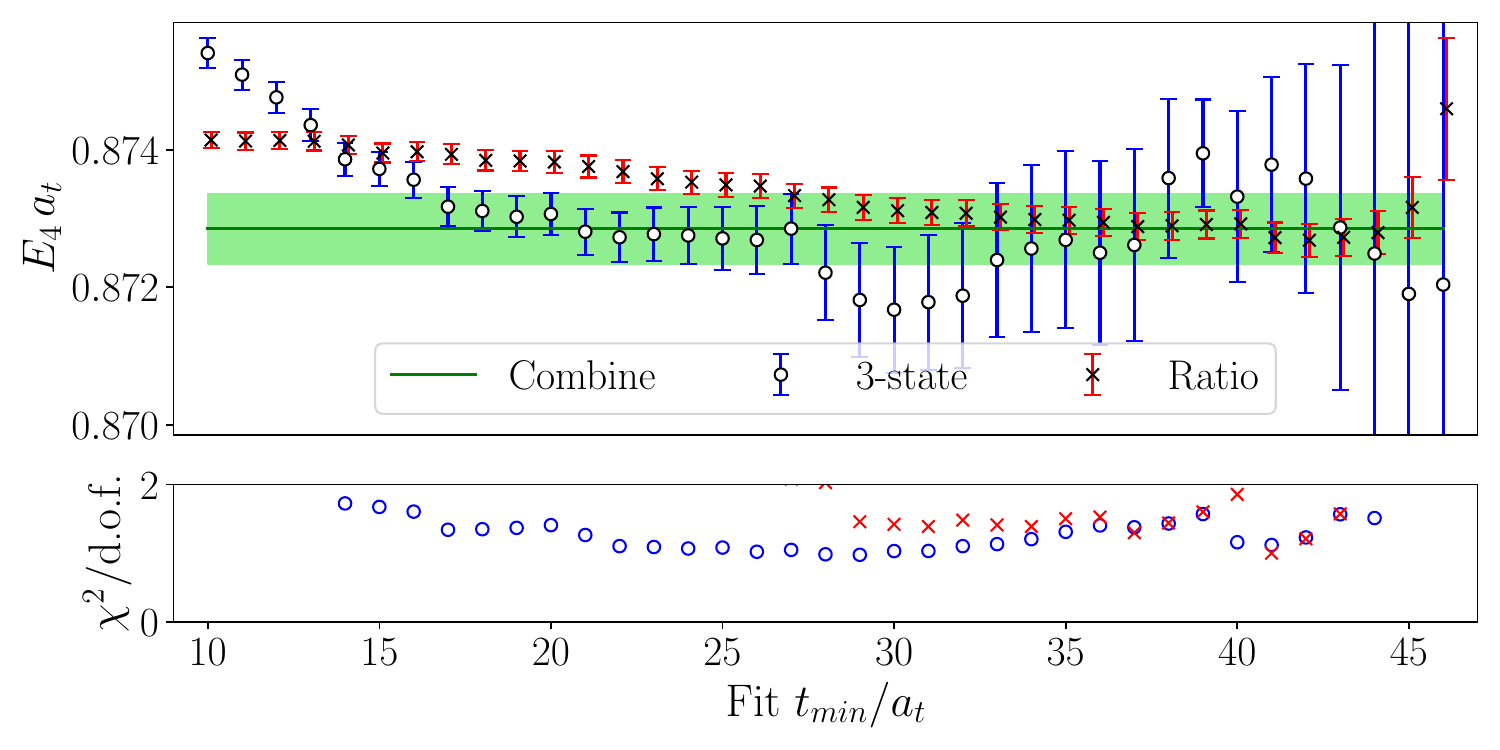}\\
	\includegraphics[width=0.45\linewidth]{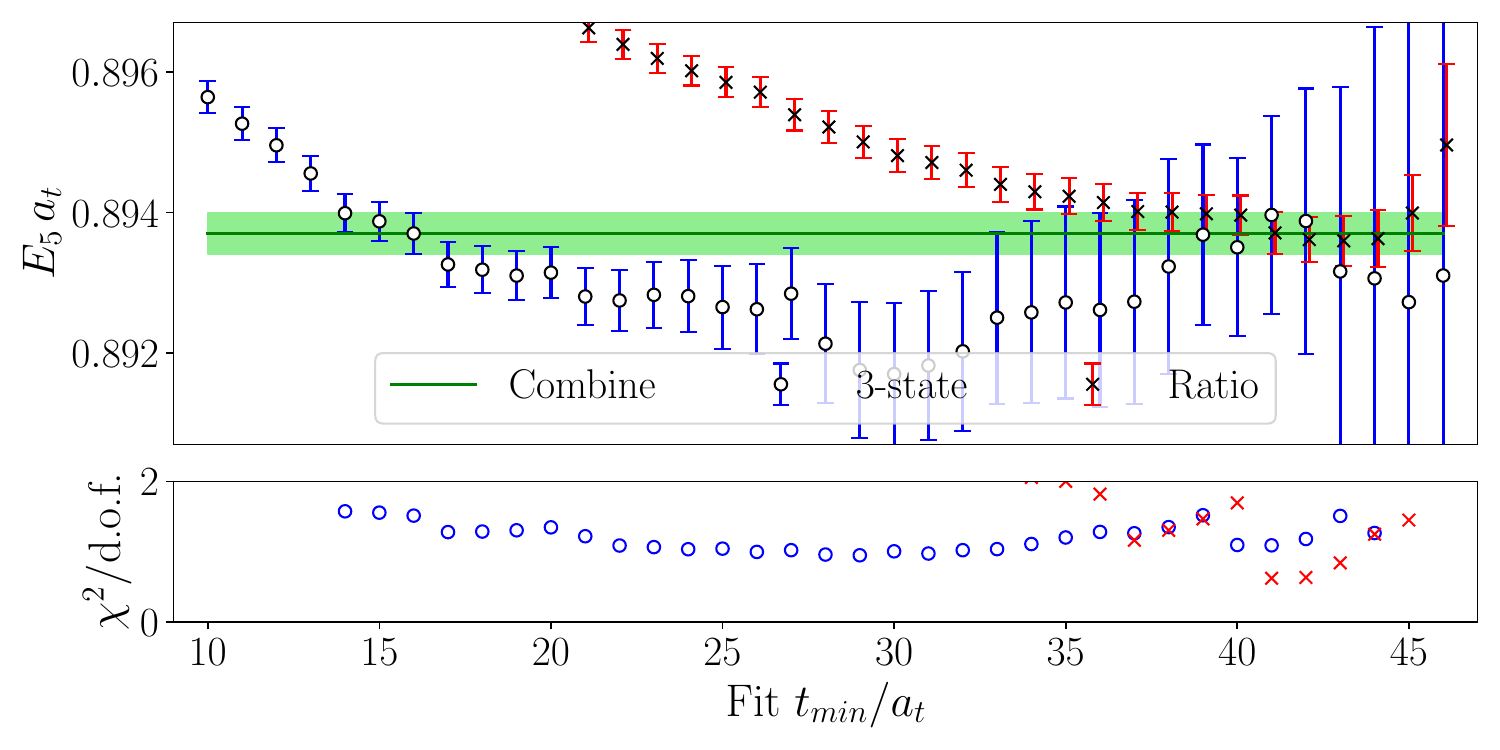}
	\includegraphics[width=0.45\linewidth]{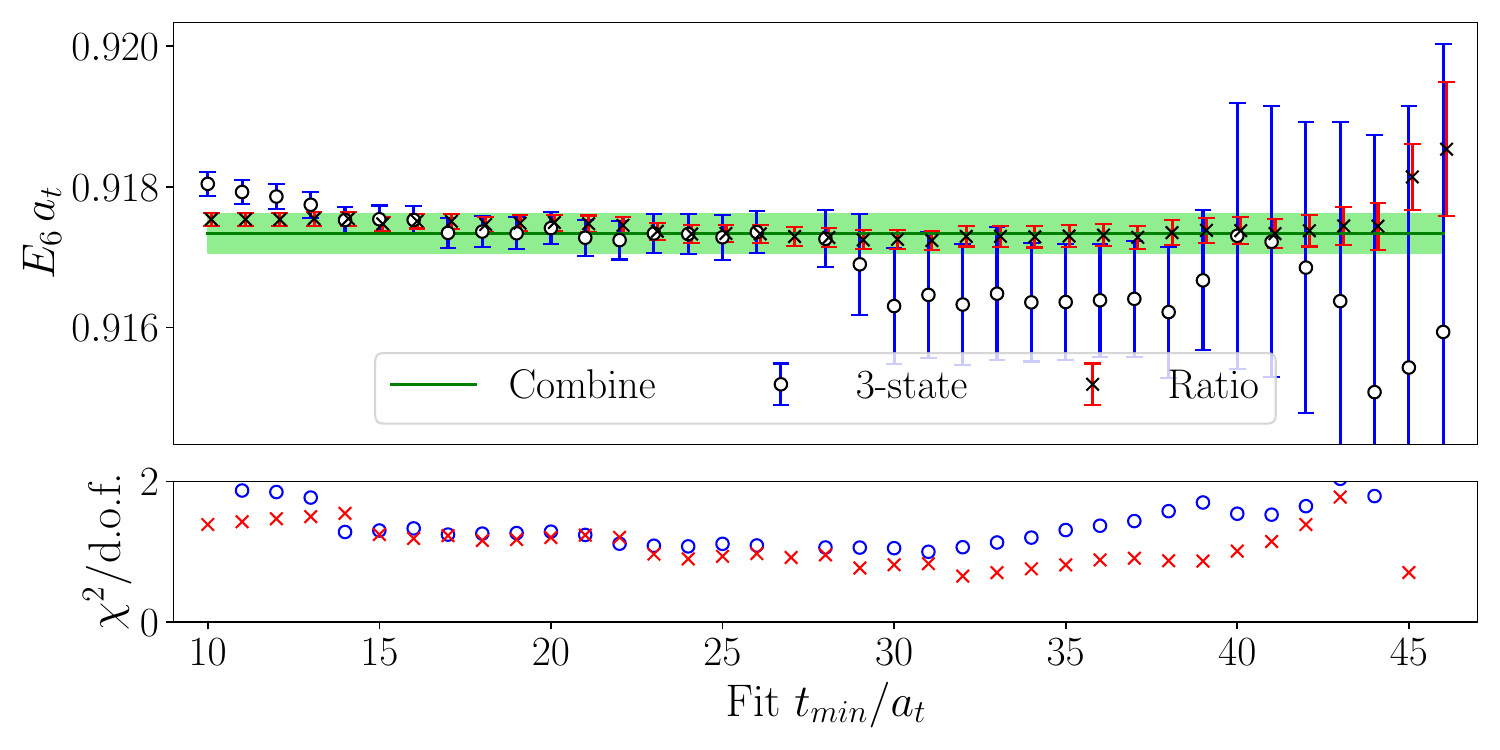}\\
	\includegraphics[width=0.45\linewidth]{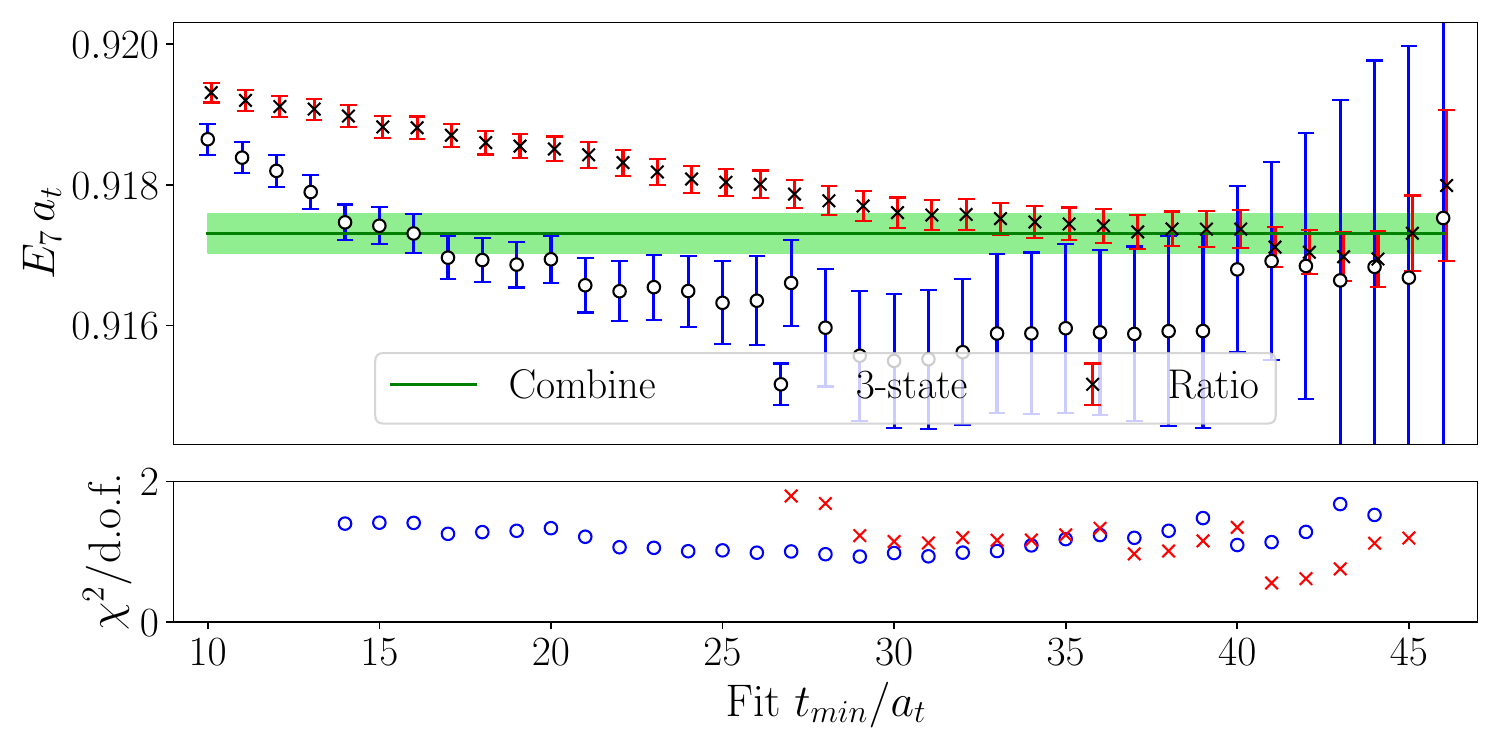}
	\includegraphics[width=0.45\linewidth]{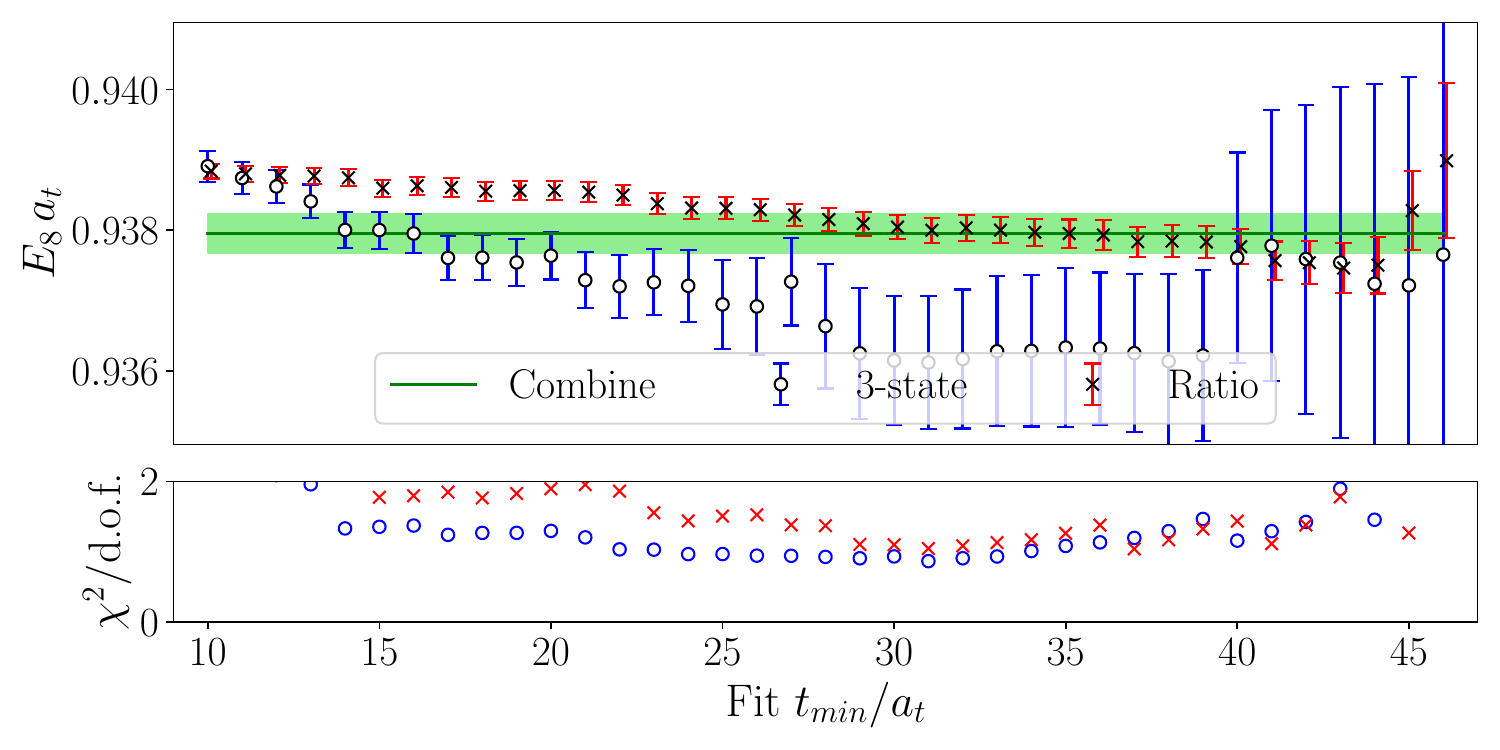}
	\caption{Stability of the two fitting methods on the L12M250 ensemble with $N_V=170$ is demonstrated by varying the minimum fitting time $t_{\rm min}$ in the $2^{++}$ system.}
	\label{fig:L12-2pp-pi-fit-tmin}
\end{figure*}
%%%%%%%%%%%%%%%%%%%%%%%%%%%

%%%%%%%%%%%%%%%%%%%%%%%%%%%
\begin{figure*}[h]
	\centering
	\includegraphics[width=0.45\linewidth]{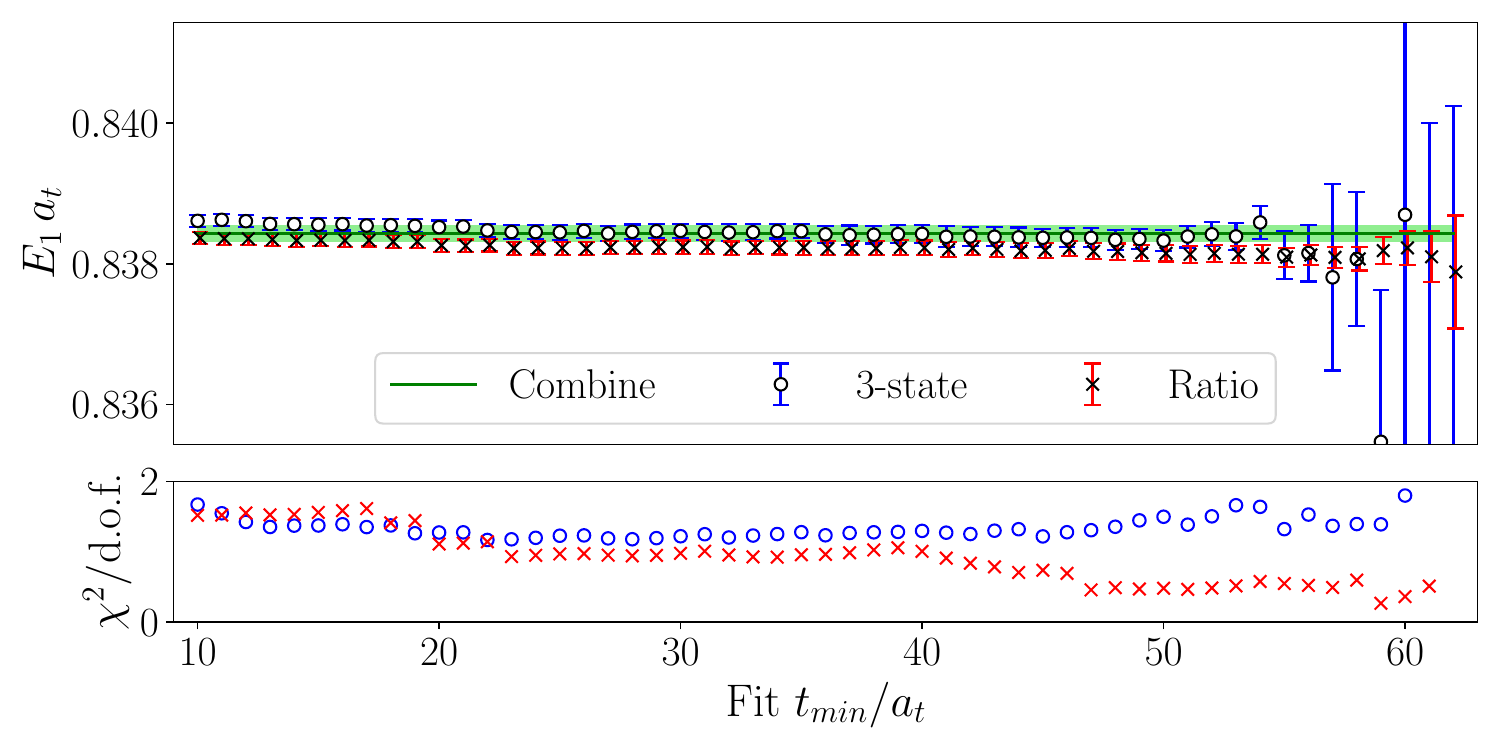}
	\includegraphics[width=0.45\linewidth]{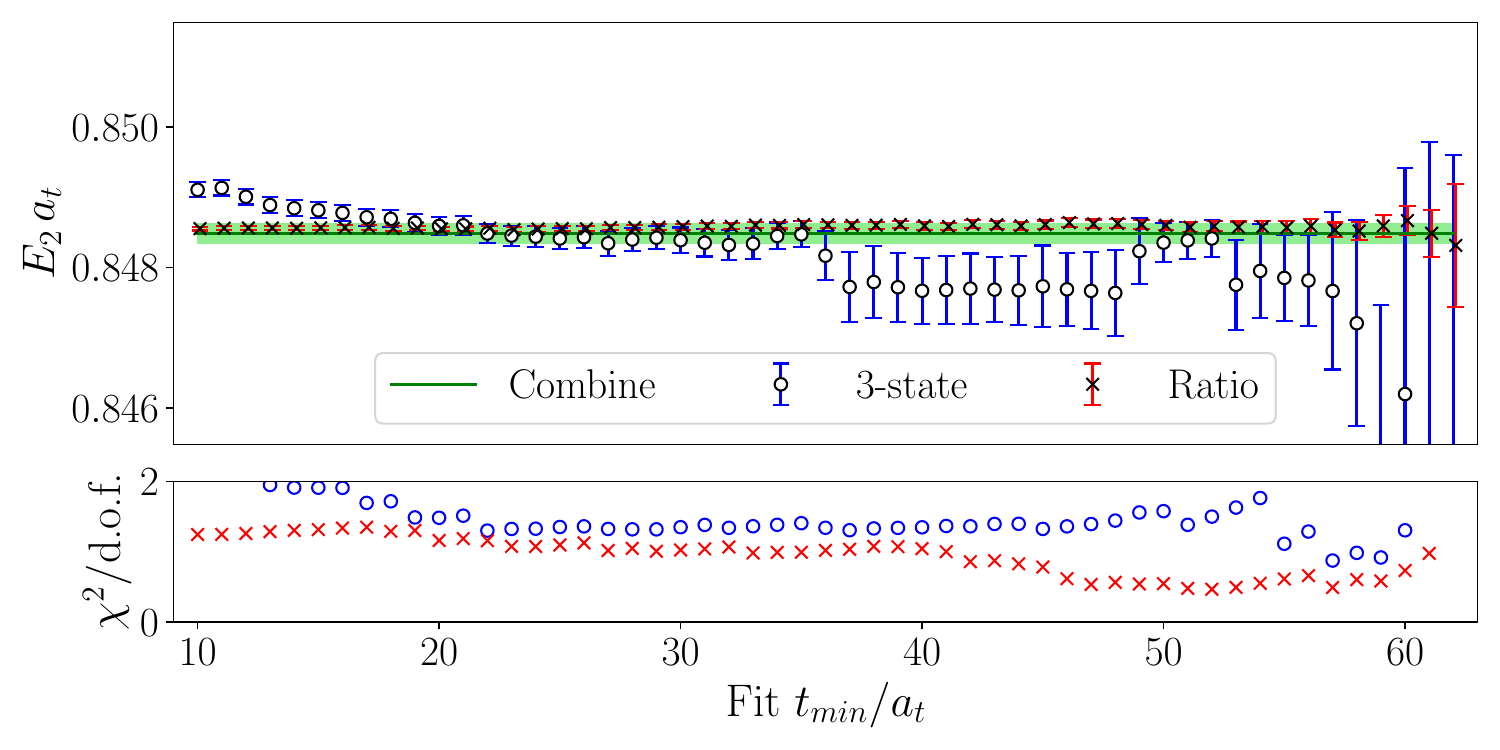}\\
	\includegraphics[width=0.45\linewidth]{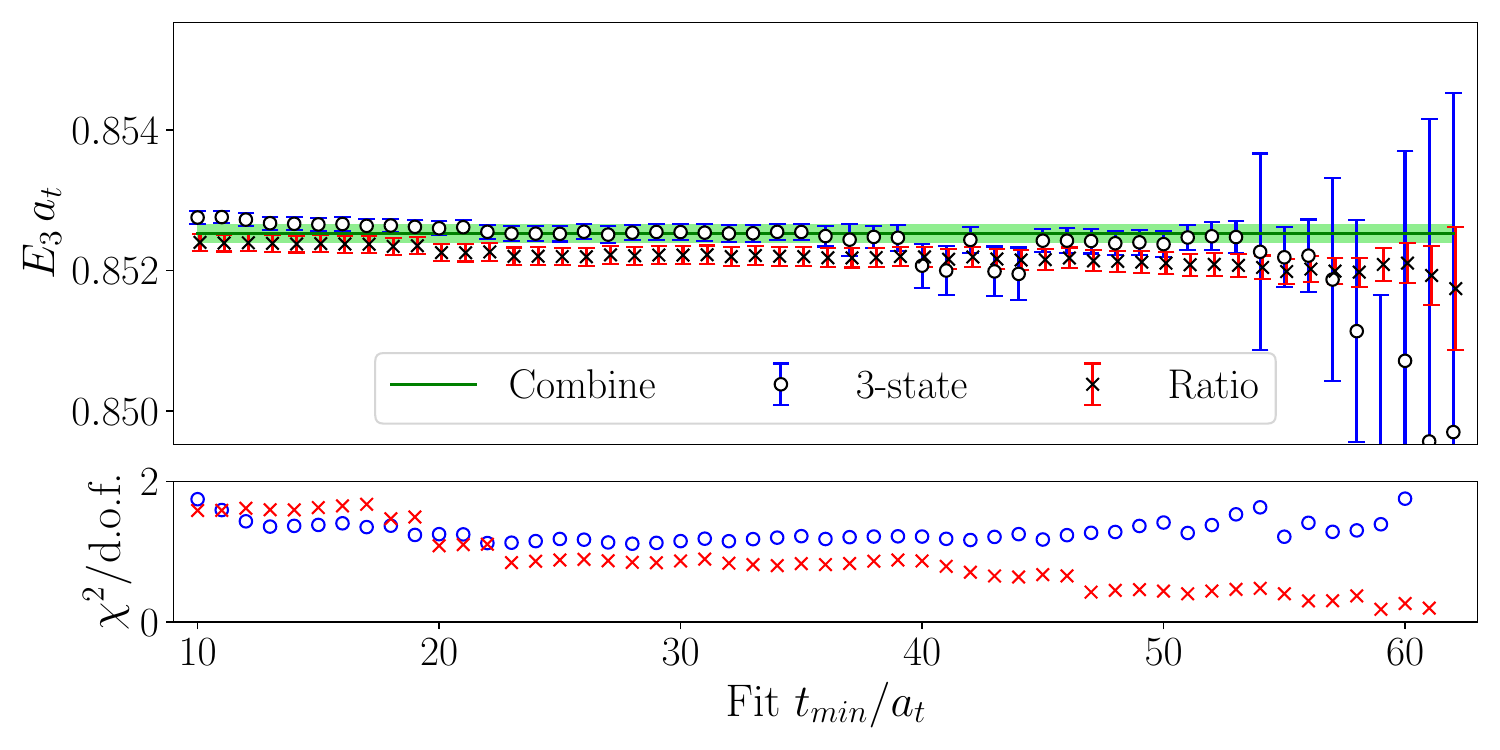}
	\includegraphics[width=0.45\linewidth]{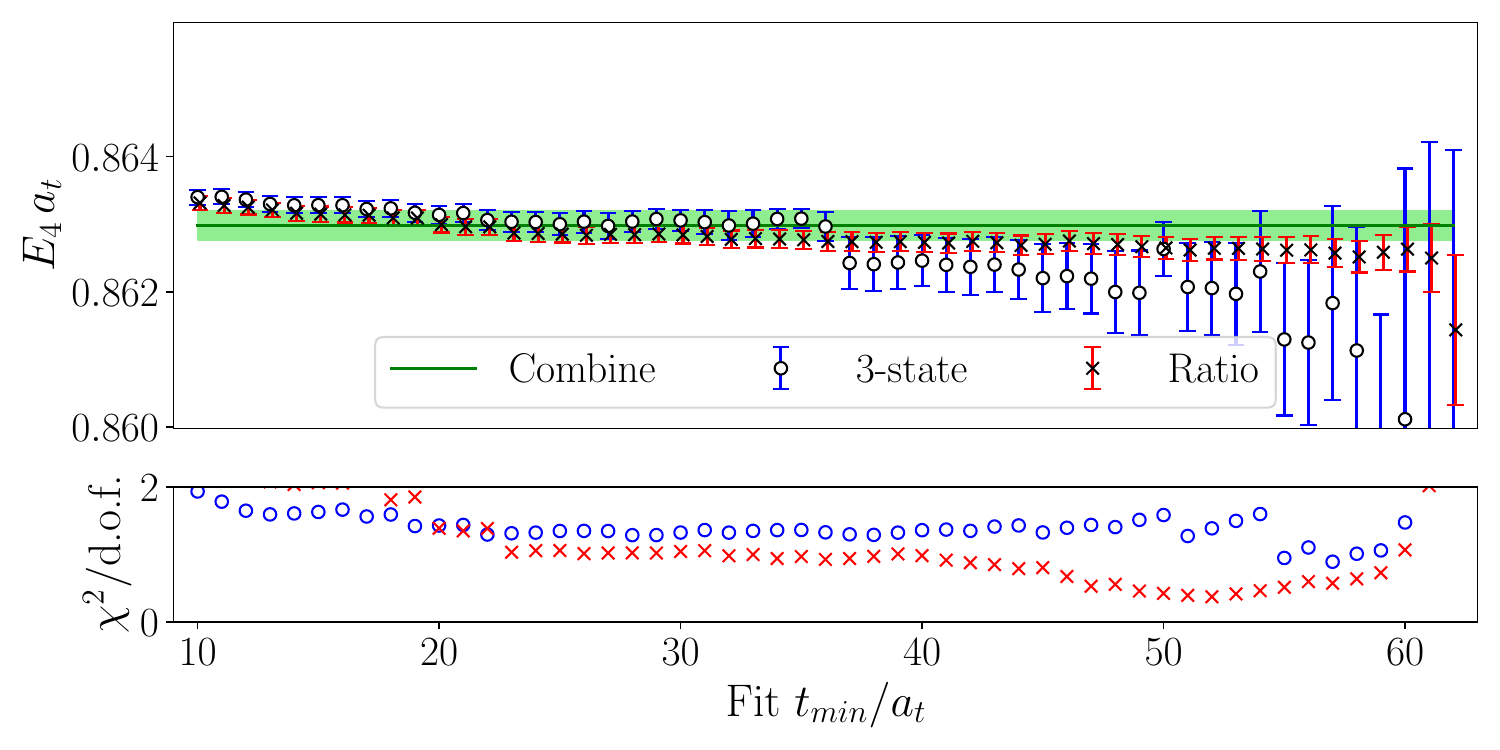}\\
	\includegraphics[width=0.45\linewidth]{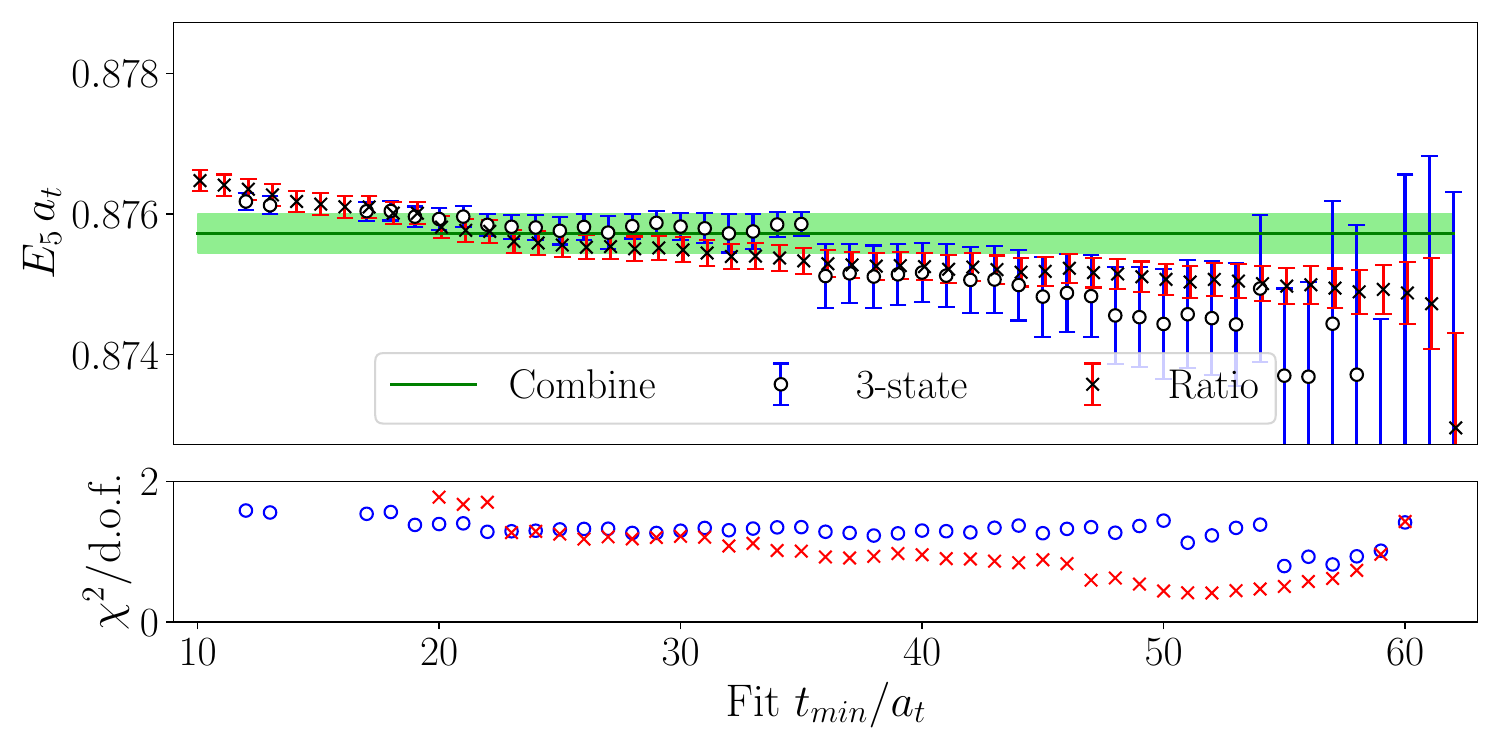}
	\includegraphics[width=0.45\linewidth]{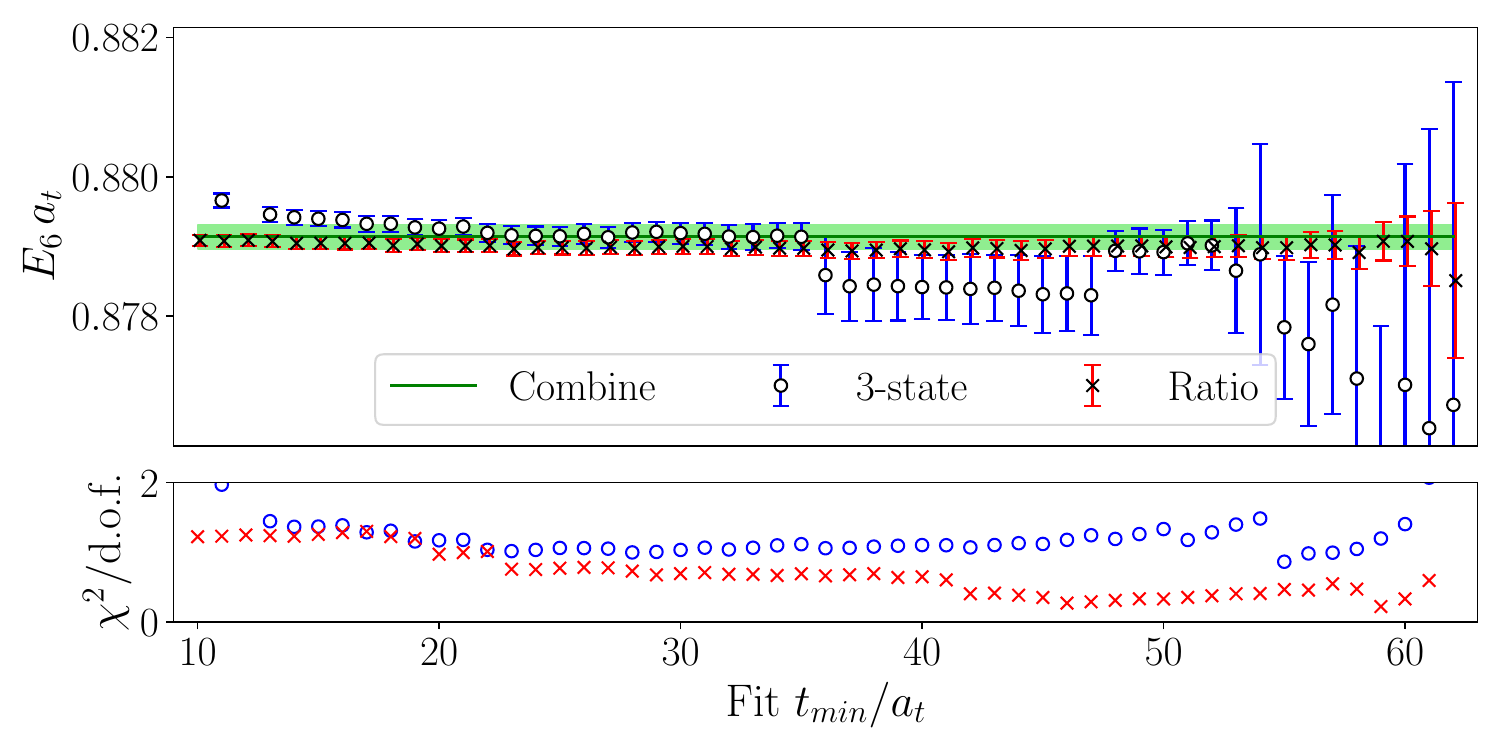}\\
	\includegraphics[width=0.45\linewidth]{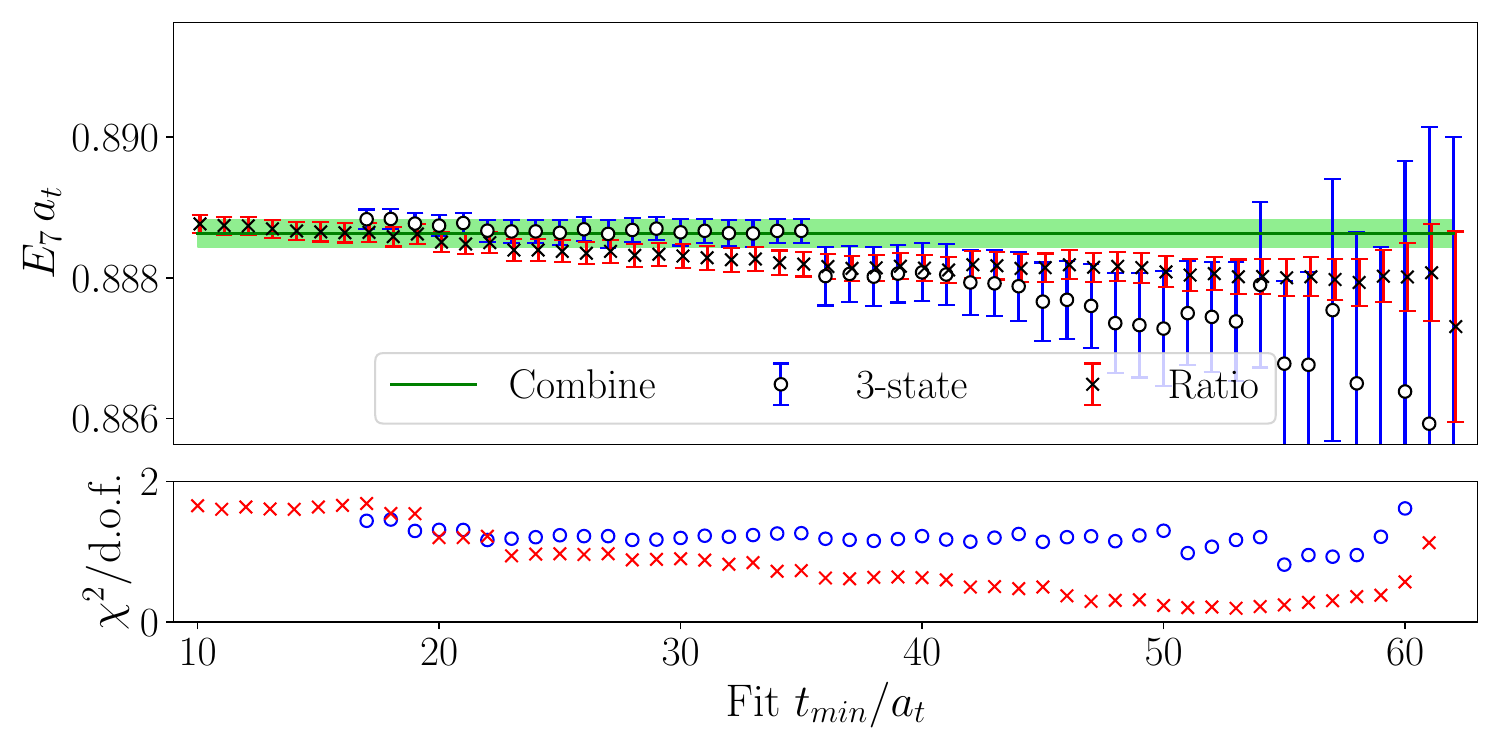}
	\includegraphics[width=0.45\linewidth]{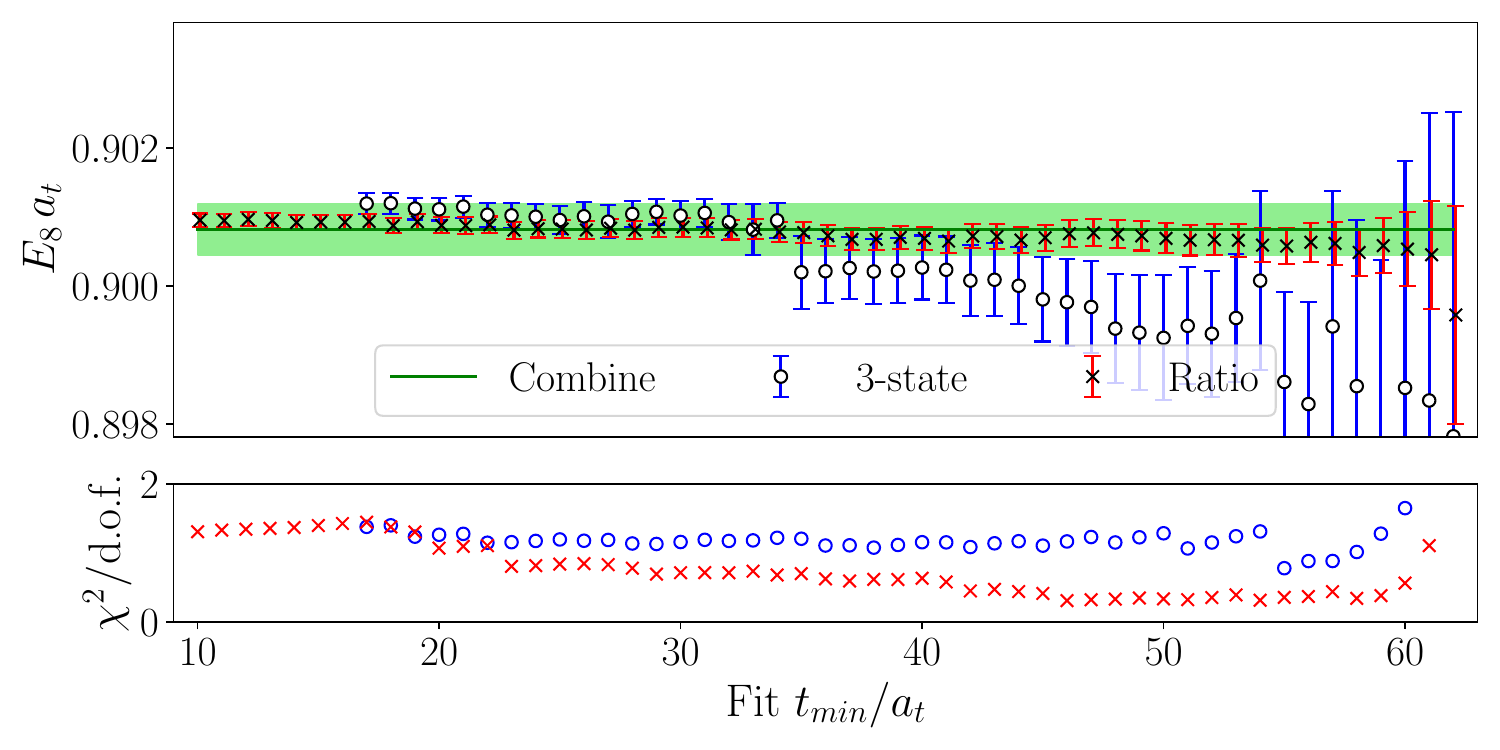}
	\caption{Stability of the two fitting methods on the L16M250 ensemble with $N_V=120$ is demonstrated by varying the minimum fitting time $t_{\rm min}$ in the $2^{++}$ system.}
	\label{fig:L16-2pp-pi-fit-tmin}
\end{figure*}
%%%%%%%%%%%%%%%%%%%%%%%%%%%

\end{document}